\documentclass[binding=0.6cm]{sapthesis}

\usepackage[nosort]{cite}
\usepackage{microtype}
\usepackage{hyperref}

\usepackage{amsmath}
\usepackage{amsfonts}
\usepackage{amssymb}

\usepackage{subfigure}

\newcommand{\dslash}{\not\!\partial}
\newcommand{\Dslash}{\not\!\! D}

\author{Natascia Vignaroli}
\title{Phenomenology of heavy fermion and vector resonances in composite Higgs models}
\submitdate{2011}
\cycle{XXIV}
\IDnumber{699314}
\advisor{Dr. Roberto Contino}

\examdate{14 February 2012}
\examiner{Prof. Giovanni Ridolfi}
\examiner{Prof. Mauro Dell'Orso}
\examiner{Prof. Antonello Polosa}

\authoremail{natascia.vignaroli@roma1.infn.it}

\begin{document}

\frontmatter
\maketitle

\begin{acknowledgements}
I would like to thank Roberto Contino for supervision and collaboration in this work, 
for having been my Ph.D. supervisor and having introduced me to the theory of composite Higgs models.
I thank Silvano Petrarca for his assistance during the first year of the Ph.D. and encouragement.
I thank Barbara Mele from whom I have learned the most of what I know about collider phenomenology and for comments on the second part of this thesis. 
I thank all my Ph.D. colleagues and friends for the stimulating environment created. Especially, I would like to thank Alessandra Mastrobuono Battisti and Francesco Sanfilippo for many useful discussions and technical suggestions. I also thank Nadia Drenska for a suggestion regarding the ROOT tool, 
that was particularly useful in the second part of this thesis.   
\end{acknowledgements}

\begin{abstract} 
In this project we study the phenomenology of models where the Higgs is a bound state of a strongly interacting dynamics at the TeV scale 
and we assess the LHC potential to discover new heavy colored states. 
In the first part of the thesis we analyze the bounds on the spectrum of Composite Higgs Models (CHM) that come from flavor observables. 
In the second part we propose a novel strategy to discover a heavy gluon ($G^*$) and heavy fermions at the LHC.  
We do so by means of simple two-site effective Lagrangians, which could also be used in further phenomenological studies on CHM.\\
From the analysis of the bounds on the CHM spectrum, we derive an important constraint on the masses of the heavy 
fermions which does not depend on the flavor structure of the sector beyond the SM.
This bound is obtained from the infrared contribution to $b \to s\gamma$ induced by the flavor-conserving effective vertex $Wt_Rb_R$.
We find that the presence of a custodial symmetry can play a role in protecting this effective coupling.
Studying the heavy colored vectors phenomenology, 
 we find that heavy composite fermions have a great impact on the phenomenology of heavy composite gluon at the LHC. 
If the composite gluon is heavier than the composite fermions, as flavor observables seem to suggest, 
the search in the channels where $G^*$ decays into one heavy fermion plus its Standard Model partner is very promising, 
with the possibility for both the $G^*$ and heavy fermions to be discovered at the early stage of the LHC. 
These channels offer also the possibility to extract important information on model's parameters, such as the top degree of compositeness.  
\end{abstract}

\tableofcontents

\mainmatter

\chapter{Introduction} 
The Standard Model (SM) of particle physics gives a very economical  
description of the spontaneous breaking
of the electroweak symmetry, but it does not explain its origin.
In fact, there are several arguments which suggest to go beyond the  
electroweak symmetry breaking (EWSB) mechanism
of the Standard Model. The main argument is that of the excessive  
fine-tuning in the Higgs mass (hierarchy problem):
the Higgs mass parameter receives large radiative corrections, the largest one from the top quark, that make unnatural the Higgs mass stabilization at the 
electro-weak scale $v\approx246$  GeV, if we admit the validity of the theory up to scales higher than the TeV. \\
A possible solution, besides Supersymmetry, is based on an analogy with the pion mass stabilization in QCD:
the Higgs, similarly to the pion, might be a composite state, generated by a new strong dynamics; as such, its mass is not sensitive
to radiative corrections above the compositeness scale, assumed to be of the order of the TeV scale. A further protection, that allows the Higgs
to be naturally lighter than the other resonances, exists if the composite Higgs is also  
the pseudo-Goldstone boson of a spontaneously broken global symmetry \cite{Georgi_Kaplan}. A pseudo-Goldstone boson Higgs is expected to be light and 
as such in agreement with the indication from the LEP electroweak precision data (EWPD).  
The new strong dynamics could be also realized by the bulk of an extra dimension. 
Indeed the composite Higgs scenario is related via the AdS/CFT correspondence \cite{Maldacena} to the Randall Sundrum framework of a warped extra dimension \cite{RS}.\\
New fermions and new gauge bosons, partners of the SM ones, are expected to emerge from the strong sector at approximately the TeV scale, 
as suggested by the naturalness principle 
(indeed, the partner sector, in particular the top one, operates in the radiative correction to the Higgs mass, stabilizing it).
The Large Hadron Collider (LHC), by probing the TeV energy range, will be able to test the EWSB mechanism and produce these new states. 
The aim of this project is studying the phenomenology of Composite Higgs Models (CHM) and the LHC potential to discover these new states. 
In this thesis we will reconsider the bounds on the CHM spectrum that come from flavor observables, 
with a special focus to $b\to s \gamma$. Guided by the results of this analysis, 
we will consider the implication of such bounds on the phenomenology of these models at the LHC 
and we design a new strategy for discovering the heavy colored resonances which appear in the spectrum.  \\
\noindent
Instead of considering a full theory we will work in an effective description valid at low energy. 
In particular, we will refer to a ``two-site'' (TS) description \cite{Sundrum,Servant}, where two sectors, the weakly-coupled sector of the elementary fields
and the composite sector, that comprises the Higgs, are linearly coupled each other through mass mixing terms \cite{Kaplan}. After diagonalization the 
elementary/composite basis rotates to the mass eigenstate one, made up of SM and heavy states that are admixture of elementary and composite modes. 
Heavier particles have larger degrees of compositeness: heavy SM particles, like the top, are more composite while the light ones are almost elementary.
In these theories the Higgs arises as a fully composite state and so it interacts with the SM and heavy particles only via their composite components. This can explain, if the right-handed 
bottom quark has a small degree of compositeness, 
the small ratio between the bottom and the top quark masses.\\
The model has to be compatible with the experimental data, in particular the LEP precision data suggest the presence of a custodial symmetry in the 
composite sector to avoid large corrections to the $\rho$ parameter. The absence of large Flavor-Changing Neutral Currents is achieved, instead, by a sort of GIM mechanism, that naturally emerges when the connection between 
the elementary and the strong sector proceeds via linear couplings. In absence of a symmetry protection, the LEP data also point toward a small degree of compositeness of the left-handed
 bottom quark (small corrections to $Z\bar{b}_{L}b_{L}$), and, by gauge invariance, of the left-handed top as well. This implies that, in order to 
obtain a heavy enough top quark, it is necessary to have an almost fully composite right-handed top quark.
It has been shown, however, that the corrections to $Z\bar{b}_{L}b_{L}$ can be suppressed if the custodial symmetry 
of the strong sector includes a Left-Right parity \cite{zbb}. This can allow for a smaller right-handed top compositeness.\\
In order to study the phenomenology at energies lower than the compositeness scale, we derive two different models,
 which we denote as TS5 and TS10. 
They describe the low-energy regime of the Minimal Composite Higgs Models (MCHM) defined in Ref. \cite{mchm, mchm2}, in the limit
in which only the leading terms in an expansion in powers of the Higgs field are retained
\footnote{see Ref. \cite{Panico-Wulzer}, for two- and three-site effective theories where the full Higgs non-linearities are
included.}. 
In MCHM the Higgs arises as the pseudo-Goldstone boson associated to the $SO(5)\to O(4)$ breaking in the composite sector; where
$O(4)$ includes $SO(4)\sim SU(2)_L\times SU(2)_R$ as well as a parity $P_{LR}$ which exchanges $SU(2)_L$ with $SU(2)_R$. 
Composite fermions are embedded in a $5=(2,2)+(1,1)$ representation of $SO(5)$ in the TS5 model and in a $10=(2,2)+(1,3)+(3,1)$ in the TS10.
In both the cases we consider, for the analysis at the LHC, the presence of a $P_{LR}$ symmetry protection to the 
$Zb_L\bar{b}_L$ coupling. \\
\newline
We find two important bounds on the masses of the heavy fermions which do not depend on the flavor structure of the sector beyond the SM (BSM).
The first comes from the measurement of the $Zb_L\bar{b}_L$ coupling, that we already mentioned and that can be suppressed assuming a $P_{LR}$ symmetry.
The second is obtained from the infrared (IR) contribution to $b \to s\gamma$ induced by the flavor conserving effective vertex $Wt_Rb_R$. 
In composite Higgs models there are two classes of effects that lead to a shift of the 
$b\rightarrow s\gamma$ decaying rate compared to the SM prediction: 
loops of heavy fermion resonances from the strong sector give a ultraviolet (UV) local contribution; 
they generate, at the compositeness scale, the flavor-violating dipole operators $\mathcal{O}_{7}$ and $\mathcal{O}^{'}_{7}$,
which define the effective Hamiltonian for the $b\rightarrow s\gamma$ decay.

\noindent
The virtual exchange of heavy resonances also generates
the effective V+A interaction of the $W$ boson and the SM quarks, $Wt_R b_R$, which in turn leads to
a shift to $b\to s\gamma$ via a loop of SM particles. This latter IR contribution is enhanced by a chiral factor $m_t/m_b$ and, since in
this case the flavor violation comes entirely from the SM V-A current, $\bar{t}_L\gamma^{\mu}s_L$, it gives a Minimal Flavor Violating (MFV) lower bound on the heavy fermion masses. \\
We also discuss the role of a parity $P_C$, which is a subgroup of the custodial $SU(2)_V$, to protect the effective coupling $Wb_R t_R$,
 which is generated by the $W$, $t_R$ and $b_R$ interactions with the composite sector.\\

In general, stronger bounds can be obtained from the UV CHM contribution to $b\to s \gamma$ and from $\epsilon^{'}/\epsilon_K$ \cite{Isidori}; 
however, these latter bounds are model dependent and in principle could be loosened by acting on the NP flavor structure.
The bound from the IR contribution to $b\to s \gamma$, on the other hand, is robust, since it is a MFV effect. 
As such it cannot be ignored in a phenomenological study on heavy fermions.\\
\newline
In the second part of the thesis we perform a study of the LHC discovery reach on a heavy gluon ($G^*$) and heavy fermions (top and bottom excitations),
 coming from a new composite sector. 
The heavy partner of the gluon has a large degree of compositeness and, as a consequence, it has
larger couplings to the heavier particles (which are also those with larger degrees of compositeness).
We find that heavy fermion resonances have a great impact on the composite gluon phenomenology:
 the $G^*$ phenomenology is strongly dependent on the ratio between $G^{*}$ mass and heavy fermions mass, $M_{G^{*}}/m_{*}$; 
according to its value we can distinguish three scenarios with different phenomenological implications for the $G^{*}$ search at the LHC. 
In the scenario where the $G^*$ is below the threshold for the production of a heavy
fermion ($M_{G^{*}}<m_{*}$), in general $G^*$ decays preferentially to top pairs. 
In this case the signal is $pp\to G^{*}\to t\bar{t}$ and it is affected 
by a huge and difficult to reduce QCD background, $pp\to g\to t\bar{t}$, 
that makes the $G^{*}$ discovery at the LHC particularly challenging. However, the study in Ref. \cite{Agashe}, 
by exploiting peculiarities of the signal, such as the large invariant mass of the $t\bar{t}$ pairs and a Left-Right asymmetry, 
shows that the discovery of a $G^{*}$ lighter than $\sim 4$ TeV is possible at 14 TeV LHC, with $100$ fb$^{-1}$. \\
If $G^*$ is heavier than fermionic resonances, the Branching Ratios
for the $G^*$ decays into one heavy fermion ($\chi$) plus its Standard Model partner ($\psi$) become relevant 
and they also increase in the case of a not fully composite right-handed top.
The analysis we perform considering these decay channels is very promising since the presence of heavy fermion resonances
 in the signal allows for a clean distinction between the signal and the background. 
We point out that there is also a third pessimistic scenario,
 corresponding to the case of a very heavy $G^*$, with a mass 
greater than heavy fermion pairs. In this case the $G^*$ total decay width becomes too large (O(TeV)) to distinguish its resonance from the background.

In our analysis we consider the channels where $G^*$ decays into one heavy fermion plus its Standard Model partner. 
The search in these channels is very promising, 
with the possibility for both the $G^*$ and heavy fermions discovery at the early stage of the LHC. 
These channels offer also the possibility to extract important information on the parameters of the model, 
such as the top degree of compositeness.  \\

Most of the results we will show in the first part of the thesis are also discussed in \cite{Vignaroli-bsGamma}, those in the second part have been presented in \cite{Bini, Vignaroli}. The importance of the heavy-light channels in the search for heavy colored vectors at the LHC has also been pointed out in Ref.~\cite{Barcelo:2011wu}, 
which appeared after completion of this work.

\chapter{Composite Higgs Models}\label{chm}
The idea behind Composite Higgs Models is that the Electro Weak Symmetry Breaking may be triggered by a new strong dynamics, 
in analogy with the chiral symmetry breaking in QCD.
In these theories a new strong sector couples to a weakly coupled sector, which coincides with that of the Standard Model without the Higgs. 
The Higgs, as the pion in QCD, is a composite state coming from the latter strong dynamics. Its composite nature  
allows for a solution to the hierarchy problem. Indeed, its mass is not sensitive
to radiative corrections above the compositeness scale, assumed to be of the order of the TeV.
The EWSB is transmitted to SM fermions by means of linear couplings \cite{Kaplan} (generated by some UV physics at the UV scale $\Lambda_{UV}$) 
between elementary fermions $\psi$ and composite fermions
\begin{equation}
\Delta\mathcal{L}=\lambda\bar{\psi}\mathcal{O}+h.c. 
\label{linCOUP}
\end{equation} 
This way to communicate the EWSB can give a natural explanation of the hierarchies in the quark
masses (through RG evolution of the composite-elementary couplings $\lambda_i$) 
and avoid the tension which occurs when trying to generate large enough quark masses and, at the same time, suppressing FCNC processes\footnote{Tension that instead affects Technicolor and Extended Technicolor Models.}. \\                                                                                                         
As a consequence of linear couplings a scenario of \textit{Partial Compositeness} of the SM particles emerges.
At energies below the compositeness scale a composite operator $\mathcal{O}$ can excite from the vacuum a tower of composite fermions of increasing mass.
Linear couplings (\ref{linCOUP}) thus turn into mass mixing terms between elementary fermions and towers of composite fermions $\chi_n$
\begin{equation}
 \langle 0|\mathcal{O}|\chi_n\rangle=\Delta_n\ \,\ \ \ \mathcal{L}_{mix}=\sum_n\Delta_n \left(\bar{\psi}\chi_n+h.c.\right )\ .
\label{mixTERM}
\end{equation} 
\begin{equation}
 \mathcal{L}=\mathcal{L}_{el}+\mathcal{L}_{com}+\mathcal{L}_{mix} 
\label{Lfull}
\end{equation} 
Because of the mass mixing terms the physical eigenstates, made up of SM and (new) heavy states, are admixture of elementary and composite modes.\\ 
The low-energy phenomenology of such theories can be exhaustively studied, and calculation can be made easier, by considering a truncation of each tower of composite fermions to the first resonance, while 
other heavy states are neglected \cite{Sundrum}. For example, the effective Lagrangian describing one elementary
chiral field $\psi_L$ and its composite partner $\chi$ is
\begin{equation}
 \Delta\mathcal{L}=\bar{\psi}_L i{\not}\partial \psi_L+\bar{\chi} (i{\not}\partial-m_*)\chi- \Delta_L \bar{\psi}_L\chi_R + h.c. \ .
\label{Lchiral}
\end{equation} 
We can rotate the fermions from the elementary/composite basis to the mass eigenstate one, the light(SM)/heavy basis, according to:
\begin{align}
\begin{split}\label{ele/compROTATION}
& \tan\varphi_{L}=\frac{\Delta_{L}}{m_*} \ \ \ \ 
	 \left\{ \begin{array}{l} 
	| light\rangle= \cos\varphi_L |\psi_L\rangle -\sin\varphi_L |\chi_L\rangle \\
	| heavy\rangle= \sin\varphi_L |\psi_L\rangle +\cos\varphi_L |\chi_L \rangle \end{array} \right.
\end{split}
\end{align}
Our eigenstate fields are thus a heavy fermion of mass $m=\sqrt{m^2_* + \Delta^2_L}$ and a light fermion, to be identified with the SM field, that will
acquire a mass after the EWSB. These fields, as we see, are superpositions of elementary and composite states. The angle $\varphi_L$ parametrizes the degree
of compositeness of the physical fields. In particular, the SM fermion has a $\sin\varphi_L\equiv\frac{\Delta_L}{\sqrt{m^2_* + \Delta^2_L}}$ degree of compositeness 
(and a $\cos\varphi_L\equiv\frac{m_*}{\sqrt{m^2_* + \Delta^2_L}}$ degree of elementarity); the mass mixing parameter $\Delta_L$ can be naturally much smaller than the mass 
$m_*$ of the composite fermion\footnote{As a result of RG evolution above the compositeness scale. 
The smallness of $\Delta$ parameters also allows for a sort of GIM mechanism that suppresses large Flavor-Changing Neutral Currents \cite{RSgim}.},
 therefore, SM fermions are in general mostly elementary with a small degree of compositeness, while heavy fermions are mostly composite with a small degree
of elementarity. We have a similar rotation, with angle $\varphi_R$, in the case of right-handed fermions.
SM fermions acquire a mass after the EWSB; since the origin of this breaking resides, by assumption, in the composite sector 
(the Higgs is a fully composite state), the SM fermion mass arises from the composite part of left-handed and right-handed SM fields:
\begin{equation}
 m_\psi = Y_* \frac{v}{\sqrt{2}} \sin\varphi_L \sin\varphi_R ,
\label{mass}
\end{equation} 
where $Y_*$ is a Yukawa coupling among composites, from which the SM Yukawa $y= Y_* \sin\varphi_L \sin\varphi_R$ originates. 
In the following we will assume that the strong sector is flavor
anarchic, so that there is no
large hierarchy between elements within each matrix $Y_*$ and the hierarchy in the masses and mixings of the SM quarks 
comes entirely from the hierarchy in the elementary/composite mixing angles 
(such `anarchic scenario' has been extensively
studied in the framework of 5D warped models, see Refs. \cite{Huber,Agashe_perez,Csaki_anarchic,Casagrande,Albrecht}).
From (\ref{mass}) we can see that heavier SM particles have larger degrees of compositeness: 
heavy SM particles, like the top, have to be quite composite while the light ones are almost elementary.\\
There are also elementary/composite mixings for the vector fields (similarly to the $\rho$-photon mixing in QCD):
\begin{equation}
	\mathcal{L}^{vect}_{mix}=\frac{M^{2}_{*}}{2}\left(\rho_{\mu}-\frac{g_{el}}{g_{*}}A_{\mu}\right)^{2}\ .
\end{equation}
Bosons rotate from the elementary/composite basis to the physical light/heavy basis according to
\begin{eqnarray}
	 \tan\theta=\frac{g_{el}}{g_{*}}\ \ \ 
        \left\{\begin{array}{l}
	A_{\mu}=\cos\theta A^{el}_{\mu}+\sin\theta\rho^{com}_{\mu}\\ 
	\rho_{\mu}=-\sin\theta A^{el}_{\mu}+\cos\theta\rho^{com}_{\mu} \end{array} \right. \ .
\end{eqnarray}
We have the following relations among SM coupling and elementary/composite couplings:
\begin{equation}
	g_{SM}=g_{*}\sin\theta \ \ g_{SM}=g_{el}\cos\theta \ .
\end{equation}
In particular, the SM gluon and the heavy gluon $G^*$ have the following superpositions with elementary and composite modes:
\begin{eqnarray}
 \tan\theta_3=\frac{g_{el3}}{g_{*3}}\ \ \ \left\{\begin{array}{l}
	g_{\mu}=\cos\theta_3 g^{el}_{\mu}+\sin\theta_3 G^{*com}_{\mu}\\ 
	G^{*}_{\mu}=-\sin\theta_3 g^{el}_{\mu}+\cos\theta_3 G^{*com}_{\mu}\ .
	\end{array} \right.
\label{GgMIX}
\end{eqnarray}

\begin{equation}
	\sqrt{4\pi \alpha_S}=g_{S}=g_{*3}\sin\theta_3 =g_{el3}\cos\theta_3\simeq g_{el3} \ .
\label{GgMIX2}
\end{equation}
\\
Experimental data give hints on the type of the new strong dynamics responsible for the EWSB. 
The LEP precision data suggest the presence of a custodial symmetry in the 
composite sector to avoid large corrections to the $\rho$ parameter. 
In order to protect $\rho$ (or equivalently the Peskin-Takeuchi T parameter) the composite sector must respect, minimally, a global symmetry:
\[
 SU(2)_L \times SU(2)_R \times U(1)_X\ ,
\]
where $SU(2)_L\times SU(2)_R$ is broken to the diagonal $SU(2)_V$ after the EWSB; 
the unbroken $SU(2)_V$ invariance acts as a custodial symmetry so that $\rho=1$ at tree level.\\
Composite gauge bosons gauge the group:
\[
 SU(3)_{C}\times SU(2)_{L}\times SU(2)_{R}\times U(1)_{X}\ .
\]
The SM electroweak group $SU(2)_{L}\times U(1)_{Y}$ can be embedded into $SU(2)_{L}\times SU(2)_{R}\times U(1)_{X}$, so that
hypercharge is realized as $Y = T^3_{R} +X$. 
The Composite Higgs transforms as a bidoublet $(2,2)$ under $SU(2)_{L}\times SU(2)_{R}$, $\mathcal{H}\equiv (H,H^c)$, 
where $H$ is the Composite Higgs doublet and $H^c=i\sigma^2 H^*$ is its conjugate.
The $\mathcal{H}$ VEV 
breaks the $SU(2)_{L}\times SU(2)_{R}\times U(1)_{X}$ group down to $SU(2)_{V}\times U(1)_{X}$ and leads to the EWSB
(($SU(2)_{V}\times U(1)_{X}$)$\cap$($SU(2)_{L}\times U(1)_{Y}$) is broken to $U(1)_{em}$). 
Therefore, we have the following relation among charges:
\begin{equation}
	Q=T^3_{L}+T^3_{R}+X=T^3_{L}+Y \ .
	\label{eq.charge}
\end{equation}

This scheme can also results from models where the Higgs arises as the pseudo-Goldstone boson associated to a $SO(5)\to SO(4)\sim SU(2)_L\times SU(2)_R$ breaking in the composite sector;
or to a $SO(5)\to O(4)$ breaking,
 where $O(4)$ includes $SO(4)\sim SU(2)_L\times SU(2)_R$ as well as a parity $P_{LR}$ which exchanges $SU(2)_L$ with $SU(2)_R$. 
This enhanced custodial symmetry can suppress the corrections to the coupling $Z\bar{b}_{L}b_{L}$, which are strongly constrained by LEP data \cite{zbb}. 

\section{$P_{LR}$ and $P_C$ symmetries}\label{PLR}
In MCHM \cite{mchm} the Higgs arises as the pseudo-Goldstone boson associated to the $SO(5)\to O(4)$ breaking in the composite sector; 
where the enhanced custodial symmetry $O(4)$ includes $SO(4)\sim SU(2)_L\times SU(2)_R$ as well as a parity 
$P_{LR}$ which exchanges $SU(2)_L$ with $SU(2)_R$. 
As shown in \cite{zbb}, this $P_{LR}$ parity, as well as the $P_C$ symmetry, subgroup of the custodial $O(4)$,
 can protect the coupling $Z\bar{b}_{L}b_{L}$ against large corrections from the composite sector. \\
Each composite operator has a definite left and right isospin quantum number, $T_{L,R}$,
and a 3rd component, $T^3_{L,R}$. 
We can also univocally assign to each SM field definite quantum
numbers, $T_{L,R}$, $T^3_{L,R}$, corresponding to those of the composite operator to which it couples. 
$P_{LR}$ and $P_C$ are symmetries of the composite sector, $P_{LR}$ exchanges $SU(2)_L$ with $SU(2)_R$ 
and $P_C$ is the subgroup of $SU(2)_V$ that transforms $\left |T_L, T_R; T^3_L, T^3_R\right\rangle \to \left |T_L, T_R; -T^3_L, -T^3_R\right\rangle$ 
($SO(3)$ vectors transform with $P_C=diag(1,-1,-1)$).
For $P_{LR}$ ($P_C$) to be a symmetry also of the interacting terms between SM fields and composite operators, 
$\Delta\mathcal{L}=\lambda\bar{\psi}\mathcal{O}+h.c.$, the SM fields $\psi$ have to be eigenstates of $P_{LR}$ ($P_C$). 
This implies:
\begin{equation}
 T_L=T_R \ \ (T^3_L=T^3_R) \ \ (P_{LR}\ invariance)
\end{equation}
\begin{equation}
 T^3_L=T^3_R=0 \ \ (P_{C}\ invariance)\ .
\end{equation} 
\noindent
If the above formulas hold, we can see that the coupling $Z\psi\bar{\psi}$, 
\begin{equation}
g_{\psi}=\frac{g}{\cos\theta_W}(Q^3_L-Q\sin^2\theta_W)\ ,
\end{equation}
 is protected against large corrections. 
Indeed, the electric charge $Q$ is conserved and the charge of the $SU(2)_L$ 3rd component, $Q^3_L$, 
is conserved by custodial invariance plus $P_{LR}$ symmetry and by $P_C$ symmetry. 
By custodial $U(1)_V$ invariance, $\delta Q^3_V=\delta Q^3_R+\delta Q^3_L=0$; if there is also a $P_{LR}$ invariance, 
$\delta Q^3_R=\delta Q^3_L$, therefore $\delta Q^3_L=0$. The same conservation, $\delta Q^3_L=0$, is obtained by $P_C$ invariance: 
the SM $W^3_L$ has an odd parity under $P_C$, $W^3_L\to -W^3_L$; if $\psi$ is a $P_C$ eigenstate it must have $T^3_L=T^3_R=0$, then 
the current $\bar{\psi}\gamma^{\mu}\psi$ is even under $P_C$ and it cannot couple to $W^3_L$, which is odd.\\
We will show (sec. \ref{PCprotect}) that the $P_C$ symmetry can also protect in a similar way the effective coupling $Wt_R b_R$ and, as a consequence, 
it can be responsible for an attenuation of the bound on heavy fermion masses, coming from the process $b\to s\gamma$. \\

In what follows we will present the Two-Site models, TS5 and TS10, which incorporate a custodial symmetry and a $P_{LR}$ parity.

\section{TS5}\label{TS5}
In the TS5 model, we consider composite fermions embedded into fundamentals $\mathbf{5_{2/3\ (-1/3)}}$ of $SO(5)\times U(1)_{X}$, 
that decompose as $\mathbf{5_{2/3\ (-1/3)}}=\mathbf{(2,2)_{2/3\ (-1/3)}}\oplus\mathbf{(1,1)_{2/3\ (-1/3)}}$ under 
$SU(2)_{L}\times SU(2)_{R}\times U(1)_{X}$. We refer to this field content, at lower energy, in the composite sector:

\begin{eqnarray}	
\nonumber
	\mathcal{Q}_{2/3}=\left[\begin{array}{cc}
	T & T_{5/3} \\ 
	B & T_{2/3} \end{array}\right]=\left(2,2\right)_{2/3} , \ \tilde{T}=\left(1,1\right)_{2/3}\\ \nonumber
	\mathcal{Q'}_{-1/3}=\left[\begin{array}{cc}
	B_{-1/3} & T' \\ 
	B_{-4/3} & B' \end{array}\right]=\left(2,2\right)_{-1/3} , \ \tilde{B}=\left(1,1\right)_{-1/3}\\ \nonumber
	\mathcal{H}=\left[\begin{array}{cc}
	\phi^{\dag}_{0} & \phi^{+} \\
	-\phi^{-} & \phi_{0} \end{array}\right]=\left(2,2\right)_{0} \\
	\label{eq.fields}
\end{eqnarray}
\noindent
We are thus introducing two classes of composite fermions, 
those filling a $\mathbf{5_{2/3}}$ representation, with $X$ charge $X=2/3$ and those in a $\mathbf{5_{-1/3}}$, with $X=-1/3$. 
We want to consider, indeed, the possibility that the SM quark doublet $(t_{L}, b_{L})$ couples to two different BSM operators,
 $\mathcal{Q}_{2/3}$ and $\mathcal{Q'}_{-1/3}$, the first responsible for generating the top mass, the second for generating the bottom mass. 
$(t_{L}, b_{L})$ is linearly coupled to $(T, B)$ through a mass mixing term we call $\Delta_{L1}$ and to $(T', B')$ through a mass mixing term
$\Delta_{L2}$. $t_{R}$ and $b_{R}$ couple respectively to $\tilde{T}$, through a mass mixing term $\Delta_{R1}$, and to $\tilde{B}$, 
through a mass mixing term $\Delta_{R2}$. The fermionic Lagrangian reads, in the elementary/composite basis:

\begin{align}
\begin{split}\label{eq.Lagrange1}
\mathcal{L}=\  & \bar{q}^i_{L}i\dslash q^i_{L}+\bar{u}^i_{R}i\dslash u^i_{R}+\bar{d}^i_{R}i\dslash d^i_{R}\\ 
& +Tr\left\{\bar{\mathcal{Q}}\left(i\dslash-M_{Q*}\right)\mathcal{Q}\right\}+\bar{\tilde{T}}\left(i\dslash-M_{\tilde{T}*}\right)\tilde{T}+Y_{*U}Tr\left\{\bar{\mathcal{Q}}\mathcal{H}\right\}\tilde{T}\\ 
& +Tr\left\{\bar{\mathcal{Q'}}\left(i\dslash-M_{Q'*}\right)\mathcal{Q'}\right\}+\bar{\tilde{B}}\left(i\dslash-M_{\tilde{B}*}\right)\tilde{B}+Y_{*D}Tr\left\{\bar{\mathcal{Q'}}\mathcal{H}\right\}\tilde{B}\\ 
& -\Delta_{L1}\bar{q}^3_{L}\left(T,B\right)-\Delta_{R1}\bar{t}_{R}\tilde{T}-\Delta_{L2}\bar{q}^3_{L}\left(T',B'\right)-\Delta_{R2}\bar{b}_{R}\tilde{B}+h.c.	\ .
\end{split}
\end{align}
\noindent
where the superscript $i$ runs over the three SM families ($i$ = 1, 2, 3), with $q^3_L\equiv(t_L , b_L )$, 
$u^3\equiv t_R$, $d^3\equiv b_R$. By construction, the elementary fields couple to the composite
ones only through the mass mixing terms, shown in the last row of (\ref{eq.Lagrange1}). This implies that the SM Yukawa
couplings arise only through the coupling of the Higgs to the composite fermions and their
mixings to the elementary fermions. We further assume that the strong sector is flavor
anarchic, so that the hierarchy in the masses and mixings of the SM quarks comes from
the hierarchy in the mixing parameters $\Delta^i_{L,R}$. In this case the mixing
parameters of the light elementary quarks can be safely neglected and one can focus on
just the third generation of composite fermions. 
\footnote{In fact, once produced, heavy fermions of the first two generations will also decay mostly to tops and
bottoms, since flavor-changing transitions are not suppressed in the strong sector, while the couplings to the
light SM quarks are extremely small, see the discussion in Ref. \cite{Sundrum}.
}\\
As a consequence of the elementary/composite mass mixings, 
the top and the bottom masses arise, after the EWSB, from the Yukawa terms in the Lagrangian (\ref{eq.Lagrange1}), 
$Y_{*U}Tr\left\{\bar{\mathcal{Q}}\mathcal{H}\right\}\tilde{T}$ and $Y_{*D}Tr\left\{\bar{\mathcal{Q'}}\mathcal{H}\right\}\tilde{B}$. 
The top mass will be proportional to $\Delta_{L1}\Delta_{R1}$ and the bottom mass to $\Delta_{L2}\Delta_{R2}$. 
The small ratio between the bottom and the top quark masses can be thus obtained both for $\Delta_{L2}\ll\Delta_{L1}$ ($\Delta_{R2}\sim\Delta_{R1}$) 
and for $\Delta_{R2}\ll\Delta_{R1}$ ($\Delta_{L2}\sim\Delta_{L1}$). \\
For $t_{R}$, $b_{R}$ and their excited states the rotation from the elementary/composite basis to the mass eigenstate one, the SM/heavy basis, 
is given by:
\begin{align}
\begin{split}\label{rotation_RR2}
&\tan\varphi_{R}=\frac{\Delta_{R1}}{M_{\tilde{T}*}}\ \ s_{R}\equiv\sin\varphi_{R} \ \ c_{R}\equiv\cos\varphi_{R}  \\ 
&\tan\varphi_{bR}=\frac{\Delta_{R2}}{M_{\tilde{B}*}}\ \ s_{bR}\equiv\sin\varphi_{bR} \ \ c_{bR}\equiv\cos\varphi_{bR} \\  
&\left\{\begin{array}{l}
t_{R}=c_{R}t^{el}_{R}-s_{R}\tilde{T}^{com}_{R}\\
\tilde{T}_{R}=s_{R}t^{el}_{R}+c_{R}\tilde{T}^{com}_{R} \end{array} \right.\ 
\left\{\begin{array}{l}
		b_{R}=c_{bR}b^{el}_{R}-s_{bR}\tilde{B}^{com}_{R}\\ 
	\tilde{B}_{R}=s_{bR}b^{el}_{R}+c_{bR}\tilde{B}^{com}_{R} \end{array}  \right.
\end{split}
\end{align}
\noindent
An analytical diagonalization of the mixing among $q^3_{L}$ and the corresponding excited states 
is possible only if we consider simplifying assumptions. In particular, we will consider the case where $\Delta_{L2}\ll\Delta_{L1}$, 
that can naturally follow, for example, from the RG flow in the full theory \cite{mchm2}. 
The first two generations of elementary quarks do not need a field rotation 
from the elementary/composite basis to the mass eigenstate basis, 
since they do not mix with the composite fermions and can thus be directly identified with the corresponding SM states.\\
We can see that in this model $t_{R}$ and $b_{R}$ are both $P_{C}$ and $P_{LR}$ eigenstates, 
since they couple to $SU(2)_{L}\times SU(2)_{R}$ singlets 
($T_L (\tilde{T},\tilde{B})=T_R (\tilde{T},\tilde{B})$, $T^3_{L} (\tilde{T},\tilde{B})=T^3_{R} (\tilde{T},\tilde{B})=0$).
Instead, $t_L$ is a $P_{LR}$ eigenstate only in the limit ($\Delta_{L1}=0$) in which 
it decouples from $T$ ($T^3_{L}(T)\neq T^3_{R}(T)$). Similarly, 
$b_L$ is a $P_{LR}$ eigenstate only for $\Delta_{L2}= 0$, in which case it decouples from $B'$ ($T^3_{L}(B')\neq T^3_{R}(B')$).\\
\noindent
So far we have made field rotations to the mass eigenstate basis before the EWSB. 
After the EWSB, the SM top and bottom quarks acquire a mass, and the heavy masses get corrections 
of order $\left(\frac{Y_{*}v}{\sqrt{2}m_{*}}\right)^2$. 
In the following, we assume $x \equiv \left(\frac{Y_{*}v}{\sqrt{2}m_{*}}\right) \ll 1$ and compute all quantities at leading order in 
$x$.

\subsection{$\Delta_{L2}\ll\Delta_{L1}$} 
In this case, since $\Delta_{L2}\ll\Delta_{L1}$, $b_{L}$ is, approximately, a $P_{LR}$ eigenstate so, approximately, we have a 
custodial symmetry protection to $Zb_{L}\bar{b}_L$.\\
The small ratio between the bottom and the top quark masses is obtained for $\Delta_{L2}\ll\Delta_{L1}$ ($\Delta_{R2}\sim\Delta_{R1}$); we have:
\begin{equation}
	m_{t}=\frac{v}{\sqrt{2}}Y_{*U}s_{1}s_{R}
\label{tmass}
\end{equation}
\begin{equation}
	m_{b}=\frac{v}{\sqrt{2}}Y_{*D}s_{2}s_{bR} \ ,
\label{bmass}
\end{equation}
\noindent
where $s_{1}=\sin\varphi_{L1}=\frac{\Delta_{L1}}{\sqrt{M^{2}_{Q*}+\Delta^{2}_{L1}}}$ and $s_{2}$ is a rotation angle proportional to $\Delta_{L2}$, 
$s_{2}=\frac{\Delta_{L2}}{M_{Q'*}}\cos\varphi_{L1}$. \\
\noindent 
The physical masses of the heavy fermions read:
\begin{align}
\left\{\begin{array}{l}
M_{\tilde{T}}=\sqrt{M^{2}_{\tilde{T}*}+\Delta^{2}_{R1}}\\
M_{\tilde{B}}=\sqrt{M^{2}_{\tilde{B}*}+\Delta^{2}_{R2}}\\
M_{T}=M_{B}=\sqrt{M^{2}_{Q*}+\Delta^{2}_{L1}} \\
M_{T5/3}=M_{T2/3}=M_{Q*}=M_T c_1 \\
M_{T'}=M_{B'}=\sqrt{M^{2}_{Q'*}+\Delta^{2}_{L2}}\simeq M_{Q'*}\\
M_{B-1/3}=M_{B-4/3}=M_{Q'*} \end{array} \right.
\label{HfMass_ts5}
\end{align}
where $c_1\equiv \cos\varphi_{L1}$. Details can be found in App. \ref{TS5A}.\\
In order for the strong sector to respect the custodial invariance, as we have shown,
 composite fermions have to fill multiplets of $SU(2)_{L}\times SU(2)_{R}\times U(1)_{X}$. 
As a consequence, the heavy partner of the SM doublet $q^3_L=(t_L,b_L)$, $Q=(T,B)$ ($=2_{1/6}$ under the SM electroweak group), is embedded in a larger multiplet, 
the bidoublet $\mathcal{Q}_{2/3}=(2,2)_{2/3}$, that includes an other doublet of heavy fermions, $Q'=(T_{5/3},T_{2/3})$($=2_{7/6}$). 
The heavy fermions $T_{5/3}$ and $T_{2/3}$ in this latter doublet are called \textit{custodians}. 
They share the same multiplet of the heavy partners of $q^3_L$ but they do not mix directly with the SM fermions. 
This implies that their masses tend to zero in the limit in which $t_L$ becomes fully composite (see for example the discussion in \cite{pomarol_serra}). 
This can be seen from eq. (\ref{HfMass_ts5}): $M_{T5/3(2/3)}$ is zero for $c_1=0$, i.e. for a fully composite $t_L$ ($s_1=1$).


\section{TS10}\label{TS10}

In TS10 we consider composite fermions embedded into a $\mathbf{10_{2/3}}$ representation 
of $SO(5)\times U(1)_{X}$, that decomposes as $\mathbf{10_{2/3}}=\mathbf{(2,2)_{2/3}}\oplus\mathbf{(1,3)_{2/3}}\oplus\mathbf{(3,1)_{2/3}}$ under 
$SU(2)_{L}\times SU(2)_{R}\times U(1)_{X}$. Therefore we refer to this field content in the composite sector:
\begin{eqnarray}	
\nonumber
	\mathcal{Q}_{2/3}=\left[\begin{array}{cc}
	T & T_{5/3} \\ 
	B & T_{2/3} \end{array}\right]=\left(2,2\right)_{2/3}\\ \nonumber 
 \mathcal{\tilde{Q}}_{2/3}=\left(\begin{array}{c}
	\tilde{T}_{5/3} \\ 
	\tilde{T}\\
        \tilde{B} \end{array}\right)=\left(1,3\right)_{2/3}\ , \  \mathcal{\tilde{Q}'}_{2/3}=\left(\begin{array}{c}
	\tilde{T}'_{5/3} \\ 
	\tilde{T}' \\
        \tilde{B}' \end{array}\right)=\left(3,1\right)_{2/3}   \\ \nonumber
	\mathcal{H}=\left[\begin{array}{cc}
	\phi^{\dag}_{0} & \phi^{+} \\
	-\phi^{-} & \phi_{0} \end{array}\right]=\left(2,2\right)_{0} \\
	\label{eq.fields_mchm10}
\end{eqnarray}
and to the following fermionic Lagrangian in the elementary/composite basis:

\begin{align}
\begin{split}\label{eq.Lagrange1_mchm10}
\mathcal{L}= \ &\bar{q}^3_{L} i\dslash q^3_{L}+\bar{t}_{R}i\dslash t_{R}+\bar{b}_{R}i\dslash b_{R} \\
& +Tr\left\{\bar{\mathcal{Q}}\left(i\dslash -M_{Q*}\right)\mathcal{Q}\right\}+Tr\left\{\mathcal{\bar{\tilde{Q}}}\left(i\dslash-M_{\tilde{Q}*}\right)\mathcal{\tilde{Q}}\right\}
+Tr\left\{\mathcal{\bar{\tilde{Q}}'}\left(i\dslash-M_{\tilde{Q}*}\right)\mathcal{\tilde{Q}}'\right\}\\ 
& +Y_{*}Tr\left\{\mathcal{H}\bar{\mathcal{Q}}\mathcal{\tilde{Q}'}\right\}+Y_{*}Tr\left\{\bar{\mathcal{Q}}\mathcal{H}\mathcal{\tilde{Q}}\right\} \\ 
& -\Delta_{L1}\bar{q}^3_{L}\left(T,B\right)-\Delta_{R1}\bar{t}_{R}\tilde{T}-\Delta_{R2}\bar{b}_{R}\tilde{B}+h.c.\ .
\end{split}	
\end{align}
\noindent
We have the following expressions for the top and bottom masses:
\begin{equation}
 m_{t}=\frac{v}{2}Y_{*}s_{1}s_{R} \ , \ \  m_{b}=\frac{v}{\sqrt{2}}Y_{*}s_{1}s_{bR}
\end{equation} 
and for the heavy fermion physical masses:
\begin{align}
\left\{\begin{array}{l}
M_{\tilde{T}}=\sqrt{M^{2}_{\tilde{Q}*}+\Delta^{2}_{R1}}\\
M_{\tilde{B}}=\sqrt{M^{2}_{\tilde{Q}*}+\Delta^{2}_{R2}}=M_{\tilde{T}} c_R / c_{bR}\simeq M_{\tilde{T}} c_R\\
M_{\tilde{T}5/3}=M_{\tilde{T}'5/3}=M_{\tilde{T}'}=M_{\tilde{B}'}=M_{\tilde{T}} c_R\\
M_{T}=M_B=\sqrt{M^{2}_{Q*}+\Delta^{2}_{L1}} \\
M_{T2/3}=M_{T5/3}= M_T c_1 \end{array} \right. \ .
\end{align}
\noindent
More details can be found in App. \ref{TS10A}.\\
Besides the custodians $T_{5/3}$ and $T_{2/3}$, which are light in the case of composite $q^3_L$, $\tilde{T}_{5/3}$ and the fermions
 in the $\mathcal{\tilde{Q}'}_{2/3}$ triplet become light for composite $t_R$ (also $\tilde{B}$ becomes light in this case).\\
In this model, both $t_{R}$ and $b_{R}$ are not $P_{LR}$ eigenstates and only $t_R$ is a $P_{C}$ eigenstate, 
as a consequence of the couplings to $\mathcal{\tilde{Q}}$ 
($T_L(\tilde{T},\tilde{B})\neq T_R(\tilde{T},\tilde{B})$; 
in particular, $b_R$ is not a $P_C$ eigenstate, since $T^3_{R}(\tilde{B})\neq 0$. Instead, $b_{L}$ is exactly a $P_{LR}$ eigenstate.


\section{$Zb_{L}\bar{b}_{L}$ in the TS Models}
Shifts in the $Z$ coupling to $b_{L}$, $g_{Lb}$, arise after the EWSB because of electroweak mixings among $b_{L}$ and heavy fermions. There is also a contribution 
from the mixing among neutral gauge bosons; however this mixing is of the order $(\frac{v}{M_{*}})^{2}\ll 1$, 
where $M_{*}$ stands for the heavy
neutral boson mass, and we will neglect it in what follows.\\
In Two-Site Models without $P_{LR}$ symmetry 
there is no custodial symmetry protection to $Zb_{L}\bar{b}_{L}$ and so the shift on $g_{Lb}$ is large. 
Naive Dimensional Analysis (NDA) \cite{NDA} gives (see, for example, \cite{Agashe_flav, SILH}):
\begin{equation}
 \frac{\delta g_{Lb}}{g_{Lb}}\sim \frac{m^{2}_{t}}{M^{2}_{Q*}s^{2}_{R}}\sim \frac{Y^2_{*} v^2 s^2_1}{M^2_{Q*}}\ .
\label{gLb_TS}
\end{equation} 
\noindent
This formula has been obtained by approximating $q^2=M^2_Z\simeq 0$. 
At $q^2=M^2_Z$ the shift receives $O\left(\frac{M^2_Z}{M^2_{Q*}}\right)$ corrections:
\begin{equation}
 \frac{\delta g_{Lb}}{g_{Lb}}\sim \frac{M^2_Z s^2_1}{M^2_{Q*}}\sim \left( \frac{v^2 Y^2_{*} s^2_1}{M^2_{Q*}}\right)\frac{g^2}{Y^2_{*}}\ . 
\label{gLb_QnonZero}
\end{equation}
When compared to (\ref{gLb_TS}), there is a suppression $\left( \frac{g}{Y_{*}}\right)^2$ (see for example \cite{contino_fcnc}), so we will neglect it in the following.

LEP and SLD experiments fix an upper bound of $0.25\%$ for the (positive) shift in the $g_{Lb}$ from its SM value. Therefore, from
the eq. (\ref{gLb_TS}), we derive the following bound for the heavy fermion mass in models without custodial symmetry protection to $Zb_{L}\bar{b}_{L}$:
\begin{equation}
M_{Q*}\gtrsim (3.2)\frac{1}{s_{R}} TeV \ .    
\label{no_cust_bound}                                    
 \end{equation} 
\noindent
In order to respect this limit without requiring too large heavy fermion masses, that would contrast with naturalness arguments, it is necessary
to have a quite composite right-handed top (i.e., a not small $s_{R}$). Instead, in models with custodial symmetry protection to $Zb_{L}\bar{b}_{L}$, 
there is no such restriction for the $t_{R}$ degree of compositeness and bounds are weaker than the one in (\ref{no_cust_bound}). Indeed, in the TS5
with $\Delta_{L2}\ll\Delta_{L1}$, where we have approximately a custodial symmetry protection to $Zb_{L}\bar{b}_L$ 
(the breaking is proportional to $\Delta_{L2}$ and is therefore small), we obtain:
\begin{align}
\begin{split} 
 \frac{\delta g_{Lb}}{g_{Lb}} =\left(\frac{Y_{*}v}{\sqrt{2}} \right)^{2}\left(\frac{s_{2}c_{bR}}{\sqrt{2}M_{\tilde{B}}}\right)^{2}[T^3_{L}(\tilde{B})-T^3_{L}(b_{L})]
=\frac{1}{2}\frac{m^{2}_{b}}{M^{2}_{Q*}}\frac{c^{4}_{bR}}{s^{2}_{bR}}\simeq\frac{1}{2}\frac{m^{2}_{t}}{M^{2}_{Q*}}\frac{s^{2}_{2}}{s^{2}_{R}} \ .
\end{split}
\end{align}
As expected, the shift is proportional to $s^{2}_{2}$ (i.e., it is proportional to $\Delta^{2}_{L2}$, the size of the
custodial symmetry breaking) and it is small (notice that is also smaller than the effect at non-zero momentum).
In the TS10, we obtain, again, a small shift:
\begin{align}
 \begin{split}
  \frac{\delta g_{Lb}}{g_{Lb}}= & \left(\frac{Y_{*}v}{\sqrt{2}}\right)^{2}\frac{s^{2}_{1}}{M^{2}_{Q*}}\left[c^{4}_{bR}(T^3_{L}(\tilde{B})-T^3_{L}(b_{L}))+(T^3_{L}(B')-T^3_{L}(b_{L}))\right]\\
  = & -\frac{m^{2}_{b}}{M^{2}_{Q*}}\frac{2-s^{2}_{bR}}{2}\simeq -\frac{m^{2}_{b}}{M^{2}_{Q*}}\ .
 \end{split}
\end{align}
Despite $b_L$ is an exact $P_{LR}$ eigenstate in the TS10, there is still a small modification that comes from the coupling of $b_{R}$, 
that explicitly breaks $P_{LR}$. Notice that $\delta g_{Lb}=0$, if we have $s_{bR}=0$.

\chapter{Bounds from flavor observables}
\section{Constraint from the process $b\rightarrow s\gamma$}\label{bsGamma}

\begin{figure}
\centering
\scalebox{0.7}{\includegraphics{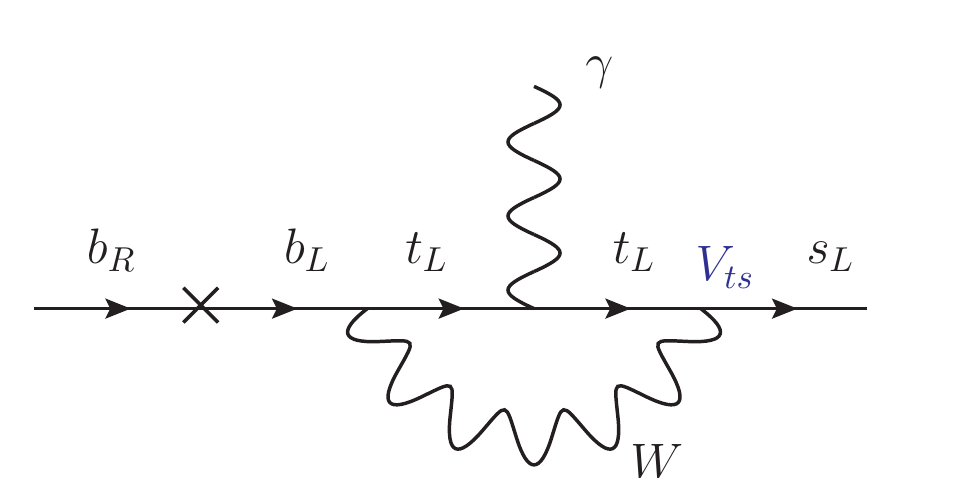}}
\caption{\textit{1 loop Infrared contribution to $C_7$ in the SM.}}
\label{C7-sm}
\end{figure}

We define, following \cite{Buras}, the effective Hamiltonian for $b\rightarrow s\gamma$:
\begin{equation}
 \mathcal{H}_{eff}=-\frac{G_{F}}{\sqrt{2}}V^{*}_{ts}V_{tb}\left[C_{7}(\mu_{b})\mathcal{O}_{7} + C^{'}_{7}(\mu_{b})\mathcal{O}^{'}_{7}\right] \ ,
\label{Heff} 
\end{equation} 
where $\mathcal{O}_{7}=\frac{e}{8\pi^{2}}m_{b}\bar{b}\sigma^{\mu\nu}F_{\mu\nu}(1-\gamma_{5})s$ and 
$\mathcal{O}^{'}_{7}=\frac{e}{8\pi^{2}}m_{b}\bar{b}\sigma^{\mu\nu}F_{\mu\nu}(1+\gamma_{5})s$.\\
In the SM the $W$ boson has a purely $V-A$ interaction to the fermions and so the contribution to the $b\rightarrow s\gamma$
process has to proceed through mass insertions in the external legs (see Fig. \ref{C7-sm}). The Wilson coefficient $C'_{7}$ is, therefore, negligible, 
because of a suppression by a factor $m_{s}/m_{b}$ in respect to the Wilson coefficient $C_{7}$,
that, evaluated at the weak scale $\mu_{w}$ is \cite{Buras}
  \begin{equation}
   C^{SM}_{7}(\mu_{w})=-\frac{1}{2}\left[ -\frac{(8x^{3}_{t}+5x^{2}_{t}-7x_{t})}{12(1-x_{t})^{3}}+\frac{x^{2}_{t}(2-3x_{t})}{2(1-x_{t})^{4}}\ln(x_{t})\right] \ ,
  \end{equation} 
with $x_{t}=\frac{m^{2}_{t}}{M^{2}_{W}}$. \\
In composite Higgs models there are two classes of effects that lead to a shift of the $b\rightarrow s\gamma$ decaying rate
compared to the Standard Model prediction. 
The first is coming from loops of heavy fermion resonances from the strong sector that generate the flavor-violating dipole operators 
$\mathcal{O}_{7}$, $\mathcal{O}^{'}_{7}$ at the compositeness scale. 
We will refer to this as the UV contribution. The second contribution comes from the tree level exchange of heavy resonances, 
which generates an effective V+A interaction of the $W $boson and the SM quarks which in turn leads to
a shift to $b\to s\gamma$ via a loop of SM particles. This latter IR contribution is enhanced by a chiral factor $m_t/m_b$. Since in
this case the flavor violation can come entirely from the SM V-A current, it gives a quite
model-independent lower bound on the heavy fermion masses.\\
By taking into account the experimental average value for the $b\rightarrow s\gamma$ branching ratio \cite{BsGamma_exp} and the 
theoretical calculation \cite{BsGamma_th}, we get, if the new physics contributions to $C_{7}$, $C^{CH}_{7}$, and to $C^{'}_{7}$, $C^{'CH}_{7}$, are considered
separately, the bounds (\ref{App_bound}):
\begin{equation}
 -0.098 \lesssim C^{CH}_{7}(m_{*}) \lesssim 0.028 
\label{C7UVlim}
\end{equation}  
\begin{equation}
 |C^{'CH}_{7}(m_{*})| \lesssim 0.37 \ ,
\label{C'7lim}
\end{equation}  
where $m_{*}$ stands for the mass of the heavy fermions in the loop (we take $m_{*}=1$ TeV). \\
The infrared contribution to $b\rightarrow s\gamma$ from the composite Higgs model is at the weak scale $\mu_{w}$ instead of $m_{*}$ (we take $\mu_W = M_W$); 
therefore, we have to account for a scaling factor
\begin{equation}
 C^{CH}_{7}(\mu_{w})=\left[ \frac{\alpha_{s}(m_{*})}{\alpha_{s}(m_{t})}\right]^{16/21} \left[\frac{\alpha_{s}(m_{t})}{\alpha_{s}(\mu_{w})} \right]^{16/23} \approx 0.79  C^{CH}_{7}(m_{*})
\end{equation}  
We get:
\begin{equation}
 -0.077\lesssim C^{CH}_{7}(\mu_{w}) \lesssim 0.023
\label{c7limIR}
\end{equation}  
\begin{equation}
 |C^{'CH}_{7}(\mu_{w})| \lesssim 0.29
\label{C'7limIR}
\end{equation}  
\noindent
While the infrared contribution to $C_7$ involves a flavor-conserving operator and brings to a MFV bound, 
the infrared contribution to $C^{'}_7$ as well as the ultraviolet contributions to $C_7$ and to $C^{'}_7$ involve 
flavor-violating operators. As a consequence, they will require assumptions on the flavor structure of the NP sector.    

\noindent
We will now evaluate the bounds on heavy masses that come from the infrared contribution to $C_7$. 
We will present first estimates of such bounds for generic composite Higgs models, which can be obtained by NDA. 
Then we will calculate the bounds for the specific two-site model TS5 and TS10, introduced in sec.s \ref{TS5} and \ref{TS10}.

\subsection{Infrared contribution to $C_7$}\label{IR_contr}

\begin{figure}
\centering
\scalebox{0.7}{\includegraphics{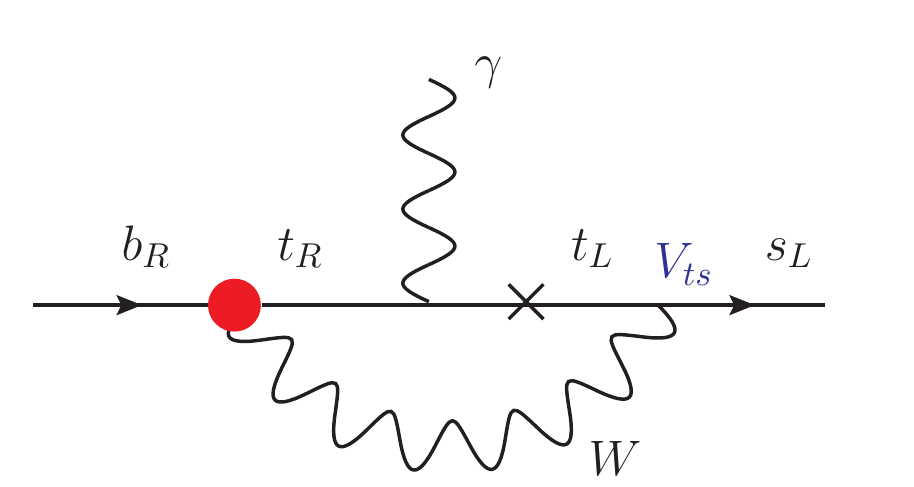}}
\caption{\textit{1 loop Infrared contribution to $C_7$. 
The red blob denotes the effective coupling $Wt_R b_R$, generated from the composite sector.}}
\label{fig:C7IR_dia}
\end{figure}

The infrared contribution to the process $b\rightarrow s\gamma$ is a one loop contribution from
the $W$ boson accompanied by top quarks, where a mass insertion in the intermediate top quark states is allowed by the presence of a 
$(V+A)$ interaction between the $W$ boson and the top and bottom quarks (Fig. \ref{fig:C7IR_dia}).
This interaction originates from a term:
\begin{equation}
\mathcal{L}\supset \mathcal{C}_{R}\mathcal{O}_{R}\ ,
\label{wright_term}
\end{equation} 
where $\mathcal{O}_{R}$ is the dimension-6 operator:
\begin{equation}
\mathcal{O}_{R}\equiv H^{c\dag}iD_{\mu}H\bar{t}_{R}\gamma^{\mu}b_{R}+h.c.\ .
\label{oWR}
\end{equation}  
At low energy, after the EWSB, the interaction in (\ref{wright_term}) gives:
\begin{equation}
\mathcal{L}\supset \frac{\mathcal{C}_{R}v^{2}}{2}\frac{g_{2}}{\sqrt{2}}\bar{b}_{R}\gamma^{\mu}t_{R}W^{-}_{\mu}\ .
\end{equation} 
\noindent
This interaction gives a contribution to the Wilson coefficient $C_{7}$ in the eq. (\ref{Heff}). We find:
\begin{equation}
C^{CH-IR}_{7}(\mu_{w})= \frac{\mathcal{C}_{R}v^{2}}{2}\frac{m_{t}}{m_{b}}f_{RH}(x_{t})
\label{CchIR}
\end{equation} 
where $x_{t}=\frac{m^{2}_{t}}{M^{2}_{W}}$ and $f_{RH}(x_{t})$ is the loop function \cite{ARH}:

\begin{align}
\begin{split}
 f_{RH}(x_{t})=\ & -\frac{1}{2}\left\{ \frac{1}{\left(1-x_{t}\right)^{3}}\frac{2}{3}\left[ -\frac{x^{3}_{t}}{2}-\frac{3}{2}x_{t}+2+3x_{t}\log(x_{t})\right]\right.\\
& + \frac{1}{\left(1-x_{t}\right)^{3}} \left[ -\frac{x^{3}_{t}}{2}+6x^{2}_{t}-\frac{15}{2}x_{t}+2-3x^{2}_{t}\log(x_{t})\right] \Bigg\} \ .
 \end{split}
\end{align}
\noindent
$f_{RH}=-0.777$, for $m_t=174$ GeV and $M_W=80.4$ GeV.\\

\begin{figure}
\centering
\scalebox{0.7}{\includegraphics{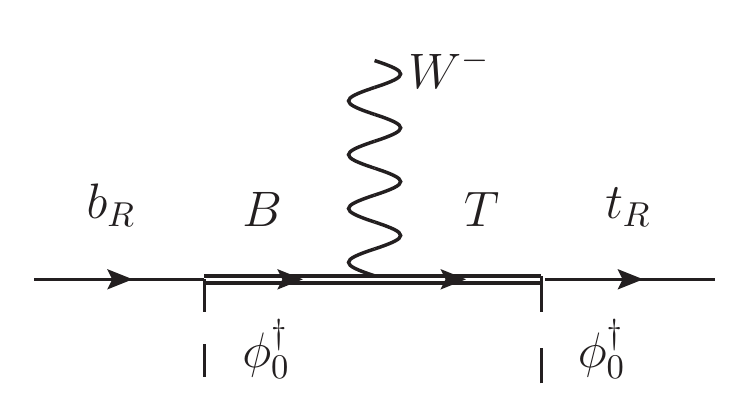}}
\caption{\textit{The CHM contribution to the effective coupling $Wt_R b_R$ (At order $\left( \frac{Y_* v}{\sqrt{2}m_*}\right)^2 $).}}
\label{wtRbR_diagram}
\end{figure}

\noindent
We point out that the bound on the CHM contributions to $b \to s\gamma$, $C^{CH}_{7}$ in eq. (\ref{c7limIR}), 
can be directly translated into a bound on the effective vertex $Wt_R b_R$, $v_R\equiv\frac{\mathcal{C}_{R}v^{2}}{2}$. 
By considering the bound in (\ref{c7limIR}) and the relation in (\ref{CchIR}), we obtain:
\begin{equation}
-0.0004<v_R<0.0013\ .
\label{bound_vR}
\end{equation} 

This bound from $b\to s \gamma$ can be compared with that from the measurement of the $Wtb$ anomalous couplings at colliders. 
Ref. \cite{saavedra_vR} reports an expected bound of $-0.012<v_R<0.024$, that can be imposed by 14 TeV LHC measurements with $30$ fb$^{-1}$. 
This latter can be obtained from studies on cross sections and top decay observables (angular distributions and asymmetries) 
in the single top production at the LHC. 
Present searches for anomalous $W$ couplings at the 7 TeV LHC \cite{LHC_vR} fix still mild bounds on $v_R$, $-0.34<v_R<0.39$, with $0.70$ fb$^{-1}$.
We can see that the bound obtained from $b\to s \gamma$ is much stronger than that from the $v_R$ measurement at collider.\\

\noindent
The CHM contribution to the effective coupling $Wt_R b_R$ is given 
by the exchange of heavy fermions that mix electro-weakly with $t_R$ and $b_R$ (fig. \ref{wtRbR_diagram}). 
At the order $x^2 $, only the $SU(2)_L$ heavy doublets, $D$, contribute to $\mathcal{C}_{R}$. 
This latter can be easily estimated by NDA \cite{NDA}:
\begin{equation}
 \mathcal{C}_{R}\sim \frac{ Y^2_{*} \xi_{bR}\xi_{tR}\xi^2_D}{M^2_D}\sim\frac{y_{b}y_{t}}{M^{2}_{D}}\frac{\xi^2_D}{\xi^2_{qL}} \ ,
\label{CR_est}
\end{equation} 
where $\xi_D$ denotes the degree of compositeness of the heavy doublet and $\xi_{\psi}$ that of the SM fermion $\psi$; 
in particular $\xi_{qL}$ is the degree of compositeness of $q^3_L=(t_L,b_L)$.\\
(\ref{CR_est}) implies:
\begin{equation}
C^{CH-IR}_{7}(\mu_{w})\sim\frac{m^{2}_{t}}{M^{2}_{D}}f_{RH}(x_{t}) \frac{\xi^2_D}{\xi^2_{qL}}\ .
\label{CchIR_est}
\end{equation}
Applying the condition in (\ref{c7limIR}) to this infrared contribution, we get the estimated bound:
\begin{equation}
M_D \gtrsim \frac{1.6 (0.87)}{\xi_{qL}} \text{TeV} \ ,
\label{C7bound_est}
\end{equation}
where the first number and the second number in parenthesis refer respectively to the case 
of a positive and of a negative $C^{CH-IR}_{7}$ contribution. 
Notice that in the case of a positive $C^{CH-IR}_{7}$ contribution we obtain a stronger bound on $M_D$, 
since the constraint in (\ref{c7limIR}) is asymmetric.\\
\noindent
We find that a subgroup of the custodial symmetry $SU(2)_V$, the $P_C$ parity, can give a suppression to the $Wt_R b_R$ coupling
and, as a consequence, to the CHM infrared contribution to $b\to s \gamma$.
The estimates we have just reported refer to generic composite Higgs models where there is not such $P_C$ protection.

\subsection{Protection by $P_C$ parity}\label{PCprotect}
The $P_C$ protection against the generation of the $Wt_R b_R$ vertex acts similarly to the $P_{LR}$ and $P_C$ protection against large corrections
to the $Zb_L b_L$ coupling, which we have discussed in sec. \ref{PLR}. 
$P_C$ is a symmetry of the sector BSM, that is respected also by the interactions of $t_R$ and $b_R$ if these latter are $P_C$ eigenstates. 
Since $P_C$ acts as $diag(1,-1,-1)$ on $SO(3)$ vectors, the $W$ is not a $P_C$ eigenstate (the composite partners of $W^1$ and $W^2$ have not the same $P_C$ eigenvalue). 
In the case in which $t_R$ and $b_R$ are both $P_C$ eigenstates,
both the $t_R$ and the $b_R$ interactions have to respect the $P_C$ parity. Then, the $Wt_R b_R$ vertex, which is $P_C$ violating,
since the $W$ is not a $P_C$ eigenstate, can arise only by paying for an additional factor, that gives a suppression. 
Whereas, in models where $t_R$ and $b_R$ are not both $P_C$ eigenstates and, as such, 
their interactions have not to respect the $P_C$ parity, the $Wt_R b_R$ vertex can be generated without suppressions.\\
 The TS5 falls into the class of models with $P_C$ protection, since in the TS5 both $t_R$ and $b_R$ are $P_C$ eigenstates. 
Considering the TS5, we can evaluate the suppression factor to $Wt_R b_R$ due to the $P_C$ protection. 
We can find it in an easy way by promoting $\Delta_{L1}$ and $\Delta_{L2}$ to spurions, 
which enforce a $SU(2)_L\times SU(2)_R$ invariance:
\[
 -\Delta_{L1}\bar{q}^3_L \left( T, B\right) \to \ -\bar{q}^3_L \mathcal{Q}_{2/3}\hat{\Delta}_{L1}
\]
\[
 -\Delta_{L2}\bar{q}^3_L \left( T', B'\right) \to \ -\bar{q}^3_L \mathcal{Q'}_{-1/3}\hat{\Delta}_{L2} \ ,
\]
where $\hat{\Delta}_{L1}=(\Delta_{L1},0)\equiv (1,2)_{1/6}$ and $\hat{\Delta}_{L2}=(0,\Delta_{L2})\equiv (1,2)_{1/6}$.
Therefore, we can write the $\mathcal{O}_R$ operator in the $SU(2)_L\times SU(2)_R$ invariant way:
\begin{equation}
 \mathcal{O}_R = \frac{1}{f^2}\bar{q}^3_R \hat{\Delta}_{L1}V_{\mu}\hat{\Delta}^{\dag}_{L2}q_R \gamma^{\mu} + h.c. \ ,
\label{OR_inv}
\end{equation}
where $f$ has the dimension of a mass, $q_R=(t_R, b_R)\equiv (1,2)_{1/6}$ 
and $V_{\mu}\equiv H^{c\dag}iD_{\mu}H$. 
Since $P_C$ is a subgroup of the custodial $SU(2)_V$, the $SU(2)\times SU(2)$ invariant operator in (\ref{OR_inv}) is also a $P_C$ invariant. 
We can notice that the $P_C$ invariance has brought to an additional factor $\frac{\Delta_{L1}\Delta_{L2}}{f^2}$ compared to (\ref{oWR}). \\
Without $P_C$ protection, the $(T,B)$ contribution to the $Wt_R b_R$ effective vertex is
\[
 s_R s_{bR} c^2_1 \left(\frac{Y_* v}{\sqrt{2} M_T} \right)^2=\frac{m_b m_t}{M^2_T}\frac{c^2_1}{s^2_1}\ ;
\]
the request for $P_C$ invariance brings to the additional factor $\frac{\Delta_{L1}\Delta_{L2}}{f^2}$. For $f^2=M_{Q*}M_{Q'*}$, we obtain
\[
  \left(\frac{Y_* v}{\sqrt{2} M_T} \right)^2 s_R s_{bR}\frac{c_1\Delta_{L1}}{M_{Q*}}\frac{c_1\Delta_{L2}}{M_{Q'*}}=\left(\frac{Y_* v}{\sqrt{2} M_T} \right)^2 s_R s_{bR}s_1 s_2=\frac{m_b m_t}{M^2_T}\ ,
\]
  that is a suppression by a factor $s^2_1/c^2_1$ in the TS5, i.e. by a factor $\frac{\xi^2_{qL}}{\xi^2_D}$.\\
\noindent

We can thus return to the estimated bounds on $M_D$ from $C^{CH-IR}_7$, and consider the case in which there is a $P_C$ protection 
to the $t_R$ and $b_R$ interactions. In such case the $\mathcal{C}_{R}$ contribution becomes:

\begin{equation}
 \mathcal{C}_{R} \sim\frac{y_{b}y_{t}}{M^{2}_{D}}  \ \ (\text{with}\ P_C) \ ,
\label{CR_est_pc}
\end{equation} 
which implies
\begin{equation}
C^{CH-IR}_{7}(\mu_{w})\sim\frac{m^{2}_{t}}{M^{2}_{D}}f_{RH}(x_{t}) \ \ (\text{with}\ P_C) \ 
\label{CchIR_est_pc}
\end{equation}
and an estimated bound:
\begin{equation}
M_D \gtrsim 1.6 (0.87) \text{TeV} \ (\text{with}\ P_C) \ .
\label{C7bound_est_pc}
\end{equation}

We will now calculate the bounds on $M_D$ from $C^{CH-IR}_7$ in the specific TS5 and TS10 models. 
As we already know, the TS5 belongs to the class of models with $P_C$ protection. 
The TS10, instead, falls in the class of models without $P_C$ protection, because in the TS10 $b_R$ is not a $P_C$ eigenstate. 
We thus expect that the bound in the TS10 will receive an enhancement factor $c_1/s_1$, compared to that in the TS5.\\
 
\noindent
In the TS5 we have a contribution to the $\mathcal{O}_R$ operator in (\ref{oWR}) both from the doublet $(T,B)$ in the $X=2/3$ representation and the doublet $(T',B')$ in the $X=-1/3$. 
We find (see App. \ref{wtrbr_calculation_TS5} and \ref{wtrbr_calculation_TS10} for details on calculation):
\begin{equation}
 \mathcal{C}^{TS5}_{R}=-\frac{y_{b}y_{t}}{M^{2}_{T}}\left( 1+\frac{M^{2}_{T}}{M^{2}_{T'}}\right) \ .
\end{equation} 
This implies:
\begin{equation}
C^{CH-IR-TS5}_{7}(\mu_{w})=-\frac{m^{2}_{t}}{M^{2}_{T}}f_{RH}(x_{t})\left( 1+\frac{M^{2}_{T}}{M^{2}_{T'}}\right) \ .
\label{CchIR-ts5}
\end{equation}
\noindent
Notice that the $\mathcal{C}^{TS5}_{R}$ contribution is negative. 
This implies a positive contribution $C^{CH-IR-TS5}_{7}$ ($f_{RH}$ is negative). 
The condition in (\ref{c7limIR}) is asymmetric and is stronger in the case of a positive $C^{CH-IR}_7$. 
Applying this condition to the Infrared contribution in (\ref{CchIR-ts5}), we get, for $r=\frac{M_{T}}{M_{T'}}=1$, 
the following bound on the $D\equiv (T,B)$ doublet mass:
\begin{equation}
 M^{TS5}_{D}\gtrsim 1.4\ TeV\ .
\label{boundIR-TS5}
\end{equation} \\

\begin{figure}[h!]
\title{\textbf{TS5 IR}}
\mbox{
{\includegraphics[width=0.7\textwidth]{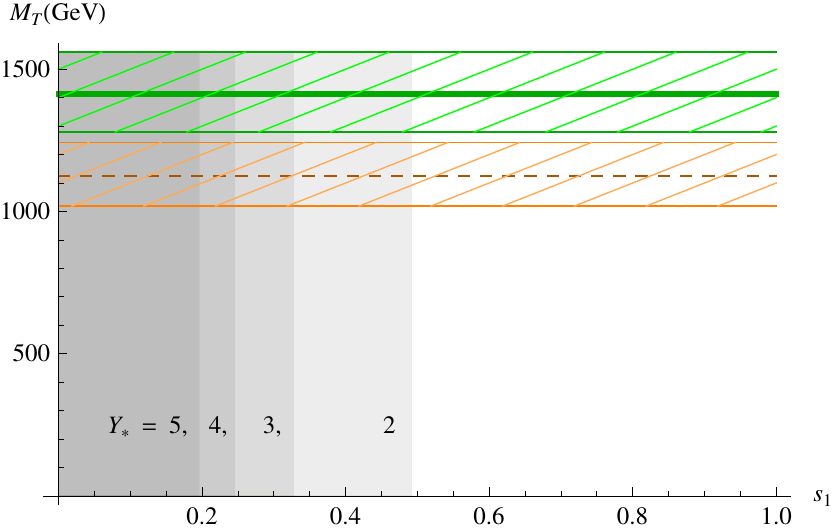}}
}
\caption{\textit{Bounds from $C^{CH-IR}_7$ (green thick line) and $C^{'CH-IR}_7$ (dashed orange line) in the TS5. 
Sketched bands define a range of variation for $M_T$ of $\pm 20\% M_{T'}$ ($r=M_T/M_{T'}\in [0.8,1.2]$). 
We also show, for different $Y_*$ values, the (grey) regions of exclusion for $s_1$, which are obtained from the condition
$s_R=\frac{\sqrt{2} m_t}{Y_* v s_1 }\leq 1$.}}
\label{fig:IR-TS5}
\end{figure}

\begin{figure}[h!]
\title{\textbf{TS10 IR}}
\includegraphics[width=0.85\textwidth]{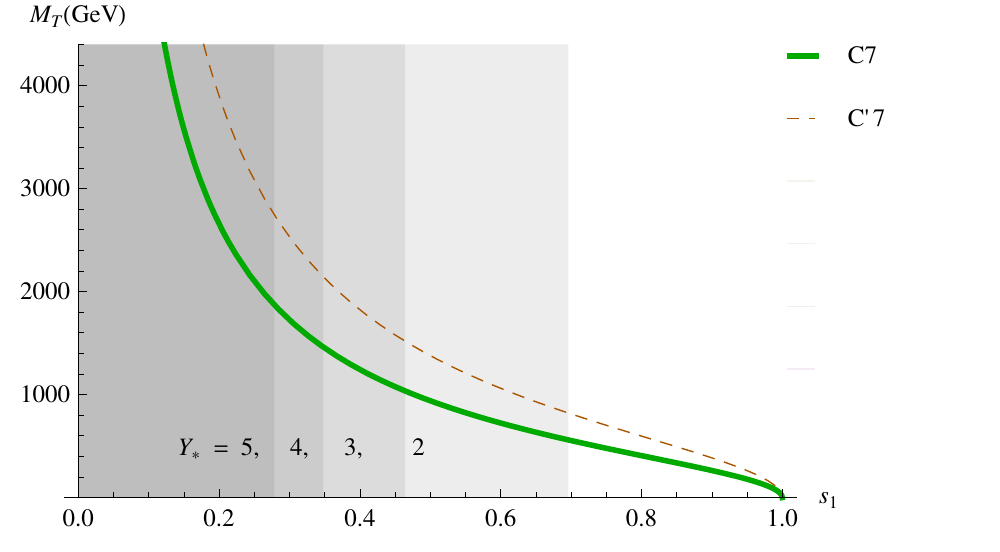}
\caption{\textit{Bounds from $C^{CH-IR}_7$ (green thick line) and $C^{'CH-IR}_7$ (dashed orange line) in the TS10. 
We also show, for different $Y_*$ values, the (grey) regions of exclusion for $s_1$, which are obtained from the condition
$s_R=\frac{2 m_t }{Y_* v s_1 }\leq 1$.}}
\label{fig:IR-TS10}
\end{figure}

\noindent
In the TS10, there is only one doublet, $D=(T,B)$, that gives a contribution to $\mathcal{C}_{R}$. 
We obtain
\begin{equation}
 \mathcal{C}^{TS10}_{R}=\frac{y_{b}y_{t}}{M^{2}_{T}}\frac{c^{2}_{1}}{s^{2}_{1}}\ ,
\end{equation}
\noindent
which implies:
\begin{equation}
C^{CH-IR-TS10}_{7}(\mu_{w})=\frac{m^{2}_{t}}{M^{2}_{T}}f_{RH}(x_{t})\frac{c^{2}_{1}}{s^{2}_{1}} \ .
\label{CchIR-ts10}
\end{equation}
\noindent
From the condition in (\ref{c7limIR}) we get finally the bound:
\begin{equation}
 M^{TS10}_{D}\gtrsim (0.54)\frac{c_1}{s_1}\ \text{TeV}\ .
\label{boundIR-TS10}
\end{equation} 
Notice that, differently from the case of the TS5 contribution, $C^{CH-IR-TS10}_{7}(\mu_{w})$ is negative. 
As such, it is constrained less strongly by the condition in (\ref{c7limIR}). 
As expected, we have found a $c_1/s_1$ enhancement of this bound, compared to (\ref{boundIR-TS5}).\\
In Fig.s \ref{fig:IR-TS5} and \ref{fig:IR-TS10} we show the value of the bound on $M_{D}$ from the IR $C_7$ contribution (thick curves) 
as function of $s_{1}$, for the TS5 (varying the $\frac{M_{T}}{M_{T'}}$ ratio in the range [0.8,1.2]) and the TS10 respectively. 
We also show, for different $Y_*$ values, the (grey) regions of exclusion for $s_1$, which are obtained from the conditions
$s_R=\frac{\sqrt{2} m_t}{Y_* v s_1 } \leq 1$ in the TS5 and $s_R=\frac{2 m_t}{Y_* v s_1 } \leq 1$ in the TS10. \\

We point out that the constraint for $s_1 \to 1$ in the TS10 is not a robust prediction but an artifact of our low-energy approximation, 
in which we have retained just the lowest-lying set of composite states. 
This is because, for $s_1 \to 1$, the couplings with Higgs of the two heavy fermions in the one doublet $D=(T,B)$ vanish and bring to a zero $C^{CH-IR-TS10}_{7}$ contribution. 
However, if we considered a second tower of heavy fermion resonances (as in a `three-site' model), we would obtain a non zero bound. 
In the TS5, for example, we have not this fake effect since we have a contribution both
from composite fermions in the $X=2/3$ representation and from those of the $X=-1/3$. \\
\noindent
We now proceed to evaluate the bounds from the $C^{'}_7$ contribution and then those from the UV contributions. 
As we already pointed out, these are contributions that involve flavor-violating operators and 
require assumptions on the flavor structure of the NP sector.
In what follows we will consider the case of flavor anarchy of the composite Yukawa matrices. 
This scenario, we remind, assumes that there is no
large hierarchy between elements within each matrix $Y_*$ and the quark mass hierarchy is completely explained by the elementary/composite mixing angles. 
We also set, for simplicity, $Y_{*U}=Y_{*D}=Y_{*}$.

\subsection{Generational mixing}\label{generation_mix}
After the EWSB, the mass eigenstate basis is obtained, as in the SM, using unitary transformations: 
$(D_{L}, D_{R})$ and $(U_{L}, U_{R})$ for down and up-type quark respectively. 
We will assume that the left rotation matrix has entries of the same order as
those of the Cabibbo-Kobayashi-Maskawa matrix:
\begin{equation}
 (D_L)_{ij}\sim (V_{CKM})_{ij}\ .
\label{VCKM}
\end{equation}
The assumption of anarchical $Y_*$ fixes the form of the rotation matrix $D_R$ to be:
\begin{equation}
 (D_R)_{ij}\sim \left( \frac{m_i}{m_j}\right)\frac{1}{(D_L)_{ij}}\ \ \text{for}\  i<j\ .
\label{DR}
\end{equation}
\noindent
Considering the estimates (\ref{VCKM}) and (\ref{DR}), 
we can evaluate the generational mixing factors in the composite Higgs model contributions to $C_{7}$ (UV) and $C^{'}_{7}$.\\
For the ultraviolet contribution to $C^{'}_{7}$, we account for the presence of a mass insertion 
that can generate the operator $\bar{b}_{L}\sigma^{\mu\nu}F_{\mu\nu}s_{R}$. 
This mass insertion brings to a factor $m_{b}(D_{R})_{23}\sim\frac{m_{s}}{(D_{L})_{23}}\sim \frac{m_{s}}{V_{ts}}$; 
where we have used first the estimate in (\ref{DR}) and then that in (\ref{VCKM}). 
The ultraviolet contribution to $C_{7}$ involves, instead, the operator $\bar{b}_{R}\sigma^{\mu\nu}F_{\mu\nu}s_{L}$ and we obtain, 
from the mass insertion, a 
generational mixing factor of $m_b (D_L)_{23}\sim m_b V_{ts}$; where the last similitude follows from the assumption in (\ref{VCKM}).\\
Evaluating, similarly, the generational mixing factor for the vertex $W t_R s_R$ in $C^{'CH-IR}_7$, one finds: 
$(D_R)_{23}\sim\frac{m_s}{m_b (D_L)_{23}}\sim \frac{m_s}{m_b V_{ts}} $, making use, again, of the estimates (\ref{DR}) and (\ref{VCKM}). 
The flavor violation in $C^{CH-IR}_7$, instead, comes entirely from the SM vertex $W t_L s_L$ and it is accounted by a factor $V_{ts}$. 
Therefore, we find that the composite Higgs model contribution to the Wilson coefficient $C^{'}_{7}$ is enhanced by a factor
\begin{equation}
 \frac{m_{s}}{m_{b}V^{2}_{ts}}\sim 8
\label{gen_mix}
\end{equation} 
compared to the contribution to $C_{7}$ both in the ultraviolet and in the infrared case.

\subsection{Infrared contribution to $C'_7$}

\begin{figure}
\centering
\scalebox{0.7}{\includegraphics{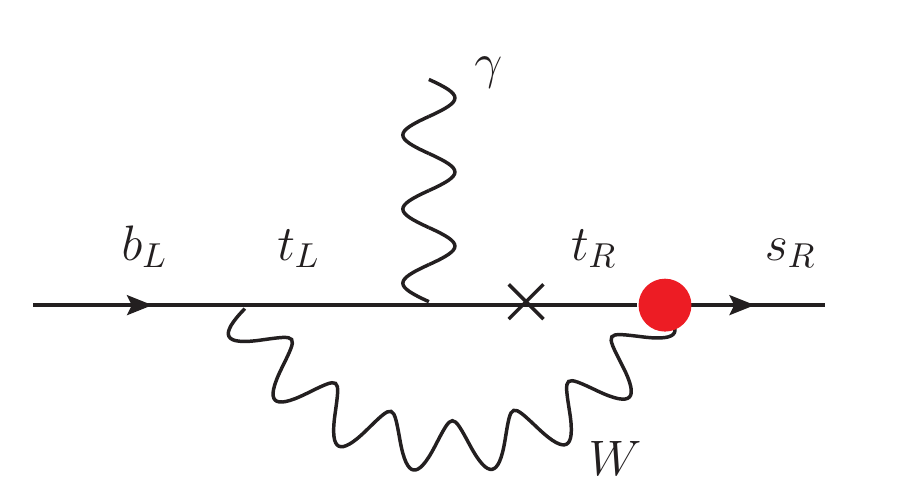}}
\caption{\textit{1 loop Infrared contribution to $C'_7$.}}
\label{fig:C7pIR_dia}
\end{figure}

Taking into account the generational mixing factor in (\ref{gen_mix}), 
the composite Higgs model contribution to the Wilson coefficient $C^{'}_{7}$ (Fig. \ref{fig:C7pIR_dia}) is given by:
\begin{equation}
 C^{'CH-IR}_{7}(\mu_{w})= \frac{\mathcal{C}_{R}v^{2}}{2}\frac{m_{s}}{m_{b}V^{2}_{ts}}\frac{m_{t}}{m_{b}}f_{RH}(x_{t})\ .
\end{equation} 
Considering the estimates for $\mathcal{C}_{R}$ in (\ref{CR_est}) and (\ref{CR_est_pc}), the condition on $C^{'CH-IR}_{7}(\mu_{w})$, eq. (\ref{C'7limIR}), 
gives thus the estimated bounds: 
\begin{equation}
 M_{D}\gtrsim 1.3\ \text{TeV}
\end{equation}
in models with $P_C$ symmetry; and
\begin{equation}
 M_{D}\gtrsim \frac{1.3}{\xi_{qL}} \text{TeV}
\end{equation}
in models without $P_C$ symmetry.\\
\noindent
Considering the specific TS5 and TS10 models, $C^{'CH-IR}_7$ gives the bounds:
\begin{equation}
 M_{D}\gtrsim 1.1\ \text{TeV}
\label{boundC7p_ts5}
\end{equation} 
in the TS5; and
\begin{equation}
 M_{D}\gtrsim\frac{c_{1}}{s_{1}}(0.80)\ \text{TeV}\ 
\label{boundC7p_ts10}
\end{equation} 
in the TS10.


Summarizing, we have found that the composite Higgs Model Infrared contribution to the process $b\to s\gamma$ 
(more specifically, the contribution to the $\mathcal{O}_7$ operator) gives a robust Minimal Flavor Violating bound on the 
mass of the bidoublet heavy fermions. 
We have found that this bound is quite strong, especially in the case of models where there is not a custodial $P_C$ symmetry 
protection to the effective coupling $Wt_R b_R$. 
We have found an estimated bound, $M_{D}\gtrsim 1.6$ TeV, in models with $P_C$ protection to the $Wt_R b_R$ vertex (where both $t_R$ and $b_R$ are $P_C$ eigenstates) 
and a bound, $M_{D}\gtrsim 1.6/\xi_{qL}$ TeV, where $\xi_{qL}$ denotes the degree of compositeness of $q^3_L=(t_L,b_L)$,
 in models without $P_C$ protection. $\xi_{qL}$ is naturally a small number, therefore the bound could be very strong in these types of models.\\ 
We can finally discuss how the bound on heavy masses can change in the case of a fully composite $t_L / t_R$:\\
 in the TS5 the bound on physical masses does not depend on the top degree of compositeness (this remains
almost true considering the full numerical calculation) and we obtain strong MFV bounds both for composite $t_L$ and composite $t_R$. 
In the TS10, because of the $P_C$ protection, we obtain very strong bound in the case of fully/almost fully composite $t_R$. 
In ref. \cite{pomarol_serra} it is found that corrections to $S$ and $T$ parameters give only weak constraints on a composite $t_R$ (both in TS5 and in TS10). 
The IR contribution to $b\to s \gamma$, instead, strongly constraint, especially in the TS10, this case.\\

One can finally discuss the validity of our results, which have been obtained `analytically' 
(i.e. by considering an expansion in $x\equiv \frac{Y_{*}v}{\sqrt{2}m_{*}}$ and retaining only the $O(x)$ terms).  
We find that the results from the numerical calculation of the bounds, obtained by diagonalizing numerically the fermionic mass matrices (see App. \ref{numeric_bound}),
do not differ more than O(1) from those we have shown, which are obtained at order $x$, in the assumption $x\ll 1$.
This can be also found by considering that the
exchange of relatively light composite fermions, that can give a contribution $\frac{Y_{*}v}{\sqrt{2}m^{CUST}_{*}} > 1$ to the effective $Wt_R b_R$ vertex, 
has to be followed by the exchange of heavier composite fermions, that reduces the overall contribution.
These relatively light composite fermions are the custodians, that become lighter than the other fermionic resonances in the limit of composite
$t_L$/$t_R$. 
By definition, however, these particles do not directly couple to SM fermions, therefore their contribution to $Wt_R b_R$ 
have to be always accompanied by the exchange of heavier composite particles. 

\subsection{Ultraviolet contribution} \label{UV_mchm5}

\begin{figure}
\centering
 \includegraphics[width=0.5\textwidth]{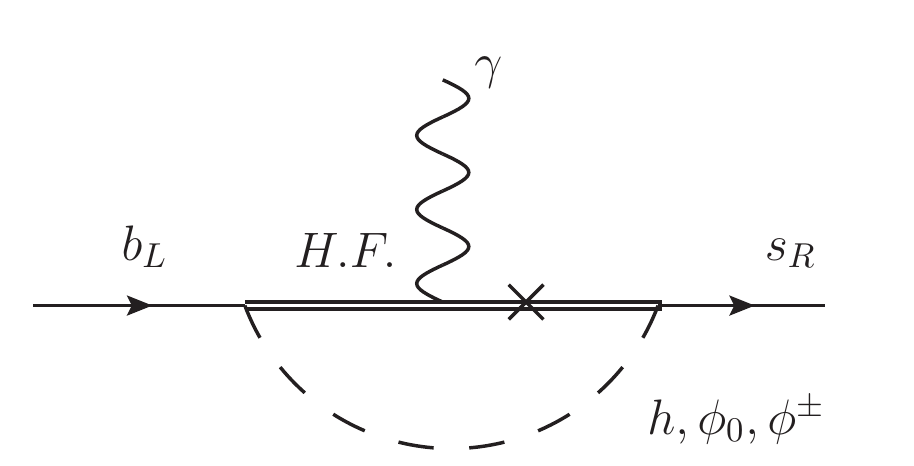}
\caption{\textit{1 loop CHM UV contribution to $C^{'}_7$.}}
 \label{UVfig}
\end{figure}

In this case the $P_{C}$ parity does not influence the bounds and we get contributions of the same size in the different models.
The leading contribution comes from diagrams with heavy fermions and would-be Goldstone Bosons in the loop\footnote{
The contribution from heavy gluon and heavy fermion exchange is suppressed. 
Indeed this contribution is approximately diagonal in the flavor space.} (Fig. \ref{UVfig}). 

\begin{equation}
C^{CH-UV}_7, C^{'CH-UV}_7 \propto s_{Li}Y_{*ik} Y_{*kl}Y_{*lj}s_{Rj}
 \label{c7UV}
\end{equation} 
The contribution (\ref{c7UV}) is not aligned with the mass matrix $m_{dij}\sim s_{Li} Y_{*ij} s_{Rj}$, therefore,
after the EWSB it remains non diagonal in the flavor space. \\

Before going on the specific TS5 and TS10 models, 
we can obtain estimated bounds from the UV contributions in generic composite Higgs models, by means of NDA.
We obtain:
\begin{equation}
 C^{CH-UV}_7 \sim \frac{\left(Y_* v \right)^2 }{M_{D}M_{\tilde{D}}}\xi_{D}\xi_{\tilde{D}} \ ,
\end{equation}
where $\tilde{D}$ denotes a heavy fermion which is a $SU(2)_L$ singlet, and 
\begin{equation}
 C^{'CH-UV}_7 \sim \frac{m_s}{m_b V^2_{ts}}\frac{\left(Y_* v \right)^2 }{M_{D}M_{\tilde{D}}}\xi_{D}\xi_{\tilde{D}} \ ,
\end{equation}
where we have taken into account the generational mixing factor in \ref{gen_mix}. 
Comparing these results with those from the IR contributions in (\ref{C7bound_est}, \ref{C7bound_est_pc}), 
we see that the UV contribution gives approximately a bound $Y_{*}/y_t$ ($\frac{Y_{*}}{y_t}\xi_{q_L}$, in the case of models without $P_C$ protection)
times stronger than the one from the IR contribution to $C_7$. 
Such UV bounds, however, are not as robust as the IR one, since they require, as we already pointed out, assumptions on the flavor structure of the sector BSM. 
In particular, we have estimated them in the scenario of flavor anarchy in the strong sector.\\

\begin{figure}[h!]
\title{\textbf{UV contribution in the TS5 ($Y_* = 3$)}}
\includegraphics[width=0.9\textwidth]{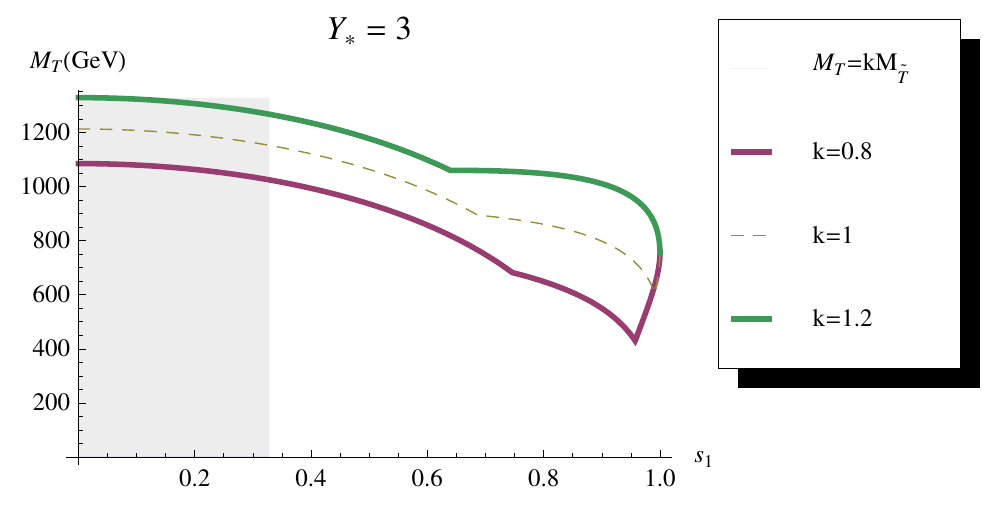}
\caption{\textit{Bounds from $C^{'CH-UV}_7$ in the TS5, for different values of $k=\frac{M_T}{M_{\tilde{T}}}$. We set $M_{\tilde{B}}=M_{\tilde{T}}$ and $M_{T'}=M_T$. 
Also shown is the exclusion region for $s_1$, obtained from the condition $s_R=\frac{\sqrt{2} m_t }{Y_* v s_1 }\leq 1$.}}
\label{fig:UV_s1_k}
\end{figure}

\begin{figure}[h!]
\title{\textbf{UV contribution in the TS5 ($k=1$)}}
\includegraphics[width=0.9\textwidth]{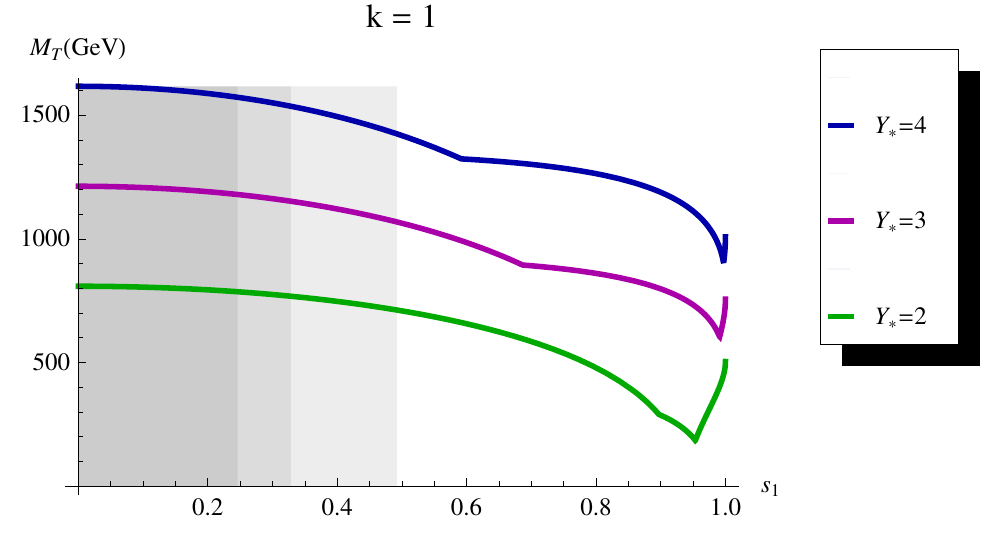}
\caption{\textit{Bounds from $C^{'CH-UV}_7$ in the TS5, for different values of $Y_*$. We set $k=\frac{M_T}{M_{\tilde{T}}}=1$, $M_{\tilde{B}}=M_{\tilde{T}}$ and $M_{T'}=M_T$.
We also show the $s_1$ exclusion regions, obtained from the condition $s_R=\frac{\sqrt{2} m_t }{Y_* v s_1 }\leq 1$. }}
\label{fig:UV_s1_Y}
\end{figure}

\noindent
In the Ref. \cite{Agashe_flav} the Ultraviolet Contribution to $b\rightarrow s\gamma$ in a Two-Site Model without a $P_{LR}$ protection to the 
$t_R$ and $b_R$ interactions is evaluated. In the following
 we will describe in details the contribution in the TS5 and we will report the results for TS10.
We can calculate the $C^{CH-UV}_{7}$ and $C^{'CH-UV}_{7}$ ultraviolet contributions by considering
the model independent analysis of Ref. \cite{Agashe_flav} and the generational mixing factor of the eq. (\ref{gen_mix}). We get the following effective Hamiltonian
\begin{equation}
 \mathcal{H}^{eff}_{neutral\ Higgs}=\frac{i\ e}{8\pi^{2}}\frac{(2\epsilon \cdot p)}{M^{2}_{w}}k_{neutral}\left[V_{ts}\bar{b}(1-\gamma_{5})s+\frac{m_{s}}{m_{b}V_{ts}}\bar{b}(1+\gamma_{5})s \right] 
\label{Heff_dipole}
\end{equation} 
where
\begin{gather}
\nonumber
 k_{neutral}\approx\sum^{4}_{i=1}\left(|\alpha^{(i)}_{1}|^{2}+|\alpha^{(i)}_{2}|^{2}\right)m_{b}\left(\frac{1}{36}\right)\frac{M^{2}_{w}}{m^{2}_{*(i)}} + \\ \nonumber
\sum^{4}_{i=1}\left(\alpha^{(i)*}_{1}\alpha^{(i)}_{2}\right)m_{*(i)}\left(\frac{1}{6}\right)\frac{M^{2}_{w}}{m^{2}_{*(i)}} \\ 
\label{kneutral1}
\end{gather} 
the index $i$ runs over the four down-type heavy fermions of the model, $\mathbf{d}^{(i)}=\tilde{B}, B', B_{-1/3}, B$, 
and the $\alpha^{(i)}_{1}$, $\alpha^{(i)}_{2}$ coefficients are defined by interaction:
\begin{equation}
 \mathcal{L}\supset \bar{\mathbf{d}}^{(i)}\left[\alpha^{(i)}_{1}(1+\gamma_{5})+\alpha^{(i)}_{2}(1-\gamma_{5})\right]bH + h.c.\ .
\end{equation} 
After the EWSB, we find the following coefficients at $O(x)$: 
\begin{gather}
\nonumber
 \alpha^{(\tilde{B})}_{1}=\frac{Y^2_{*}v}{2}s_{bR}\left[\frac{1}{M_{B'}}+\frac{M_{B'}+c_{bR}M_{\tilde{B}}}{M^2_{B'}-M^2_{\tilde{B}}} \right]\\ \nonumber
 \alpha^{(\tilde{B})}_{2}=-\frac{Y_{*}}{2\sqrt{2}}s_2 c_{bR}\\ \nonumber
 \alpha^{(B')}_{1}=\alpha^{(B_{-1/3})}_{1}=-\frac{Y_{*}}{2\sqrt{2}}s_{bR}\\ \nonumber
 \alpha^{(B')}_{2}=\alpha^{(B_{-1/3})}_{2}=-\frac{Y^2_{*}v}{4}s_{2}\left[ \frac{M^2_{B'}M_{\tilde{B}}-s^2_{bR}M^3_{\tilde{B}}-c_{bR}M^3_{B'}+2c_{bR}M_{B'}M^2_{\tilde{B}}}{M_{B'}M_{\tilde{B}}(M^2_{B'}-M^2_{\tilde{B}})}\right]  \\  
\label{alfa}
\end{gather} 
the heavy fermion $B$ gives a contribution of $O(s^{2}_{2})$ to $k_{neutral}$ and we neglect it.\\
Considering the eq. (\ref{kneutral1}) and the coefficients in (\ref{alfa}), neglecting again $O(x^{2})$ terms, we obtain:
\begin{equation}
  k_{neutral}\approx-m_b M^2_W Y^2_{*}\frac{1}{8}\left(\frac{c_{bR}}{M_{B'}M_{\tilde{B}}}-\frac{7}{18}\frac{s^{2}_{bR}}{M^2_{B'}}\right)   \ .
\label{kneutral}
\end{equation} 
From this expression of $k_{neutral}$ we obtain the following TS5 ultraviolet contributions to the Wilson coefficient of the effective Hamiltonian in (\ref{Heff}):

\begin{align}
\begin{split}
 & C^{CH-UV}_{7}(m_{*})=\frac{1}{16}\frac{\sqrt{2}}{G_{F}}Y^{2}_{*}\left(\frac{c_{bR}}{M_{B'}M_{\tilde{B}}}-\frac{7}{18}\frac{s^{2}_{bR}}{M^2_{B'}}\right)\ ;\\
& C^{'CH-UV}_{7}(m_{*})=\frac{1}{16}\frac{\sqrt{2}}{G_{F}}Y^{2}_{*}\left(\frac{c_{bR}}{M_{B'}M_{\tilde{B}}}-\frac{7}{18}\frac{s^{2}_{bR}}{M^2_{B'}}\right)\frac{m_{s}}{m_{b}V^{2}_{ts}} \ .
\end{split}
\end{align}
Assuming $s_{bR}$ small, the above formulas become:
\begin{equation}
 C^{CH-UV}_{7}(m_{*})=\frac{1}{16}\frac{\sqrt{2}}{G_{F}}\frac{Y^{2}_{*}}{M_{B'}M_{\tilde{B}}}\ ;\ 
C^{'CH-UV}_{7}(m_{*})=\frac{1}{16}\frac{\sqrt{2}}{G_{F}}\frac{Y^{2}_{*}}{M_{B'}M_{\tilde{B}}}\frac{m_{s}}{m_{b}V^{2}_{ts}} \ .
\end{equation}
\noindent
Finally, the condition on $C^{'CH-UV}_{7}$ in the eq. (\ref{C'7lim}) gives the bound:
\begin{equation}
 \sqrt{M_{B'}M_{\tilde{B}}}\gtrsim(0.40)\ Y_{*}\ \text{TeV}\ ;
\label{UVbound}
\end{equation} 
where, for simplicity we have set $s_{bR}=0$.
The condition (\ref{C7UVlim}) on $C^{CH-UV}_{7}$ gives a stronger bound, 
\begin{equation} 
\sqrt{M_{B'}M_{\tilde{B}}}\gtrsim(0.52)\ Y_{*} \text{TeV} \ ,
\end{equation}
 if $C^{CH-UV}_{7}(m_{*})$ is a negative contribution.\\ 
There is also a contribution to $b\rightarrow s\gamma$ from diagrams with heavy fermions and charged Higgs in the loop. 
Following a similar procedure as the one used before (\ref{kchargedApp}) we find, neglecting $O(x^{2})$ terms:
\begin{equation}
  k_{charged}\approx m_b M^2_W Y^2_*\frac{5}{48}\frac{1}{M_{B'}M_{\tilde{B}}}+O(s^2_1)+O(s^2_{bR}) \ .
\end{equation} 
If we can neglect $O(s^{2}_{1})$ and $O(s^{2}_{bR})$ terms, $k_{charged}$ gives a weaker bound than the one from $k_{neutral}$. 
The full expression of $k_{charged}$ can be found in App. \ref{AppUVcontrib}, here we have just reported, for simplicity, the result for small $s_1$ and $s_{bR}$ angles. \\
In Fig. \ref{fig:UV_s1_k} we show the bound on the doublet mass $M_{T}$ as function of $s_{1}$ from the condition on $C^{'CH-UV}_{7}$, 
for different values of the ratio $k=\frac{M_T}{M_{\tilde{T}}}$ between doublet and singlet mass and for a value $Y_{*}=3$. 
We set $M_{\tilde{B}}=M_{\tilde{T}}$ and $M_{T'}=M_T$. These values are obtained
taking into account the strongest values between the neutral Higgs contribution and the charged Higgs one. We set $s_{bR}=s_{1}$.\\
In Fig. \ref{fig:UV_s1_Y} we show the bound on the doublet mass $M_{T}$ from the condition on $C^{'CH-UV}_{7}$ as function of $s_{1}$, 
for different values of $Y_{*}$. We set $k=\frac{M_T}{M_{\tilde{T}}}=1$, $M_{\tilde{B}}=M_{\tilde{T}}$ and $M_{T'}=M_T$.

\subsubsection{Ultraviolet contribution in the TS10}
\begin{figure}[h!]
\title{\textb{UV contribution in the TS10 ($Y_* = 3$)}}
\includegraphics[width=0.9\textwidth]{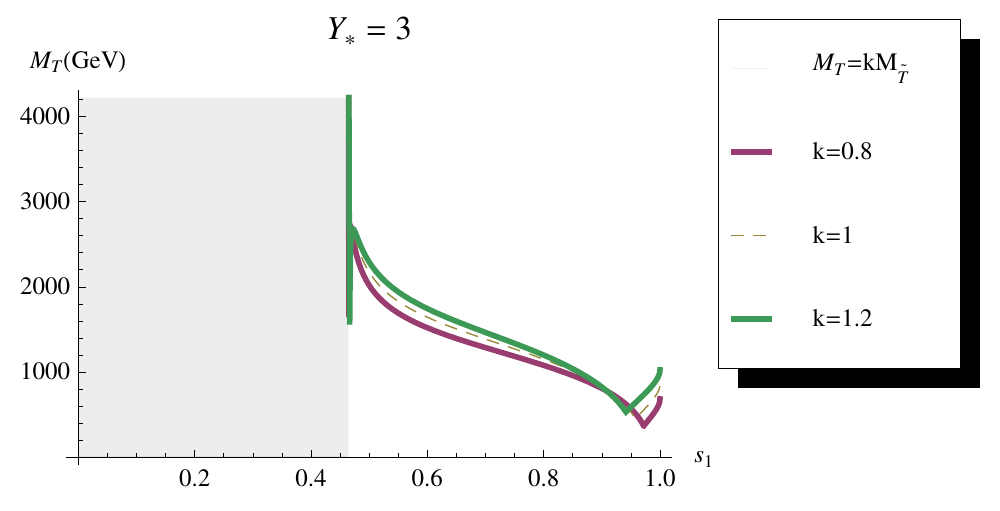}
\caption{\textit{Bounds from $C^{'CH-UV}_7$ in the TS10, for different values of $k=\frac{M_T}{M_{\tilde{T}}}$ ($M_{\tilde{B}}\simeq c_R M_{\tilde{T}}$, $M_{\tilde{B}'}=M_{\tilde{T}'}= c_R M_{\tilde{T}}$).
Also shown is the exclusion region for $s_1$, obtained from the condition $s_R=\frac{2 m_t }{Y_* v s_1 }\leq 1$.}}
\label{fig:UV_s1_k_ts10}
\end{figure}

\begin{figure}[h!]
\title{\textbf{UV contribution in the TS10 ($k = 1$)}}
\includegraphics[width=0.9\textwidth]{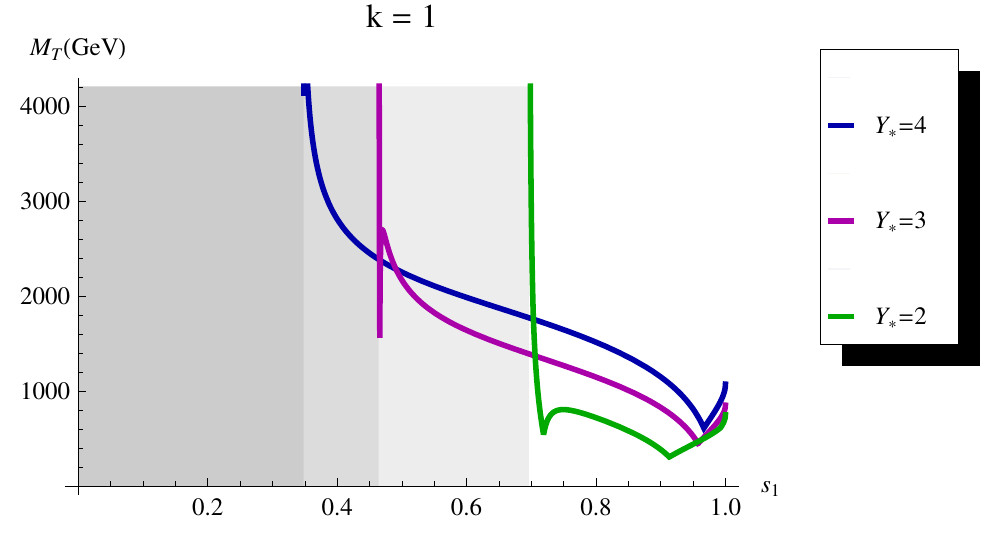}
\caption{\textit{Bounds from $C^{'CH-UV}_7$ in the TS10, for different values of $Y_*$. We set $k=\frac{M_T}{M_{\tilde{T}}}=1$.
We also show the $s_1$ exclusion regions, obtained from the condition $s_R=\frac{2 m_t }{Y_* v s_1 }\leq 1$.}}
\label{fig:UV_s1_Y_ts10}
\end{figure}

For the TS10 model, applying the same procedure as for the case of TS5, we get:
\begin{align}
\begin{split}\label{kneutral_ts10}
& k_{neutral} =  m_b M^2_W Y^2_*\\
&\times \frac{7 M_T M^2_{T'} s^2_1 - 18 M_{\tilde{B}} M^2_{\tilde{B}'} \sqrt{1-s^2_1}+ M^2_{\tilde{B}} \left( 7 M_B s^2_1 -18 M_{\tilde{B}'} \sqrt{1-s^2_1}\right)}{288 M^2_{\tilde{B}} M_B M^2_{\tilde{B}'}} +O(s_{bR})\\
 & = - m_b M^2_W Y^2_{*}\frac{1}{16}\left( \frac{1}{M_{B}M_{\tilde{B}}}+\frac{1}{M_{B}M_{\tilde{B}'}}\right) +O(s^2_1)+O(s_{bR})
\end{split}
\end{align}

\begin{equation}
 k_{charged}= m_b M^2_W Y^2_* \left(\frac{5}{48}\frac{1}{M_{B}M_{\tilde{B}}}+\frac{5}{48}\frac{1}{M_{B}M_{\tilde{B}'}}+ \frac{5}{96}\frac{s^2_R}{M^2_{B}}\right) +O(s^2_1)+O(s^2_{bR})
\label{kcharged_ts10}
\end{equation} 
If the left-handed bottom quark has a small degree of compositeness, we can neglect $O(s^{2}_{1})$ 
(while $s_{bR}$ is naturally very small in the TS10, in order to account for the ratio $m_b/m_t\ll 1$). 
The charged contribution, in this case, gives a stronger bound than the one from $k_{neutral}$: 
\begin{equation}
 \sqrt{M_{B}M_{\tilde{B}}}\gtrsim(0.58)\ Y_{*}\ TeV\ ,
\label{UVbound-ts10}
\end{equation} 
from the condition (\ref{C'7lim}) on $C^{'CH-UV}_{7}$. A stronger bound, 
\begin{equation}
\sqrt{M_{B}M_{\tilde{B}}}\gtrsim(0.75)\ Y_{*} \text{TeV} \ ,
\end{equation} 
comes from the condition (\ref{C7UVlim}) on $C^{CH-UV}_{7}$, if this last contribution has a negative sign.\\
\noindent
In Fig. \ref{fig:UV_s1_k_ts10} we show the bound on the doublet mass $M_{T}$ as function of $s_{1}$ from the condition on $C^{'CH-UV}_{7}$, 
for different values of the ratio $k=\frac{M_T}{M_{\tilde{T}}}$ between doublet and $\tilde{T}$ singlet mass and for a value $Y_{*}=3$. 
The custodian singlet masses have the following relations with $M_{\tilde{T}}$:
$M_{\tilde{B}}\simeq c_R M_{\tilde{T}}$, $M_{\tilde{B}'}=M_{\tilde{T}'}= c_R M_{\tilde{T}}$. \\
In Fig. \ref{fig:UV_s1_Y_ts10} we show the bound on the doublet mass $M_{T}$ as function of $s_{1}$ from the condition on $C^{'CH-UV}_{7}$, 
for different $Y_{*}$ values. We set $k=\frac{M_T}{M_{\tilde{T}}}=1$. \\
All these bounds are obtained taking into account the strongest values between the neutral Higgs contribution and the charged Higgs one.\\ 
We can see that in the TS10 model, the UV bounds are particularly strong in the case of fully composite $t_R$. This is an effect caused by the exchange of the custodians
$\tilde{T}'$, $\tilde{B}'$ and of the $\tilde{B}$, that are light in the limit of a composite $t_R$. 
In particular, when $t_R$ is fully composite ($s_R=1$), 
$M_{\tilde{B}}(\simeq c_R M_{\tilde{T}})$ and $M_{\tilde{B}'}=M_{\tilde{T}'}(= c_R M_{\tilde{T}})$ vanish. 
This causes the divergence of the bounds for $s_R \to 1$. 
Such divergences can be seen in the curves in Fig.s \ref{fig:UV_s1_k_ts10} and \ref{fig:UV_s1_Y_ts10}, 
when they approach the (grey) exclusion regions for $s_1$ (indeed, the minimum value of $s_1$ 
allowed by the condition $s_R=\frac{2 m_t }{Y_* v s_1 }\leq 1$ is obviously obtained in the case $s_R=1$).\\

\noindent
We summarize in Tab. \ref{tab:estimates_calc} the bounds on heavy fermion masses that can be obtained from the process $b\to s \gamma$. 
We report the estimated bounds for generic Composite Higgs Models (with or without $P_C$ protection) that we have found by means of NDA, 
and the bounds in the specific Two-Site models TS5 and TS10. 
We denote by $\xi_{\psi/\chi}$ the degree of compositeness of a SM/Heavy fermion, in a generic Composite Higgs Model. 
$D$ ($\tilde{D}$) denotes a $SU(2)_L$ heavy fermion doublet (singlet). 
For the estimated bounds from $C^{CH}_7$ and for the bounds from $C^{CH-UV}_7$, we report both the values that can be obtained in the case of a positive (the first number) or of a negative 
(the second number in parenthesis) contribution.  

\begin{table}[h!]
\rotatebox{90}{\scalebox{0.8}{\begin{tabular}{lcccc}
\hline
\multicolumn{1}{l}{$C^{CH-IR}_{7}(\mu_w)$} &\multicolumn{2}{|c}{$w/\ P_C$ \ \ $\sim\frac{\left(y_t v \right)^2 }{M^{2}_{D}}\xi^{2}_{D}$}&\multicolumn{2}{|c|}{$w/o\ P_C$ \ \ $\sim\frac{\left(y_t v \right)^2 }{M^{2}_{D}}\left( \frac{\xi_D}{\xi_{qL}}\right)^2 $}\\
\cline{2-5}
&\multicolumn{1}{|c}{ESTIMATED} &\multicolumn{1}{|c}{TS5}&\multicolumn{1}{|c}{ESTIMATED}&\multicolumn{1}{|c|}{TS10}\\
&\multicolumn{1}{|c}{$M_D \gtrsim 1.6 (0.87)\ \text{TeV}$}&\multicolumn{1}{|c}{$M_D \gtrsim 1.4\ \text{TeV}$}&\multicolumn{1}{|c}{$M_D \gtrsim \frac{1.6 (0.87)}{\xi_{qL}}$ TeV}&\multicolumn{1}{|c|}{$M_D \gtrsim \frac{0.54}{s_{1}}$ TeV}\\
\hline \hline
\multicolumn{1}{l}{$C^{'CH-IR}_{7}(\mu_w)$} &\multicolumn{2}{|c}{$w/\ P_C$ \ \ $\sim\frac{\left(y_t v \right)^2 }{M^{2}_{D}}\xi^{2}_{D}\frac{m_s}{m_b V^{2}_{ts}}$}&\multicolumn{2}{|c|}{$w/o\ P_C$ \ \ $\sim\frac{\left(y_t v \right)^2 }{M^{2}_{D}}\left( \frac{\xi_D}{\xi_{qL}}\right)^2\frac{m_s}{m_b V^{2}_{ts}} $}\\
\cline{2-5}
&\multicolumn{1}{|c}{ESTIMATED} &\multicolumn{1}{|c}{TS5}&\multicolumn{1}{|c}{ESTIMATED}&\multicolumn{1}{|c|}{TS10}\\
&\multicolumn{1}{|c}{$M_D \gtrsim 1.3$ TeV}&\multicolumn{1}{|c}{$M_D \gtrsim 1.1$ TeV}&\multicolumn{1}{|c}{$M_D \gtrsim \frac{1.3}{\xi_{qL}}$ TeV [$(1.3)Y_{*}$TeV]}&\multicolumn{1}{|c|}{$M_D \gtrsim \frac{0.80}{s_{1}}$ TeV [$(0.56)Y_{*}$TeV]}\\
\hline \hline
\multicolumn{1}{l}{$C^{CH-UV}_{7}(m_{*})$} &\multicolumn{3}{|c|}{$\sim\frac{\left(Y_* v \right)^2 }{M_{D}M_{\tilde{D}}}\xi_{D}\xi_{\tilde{D}}$}& \\
\cline{2-4}
&\multicolumn{1}{|c}{ESTIMATED} &\multicolumn{1}{|c}{TS5}&\multicolumn{1}{|c|}{TS10}\\
&\multicolumn{1}{|c}{$\sqrt{M_D M_{\tilde{D}}} \gtrsim 1.5 (0.79)Y_*$ TeV}&\multicolumn{1}{|c}{$\sqrt{M_D M_{\tilde{D}}} \gtrsim 0.52 (0.28)Y_*$ TeV}&\multicolumn{1}{|c|}{$\sqrt{M_D M_{\tilde{B}}} \gtrsim 0.75 (0.40)Y_*$ TeV}\\
\cline{1-4}
\multicolumn{1}{l}{$C^{'CH-UV}_{7}(m_*)$}&\multicolumn{3}{|c|}{$\sim\frac{\left(Y_* v \right)^2 }{M_{D}M_{\tilde{D}}}\xi_{D}\xi_{\tilde{D}}\frac{m_s}{m_b V^{2}_{ts}}$}&\\
\cline{2-4}
&\multicolumn{1}{|c}{ESTIMATED} &\multicolumn{1}{|c}{TS5}&\multicolumn{1}{|c|}{TS10}\\
&\multicolumn{1}{|c}{$\sqrt{M_D M_{\tilde{D}}} \gtrsim (1.1)Y_*$ TeV}&\multicolumn{1}{|c}{$\sqrt{M_D M_{\tilde{D}}} \gtrsim (0.40)Y_*$ TeV}&\multicolumn{1}{|c|}{$\sqrt{M_D M_{\tilde{B}}} \gtrsim (0.58)Y_*$ TeV}\\
\cline{1-4}
\end{tabular}}}
\caption{\textit{Estimated bounds for general Composite Higgs Models and for the specific TS5 and TS10 at small elementary/composite mixing angles $s_1$ and $s_{bR}$. For the estimated bounds from $C^{CH}_7$ and for the bounds from $C^{CH-UV}_7$,
 we report both the values that can be obtained in the case of a positive (the first number) or a negative 
(the second number in parenthesis) contribution.  
}}
\label{tab:estimates_calc}
\end{table}

\section{Constraint from $\epsilon^{'}/\epsilon_{K}$}
The bound on the mass of the heavy fermions that comes from the direct CP violating observable
of the $K^0\to 2\pi$ system, $Re(\epsilon '/\epsilon)$, can be even stronger, in the assumption of anarchic $Y_*$, 
than those obtained from $b\to s \gamma$, as already found in \cite{Isidori}. 
As we already pointed out, however, it is a bound that strongly depends on the assumptions made on the flavor structure of the new physics sector and,
 as a consequence, it is non minimal flavor violating.\\
As in the UV contribution to $b\rightarrow s\gamma$, the custodial symmetry does not influence the bound and we obtain contributions of the same size 
for the different models. In what follows we describe the bound in the TS5 and in the TS10.\\ 
\noindent
New Physics contribution can be parametrized at low energy by chromo-magnetic operators:
\begin{equation}
 \mathcal{O}_{G}=\bar{s}\sigma^{\mu\nu}T^{a}G^{a}_{\mu\nu}\left(1-\gamma_{5} \right)d \ \ , \ \  \mathcal{O'}_{G}=\bar{s}\sigma^{\mu\nu}T^{a}G^{a}_{\mu\nu}\left(1+\gamma_{5} \right)d\ .
\end{equation} 
As for the UV contribution to $b\rightarrow s\gamma$, the leading contribution to $\epsilon^{'}/\epsilon_{K}$ comes from diagrams with heavy fermions and Higgs in the loop, 
that generate the $\mathcal{O}_{G}$ and $\mathcal{O'}_{G}$ operators (1 loop diagrams are the same as for the UV contribution to $b\to s \gamma$, Fig. \ref{UVfig}, 
with the replacements $\gamma \to g$, $b \to s$ and $s \to d$). \\
The related coefficients $\mathcal{C}_{G}$ and $\mathcal{C'}_{G}$, in analogy with $\mathcal{C}_{7}$ and $\mathcal{C'}_{7}$ of the UV contribution to $b\rightarrow s\gamma$,
differ by a generational mixing factor that, in the assumption of anarchic $Y_{*}$, we estimate to be $\sim \frac{m_{d}}{m_{s}V^{2}_{us}}$. 
We consider only the generation mixing $(1-3)\times (2-3)$, via 3rd generation.\\
In analogy with (\ref{Heff_dipole}), we define:
\begin{equation}
 \mathcal{A}^{eff-chromo}_{neutral\ Higgs}=\frac{i\ g_{s}}{8\pi^{2}}\frac{(2\epsilon \cdot p)}{M^{2}_{w}}k^{G}_{neutral}\left[V_{us}\bar{s}(1-\gamma_{5})d+\frac{m_{d}}{m_{s}V_{us}}\bar{s}(1+\gamma_{5})d \right]\ , 
\label{Heff_chromo}
\end{equation} 
where
\begin{gather}
\nonumber
 k^{G}_{neutral}\approx\sum^{4}_{i=1}\left(|\alpha^{(i)}_{1}|^{2}+|\alpha^{(i)}_{2}|^{2}\right)m_{s}\left(-\frac{1}{12}\right)\frac{M^{2}_{w}}{m^{2}_{*(i)}} + \\ \nonumber
\sum^{4}_{i=1}\left(\alpha^{(i)*}_{1}\alpha^{(i)}_{2}\right)m_{*(i)}\left(-\frac{1}{2}\right)\frac{M^{2}_{w}}{m^{2}_{*(i)}} \\ 
\label{kGneutral1}
\end{gather} 
the index $i$ runs over the four down-type heavy fermions of the model, $\mathbf{d}^{(i)}$, 
and the $\alpha^{(i)}_{1}$, $\alpha^{(i)}_{2}$ coefficients are defined by interaction:
\begin{equation}
 \mathcal{L}\supset \bar{\mathbf{d}}^{(i)}\left[\alpha^{(i)}_{1}(1+\gamma_{5})+\alpha^{(i)}_{2}(1-\gamma_{5})\right]bH + h.c.\ .
\end{equation} 
\noindent
After the EWSB, neglecting $O(x^{2})$ terms, we find in the TS5:
\begin{equation}
  k^{G}_{neutral}= \frac{3}{8}m_{s}M^{2}_{w}\frac{Y^{2}_{*}}{M_{B'}M_{\tilde{B}}} + O(s^2_{sR})\ ,
\label{kGneutral2}
\end{equation} 
\noindent
where $s_{sR}$ defines the degree of compositeness of the right-handed strange quark and has naturally a small value. 
In the limit in which $s_{sR}=0$, we obtain the same result also in the TS10. \\
\noindent
We can therefore calculate the $\mathcal{C}_{G}$ and $\mathcal{C'}_{G}$ contributions:
\begin{equation}
 \mathcal{C}_{G}=-\frac{1}{16 \pi^2}\frac{k^{G}_{neutral}}{M^{2}_{w}m_{s}}V_{us} \ , \ \mathcal{C'}_{G}=\frac{m_{d}}{m_{s}V^{2}_{us}}\mathcal{C}_{G}\ .
\end{equation} 
\noindent
Defining
\begin{equation}
 \delta_{\epsilon^{'}}=\frac{Re(\epsilon^{'}/\epsilon)_{CH}-Re(\epsilon^{'}/\epsilon)_{SM}}{Re(\epsilon^{'}/\epsilon)_{exp}}
\end{equation} 
\noindent
we obtain
\begin{equation}
 |\delta_{\epsilon^{'}}|\approx \left( 58\ TeV\right)^{2}B_{G}|\mathcal{C}_{G}-\mathcal{C'}_{G}|<1 \ ,
\label{epsPbound}
\end{equation} 
\noindent
where $Re(\epsilon^{'}/\epsilon)_{SM}$ has been estimated as in Ref. \cite{Isidori}; $B_G$ denotes the hadronic bag-parameter, 
$\left\langle 2\pi_{I=0}| y_s \mathcal{O}_{G} | K^0 \right\rangle $. 
We take $B_{G}=1$\footnote{ That corresponds to the estimate of the hadronic matrix element $\left\langle 2\pi_{I=0}| y_s \mathcal{O}_{G} | K^0 \right\rangle $
 in the chiral quark
model and to the first order in the chiral expansion.} and we take into account separately the contribution from $\mathcal{C}_{G}$ and $\mathcal{C'}_{G}$.\\
In the limit $s_{sR}=0$ we obtain from (\ref{epsPbound}):

\begin{equation}
 \sqrt{M_{B'}M_{\tilde{B}}} \gtrsim  (1.3) Y_*\  TeV\ ,
\label{epsPbound_2}
\end{equation} 
\noindent
which is in agreement with the result in \cite{Isidori}.
The contribution from the charged Higgs interactions gives weaker bounds than those from the neutral Higgs contribution.

\section{Constraints from $\epsilon_{K}$}\label{KK}
The strongest constraint on the $G^*$ mass comes from the data on $K-\bar{K}$ mixing and it is not a Minimal Flavor Violating bound. 

\noindent
It is well-known \cite{kkgBound} that the dominant contribution to the CP violation in $K-\bar{K}$ mixing, $\epsilon_K$,
comes from the heavy gluon exchange between left-handed and right-handed down-type quark
currents. Such exchange gives a contribution to the operator:
\[
 \mathcal{O}_{4}=\bar{d}^{\alpha}_{R}s^{\alpha}_{L}\bar{d}^{\beta}_{L}s^{\beta}_{R}\ .
\]
This contribution includes two terms, one from
the ``direct'' mixing of the 1-2 generations and another from the $(1-3)\times(2-3)$
mixing, via 3rd generation.
If one calculates this latter contribution, i.e. by considering the mixing via 3rd generation, finds:
\begin{equation}
 \mathcal{C}_{4}\sim\frac{g^{2}_{*3}}{M^{2}_{G*}}(D^{\dag}_{L})_{13}(D_{L})_{23}s^{2}_{1}(D_{R})_{23}(D^{\dag}_{R})_{13}s^{2}_{bR}\sim s^{2}_{1}s^{2}_{bR}\frac{g^{2}_{*3}}{M^{2}_{G*}}\frac{m_{s}m_{d}}{m^{2}_{b}}\ .
\label{C4_1}
\end{equation} 
\noindent
In the Two-Site Model without a $P_{LR}$ protection to $Wt_R b_R$ of ref. \cite{Agashe_flav, Sundrum}, in the TS10 and in the TS5 with $\Delta_{L2}=\Delta_{L1}$, 
where the bottom mass is proportional to the same angle that defines the $b_{L}$ composite/elementary mixing, the eq. (\ref{C4_1}) reduces to:
\begin{equation}
 \mathcal{C}_{4}\sim\frac{g^{2}_{*3}}{M^{2}_{G*}}\frac{m_{s}m_{d}}{\left(\frac{Y_{*}v}{\sqrt{2}}\right)^{2}}\ ;
\end{equation} 
from which the bound \cite{Agashe_flav}:
\begin{equation}
 M_{G*}\gtrsim (11)\frac{g_{*3}}{Y_{*}} \text{TeV} \ .
\end{equation}
\noindent
In the TS5 (with $\Delta_{L2}\ll\Delta_{L1}$), instead, there is a $s^{2}_{1}/s^{2}_{2}$ enhancement:
\begin{equation}
 \mathcal{C}_{4}\sim \frac{s^{2}_{1}}{s^{2}_{2}}\frac{g^{2}_{*3}}{M^{2}_{G*}}\frac{m_{s}m_{d}}{\left(\frac{Y_{*}v}{\sqrt{2}}\right)^{2}}\ ,
\end{equation} 
so that the bound becomes:
\begin{equation}
 M_{G*}\gtrsim (11)\frac{s_1}{s_2}\frac{g_{*3}}{Y_{*}} \text{TeV} \ .
\end{equation}
We have a strong bound, especially in the TS5 (with $\Delta_{L2}\ll\Delta_{L1}$), though it could still allow for a relatively light $G^*$,
 if we had a ratio $\frac{g_{*3}}{Y_{*}}\ll 1$ \footnote{In this case, however, 
a tension between the bound from $\epsilon_{K}$ and the one from the UV contribution to $b\rightarrow s\gamma$
exists, as pointed out in \cite{Agashe_flav}, because of the opposite dependence on the composite Yukawa coupling. 
So it is not possible to simultaneously soften these bounds in the scenario of anarchic composite Yukawa coupling.}.
Most importantly, this is a bound that strongly depends on the flavor structure of the sector Beyond the SM. 
If we consider BSM flavor scenarios less minimal than 
that of anarchic $Y_*$, the constraint can be much softened. 
For example, recent studies show that, if the strong sector is invariant under additional flavor 
symmetries \cite{Fitzpatrick:2007sa, Santiago:2008vq, Csaki, Csaki:2008eh, Csaki:2009wc}
or preserves CP~\cite{Redi_Weiler}, the heavy gluon can have a mass as light as a few TeVs.

\chapter{Heavy colored vectors at the LHC}

In this part of the project we perform a study of the LHC discovery reach on a heavy gluon ($G^*$) and heavy fermions (top and bottom excitations), 
coming from a new composite sector. 
We consider the ``two-site'' descriptions (TS5 and TS10), which we have derived in the previous part. 
The results obtained and the strategies we applied, however, can be generalized to Randall-Sundrum models and to the
 wider class of models that address the hierarchy problem by considering a new strong sector responsible for the EWSB.\\
We start by considering the TS5 model of sec. \ref{TS5} with the following assumptions:
$Y_{*U}=Y_{*D}=Y_*$ and $s_{bR}=s_{1}$ (i.e. $b_L$ as much composite as $b_R$).
 We remind that the physical mass $M_T=M_B$ \footnote{We are working at order $x\equiv\frac{Y_* v}{\sqrt{2}m_*}$ and we are thus neglecting $O(x^2)$ corrections in the heavy masses.} of the $(T,B)$ doublet of heavy fermions 
(partner of the SM $q^3_L=(t_L, b_L)$ doublet) is related to the other heavy fermion masses as:
\[
M_{T5/3}=M_{T2/3}=M_{T}c_1 \ \ M_{\tilde{T}}=k M_T \ \  M_{\tilde{B}}=k' M_T \ .
\]
We set, for simplicity, $k=k'=1$.\\
We have also the following relations with the heavy fermions in the $\mathcal{Q'}$ bidoublet,
\[
 M_{T'}=M_{B'}=rM_T\ \ \ M_{B-1/3}=M_{B-4/3}\simeq M_{T'}=rM_T\ ,
\]
we will set $r=1$ in the calculation of the BRs and of the $G^*$ width. 
However, these heavy fermions are not much relevant in the analysis we perform in the $\psi\chi$ channel, 
since they can only be produced in pairs from the decay of the $G^*$ (in particular, the process $G^*\to T't (B'b)$ receives a suppression by $s^2_2$).

\noindent
In our model $G^*$ transforms as $(\mathbf{8}, \mathbf{1},\mathbf{1})_0$ under 
$SU(3)_c \times  SU(2)_L \times SU(2)_R \times U(1)_X$. 
Composite fermions are $SU(3)_c$ triplets and transform as shown in (\ref{eq.fields}) under $SU(2)_L \times SU(2)_R \times U(1)_X$. 
The Lagrangian which describes $G^*$ and fermion interactions in the TS5 reads, in the elementary/composite basis
  (we work in the
electroweak gauge-less limit, and omit the terms involving the $SU(2)_L \times U(1)_Y$ elementary gauge fields, which play no role in our analysis):
%
\begin{align}  \label{eq:Ltotal}
{\cal L} =&  \, {\cal L}_{elementary} + {\cal L}_{composite} + {\cal L}_{mixing} \\[0.7cm]
\label{eq:Lelem}
 {\cal L}_{elementary} =& \,\bar q^i_L i\!\Dslash\, q^i_L + \bar u^i_R i\!\Dslash\, u^i_R +  \bar d^i_R i\!\Dslash\, d^i_R 
                                    -\frac{1}{4g_{el3}^2} \, G_{\mu\nu} G^{\mu\nu}  \\[0.5cm]
\label{eq:Lcomp}
\begin{split} 
 {\cal L}_{composite}  =&  
  \, \text{Tr} \left\{ \bar{\cal Q} \left( i\!\!\dslash\, - \!\not\! G^*\! - M_{Q*} \right) {\cal Q}  \right\} + 
  \bar{\tilde T}  \left( i\!\!\dslash\, - \!\not \! G^*\! - M_{\tilde T*} \right) \tilde T   \\[0.2cm]
  & + \text{Tr} \left\{ \bar{\cal Q^\prime} \left( i\!\!\dslash\, - \!\not\! G^*\! - M_{Q^\prime *} \right) {\cal Q^\prime}  \right\} + 
  \bar{\tilde B}  \left( i\!\!\dslash\, - \!\not \! G^*\! - M_{\tilde B *} \right) \tilde B  \\[0.2cm]
  & + Y_{*U} \, \text{Tr} \{ \bar{{\cal Q}} \,  {\cal H} \} \, \tilde{T}  + Y_{*D} \, \text{Tr} \{ \bar{{\cal Q^\prime}} \,  {\cal H} \} \, \tilde{B}
      -\frac{1}{4g_{*3}^2} \, G^*_{\mu\nu} G^{*\, \mu\nu}  
\end{split} \\[0.5cm]
\begin{split} \label{eq:Lmixing}
{\cal L}_{mixing} =&  - \Delta_{L1}\, \bar q^3_L \left( T,B\right) - \Delta_{L2}\, \bar q^3_L \left( T^\prime,B^\prime\right)  
 - \Delta_{tR}\, \bar t_R \tilde T  - \Delta_{bR}\, \bar b_R \tilde B + h.c. \\[0.2cm]
  & + \frac{1}{2}\, \frac{{\bar M}^2_{G_*}}{g_{*3}^2} \left( g_\mu - G_\mu^* \right)^2\, .
\end{split}
\end{align}

The derivative $D_\mu$ is covariant under $SU(3)_c$ transformations.
The superscript $i$ in eq.(\ref{eq:Lelem}) runs over the three SM families ($i =1,2,3$), with $q^3_L \equiv (t_L, b_L)$, $u^3_R \equiv t_R$, 
$d^3_R \equiv b_R$.
Lagrangian (\ref{eq:Ltotal}) can be diagonalized, before EWSB, by performing a field rotation 
from the composite/elementary to the mass eigenstate basis (the resulting Lagrangian can be found in the appendix \ref{Gstar_L}, eq. \ref{eq:Lrotated}).
 In this latter basis the couplings of $G^*$ to the fermions read:
\begin{align}
\label{eq:Gqq}
g_{G^* qq} =& g_S \tan\theta_3\, , && q = u,d,c,s \\[0.3cm]
\label{eq:Gpsipsi}
g_{G^* \psi \psi} =& 
  g_S \left( \tan\theta_3 c^2_{\psi} - \cot\theta_3 s^2_{\psi} \right) \, , &&   \psi = t_L, b_L , t_R, b_R\\[0.3cm]
\label{eq:Ghl}
g_{G^* \psi\chi} =& 
  g_S \, \frac{s_{\psi}c_{\psi}}{\sin\theta_3 \cos\theta_3}\, , && \psi\chi = Tt_L , Bb_L, \tilde T t_R, \tilde B b_R\\[0.3cm]
\label{eq:Ghh}
g_{G^* \chi\chi} =& 
  g_S \, \left( \tan\theta_3 s^2_{\psi} - \cot\theta_3 c^2_{\psi} \right)\, , && \chi = \text{any of the heavy fermions}\, ,
\end{align}

$s_{\psi}\equiv \sin\varphi_{\psi}$ is the sine of the angle that defines the rotation
from the elementary/composite basis $\psi^{ele}/ \chi^{com}$ to the basis of the mass eigenstates $\psi/\chi$
 (see, for example, eq.(\ref{ele/compROTATION})).
$s_{\psi}$ denotes thus the degree of compositeness of SM fermions
 (the relating cosine, $c_{\psi}$, represents the degree of elementarity of SM fermions).
Referring to TS models of sections \ref{TS5} and \ref{TS10}, $s_{\psi} =$ $s_{1}$ for $\psi=t_L, b_L$, $s_{\psi} =s_{bR}$ for $\psi=b_R$, $s_{\psi} =s_{R}$ for $\psi=t_R$.
For light quarks $s_{\psi}\simeq0$ ($c_{\psi}\simeq1$); this follows from the assumption of flavor anarchy in the strong sector (see the discussion in sec. \ref{TS5})
\footnote{If the strong sector is not anarchic, the phenomenology
can change significantly and additional signatures, like for example the unsuppressed decay of $G^*$ to light quarks, can appear, see for example
Ref.~\cite{Redi_Weiler}.
Our analysis, on the other hand, can still be relevant although the decay modes considered here might not be the dominant ones.}.
As we know from chapter \ref{chm}, $s_{\psi}$ represents also the degree of elementarity ($c_{\psi}$ the degree of compositeness) of the heavy fermion $\chi$,
 partner of the SM fermion $\psi$. Possible $\psi\chi$ partners are the following:
\begin{equation}
\chi\psi \equiv\  Tt_L , Bb_L, \tilde T t_R, \tilde B b_R
\end{equation}
As we already pointed out in the previous part of the project, there are also heavy fermions that do not mix directly with elementary fermions, which are
called custodians. For the custodians $s_{\psi}=0$ ($c_{\psi}=1$).\\
We remind that $\theta_3$ parametrizes the $G^*$ degree of compositeness (see eq.(\ref{GgMIX})) and is related to the coupling $g_{*3}$ of the 
$SU(3)_c$ interaction of the composite sector as $\tan\theta_3=\frac{g_{el3}}{g_{*3}}\simeq\frac{\sqrt{4\pi\alpha_S}}{g_{*3}}$.
 In the analysis we consider initially a reference value for the model, $\tan\theta_3= 0.44$, that corresponds to a value $g_{*3}=3$.
 We also set $Y_*=g_{*3}$.\\ 

\section{Remarks on the $G^{*}$ and the heavy fermions search at the LHC }

\subsection{$G^*$ Production}

\begin{figure}[t]
\mbox{
\subfigure[$G^{*}$ production.]{\includegraphics[width=0.3\textwidth]{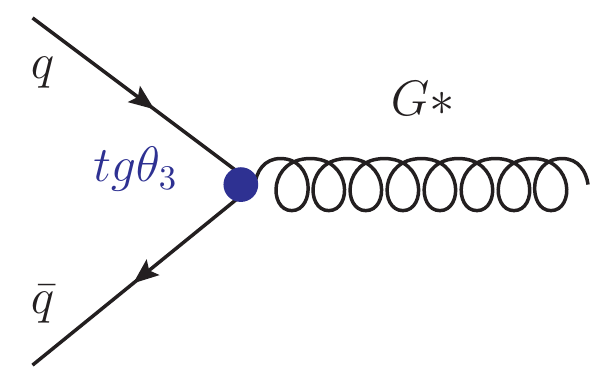}}} 
\subfigure[$G^{*}$ production cross section at the LHC with $\sqrt{s}=14$ TeV and $\sqrt{s}=7$ TeV]{\includegraphics[width=0.8\textwidth]{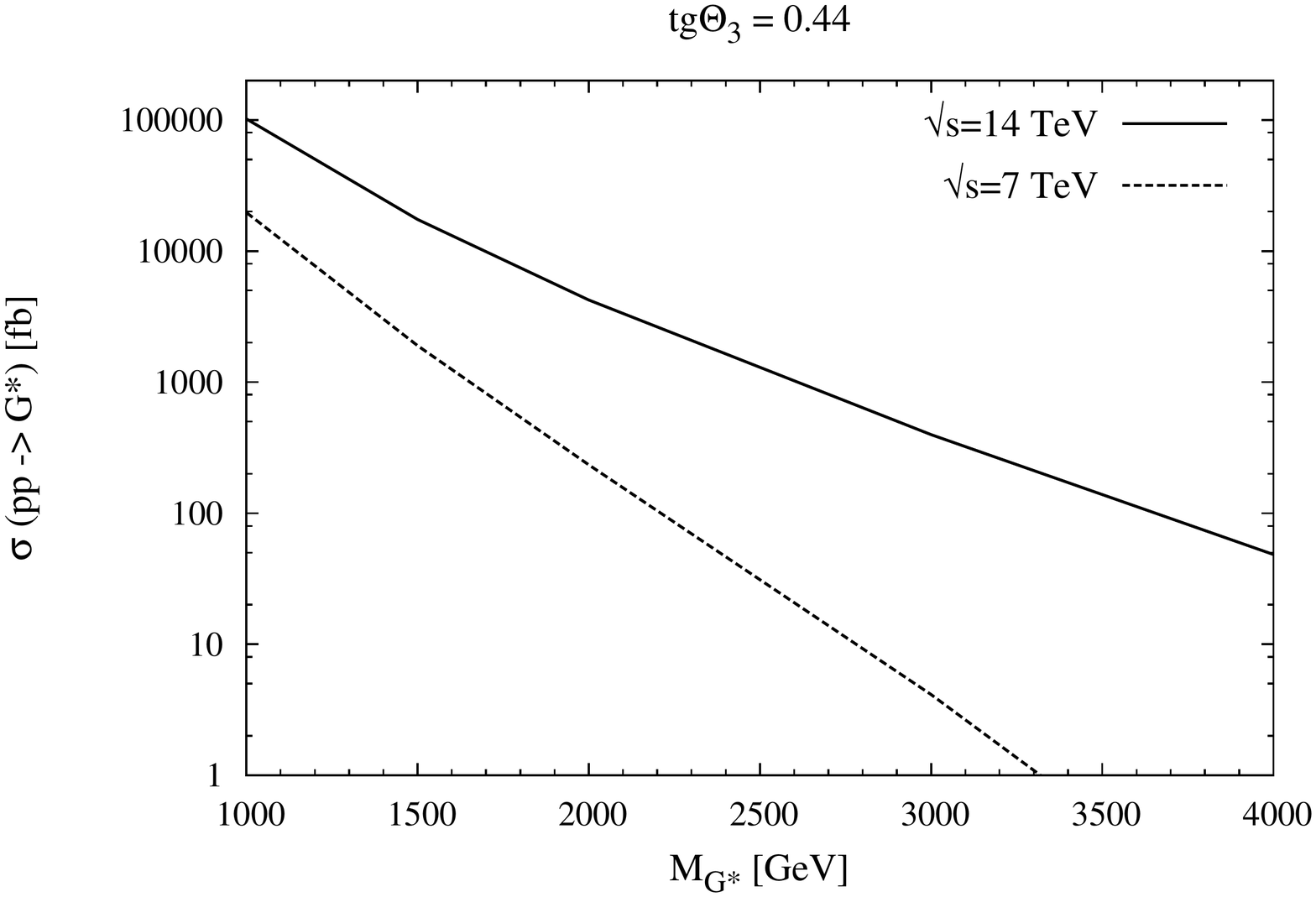}
}
\caption{\textit{$G^{*}$ production.}}
\label{Gprod_fig}
\end{figure}

The $G^*$ is produced in $pp$ collision by quark-antiquark annihilation, the production by gluon fusion being forbidden by gauge invariance.\\
The $G^*$ couples to the light quarks $u,d$ in the proton with a coupling:

\begin{equation}
 g_{G^* u\bar{u}}=g_{G^* d\bar{d}}\simeq g_{S}\tan\theta_3\ .
\label{Gprod}
\end{equation} 

Light quarks, indeed, are almost fully elementary and, as a consequence, they couple, with a coupling $g_{el3}$, to the elementary component of the heavy gluon, 
$G^* \sin\theta_3$ (see eq.s (\ref{GgMIX},\ref{GgMIX2})).
The cross section for the production of the $G^*$ is therefore proportional to $\tan^2\theta_3$. \\
The cross section for the production of the $G^*$ drops exponentially with the $G^*$ mass, as we can see from Fig. \ref{Gprod_fig}. 
This makes the discovery at the LHC difficult for a mass of $G^*$ above $4$ TeV (considering a value $\tan\theta_3 =0.44$). 
We have calculated the cross section for the $G^*$ production by using MADGRAPH/MADEVENT v4 \cite{Madgraph}; we have implemented 
our model for the $G^*$ interactions in MADGRAPH, making use of the FeynRules package \cite{Feynrules}. 
We have used the CTEQ6L1 pdf set and the factorization and renormalization scales, $Q=M_{G*}$.

\subsection{$G^*$ Branching Ratios and total decay width}\label{Sec_BRs}

\begin{figure}[h!]
\centering
\mbox{
\subfigure[$G^{*}$ decay Branching Ratios ($s_{R}=1$)]{
\includegraphics[width=0.77\textwidth]{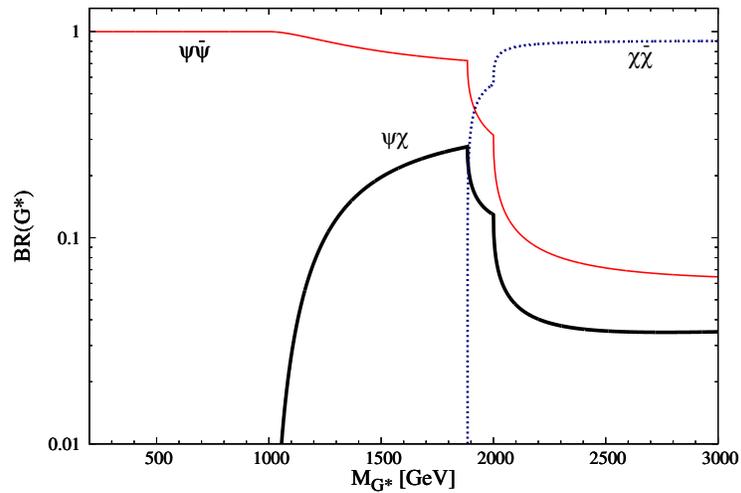}}}\\
\mbox{\subfigure[$G^{*}$ total decay width ($s_{R}=1$)]{
\includegraphics[width=0.77\textwidth]{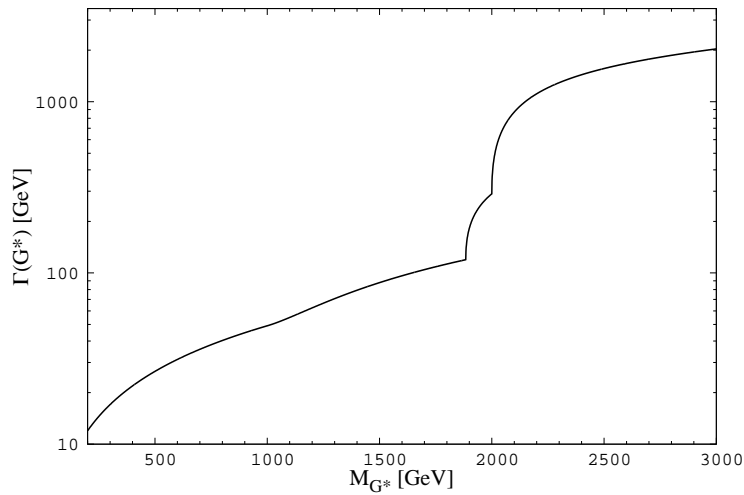}}
}
\caption{\textit{$G^{*}$ decay Branching Ratios and $G^{*}$ total decay width as functions of the $G^{*}$ mass in the TS5,
 with a fully composite $t_R$, $s_R=1$. $\tan\theta_3$ and $Y_*$ parameters are set at the reference values: $\tan\theta_3=0.44$, $Y_*=3$. 
The $(T,B)$ heavy fermions (partners of $q^3_L\equiv(t_L,b_L)$) mass has been set to $M_T=M_B=1$ TeV. 
$\psi\bar{\psi}$ denotes the BR for the $G^{*}$ decays into SM fermion pairs [red curve], 
$\psi\chi$ those for the $G^{*}$ decays into one heavy ($\chi$) plus one SM ($\psi$) fermion [thick curve] and $\chi\bar{\chi}$ 
those for the $G^{*}$ decays into a pair of heavy fermions [dotted curve].}}
\label{BR_sR1}
\end{figure}

In order to study the $G^*$ phenomenology, we look first at its decay modes and we calculate its total decay width.\\
We can distinguish three classes of decays for the $G^*$: the decays into SM fermion pairs, $G^* \to \psi\bar{\psi}$, 
those into one heavy ($\chi$) plus one SM fermion, $G^* \to \chi\psi$, and the decays into heavy fermion pairs, $G^* \to \chi\bar{\chi}$.
Decay widths are as follows: 
\begin{align}
 \Gamma\left(G^{*}\rightarrow\psi\bar{\psi}\right)= &\frac{\alpha_{S}}{12}M_{G*}\left(s^{2}_{\psi}\cot\theta_{3}-c^{2}_{\psi}\tan\theta_{3}\right)^{2}  \\[0.5cm]
 \Gamma\left(G^{*}\rightarrow\chi\bar{\psi}+\psi\bar{\chi}\right)= &\frac{\alpha_{S}}{6}M_{G*}\frac{s^{2}_{\psi}c^{2}_{\psi}}{\sin^{2}\theta_{3}\cos^{2}\theta_{3}}
\left(1-\frac{m^{2}_{*}}{M^{2}_{G*}}\right)\left(1-\frac{1}{2}\frac{m^{2}_{*}}{M^{2}_{G*}}-\frac{1}{2}\frac{m^{4}_{*}}{M^{4}_{G*}}\right)  \\[0.5cm]
\begin{split}
\Gamma\left(G^{*}\rightarrow \chi\bar{\chi}\right)= &\frac{\alpha_{S}}{12}M_{G*} \left\{\left[\left(c^{2}_{\psi}\cot\theta_{3}-s^{2}_{\psi}\tan\theta_{3}\right)^{2}+\cot^{2}\theta_{3}\right]\left(1-\frac{m^{2}_{*}}{M^{2}_{G*}}\right)\right.\\
& +6\left( c^2_{\psi}\cot^2\theta_3 -s^2_{\psi}\right)\frac{m^{2}_{*}}{M^{2}_{G*}} \Bigg\} \sqrt{1-4\frac{m^{2}_{*}}{M^{2}_{G*}}}\  ,
 \end{split}
\end{align}

\noindent
where $m_*$ denotes the heavy fermion physical mass. \\
We show in Fig. \ref{BR_sR1} the BRs and the total decay width of the $G^*$, as functions of the $G^{*}$ mass.
Fig. \ref{BR_sR1} refers to the case of a fully composite $t_R$ ($s_R =1$). 
The mass of the $(T,B)$ heavy fermions (partners of $q^3_L\equiv(t_L , b_L)$) has been set to $1$ TeV.

\subsection{Three scenarios for the $G^*$ search at the LHC }

\begin{figure}[h!]
\mbox{
\subfigure[$G^{*}$ decay Branching Ratios ($s_{R}=0.8$)]{
\includegraphics[width=0.52\textwidth]{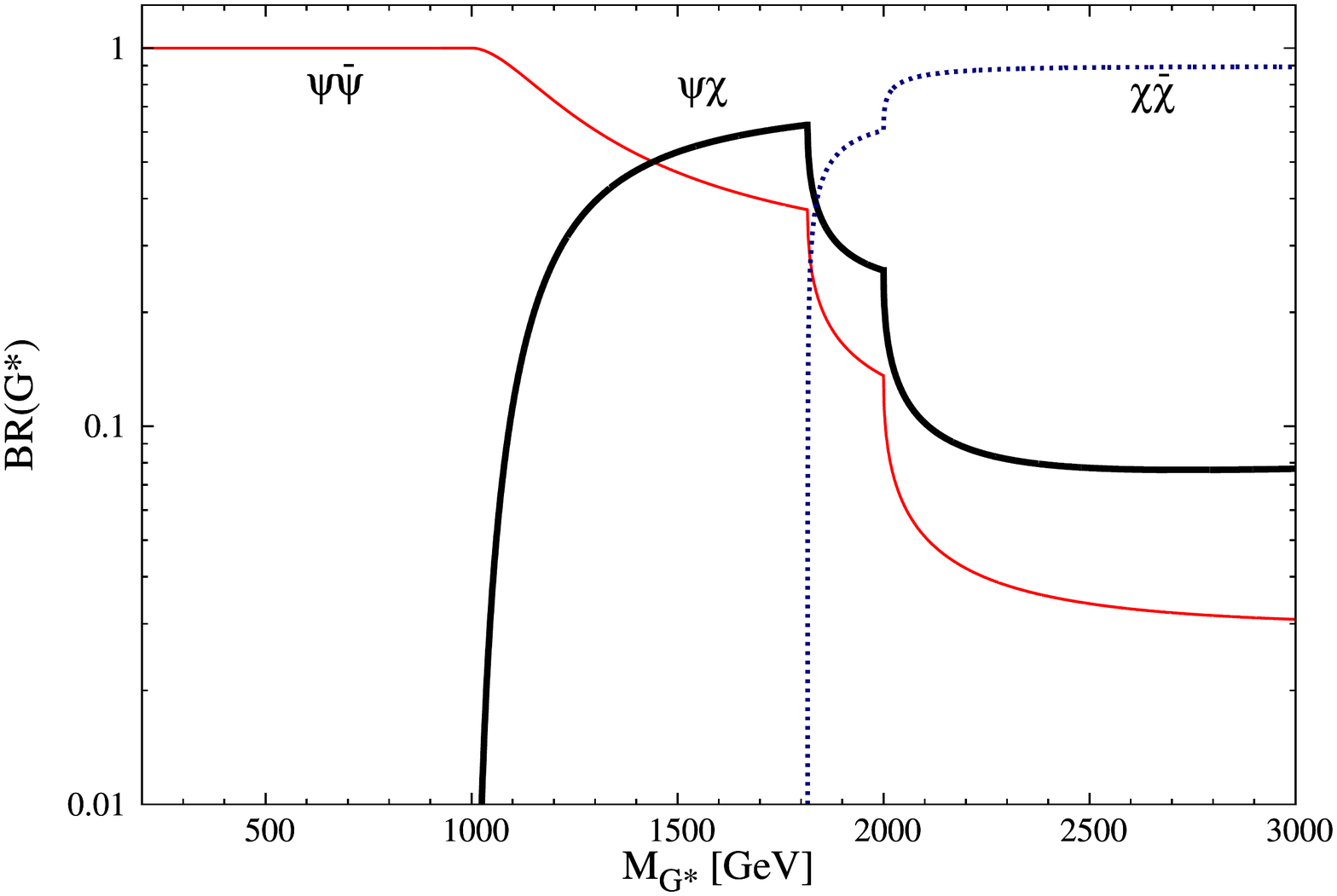}}
\subfigure[$G^{*}$ total decay width ($s_{R}=0.8$)]{
\includegraphics[width=0.52\textwidth]{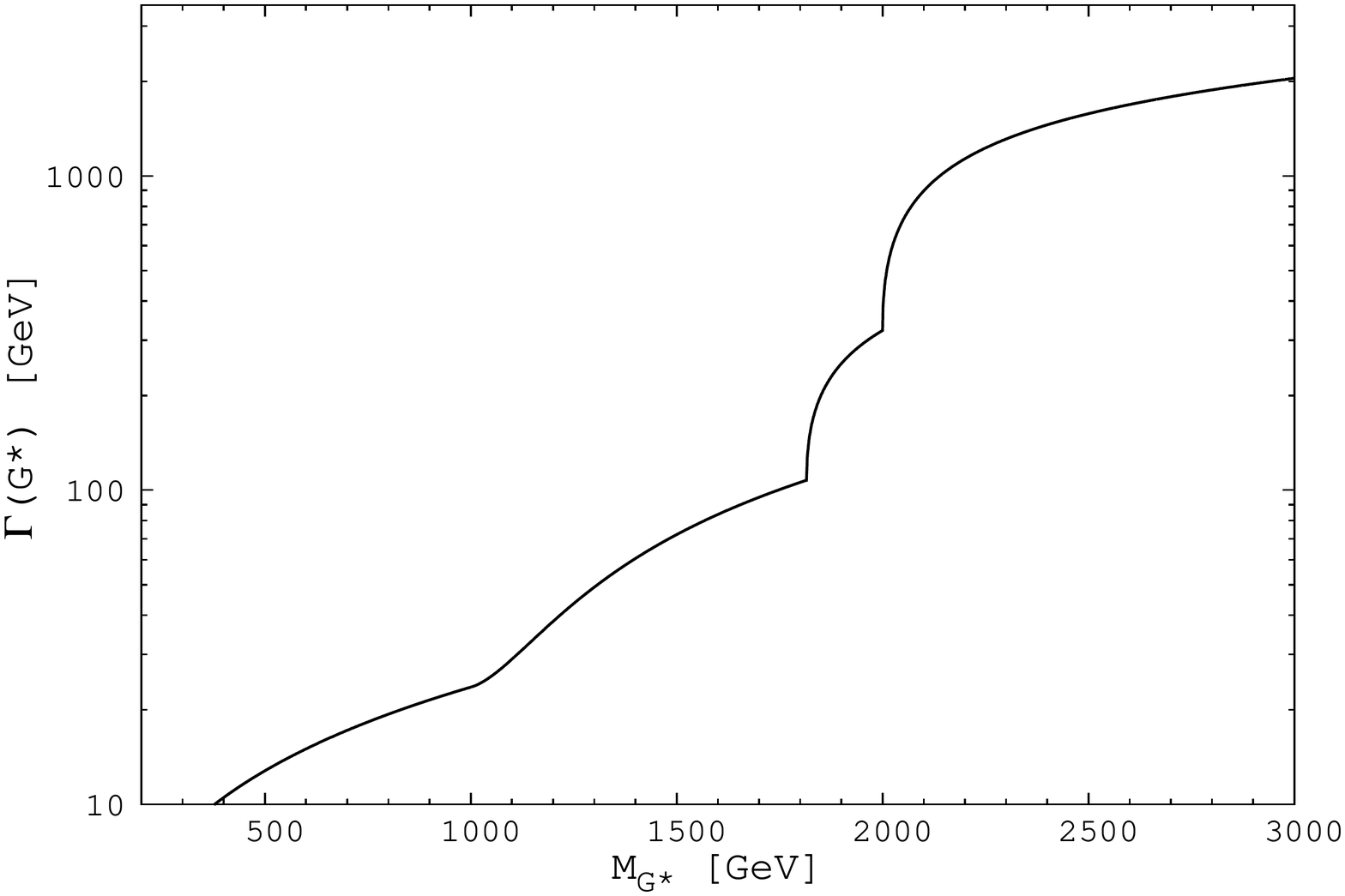}}
}
\caption{\textit{$G^{*}$ decay Branching Ratios and $G^{*}$ total decay width as functions of the $G^{*}$ mass,
 for $s_{R}=0.8$. We also set: $M_T=1$ TeV, $\tan\theta_3=0.44$, $Y_*=3$.}}
\label{BR_sR08}
\end{figure}

\begin{figure}[h!]
\mbox{
\subfigure[$G^{*}$ decay Branching Ratios ($s_{R}=0.6$)]{
\includegraphics[width=0.52\textwidth]{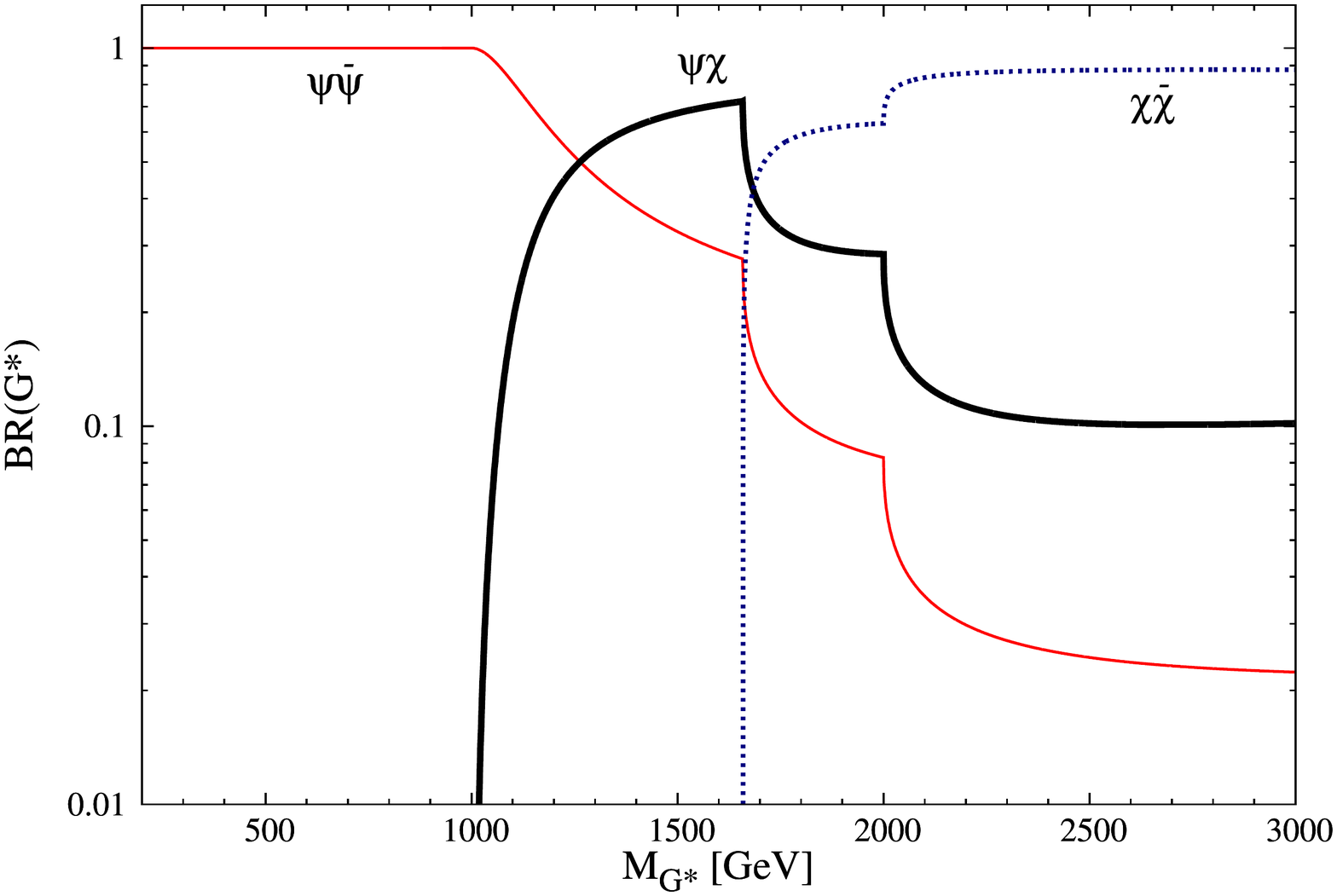}}
\subfigure[$G^{*}$ total decay width ($s_{R}=0.6$)]{
\includegraphics[width=0.52\textwidth]{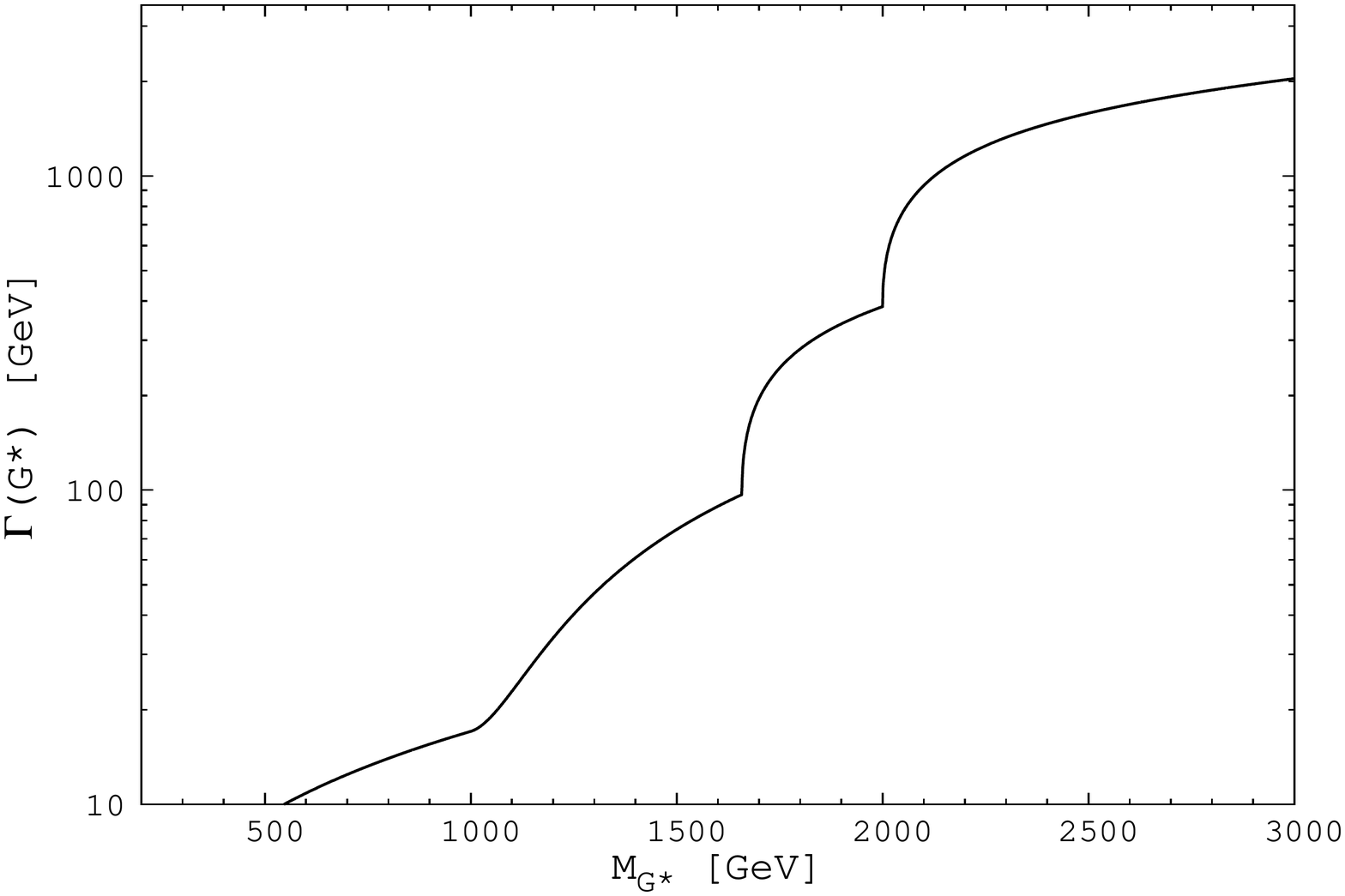}}
}
\caption{\textit{$G^{*}$ decay Branching Ratios and $G^{*}$ total decay width as functions of the $G^{*}$ mass, for $s_{R}=0.6$. 
We also set: $M_T=1$ TeV, $\tan\theta_3=0.44$, $Y_*=3$.}
}
\label{BR_sR06}
\end{figure}

As Fig. \ref{BR_sR1} shows, we can clearly distinguish three scenarios with different phenomenological implications for the $G^*$ search at the LHC,
 depending on the ratio between $G^*$ mass and the heavy fermion mass $M_T$.\\
 When the $G^*$ mass is below the threshold for the production of a heavy fermion ($M_{G^*}<M_T$), 
$G^*$ decays completely into pairs of SM quarks: $t\bar{t}$ pairs or dijets. 
The relative importance of these decay channels is controlled by $\tan\theta_3$ and the top degrees of compositeness. 
For small values of $\tan\theta_3$,
which are naturally implied by the hierarchy of couplings $g_{el3} \ll g_{*3}$, or large top degree of compositeness,
the dominant channel is $t\bar t$. For example, for a fully composite $t_R$, $s_{R} =1$, and $\tan\theta_3 =0.2$, $Y_* =3$, 
one has $BR(G^* \to t\bar t) = 0.98$, $BR(G^* \to q\bar q) = 0.012$.
On the other hand, large branching ratios to pairs of light quarks can be  obtained even for moderate top degrees of compositeness
if $\tan\theta_3$ is not too small.
For example, for $s_{R} =0.6$, $\tan\theta_3 =0.44$, $Y_* =3$  one has $BR(G^* \to t\bar t) = 0.18$, $BR(G^* \to q\bar q) = 0.69$.
In this case the strongest discovery reach (or exclusion power) comes from the dijet searches.
This first scenario, where the decays to one heavy fermion are kinematically forbidden, 
is the only one studied in the literature on the $G^{*}$ search at the LHC. In particular, 
the searches for a heavy gluon have been focussed on the channel $pp\to G^{*}\to t_R\bar{t}_R$ \cite{Agashe, Randall}, 
assuming the case of a fully composite top right and a small value of $\tan\theta_3$.
%
The signal $G^{*}\to t_R\bar{t}_R$ is affected by huge and difficult to reduce QCD background, $pp\to g\to t\bar{t}$,  
that makes the $G^{*}$ discovery at the LHC particularly challenging. However, the study in Ref. \cite{Agashe}, 
by exploiting peculiarities of the signal, such as the large invariant mass of $t\bar{t}$ pairs and a Left-Right asymmetry, 
shows that a discovery of a $G^{*}$ lighter than $\sim 4$ TeV is possible with $100$ fb$^{-1}$ at the 14 TeV LHC. 
\\
When the $G^*$ reaches the threshold for the decay into one heavy fermion, the $G^* \to \psi\chi$
decays become relevant, while the BR for the $G^*$ decay into SM quark pairs decreases. 
The $G^*$ width remains quite narrow, below $\sim 100$ GeV. 
When the $G^*$ reaches the threshold for the production of pairs of custodians ($T_{5/3}$ and $T_{2/3}$), 
$BR(G^* \to \psi\chi)$ slightly decreases. The intermediate scenario where $M_{T}<M_{G^{*}}<2M_{T}$ and, especially, 
the scenario where $M_{T}<M_{G^{*}}<2M_{T5/3}$ seem to be very promising for the $G^{*}$ search at the LHC. 
We identify the $G^{*}$ decay into a top (bottom) and its heavy partner as the best search channel
\footnote{Also the study in Ref. \cite{Dobrescu}, that considers a model with gluon-prime and top-prime, similar to the TS model that we are analyzing, 
has suggested the production of a single top-prime
in association with a top as a promising channel for an observation at CDF (they consider a different parameter space, with lighter heavy colored vectors).}.
Since SM gluon interactions with one heavy and one SM fermion are forbidden, these search channels, differently from the cases in the other scenarios, 
are not overwhelmed by irreducible backgrounds with gluon mediations. 
The presence of heavy fermion resonances only in the signal turns out to be crucial to reduce backgrounds. \\
 Finally, we can recognize a third scenario, 
which corresponds to the case of a $G^*$ heavier than heavy fermion pairs ($M_{G^{*}}>2M_{T}$). In this scenario $G^*$ decays completely
into pairs of heavy fermions. Due to the large number of available channels and the large couplings, 
its total decay width has a rapid increase; it grows up to $\sim 1$ TeV. 
Such a width is too large to be able to distinguish the $G^{*}$ resonance from the background, 
in which the same production of heavy fermion pairs is mediated by the SM gluon
instead of the $G^{*}$. So, if the $G^{*}$ is sufficiently heavy to decay into heavy fermion pairs, 
we argue that it will not be easily discovered at the LHC.\\
We do not have a model independent hint from experimental data on which of the three scenarios could really exists.
The data, however, give generally stronger constraints on the $G^*$ mass than on the heavy fermion masses.
In particular, the bound on the $G^*$ mass from the data on $K-\bar{K}$ mixing, that we discussed in sec.\ref{KK}, is a very strong bound,
even though it depends on the flavor structure of the sector Beyond the SM. The first scenario seems thus to be not the favorite one.  \\
In consideration of all we discussed, an analysis focused on the intermediate scenario seems to be needed for the $G^*$ search at the LHC. \\
In Fig. \ref{BR_sR1}, we have shown the $G^*$ BRs in the case of a fully composite $t_R$ (which is the scenario considered in the analysis of Ref.\cite{Agashe}).
$t_R$, however, is not forced to be fully composite. As found in the previous part of the project 
a not fully composite top seems, instead, to be the preferred case by the data on $b \to s \gamma$ for TS10 model 
(and, in general, for models without $P_C$ protection to the $Wt_R b_R$ coupling). 
We calculate $G^*$ Branching Ratios and total decay width for different values of the $t_R$ degree of compositeness. 
We show the results in Fig.s \ref{BR_sR08},\ref{BR_sR06}.
 We see that the $G^* \to \chi\psi$ decays become more important in the case of a not fully composite $t_R$ 
(until we reach the case of a fully composite $t_L$. We discuss this latter case in Sec. \ref{tCOM}).
  
\section{Search channels}

We focus our search on the promising channels where the heavy gluon $G^{*}$ (with $M_T<M_{G^{*}}<2M_{T5/3}$) decays into one heavy fermion ($\chi$) 
plus its SM partner ($\psi$)\footnote{We consider both the charge, $G^* \to \bar{\chi}\psi+\chi\bar{\psi}$.}:
\[
 pp\to G^* \to \chi\psi \ \equiv Tt_L , Bb_L, \tilde T t_R, \tilde B b_R \ .
\]
These fermions can be
a left-handed top with its heavy partner $T$, a left-handed bottom with its heavy partner $B$, 
a right-handed bottom (if not fully elementary) with the heavy fermion $\tilde{B}$ or 
(in the case of a not fully composite $t_R$) a right-handed top with its heavy partner $\tilde{T}$.
We find that cross sections for the heavy fermions production via the $G^{*}$ exchange can be even larger than those of
single and pair production mechanisms, those typically considered in the literature (see ref.s \cite{Servant, Carena:2007tn, AguilarSaavedra:2005pv, Skiba:2007fw, AguilarSaavedra:2009es}, for analyses on pair production, and \cite{Wulzer, Azuelos:2004dm, Atre:2008iu, Atre:2011ae}, for single production). 
For example, considering $M_{G*}=3$ TeV and $M_T=2$ TeV, we find a $b\bar{B}+\bar{b}B$ (or $t\bar{T}+\bar{t}T$) cross section of $58\div 84\ fb$ (at 14 TeV LHC\footnote{Applying only minimal cuts: $p_{Tb}> 20$ GeV.}), varying 
the value of the $G^*$ coupling to light quarks, $g_S\tan\theta_3$, from the value $0.2 g_S$ to $0.6 g_S$ 
and considering an intermediate value, $s_R=0.6$, for the top degree of compositeness.
Therefore, these channels are also promising for the heavy fermions discovery at the LHC.\\

Depending on the type of heavy fermion decays, we can distinguish search channels with different final states.

\subsection{Heavy fermions decay}\label{BRsHf}

Heavy fermions are essentially composite states, therefore they couple strongly to composite modes.
Heavy fermions ($\chi$) decay thus into longitudinally polarized bosons or into the Higgs (plus a SM fermion $\psi$);
 the widths for these decays are as follows:

\[
\Gamma\left(\chi\rightarrow W_L\psi\right)=\frac{\lambda^2_{W\chi}}{32 \pi}M_{\chi}
\left[\left( 1+\frac{m^2_{\psi}-M^2_W}{M^2_{\chi}}\right)\left( 1+\frac{m^2_{\psi}+2M^2_W}{M^2_{\chi}}\right)-4\frac{m^2_{\psi}}{M^2_{\chi}}\right] 
\]
\[
 \times\sqrt{1-2\frac{m^2_{\psi}+M^2_W}{M^2_{\chi}}+\frac{\left(m^2_{\psi}-M^2_W\right)^2 }{M^4_{\chi}}}
\]

\[
\Gamma\left(\chi\rightarrow Z_L\psi\right)=\frac{\lambda^2_{Z\chi}}{64 \pi}M_{\chi}
\left[\left( 1+\frac{m^2_{\psi}-M^2_Z}{M^2_{\chi}}\right)\left( 1+\frac{m^2_{\psi}+2M^2_Z}{M^2_{\chi}}\right)-4\frac{m^2_{\psi}}{M^2_{\chi}}\right] 
\]
\[
 \times\sqrt{1-2\frac{m^2_{\psi}+M^2_Z}{M^2_{\chi}}+\frac{\left(m^2_{\psi}-M^2_Z\right)^2 }{M^4_{\chi}}}
\]

\[
\Gamma\left(\chi\rightarrow h\psi\right)=\frac{\lambda^2_{h\chi}}{64 \pi}M_{\chi}
\left( 1+\frac{m^2_{\psi}}{M^2_{\chi}}-\frac{M^2_{h}}{M^2_{\chi}}\right)
\sqrt{\left(1-\frac{m^2_{\psi}}{M^2_{\chi}}+\frac{M^2_{h}}{M^2_{\chi}}\right)^2-4\frac{M^4_{h}}{M^4_{\chi}}}\ .
\]
\noindent
Vertices $\lambda_{W/Z/h\chi}$ are the following (with the exception of $\tilde{B}$, 
we can calculate them by diagonalizing the fermionic mass matrices, (\ref{Mup}), (\ref{Mdown}), (\ref{Mup_mchm10}) and (\ref{Mdown_mchm10}),
 at the order $\frac{Y_* v}{\sqrt{2}M_{\chi}}$, 
neglecting further terms of the order $\left( \frac{Y_* v}{\sqrt{2}M_{\chi}}\right)^2$, coming from the electroweak mixing among heavy fermions):

\[
 \lambda_{WT}\simeq 0 \ , \ \lambda_{ZT}=\lambda_{hT}=Y_* c_1s_R
\]
\noindent
for the $T$ decays ($T\to Wb$, $T\to Zt$, $T\to ht$). 
$\lambda_{WT}$ is of the order $s^2_2$ in the TS5 and of the order $s^2_{bR}$ in the TS10 
(we remind that in the TS10 $s_{bR}$ has to be small, in order to account for the ratio $m_b/m_t\ll 1$).

\[
 \lambda_{WB}= Y_* c_1s_R \ , \ \  \lambda_{ZB}\simeq 0 \ ,\ \  \lambda_{hB}\simeq 0
\]
\noindent
for the $B$ decays ($B\to Wt$, $B\to Zb$, $B\to hb$). $\lambda_{Z/hB}$ is of the order $s^2_2$ in the TS5 and of the order $s^2_{bR}$ in the TS10.

\[
 \lambda_{W\tilde{T}}=\lambda_{Z\tilde{T}}=\lambda_{h\tilde{T}}= Y_* s_1 c_R 
\]
\noindent
for the $\tilde{T}$ decays ($\tilde{T}\to Wb$, $\tilde{T}\to Zt$, $\tilde{T}\to ht$).\\
For the $\tilde{B}$ decays:
\[
 \lambda_{Z\tilde{B}}=\lambda_{h\tilde{B}}\sim Y_* s_{bR}\times O(x)\ , \ \  \lambda_{W\tilde{B}}\sim Y_* s_2 c_{bR} \ \ \text{(in the TS5)}
\]
\[
 \lambda_{W\tilde{B}}=\lambda_{Z\tilde{B}}=\lambda_{h\tilde{B}}=Y_* s_1 c_{bR} \ \ \text{(in the TS10)}
\]
In the TS5, $\tilde{B}$ can decay into $Z(/h)b$ via the electroweak mixing with $B'$ and with $B_{-1/3}$, without paying for a suppression by $s_2$.
The decay of $\tilde{B}$ into $Wtb$, instead, is suppressed by $s_2$. We have 
$\lambda_{Z(/h)\tilde{B}}\sim Y_*\left( \frac{Y_* v}{\sqrt{2}} \frac{c_{bR}M_{\tilde{B}}+M_{B'}}{M^2_{\tilde{B}}-M^2_{B'}}\right)s_{bR}\gg \lambda_{W\tilde{B}}\sim Y_* s_2 c_{bR}$.
Therefore, in the TS5, $BR\left(\tilde{B}\rightarrow Z(/h)b\right)\gg BR\left(\tilde{B}\rightarrow Wt\right)$ 
and we will neglect the $\tilde{B}$ contribution to the $G^*\to\psi\chi\to Wtb$ signal. \\

\begin{figure}[h!]
\mbox{\subfigure[$\Gamma(T) $]{\includegraphics[width=0.53\textwidth]{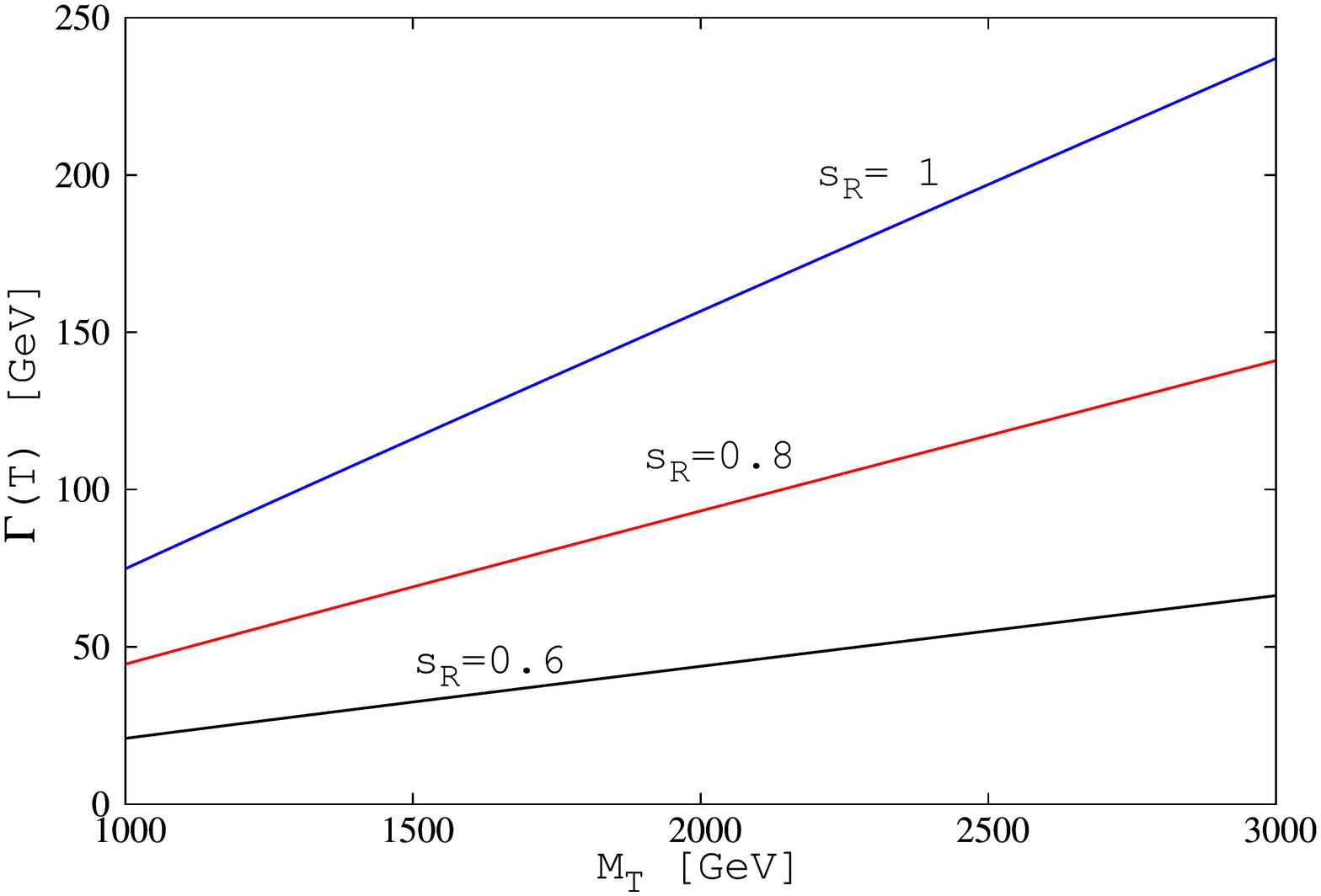}}
\subfigure[$\Gamma(\tilde{T})$]{\includegraphics[width=0.53\textwidth]{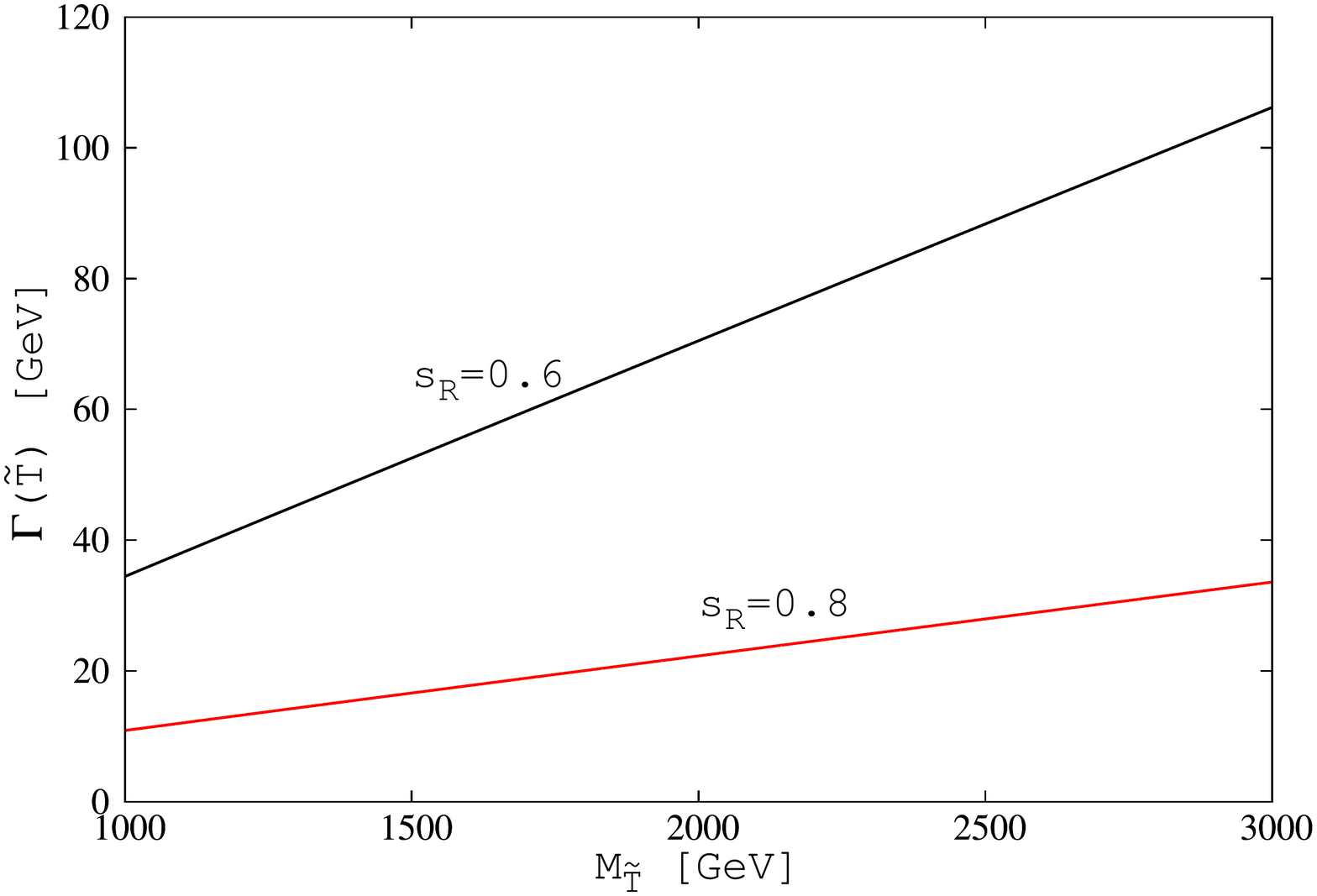}}
}
\caption{\textit{Heavy Fermion total decay widths in the TS5, as functions of the heavy fermions mass, for different values of the top degree of compositeness, $s_R$, and for $Y_*=3$. 
$(\Gamma(T)\simeq\Gamma(B))$.}}
\label{HFwidths}
\end{figure}

\noindent
For the Branching Ratios in the TS5, we find:
\[ 
 BR\left(T\rightarrow Z_Lt\right)\simeq 50\% \ \, \ \  BR\left(T\rightarrow ht\right)\simeq 50\% \ .
\]
\[ 
 BR\left(B\rightarrow W_Lt\right)\simeq 100\% 
\]
\[ 
 BR\left(\tilde{T}\rightarrow W_Lb\right)\simeq 50\% \ \, \ \ BR\left(\tilde{T}\rightarrow Z_Lt\right)\simeq 25\% \ \, \ \ BR\left(\tilde{t}\rightarrow ht\right)\simeq 25\%
\]
\[ 
 BR\left(\tilde{B}\rightarrow Z_Lb\right)\simeq 50\% \ \, \ \ BR\left(\tilde{B}\rightarrow hb\right)\simeq 50\%
\]

\noindent
The Branching Ratios in the TS10 are the same as in the TS5, apart from those of $\tilde{B}$, for which we have:
\[ 
 BR\left(\tilde{B}\rightarrow W_Lt\right)\simeq 50\% \ \, \ \ BR\left(\tilde{B}\rightarrow Z_Lb\right)\simeq 25\% \ \, \ \ BR\left(\tilde{B}\rightarrow hb\right)\simeq 25\%  \ (TS10)
\]

\noindent
Since the heavy fermions are much heavier than their decay products, these BRs are essentially independent on the mass of the heavy fermions.\\

\begin{figure}[]
\centering
\mbox{\subfigure[$Wtb$]{\includegraphics[width=0.7\textwidth]{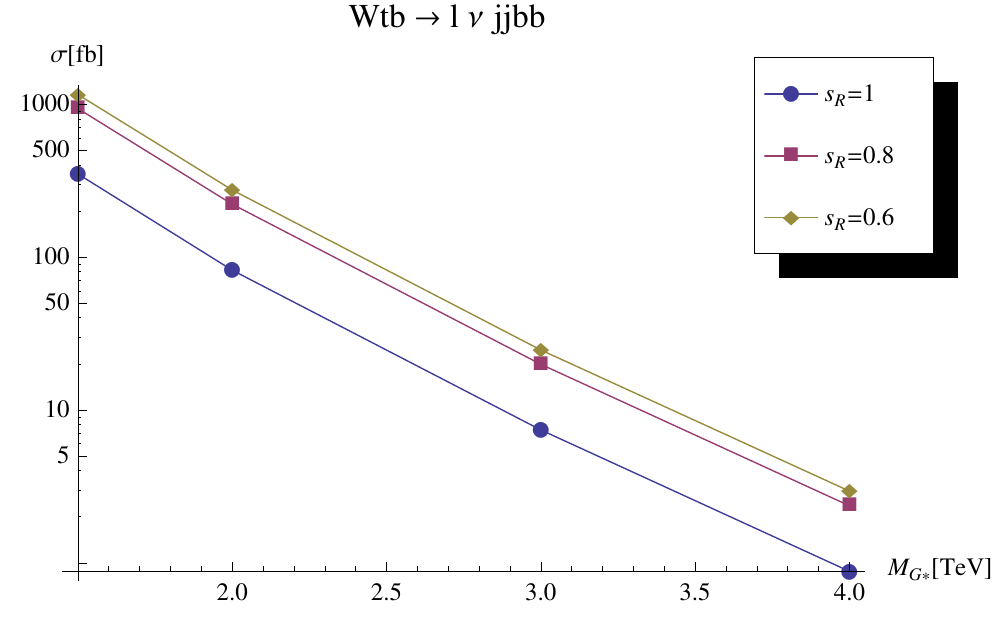}}}\\
\mbox{\subfigure[$Z(/h)t\bar{t}$]{\includegraphics[width=0.7\textwidth]{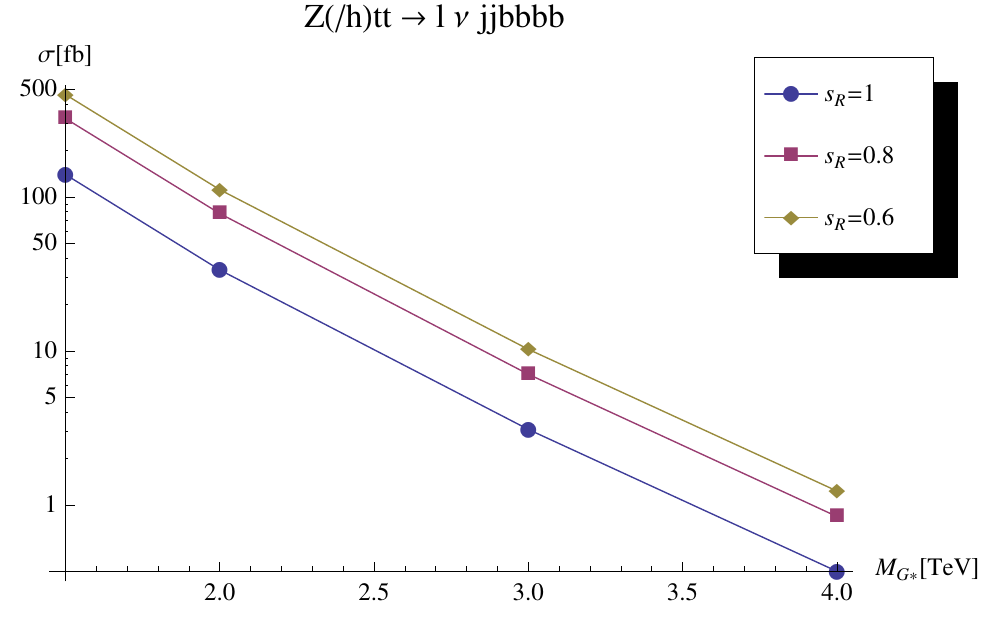}}}\\
\mbox{\subfigure[$Z(/h)b\bar{b}$]{\includegraphics[width=0.7\textwidth]{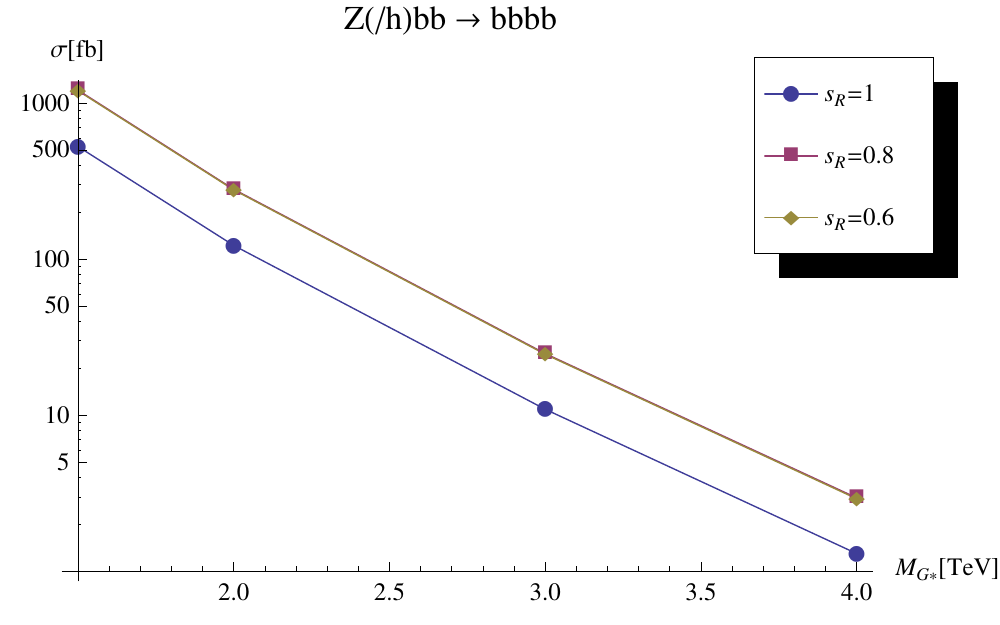}}}
\caption{\textit{Cross Section values (with only minimal cuts) for the three Search channels at the $14$ TeV LHC. We set $M_{G*}/M_T=1.5$. $l\equiv e/\mu$.}}
\label{xsec_channel}
\end{figure}

Considering this pattern of decays, we can identify three search channels: $Wtb$, $Z(/h)t\bar{t}$, $Z(/h)b\bar{b}$.
\begin{figure}[h!]
\begin{center}
\mbox{\subfigure[$Wtb$]{\includegraphics[width=0.6\textwidth]{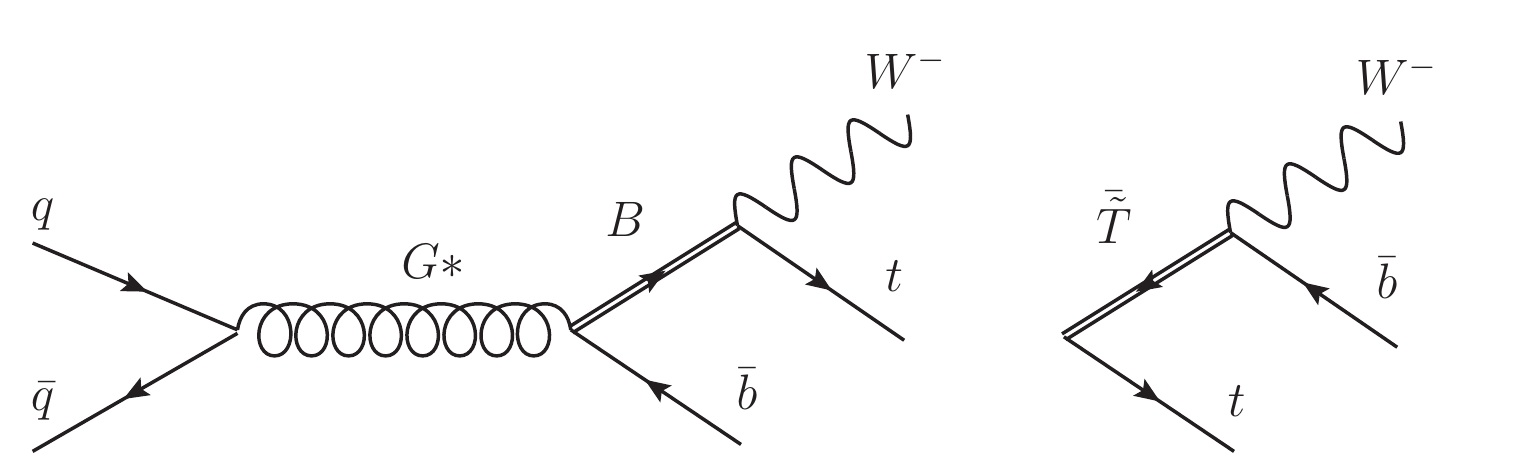}}}\\
\mbox{\subfigure[$Z(/h)t\bar{t}$]{\includegraphics[width=0.4\textwidth]{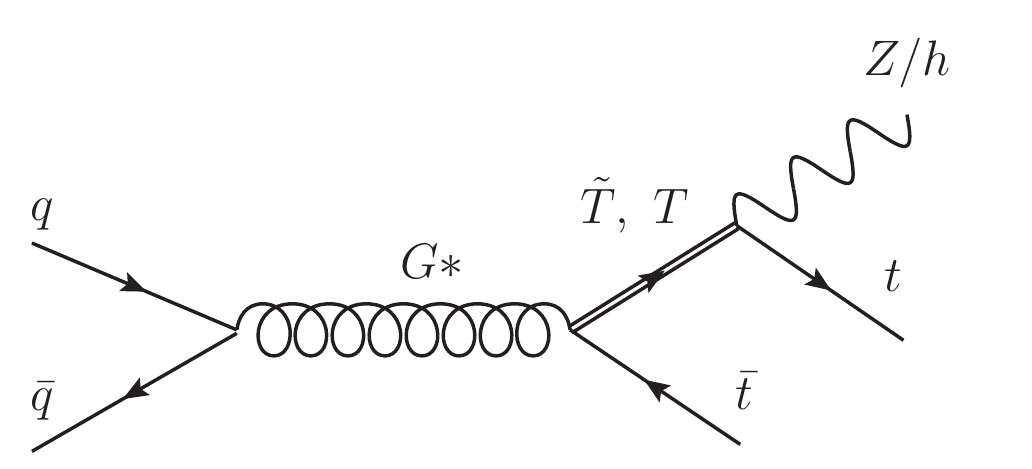}}
\subfigure[$Z(/h)b\bar{b}$]{\includegraphics[width=0.4\textwidth]{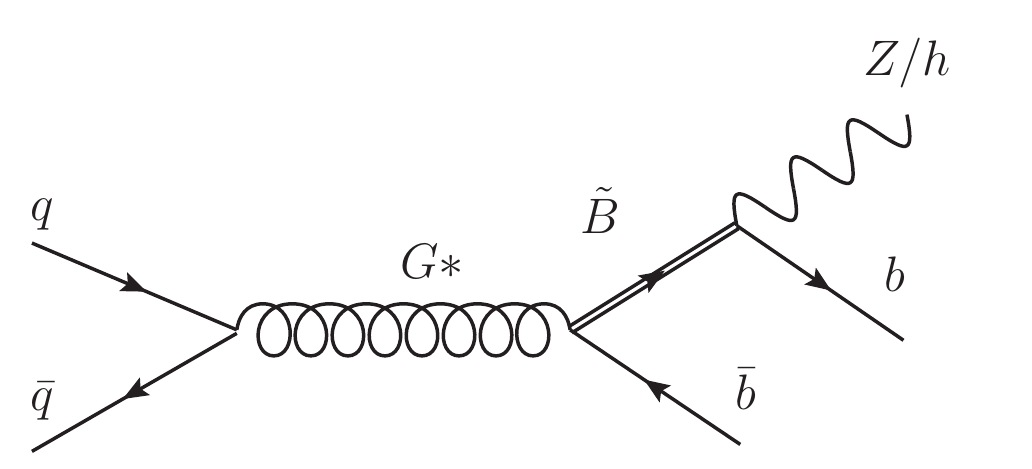}}}
\end{center}
\caption{\textit{The three $G^*\to \psi\chi$ Search channels.}}
\end{figure}\\
\noindent
The $Z(/h)b\bar{b}$ channel has a signal from $G^{*}$ only if $b_R$ is not too much elementary, therefore it could be a very interesting test to
establish the $b_R$ degree of compositeness. $Wtb$ and $Z(/h)t\bar{t}$ channels, besides being promising for extracting information on the model parameters,
 prove to be very good channels for the $G^{*}$ discovery. Indeed, the presence in the signal of heavy fermion resonances allows for a clean distinction
between the signal and the background; the $Z(/h)b\bar{b}$ and $Z(/h)t\bar{t}$ channels could be also promising for the Higgs searches. \\
An analysis of all these channels should be very interesting. We will focus our analysis on one of them. In order to select
the most promising one for a quick discovery at the LHC, we look first at the cross section values for these search channels at the LHC. \\
We show in Fig. \ref{xsec_channel} the cross section values for the three search channels at the $14$ TeV LHC.
These values have been obtained from a Monte Carlo simulation, using MADGRAPH/MADEVENT v4 (with only minimal cuts\footnote{
We apply only a cut on the bottom transverse momentum, $p_{Tb}>20$ GeV. We set factorization and renormalization scales, $Q=M_{G*}$.
}). We implemented our TS5 model in MADGRAPH, making use of the FeynRules package. 
We have calculated the cross section values of the $pp\to G^* \to \chi\psi \to$ $Wtb$, $Z(/h)tt$, $Z/(h)bb$ processes, then we have multiplied them by the theoretical
values for the BRs of the decays: $Wtb\to l\nu jjbb$, $Z(/h)tt\to(Z(/h)\to bb)l\nu jjbb$, $Z(/h)bb\to(Z(/h)\to bb)bb$.
These decays are those we guess to be the most suitable for a search analysis at the LHC. the $Z$ decays into leptons would bring to cleanest final states,
but to rather low values of the cross section, $\frac{BR(Z\to l^+l^-)}{BR(Z\to b\bar{b})}\simeq 0.44$. 
The semileptonic decay channel $Wtb\to l\nu jjbb$ is considered the gold channel in the studies of $t\bar{t}$ resonances; 
the cross section for semileptonic decay is lower than the one from a fully hadronically decay; 
the semileptonic channel, however, differently from the all-hadronic one, has not to confront the large dijet background. 
A study of the all-hadronic channel could be interesting, even though a ``top-tagging'' procedure, as the one described in \cite{topTag}, should be needed.
The cross section values in Fig. \ref{xsec_channel} are shown as function of the heavy gluon mass.
 They are evaluated in the intermediate scenario where $M_T < M_{G*} < 2 M_T$; we fix the ratio between heavy gluon and heavy (doublet) fermion mass as $M_{G*}/M_T=1.5$.\\ 
 Fig. \ref{xsec_channel} shows an exponential decrease of the cross section with the $G^*$ mass.
 This is a consequence of the exponential decrease of the $G^*$ production cross section with $M_{G*}$. 
We remind that the heavy gluon is produced in a Drell-Yan process; the production by gluon fusion is forbidden at tree level, by gauge invariance.\\
As we can see from Fig. \ref{xsec_channel}, the $Wtb$ channel has the highest cross section values. 
An analysis of the $Wtb$ channel is simpler than an analysis of the $Z(/h)tt$ channel; 
in particular, there is a lower number of final states and of jets. 
Moreover, the $Wtb$ channel has the advantage to include the decay of the $G^*$ into top pairs, which is the usual search channel for the $G^*$ discovery.
This, as we will discuss afterwards, could provide hints on the $t_R$ degree of compositeness.   
The $Z(/h)bb$ channel has also high cross section values, but, as already said, it strongly depends on the $b_R$ degree of compositeness 
($BR(G^* \to \tilde{B}b_R)\propto s^2_{bR}$). \\

\begin{figure}[]
\centering 
\subfigure[$s_R=1$]{\includegraphics[width=0.65\textwidth]{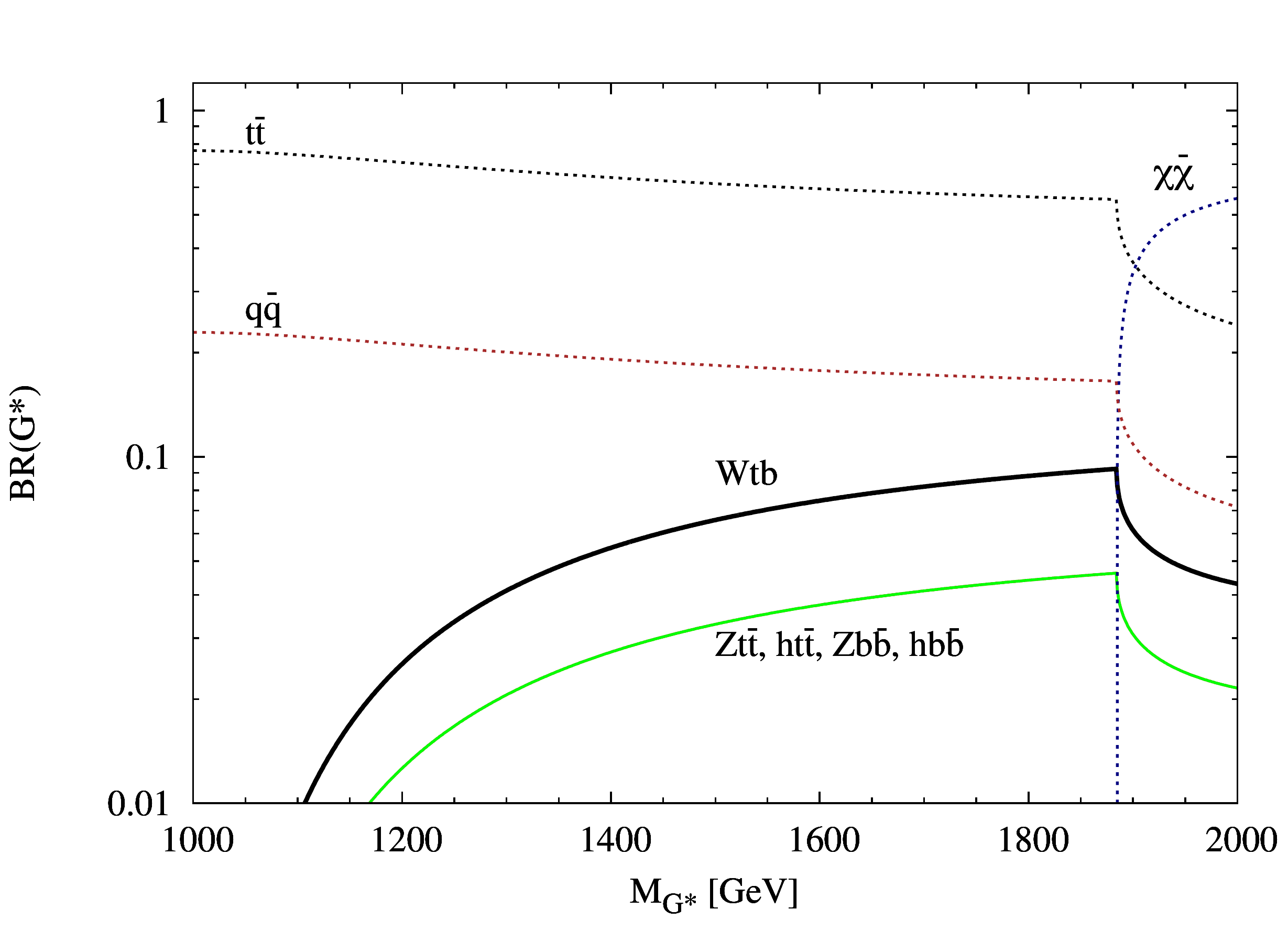}}\\
\subfigure[$s_R=0.8$]{\includegraphics[width=0.65\textwidth]{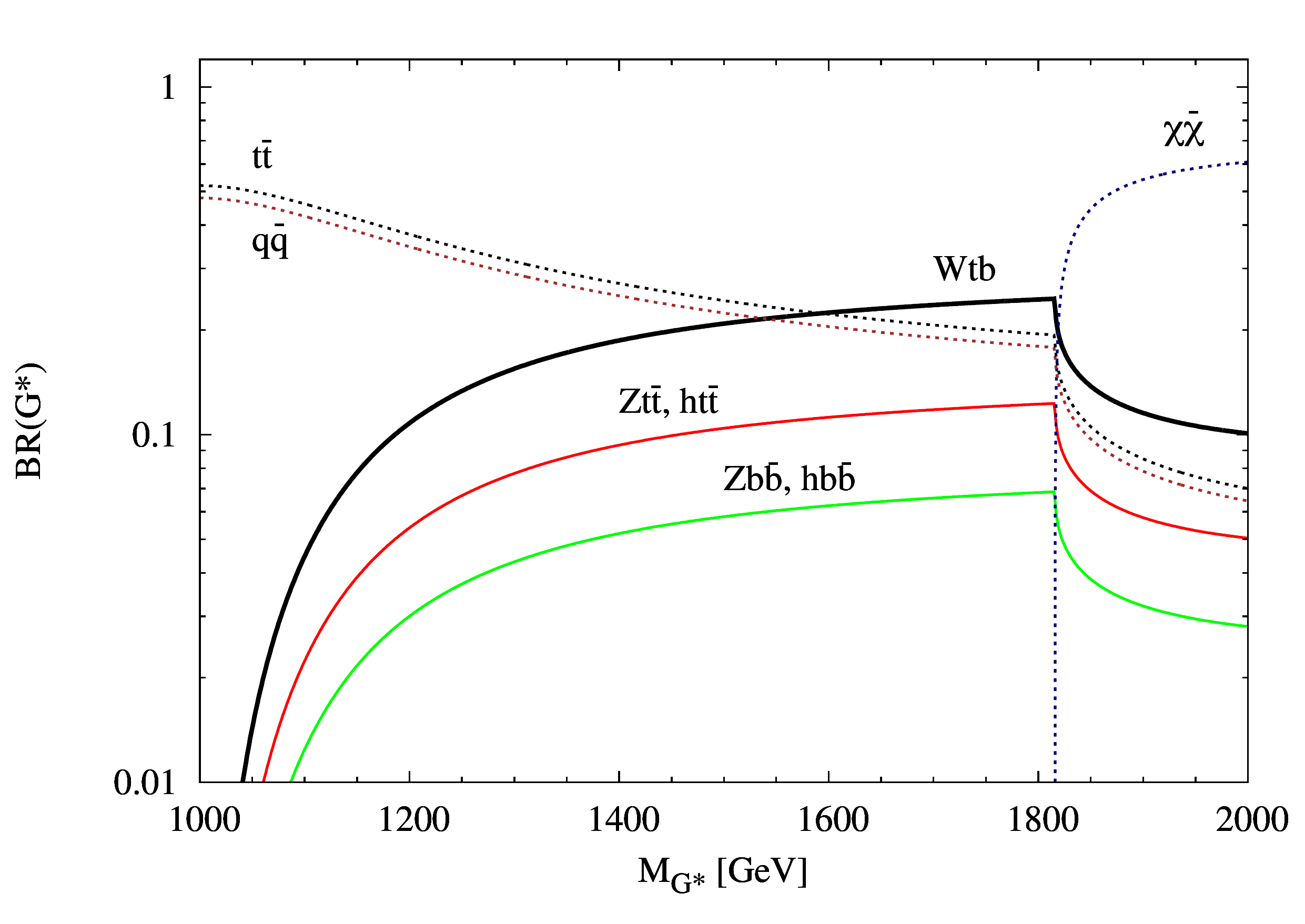}}\\
\subfigure[$s_R=0.6$]{\includegraphics[width=0.65\textwidth]{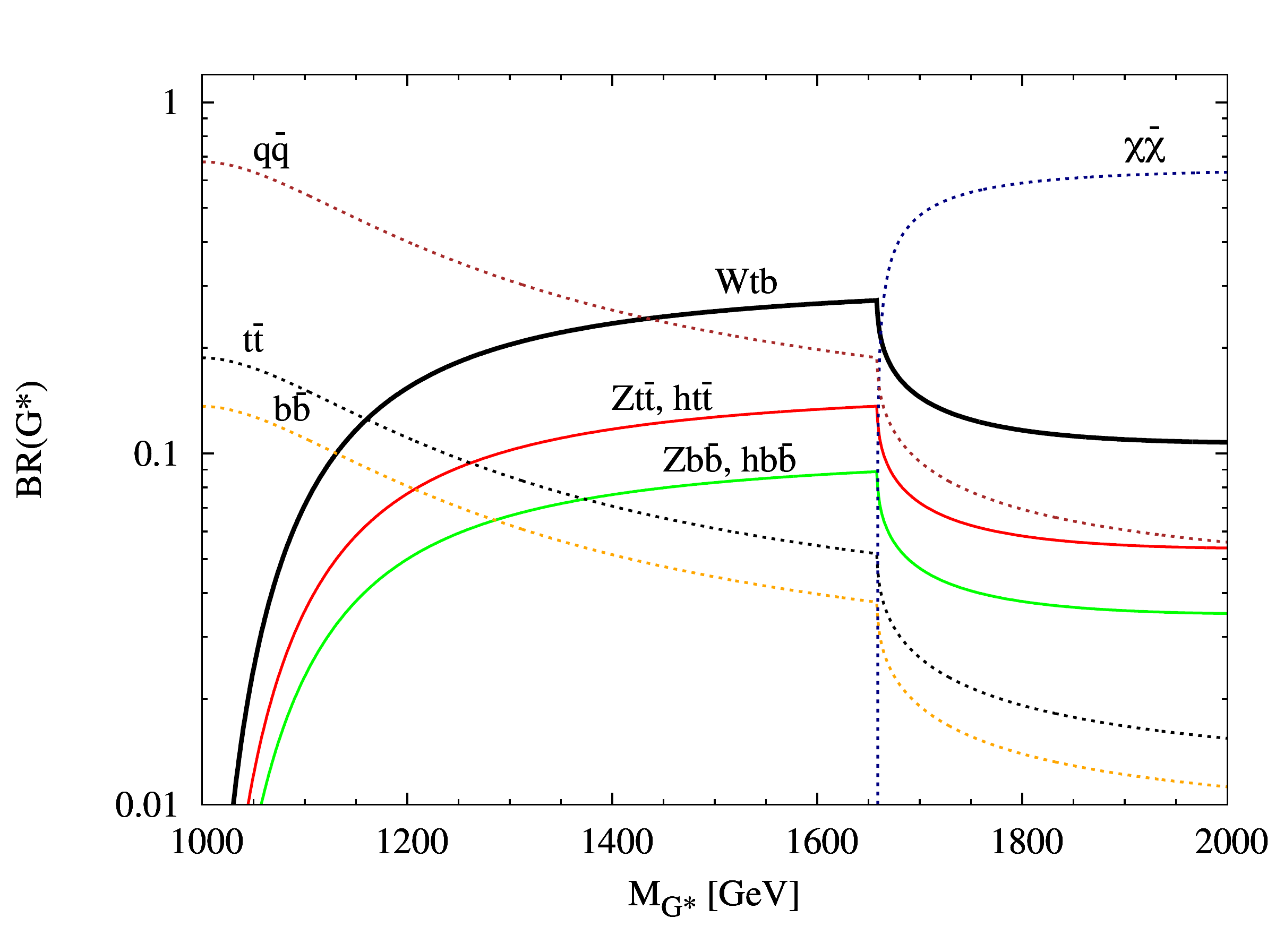}}
\caption{\textit{BRs for the $G^*$ decays in the different search channels, as functions of $M_{G^*}$, for several values of the top degrees of compositeness, 
considering the intermediate scenario where $M_T<M_{G*}<2 M_T$. We set $M_T=1$ TeV, $\tan\theta_3=0.44$, $Y_*=3$. 
The decay channel $ht\bar t$ ($hb\bar b$) has the same BR of $Zt\bar t$ ($Zb\bar b$).}}
\label{BR_channels}
\end{figure}

\begin{figure}[tbp]
\begin{center}
\includegraphics[width=0.495\textwidth,clip,angle=0]{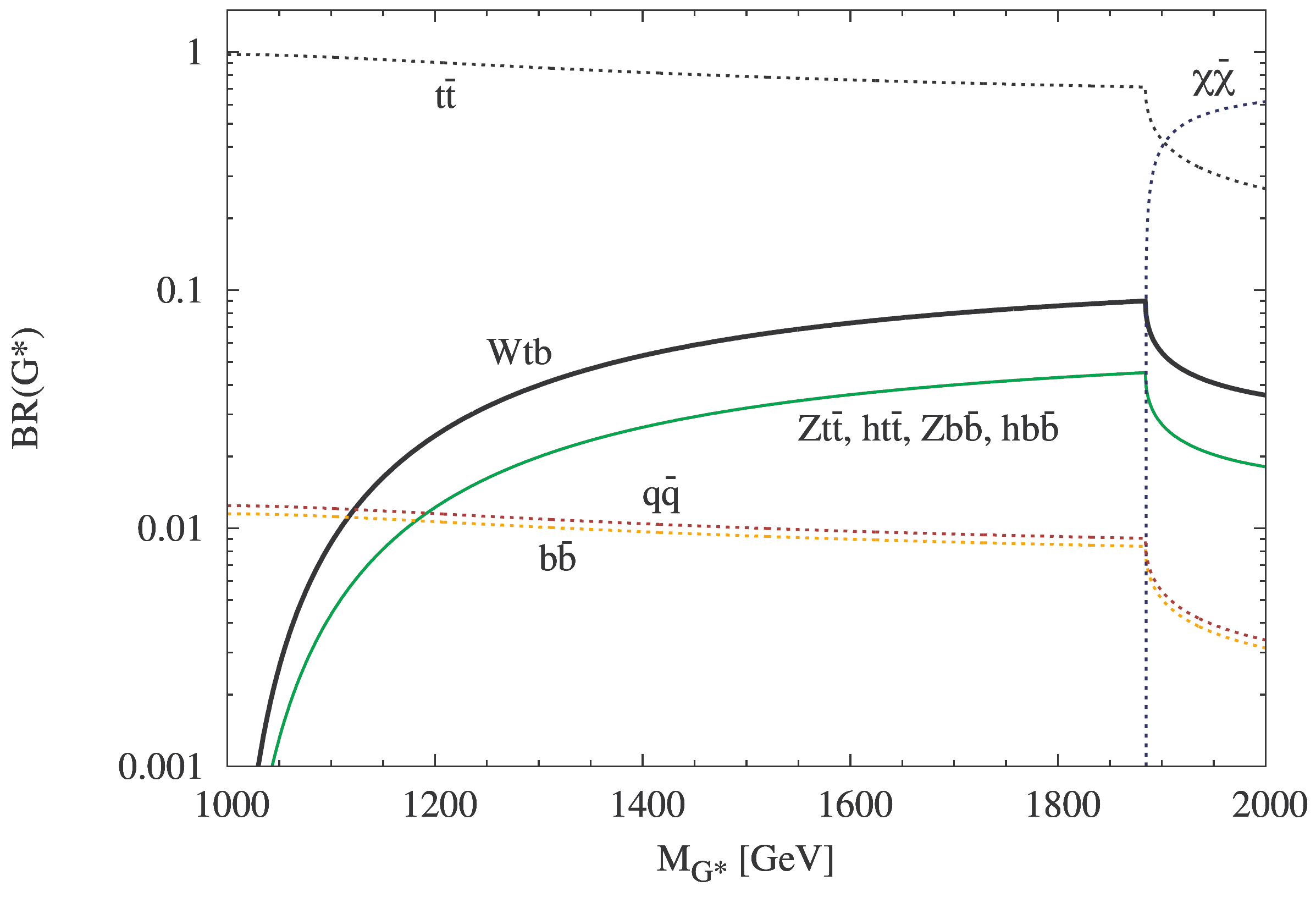}
\includegraphics[width=0.495\textwidth,clip,angle=0]{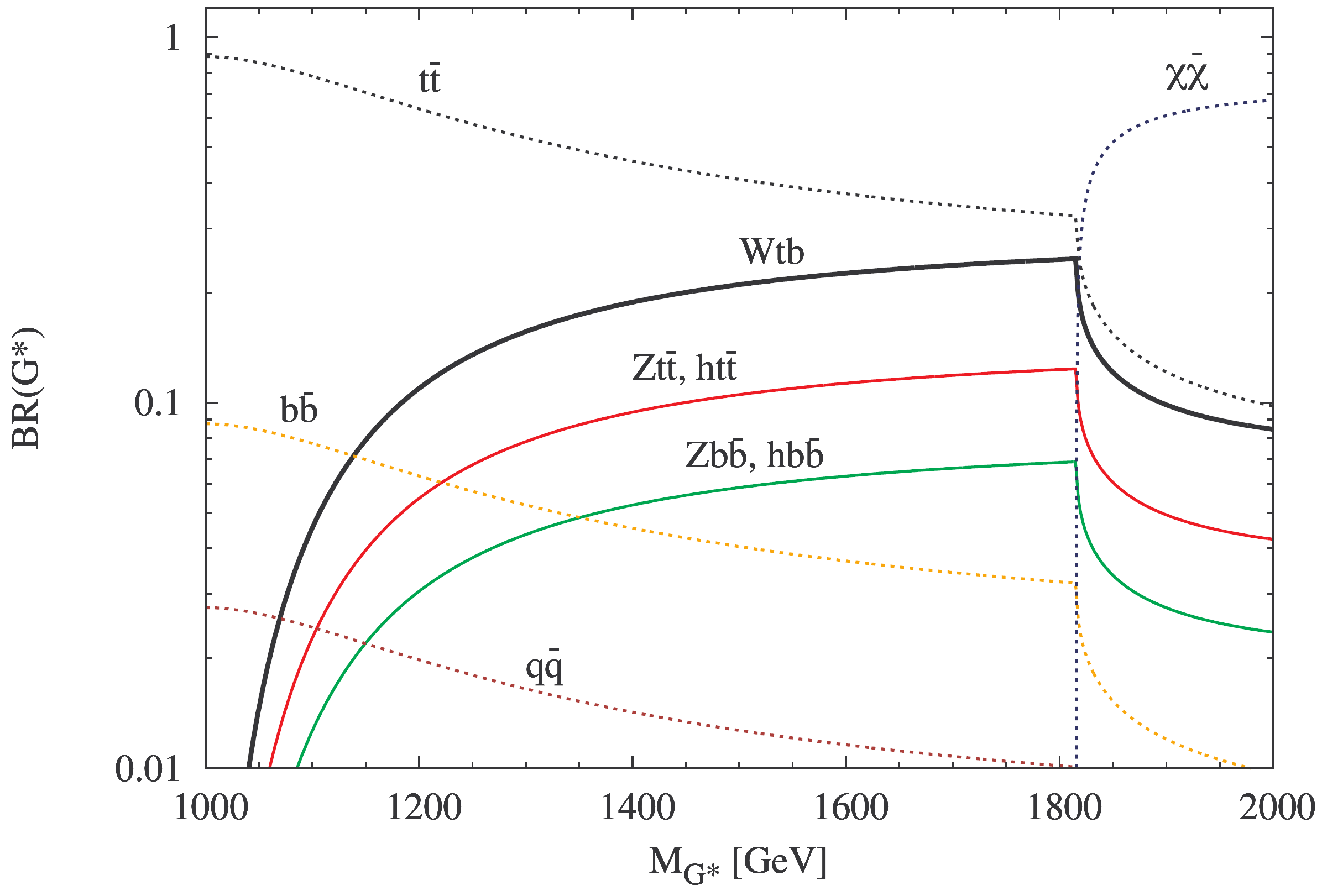}
\caption[]{
\label{fig:hldecays2}
\textit{
BRs of $G^*$ to the various final channels as functions of $M_{G^*}$.
Compared to Fig.s~\ref{BR_channels}, we set a smaller $\tan\theta_3$ value, $\tan\theta_3=0.2$. 
The plot on the left is obtained for a fully composite $t_R$, $s_R=1$, that on the right for $s_R=0.8$.
Both plots are done fixing $M_{T} =1\,$TeV and $Y_* =3$.}
}
\end{center}
\end{figure}

We show in Fig. \ref{BR_channels} the BRs for the $G^*$ decays in all the possible channels (in the TS5 model), for several top degrees of compositeness. 
We consider the intermediate scenario ($M_T<M_{G*}<2 M_T$) and we set: $\tan\theta_3=0.44$ and $Y_*=3$. This clarifies the yield for the different search channels. 
In the figure, $t\bar{t}$ denotes the $G^*$ decay to top pairs and is not included in $Wtb$, 
which denotes only the heavy-light decays $G^*\to\psi\chi\to Wtb$. 
We will further discuss in sec. \ref{Hint_sR} the relative amount of the $\psi\chi$ and $t\bar{t}$ components in the $G^*\to Wtb$ signature. 
$q\bar{q}$ denotes the BR for the $G^*$ decays into pairs of light quarks, $q=u,d,c,s$.
$\chi\chi$ those for the $G^*$ decays into pairs of custodian heavy fermions; these heavy fermions, we remind, 
become increasingly lighter when the degree of compositeness of the left-handed top increases (i.e., when $s_R$ decreases). \\  
Finally, Fig.s \ref{fig:hldecays2}, obtained for $\tan\theta_3=0.2$, clearly illustrate that, while large branching fractions for the heavy-light decays
 are mainly implied by the kinematics and as such are a robust prediction,
the value of $BR(G^* \to q\bar q)$ is strongly dependent on $\tan\theta_3$ and can thus be easily made small. 
The left plot refers to the same benchmark point adopted in Ref.~\cite{Agashe}, $s_{R} = 1$, $\tan\theta_3 =0.2$, 
$Y_* =3$ ($\sin\varphi_{L1} = 0.33$); in this case the $t\bar t$ channel largely dominates over the others until the threshold 
$M_{G^*} = 2 M_{T_{5/3}}$, as a consequence of the full degree of compositeness of the right-handed top. 
The right plot, obtained for a slightly smaller degree of compositeness of $t_R$, $s_R=0.8$, and for $\tan\theta_3 =0.2$, 
$Y_* =3$, however, shows that, with just a small variation from the case of a fully composite $t_R$, 
the $t\bar t$ branching ratio is substantially reduced while those of the heavy-light channels, especially $Wtb$, are sizable.\\

Taking into account all these considerations, we focus our analysis on the $Wtb$ channel.
An analysis of the other heavy-light channels, anyway, remains very interesting, especially if heavy colored vectors are discovered in the $Wtb$ channel.
 It can provide further information on the model and a check of the model predictions. \\

\captionsetup[subfigure]{position=bottom}

\section{Analysis of the $Wtb$ channel}
We analyze the $Wtb$ channel in the intermediate scenario (and in particular in the region $M_T<M_{G*}<2 M_{T5/3}$). 
We fix the ratio between the $G^*$ and the heavy fermions mass at the value $M_{G^*}/M_{T}=1.5$ and we consider several $G^*$ masses.
Because of the Minimal Flavor Violating bound of $\sim1$ TeV on the mass of the $D\equiv (T,B)$ heavy fermions that we have obtained from $b\to s \gamma$ in the first part of this project,
we do not consider $M_{T}$ values lower than $1$ TeV. Therefore, we focus our analysis on a region $M_{G*}\gtrsim 1.5$ TeV. \\ 
The constraint on $G^*$ mass coming from $K-\bar{K}$ mixing, that we have discussed on section \ref{KK}, is quite strong, $M_{G*}\gtrsim (11)\frac{g_{*3}}{Y_{*}}$ TeV
\footnote{in the TS5, we remind, there is also a $s_1/s_2$ enhancement of this bound.}, though it could still allow for a relatively light $G^*$,
 if we had a ratio $\frac{g_{*3}}{Y_{*}}\ll 1$.
Most importantly, this is a bound that strongly depends on the flavor structure of the sector Beyond the SM. If we consider BSM flavor scenario different from
the one of anarchic $Y_*$ (where it is assumed that there is no
large hierarchy between elements within each matrix $Y_*$ and the quark mass hierarchy is completely explained by the elementary/composite mixing angles), 
the constraint can be much softened. 
For example, recent studies show that, if the strong sector is invariant under additional flavor 
symmetries \cite{Fitzpatrick:2007sa, Santiago:2008vq, Csaki, Csaki:2008eh, Csaki:2009wc}
or preserves CP~\cite{Redi_Weiler}, the heavy gluon can have a mass as light as a few TeVs and its phenomenology is not qualitatively
modified. In the following we will assume that some mechanism is at work to alleviate the flavor bounds on $M_{G*}$.\\
We will consider in the analysis an intermediate degree of compositeness for the top, $s_R=0.6$, and we will fix $Y_{*}=3$ and $\tan\theta_3 = 0.44$, 
this latter corresponds to $g_{*3}\simeq3$ for $\alpha_S (M_Z)=0.118$. 
$g_{*3}=Y_{*}=3$ are reference values for composite models. Summing up, we will consider in the analysis the following set of parameters and assumptions:
\begin{gather}
 \tan\theta_3 = 0.44 \ \ \  s_R=0.6 \ \ \ Y_*=3 \nonumber \\
M_{G*}/M_{T}=1.5\ \ \ M_T=M_{\tilde{B}}=M_{\tilde{T}}
\label{SetPar}
\end{gather}
Our final results will be quite independent of the specific values of $Y_*$, since the 
latter determine basically only the decay width of the heavy fermions
\footnote{We point out, however, that a dependence on $Y_*$ is implicit in the evaluation of the $q^3_L$ degree of compositeness, $s_1$. 
In our model, $s_1=\frac{\sqrt{2}m_t}{Y_* v s_R}$. This implies a dependence of the $G^*$ decay BRs on $Y_*$, 
that is anyway soft, if we remain in a scenario of a not fully composite $q^3_L$.}. 
 We will discuss on Sec. \ref{ParSpace} the dependence on the $\tan\theta_3$ and $s_R$ values. \\
\begin{figure}[t]
\centering
\includegraphics[width=0.8\textwidth]{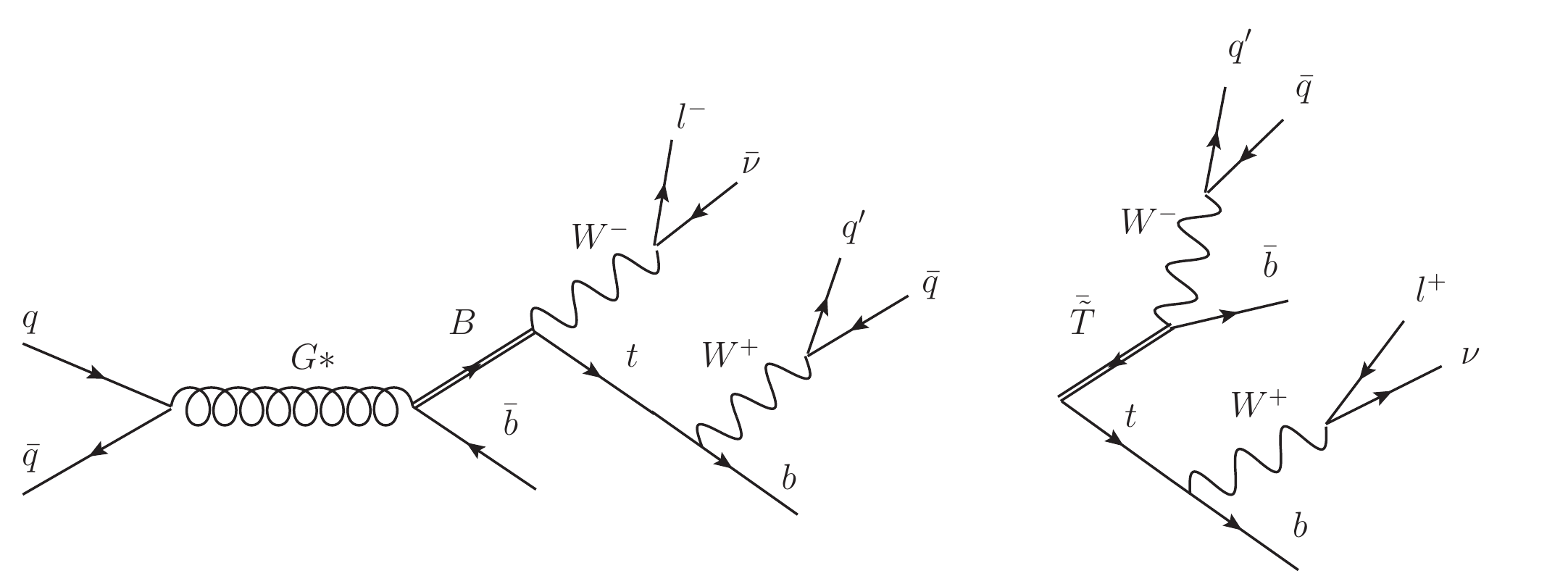}
\caption{\textit{$\chi\psi$ Signal in the $Wtb$ channel.}}
\label{wtb_S}
\end{figure}
The $\chi\psi$ component of the $Wtb$ signal is made for two third of a left-handed bottom with its heavy partner $B$ and for one third
 of a right-handed top with its heavy partner $\tilde{T}$.   
\[
 pp \to G^* \to \chi\psi \equiv \ b_L B_L \ (67 \%) \ , \ \ t_R \tilde{T}_R \ (33 \%)\ \to Wtb
\]
The percentages
we indicate refer to the TS5 model with the assumptions: $s_{bR}=s_{1}$, $s_R =0.6$. 
The $\psi\chi$ composition is model dependent and can vary with the different models.
The other component of the $Wtb$ signal comes from the $G^* \to \bar{t}t$ decays. The importance of this component depends on the top degree of compositeness 
(we will deepen this question on sec. \ref{Hint_sR}). For $s_R=0.6$ the $G^*$ decays into $t\bar{t}$ pairs constitutes about the $19\%$ of the $Wtb$ signal (before the acceptance cuts). 

\subsection{Signal and Background Simulation}

We look for the Signal
\begin{equation}
 q\bar{q} \to G^* \to \chi\bar{\psi}+\psi\bar{\chi} \ (+t\bar{t}) \to l^{\pm}\nu b\bar{b}q\bar{q}'
\end{equation} 
at the hard scattering level. The physical final state observed in the ATLAS or CMS detector is of the form:
\begin{equation}
 l^{\pm}+ n\ jets+ \not{E}_T\ \ , l=e,\mu\ .
\label{final_states}
\end{equation}   
We generate both the signal and the SM background events at the partonic level. We do not include detector effects in our analysis, except for a gaussian smearing on the jets;
 we smear both the jet energy and momentum absolute value by $\Delta E/E = 100\% / \sqrt{E/GeV}$, and the jet momentum direction using an angle resolution $\Delta\phi= 0.05$
radians and $\Delta\eta = 0.04$.\\
Signal events have been generated with MADGRAPH. The background to our signal comes mostly from $WWb\bar{b}$, 
which includes the resonant sub-processes $Wtb\to WWbb$ (single top) and $t\bar{t}\to WWbb$; the latter, in particular, 
gives the largest contribution after the acceptance cuts of eq. (\ref{AcceptanceCut}).
other relevant backgrounds are $W(\to l\nu)b\bar{b}+jets$ and $W(\to l\nu)+jets$. We do not include other reducible backgrounds which are expected to be subdominant,
in particular: $b\bar{b} + jets$, where one light jets is misreconstructed as a lepton (it should be
possible to reduce it down to a negligible level by requiring enough missing energy in the
event); single-top processes, $t + jets$, $tb + jets$, $Wt + jets$ (after the request of two $b$-tags all
these backgrounds are expected to be much smaller than $Wtb$ \footnote{See for example Table 32 at page 32 of Ref. \cite{Alpgen}}, 
which is included in our
analysis)
\footnote{Other backgrounds as $WW+jets$ and $WZ+jets$ can be neglected as well, once a b-tagging procedure is applied. 
The cross section for $WW+jets$ is only the $\sim 2\%$ of the cross section for $W+jets$. 
The cross section for $WZ+jets$ is less than $1/2$ of the $WW+jets$ cross section. 
The component $WZ(\to b\bar{b})+jets$ is not suppressed by the b-tagging,
 but it has a low cross section ($\sim 950 fb$ after the acceptance cuts) compared with the other backgrounds. Moreover it can be easily suppressed 
by a cut on the $b$ pairs invariant mass, $M_{bb}\gtrsim M_Z$, that has practically no effect on the signal.}. 
We have simulated the background $WWb\bar{b}$ with MADGRAPH and the remaining background samples with ALPGEN \cite{Alpgen}\footnote{
We have used the CTEQ6L1 pdf set. 
The samples for the $Wtb$ signal and the background $WWbb$ have been generated in MADGRAPH using respectively the factorization and renormalization scales, $Q=M_{G*}$ and $Q=\sqrt{M^2_W+\sum_b p^2_T}$, where the last sum is over the $p_T$ of the $b$ jets. 
The samples for $Wbb+jets$ and $W+jets$ have been generated in ALPGEN using the scale $Q^2=M^2_W + p^2_{T,W}$.}. 
The background $Wbb+jets$ includes the samples $Wbb+1J$, $Wbb+2J$, $Wbb+3J$ and the $W+jets$ the samples $W+3J$ and $W+4J$. 
Including all these samples with increasing multiplicity of light jets in the final state is redundant, and in
principle leads to a double counting of kinematic configurations. A correct procedure would
be resumming soft and collinear emissions by means of a parton shower, and follow some
matching technique to avoid double counting. In our analysis we retain all the $Wbb+nJets$ and $W+nJets$
samples, but the cuts we will impose suppress the events with larger number of jets and
thus strongly reduce the amount of double counting. Moreover, the $W + jets$ background will turn out to be much smaller
than the others at the end of our analysis, so that in this case the issue of double counting
can be safely ignored.\\
The number of reconstructed jets in the detector ($n$ in the eq. \ref{final_states}) depends in general on the adopted jet algorithm and on its parameters. 
In our simple parton-level analysis jets are identified by matching them directly on the quarks and gluons from the hard scattering and imposing
a set of acceptance and isolation cuts. We will require $n \ge 3$ and two among the jets to be $b$-jets. This choice is motivated by the following analysis. \\
We calculate the number of reconstructed jets in the signal events for two different jet isolation values: $\Delta R_{jj'}\gtrsim 0.4$ and $\Delta R_{jj'}\gtrsim0.7$. 
If two jets are
closer than the separation $\Delta_R$, they are merged into a single jet whose four-momentum
is the vectorial sum of the jet momenta. 
The results in table \ref{jetNum} show that in the most part of the signal events (about the $50\%$), the two light-jets, coming from the $W$,
 merge into a single `multiple' jet. 
This is a feature of processes where a particle, as the hadronically decaying $W$ of our signal, comes from the decays of much heavier particles, such as the $G^*$ or the heavy fermions.
As a consequence, this particle is very boosted and it decays to collimated particles that look like a single jet.
We can notice that, as naively expected due to the increasing boost of all the decay products, the number of
events with four jets decreases for larger $G^*$ masses, while the fraction of events in which
the hadronic $W$ is reconstructed as a single jet increases.
 We see also that for the cases of heavier $G^*$ and composite fermions, a relevant part of the b-jets merges with the two light-jets in a single multiple-jet.
This situation corresponds to having a very boosted top that decays into collimated particles. 
In order to analyze this type of events a specific strategy of top-tagging, as the one described in \cite{topTag}, is needed. 
The use of boosted jet techniques can lead to a better sensitivity on events
with three jets, and allow the study of the extreme case, most probable at high $G^*$ masses,
in which the hadronically decayed top is reconstructed as a single fat jet. On the other hand, 
while the SM (top or bottom) quark originating from the $G^*$ decay is always
highly boosted, for $M_{G*}/M_\chi =1.5$ (as we will consider in our analysis) the heavy quark is not 
(we will return to this aspect in sec. \ref{ParSpace}). 
As a consequence, an event selection strategy which relies on a large boost of all the decay
products, adopted for example by some of the LHC searches for heavy resonances decaying
to $t\bar{t}$ \cite{LHC_tt}, might have a poor efficiency on our topology of signal events.\\
Considering all we have discussed, in order to catch the most part of the signal, we need a semi-inclusive analysis: we require at least three jets, reconstructed by considering a jet 
cone size $\Delta R=0.4$. We also require a two $b$-jets tagging:
\begin{equation}
 pp\to l^{\pm}+ n\ jets+ \not{E}_T\ \ (l=e,\mu)\ \ , n\geq 3\ ,\ 2\ b\ \text{tags}
\label{final_states_2}
\end{equation}   
The 2 $b$ tags requirement allows for a very effective reduction 
of large backgrounds, such as the $W+jets$. 
The requirement is also important to being able to identify the $Wtb$ final states of our signal and, therefore, to recognize, 
as we will show in the next sections, the heavy fermion resonances. 

\begin{table}
\begin{tabular}[]{cccc}
\multicolumn{1}{c}{}&\multicolumn{1}{c}{}&\multicolumn{1}{c|}{$\Delta R=0.4$}&\multicolumn{1}{c}{$\Delta R=0.7$}\\
\hline
\hline
\multicolumn{1}{c|}{$M_{G*}=1.5$ TeV}&\multicolumn{1}{c|}{2j+2b}&\multicolumn{1}{c|}{$42\%$}&\multicolumn{1}{c|}{$12\%$}\\
\cline{2-4}
\multicolumn{1}{c|}{}&\multicolumn{1}{c|}{1j+2b}&\multicolumn{1}{c|}{$31\%$ (1Mj+2b $18\%$)}&\multicolumn{1}{c|}{$49\%$ (1Mj+2b $42\%$)}\\
\cline{2-4}
\multicolumn{1}{c|}{}&\multicolumn{1}{c|}{1j+1b}&\multicolumn{1}{c|}{$7.0\%$}&\multicolumn{1}{c|}{$19\%$}\\
\hline
\hline
\multicolumn{1}{c|}{$M_{G*}=2$ TeV}&\multicolumn{1}{c|}{2j+2b}&\multicolumn{1}{c|}{$29\%$}&\multicolumn{1}{c|}{$6\%$}\\
\cline{2-4}
\multicolumn{1}{c|}{}&\multicolumn{1}{c|}{1j+2b}&\multicolumn{1}{c|}{$42\%$ (1Mj+2b $33\%$)}&\multicolumn{1}{c|}{$49\%$ (1Mj+2b $45\%$)}\\
\cline{2-4}
\multicolumn{1}{c|}{}&\multicolumn{1}{c|}{1j+1b}&\multicolumn{1}{c|}{$10\%$}&\multicolumn{1}{c|}{$27\%$}\\
\hline
\hline
\multicolumn{1}{c|}{$M_{G*}=3$ TeV}&\multicolumn{1}{c|}{2j+2b}&\multicolumn{1}{c|}{$13\%$}&\multicolumn{1}{c|}{$2.7\%$}\\
\cline{2-4}
\multicolumn{1}{c|}{}&\multicolumn{1}{c|}{1j+2b}&\multicolumn{1}{c|}{$52\%$ (1Mj+2b $46\%$)}&\multicolumn{1}{c|}{$47\%$ (1Mj+2b $45\%$)}\\
\cline{2-4}
\multicolumn{1}{c|}{}&\multicolumn{1}{c|}{1j+1b}&\multicolumn{1}{c|}{$17\%$}&\multicolumn{1}{c|}{$37\%$}\\
\hline
\hline
\multicolumn{1}{c|}{$M_{G*}=4$ TeV}&\multicolumn{1}{c|}{2j+2b}&\multicolumn{1}{c|}{$7.5\%$}&\multicolumn{1}{c|}{$1.5\%$}\\
\cline{2-4}
\multicolumn{1}{c|}{}&\multicolumn{1}{c|}{1j+2b}&\multicolumn{1}{c|}{$53\%$ (1Mj+2b $49\%$)}&\multicolumn{1}{c|}{$46\%$ (1Mj+2b $45\%$)}\\
\cline{2-4}
\multicolumn{1}{c|}{}&\multicolumn{1}{c|}{1j+1b}&\multicolumn{1}{c|}{$25\%$}&\multicolumn{1}{c|}{$42\%$}\\
\hline
\hline
\end{tabular}
\caption{ \textit{Percentages of the number of $G^*\to\psi\chi$ signal events where the jet content is reconstructed to be respectively $2j+2b$, $1j+2b$ and $1j+1b$, 
as a function of the $G^*$ mass for $\sqrt{s}=14$ TeV. $Mj$ denotes a `multiple' jet. The reconstructed jet has to be 
inside the hadronic calorimeter, we require $|\eta_j|<5$ ($|\eta_b|<2.5$, for the b-tagging), and it has to be not too soft, we require $p_{Tj}>30$ GeV. }
}
\label{jetNum}
\end{table}

We will introduce first the strategy of analysis and the results for the nominal value of the LHC center-of-mass energy, $\sqrt{s}=14$ TeV.
Then (in sec. \ref{7TEV}), we will repeat the analysis considering the center-of-mass energy of the current runs at the LHC, $\sqrt{s}=7$ TeV.

\subsection{Results after acceptance cuts}

Our final states (\ref{final_states}) have to be detectable with the ATLAS and CMS experiments;
 the lepton (an electron or a muon) has to go inside the electromagnetic calorimeter or the muon chambers ($|\eta|<2.5$), otherwise it is missed. 
It has also to be enough energetic to activate the lepton trigger. The jets, as well, have not to go outside the hadronic calorimeter ($|\eta|<5$) 
and they have not to be soft ($p_{Tj}>30$ GeV), in order to be distinguished from the jets in the Initial State Radiation (ISR) or in underline events. 
Following the previous arguments, we require at least three jets, reconstructed by considering a jet cone size $\Delta R=0.4$. 
This means that the jets have to be separated from each other by an angular distance $\Delta R>0.4$; they have also to be isolated from the lepton ($\Delta R_{lj} > 0.4$).
In the case of b-jets, a rapidity value in the central rapidity region ($|\eta|<2.5$) is required for the b-tagging procedure.
Summing up, we do the following requirements. We require 1 lepton obeying 
\[
|\eta_l|<2.5 , \  p_{T l}> 20\ GeV
\]
and at least 3 jets, among which we require 2 b-jets, obeying
\begin{equation}
 \Delta R_{jj'} > 0.4 , \ \Delta R_{lj} > 0.4 , \  |\eta_j|<5 \ (|\eta_b|<2.5\ for\ the\ b-TAG), \ p_{T j}> 30\ GeV  .
\label{AcceptanceCut}
\end{equation}

\begin{figure}[h]
\centering
\includegraphics[width=0.4\textwidth, angle=90]{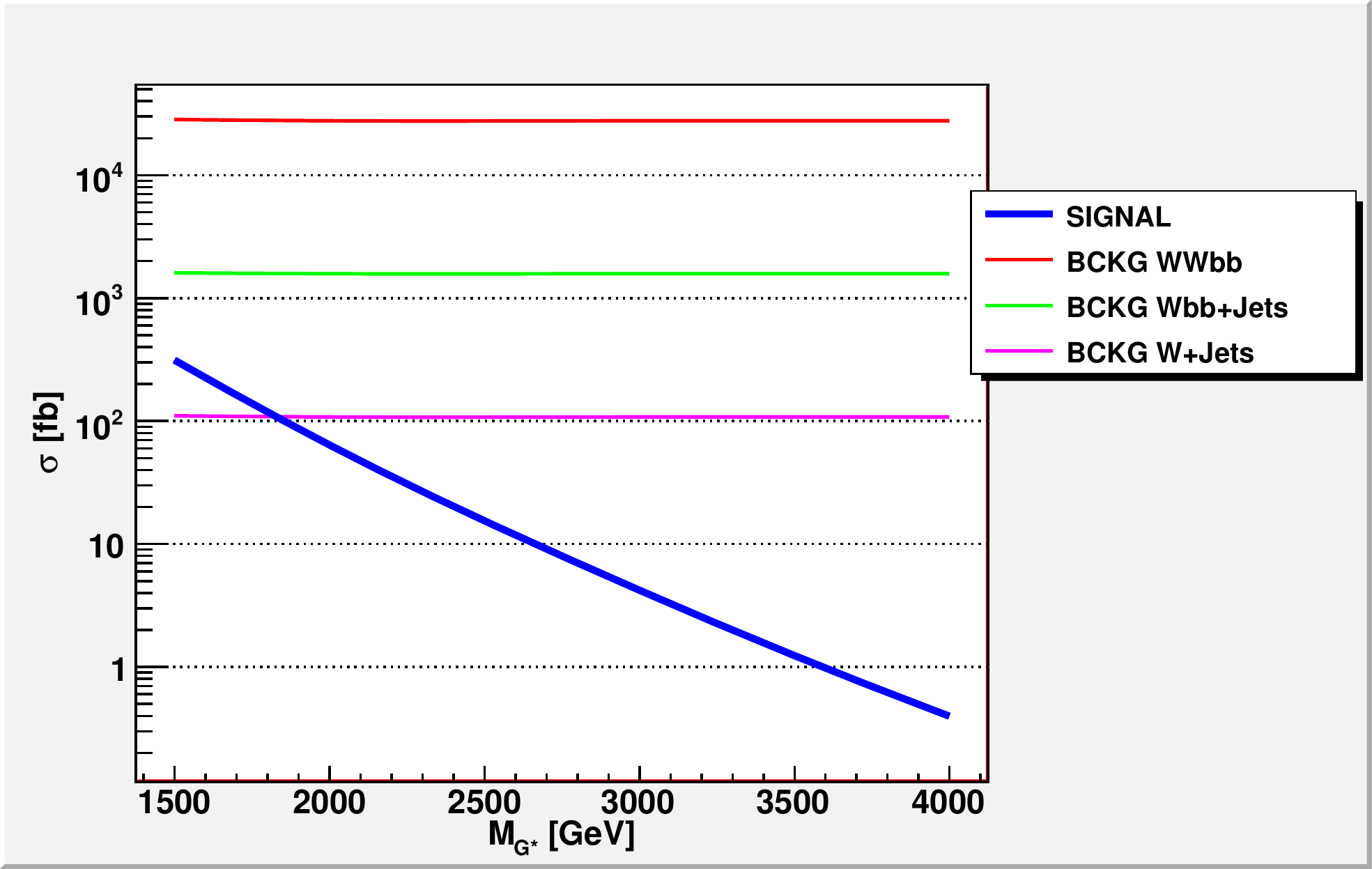}
\caption{\textit{Cross Section values after the Acceptance Cuts and the b-tagging ($\sqrt{s}=14$ TeV).}
}
\label{fig_XSEC_AC_Btag}
\end{figure}

The cross section values for the signal and the background after the application of the acceptance and isolation cuts in (\ref{AcceptanceCut}) are shown in table \ref{XSEC_AC-bTAG}.
We also estimate the cross section values after a procedure of b-tagging. 
We take into account a b-tagging efficiency of $60\%$ (that is of $36\%$ for the tagging of both the 2 b-jets) and a misidentification factor of $1/100$ for the light jets to be tagged as $b$ jets
\footnote{We also consider combinatorial factors in the light-jets misidentification.}. \\
We see that after b-tagging the main background is the $WWbb$, which is made up for the most part of $t\bar{t}$ events. 
$Wbb+jets$ events are also important, while the $W+jets$ events have a quite low cross section. The cross section for the signal, as expected, 
decrease exponentially with the increasing of the $G^*$ mass (fig. \ref{fig_XSEC_AC_Btag}).  

\begin{table}[t]
\subtable[SIGNAL]{
\begin{tabular}[]{ccccccc}
\multicolumn{1}{c}{}&\multicolumn{3}{||c}{ $\sigma\ [fb]$ After Acceptance Cuts} &\multicolumn{3}{||c}{ $\sigma\ [fb]$ After b-TAG}\\
\hline
\hline
\multicolumn{1}{c}{$M_{G*}=1.5$ TeV}&\multicolumn{1}{||c}{$\chi\psi$} &\multicolumn{1}{c}{728}&\multicolumn{1}{c}{846}&\multicolumn{1}{||c}{$\chi\psi$} &\multicolumn{1}{c}{262}&\multicolumn{1}{c}{305}\\
\cline{2-3}\cline{5-6}
\multicolumn{1}{c}{}&\multicolumn{1}{||c}{$t\bar{t}$} &\multicolumn{1}{c}{118}&\multicolumn{1}{c}{}&\multicolumn{1}{||c}{$t\bar{t}$} &\multicolumn{1}{c}{42.6}&\multicolumn{1}{c}{}\\
\hline
\hline
\multicolumn{1}{c}{$M_{G*}=2$ TeV}&\multicolumn{1}{||c}{$\chi\psi$} &\multicolumn{1}{c}{158}&\multicolumn{1}{c}{178}&\multicolumn{1}{||c}{$\chi\psi$} &\multicolumn{1}{c}{56.9}&\multicolumn{1}{c}{63.9}\\
\cline{2-3}\cline{5-6}
\multicolumn{1}{c}{}&\multicolumn{1}{||c}{$t\bar{t}$} &\multicolumn{1}{c}{19.5}&\multicolumn{1}{c}{}&\multicolumn{1}{||c}{$t\bar{t}$} &\multicolumn{1}{c}{7.00}&\multicolumn{1}{c}{}\\
\hline
\hline
\multicolumn{1}{c}{$M_{G*}=3$ TeV}&\multicolumn{1}{||c}{$\chi\psi$} &\multicolumn{1}{c}{10.4}&\multicolumn{1}{c}{11.7}&\multicolumn{1}{||c}{$\chi\psi$} &\multicolumn{1}{c}{3.80}&\multicolumn{1}{c}{4.26}\\
\cline{2-3}\cline{5-6}
\multicolumn{1}{c}{}&\multicolumn{1}{||c}{$t\bar{t}$} &\multicolumn{1}{c}{1.29}&\multicolumn{1}{c}{}&\multicolumn{1}{||c}{$t\bar{t}$} &\multicolumn{1}{c}{0.463}&\multicolumn{1}{c}{}\\
\hline
\hline
\multicolumn{1}{c}{$M_{G*}=4$ TeV}&\multicolumn{1}{||c}{$\chi\psi$} &\multicolumn{1}{c}{0.870}&\multicolumn{1}{c}{1.10}&\multicolumn{1}{||c}{$\chi\psi$} &\multicolumn{1}{c}{0.320}&\multicolumn{1}{c}{0.404}\\
\cline{2-3}\cline{5-6}
\multicolumn{1}{c}{}&\multicolumn{1}{||c}{$t\bar{t}$} &\multicolumn{1}{c}{0.233}&\multicolumn{1}{c}{}&\multicolumn{1}{||c}{$t\bar{t}$} &\multicolumn{1}{c}{0.0837}&\multicolumn{1}{c}{}\\
\hline
\hline
\end{tabular}
}
\subtable[BACKGROUND]{
 \begin{tabular}[]{ccccccc}
\multicolumn{1}{c}{}&\multicolumn{3}{||c}{ $\sigma\ [pb]$ After Acceptance Cuts} &\multicolumn{3}{||c}{ $\sigma\ [pb]$ After b-TAG}\\
\hline
\hline
\multicolumn{1}{c}{$WWbb$}&\multicolumn{1}{||c}{} &\multicolumn{1}{c}{}&\multicolumn{1}{c}{76.9}&\multicolumn{1}{||c}{} &\multicolumn{1}{c}{}&\multicolumn{1}{c}{27.7}\\
\hline
\hline
\multicolumn{1}{c}{$Wbb+jets$}&\multicolumn{1}{||c}{$Wbb+1J$} &\multicolumn{1}{c}{2.19}&\multicolumn{1}{c}{4.40}&\multicolumn{1}{||c}{$Wbb+1J$} &\multicolumn{1}{c}{0.794}&\multicolumn{1}{c}{1.58}\\
\cline{2-3}\cline{5-6}
\multicolumn{1}{c}{}&\multicolumn{1}{||c}{$Wbb+2J$} &\multicolumn{1}{c}{1.61}&\multicolumn{1}{c}{}&\multicolumn{1}{||c}{$Wbb+2J$} &\multicolumn{1}{c}{0.574}&\multicolumn{1}{c}{}\\
\cline{2-3}\cline{5-6}
\multicolumn{1}{c}{}&\multicolumn{1}{||c}{$Wbb+3J$} &\multicolumn{1}{c}{0.599}&\multicolumn{1}{c}{}&\multicolumn{1}{||c}{$Wbb+3J$} &\multicolumn{1}{c}{0.215}&\multicolumn{1}{c}{}\\
\cline{2-3}\cline{5-6}
\hline
\hline
\multicolumn{1}{c}{$W+jets$}&\multicolumn{1}{||c}{$W+3J$} &\multicolumn{1}{c}{322}&\multicolumn{1}{c}{417}&\multicolumn{1}{||c}{$W+3J$} &\multicolumn{1}{c}{0.0675}&\multicolumn{1}{c}{0.108}\\
\cline{2-3}\cline{5-6}
\multicolumn{1}{c}{}&\multicolumn{1}{||c}{$W+4J$} &\multicolumn{1}{c}{95.1}&\multicolumn{1}{c}{}&\multicolumn{1}{||c}{$W+4J$} &\multicolumn{1}{c}{0.0412}&\multicolumn{1}{c}{}\\
\cline{2-3}
\hline
\hline
\multicolumn{1}{c}{Total BCKG}&\multicolumn{1}{||c}{} &\multicolumn{1}{c}{}&\multicolumn{1}{c}{498}&\multicolumn{1}{||c}{} &\multicolumn{1}{c}{}&\multicolumn{1}{c}{29.4}\\
\hline
\hline
\end{tabular}
}
\caption{\textit{Cross Section values after the Acceptance Cuts and the b-tagging ($\sqrt{s}=14$ TeV).}
}
\label{XSEC_AC-bTAG}
\end{table}

\subsection{Reconstruction procedure} \label{reco}

The strong point of our analysis is the presence in the signal of heavy resonances, both from the $G^*$ and the heavy fermions, that allow for a clean distinction between the signal and the background.
In order to distinguish with precision the resonance of the heavy gluon in the total invariant mass distribution, $M_{all}$, 
we need to reconstruct the momentum of the neutrino from the missing energy. 
Once we have reconstructed the neutrino momentum, our aim is to recognize the resonances of the heavy fermions. To do this, we need to
tag the $Wtb$ final states of our signal (see fig. \ref{wtb_S}).
 The heavy partners of the bottom decay into a $W$ plus the top, therefore they will produce a peak in the $Wt$ invariant mass distribution, $M_{Wt}$.
The heavy partner of the top decays into a $W$ plus a bottom and it will give a peak in the $Wb$ invariant mass distribution, $M_{Wb}$. \\

The transverse momentum of the neutrino can be reconstructed from the transverse missing momentum;
this latter can be estimated, considering a $p^{TOT}_T=0$ hypothesis, as $p^{miss}_T=-\sum p_T$, where $\sum p_T$ is the sum over the $p_T$ of all the detected final states.
We compute the transverse missing momentum by including a gaussian resolution $\sigma(p^{miss}_{x/y})=a\cdot\sqrt{\sum E_T/GeV}$, 
where $\sum E_T$ is the total transverse energy deposited in the
calorimeters (from electrons, muons and jets). We choose $a = 0.49$.~\footnote{This numerical value,
as well as the $b$-tagging efficiency and rejection rate and the resolution parameters considered in the jet smearing, have been chosen according 
to the performance of the ATLAS detector~\cite{Aad:2008zzm}.} 
Once we have estimated the neutrino transverse momentum, we can derive the neutrino longitudinal momentum, $p_z$, by requiring that the
neutrino and the lepton reconstruct an on-mass-shell $W$, $M_{l\nu} = 80.4$ GeV. The condition
\begin{equation}
 (E^{l}+E^{\nu})^{2}-(p^{l}_x+p^{\nu}_x)^{2}-(p^{l}_y+p^{\nu}_y)^{2}-(p^{l}_z+p^{\nu}_z)^{2}=M^2_W
\label{neutrino}
\end{equation}
 gives two solutions for $p^{\nu}_z$. For the events where the neutrino comes from the decay of a top, as we will explain afterwards,
 we consider the invariant mass of the top decay products, $M_{W(\to l\nu)b}$, and 
 we select the solution that gives the $M_{W(\to l\nu)b}$ value closest to the top mass, $m_t=174$ GeV (we do this selection during the subsequent top-tagging procedure). 
We take both of the two solutions otherwise (obviously we take into account a weighting factor of 0.5 for these latter events). 
In this case, we will require that both of the two reconstructed events, one for each of the two solutions for the neutrino, have to respect the conditions that we will further impose.\\
We find that in the $\simeq 20 \%$ of the events, both for the signal and the background, the eq. (\ref{neutrino}) has imaginary solutions 
(this corresponds to the case of a quite off-shell leptonically decayed $W$). In this case we decide to throw out the event. 
Our neutrino reconstruction procedure has, therefore, an efficiency of about the $80\%$.\\

\begin{figure}
\centering
\includegraphics[width=0.5\textwidth, angle=90]{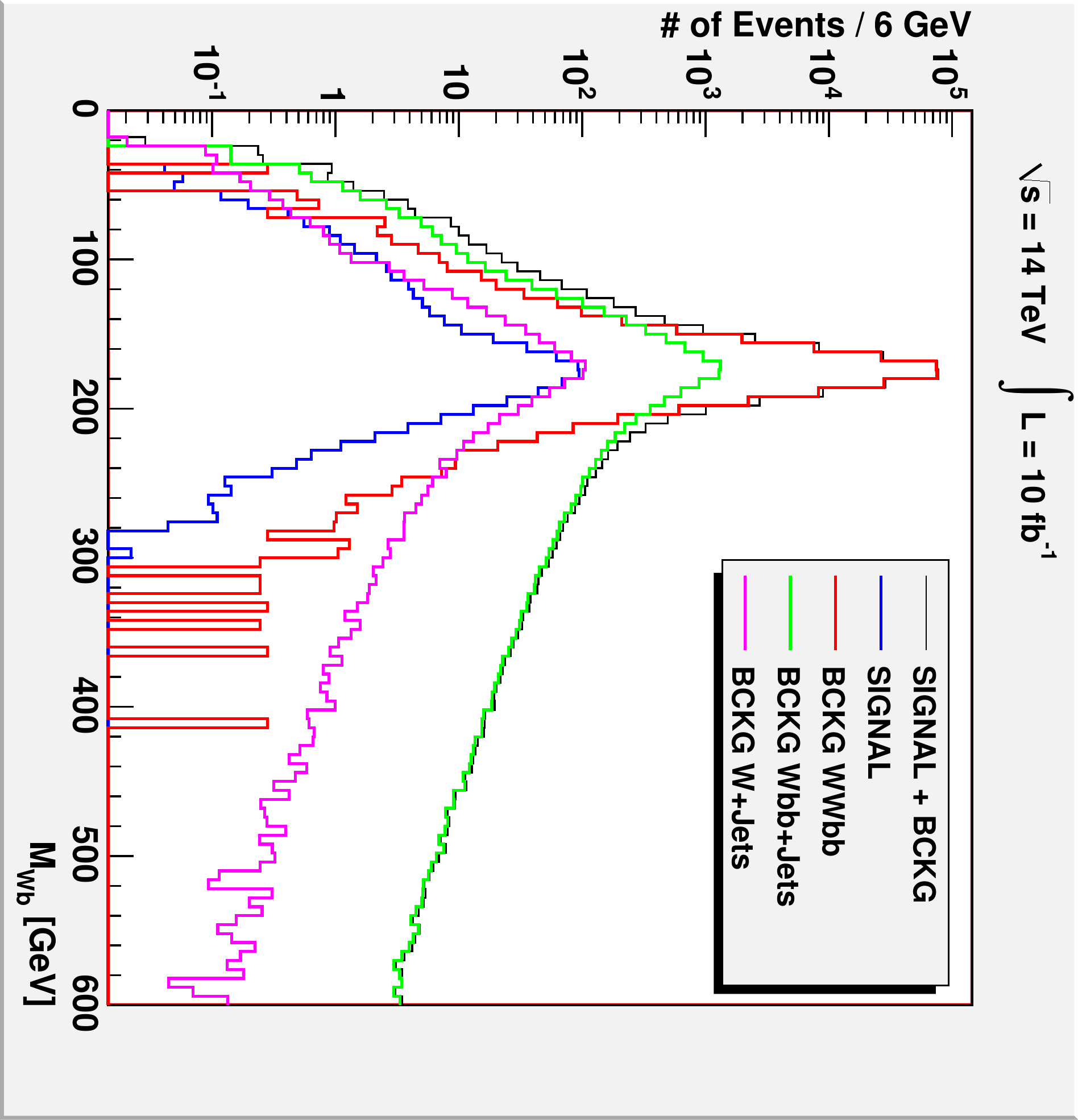}
\caption{\textit{Distribution of the invariant mass of the $Wb$ pair we select as the one coming from the decay of the top (by using the procedure explained in the text), for signal (with $M_{G^*}=2$ TeV) and backgrounds.}
}
\label{top_reco}
\end{figure}

Once we have reconstructed the momentum of the neutrino, we want to recognize the top which is in our $Wtb$ signal. 
To do this, we first reconstruct the leptonically and hadronically decayed $W$s and then we consider all the possible $Wb$ combinations.
The $Wb$ pair that gives the $M_{Wb}$ invariant mass closest to the top mass, $m_t=174$ GeV, is selected as the pair coming from the decay of the top.
The top quadri-momentum is then reconstructed by summing on the quadri-momentum of the $W$ and of the $b$, that form the selected pair. \\
More in detail, we reconstruct the hadronically decayed $W$ by summing on the quadri-momentum of all the light jets. 
For the signal and the $WWbb$ background, we will reconstruct a `true' $W$, for the $W+jets$ and the $Wbb+jets$ backgrounds, we will obtain a fictitious unphysical $W$. 
We reconstruct two leptonically decayed $W$s, one for each of the two reconstructed neutrinos, that we obtained after the procedure of neutrino reconstruction we have previously explained (therefore, considering that we have two $b$s in the final state, we have to take into account six possible $Wb$ combinations). 
If the selected $Wb$ pair is formed by a leptonically decayed $W$, we select as the `true' neutrino the one which comes from the $W$ in the selected pair 
(the other neutrino, with the leptonically decayed $W$ from which this latter originates, is thrown out). \\
In fig. \ref{top_reco} we show the distribution of the invariant mass of the selected $Wb$ pair, 
for a signal referred to a $G^*$ with a mass of $2$ TeV and for the different backgrounds. 
The peak we see next to $174$ GeV corresponds, for the signal and the $t\bar{t}$ background, to a physical top resonance; 
the peak for the $Wbb+jets$ and $W+jets$ backgrounds does not correspond, instead, to a `true' physical top; it is a consequence of the procedure of top-tagging we used.
At this level of the analysis, we do not want yet to distinguish between the signal and the background. Our aim is to recognize the top we know to be in the signal.
Therefore, we choose to keep all the events in the $M_{Wb}$ range $[80\ GeV, 250\ GeV]$, in order to catch the most part of the signal. \\

We show in table \ref{NeuTopEff} the efficiencies of the neutrino reconstruction ($\epsilon_{\nu}$) and the top-tagging ($\epsilon_t$) procedures, for signal and backgrounds. \\

\begin{table}[t!]
\subtable[SIGNAL]{
\begin{tabular}[]{cccc}
 \multicolumn{2}{c}{}& \multicolumn{1}{c}{$\epsilon_{\nu}$} &\multicolumn{1}{c}{$\epsilon_t$}\\
\hline
\multicolumn{1}{c}{$M_{G*}=1.5$ TeV}&\multicolumn{1}{||c}{$\chi\psi$} &\multicolumn{1}{c}{0.81}&\multicolumn{1}{c}{0.99}\\
\cline{2-4}
\multicolumn{1}{c}{}&\multicolumn{1}{||c}{$t\bar{t}$} &\multicolumn{1}{c}{0.82}&\multicolumn{1}{c}{1.0}\\
\hline
\multicolumn{1}{c}{$M_{G*}=2$ TeV}&\multicolumn{1}{||c}{$\chi\psi$} &\multicolumn{1}{c}{0.80}&\multicolumn{1}{c}{0.99}\\
\cline{2-4}
\multicolumn{1}{c}{}&\multicolumn{1}{||c}{$t\bar{t}$} &\multicolumn{1}{c}{0.83}&\multicolumn{1}{c}{1.0}\\
\hline
\multicolumn{1}{c}{$M_{G*}=3$ TeV}&\multicolumn{1}{||c}{$\chi\psi$} &\multicolumn{1}{c}{0.79}&\multicolumn{1}{c}{0.99}\\
\cline{2-4}
\multicolumn{1}{c}{}&\multicolumn{1}{||c}{$t\bar{t}$} &\multicolumn{1}{c}{0.82}&\multicolumn{1}{c}{1.0}\\
\hline
\multicolumn{1}{c}{$M_{G*}=4$ TeV}&\multicolumn{1}{||c}{$\chi\psi$} &\multicolumn{1}{c}{0.77}&\multicolumn{1}{c}{0.99}\\
\cline{2-4}
\multicolumn{1}{c}{}&\multicolumn{1}{||c}{$t\bar{t}$} &\multicolumn{1}{c}{0.82}&\multicolumn{1}{c}{1.0}\\
\hline
\end{tabular}
}
\subtable[BACKGROUND]{
\begin{tabular}[]{cccc}
 \multicolumn{1}{c}{}& \multicolumn{1}{c}{$\epsilon_{\nu}$} &\multicolumn{1}{c}{$\epsilon_t$}\\
\hline
\multicolumn{1}{c}{$WWbb$} &\multicolumn{1}{c}{0.82}&\multicolumn{1}{c}{1.0}\\
\hline
\multicolumn{1}{c}{$Wbb+jets$} &\multicolumn{1}{c}{0.79}&\multicolumn{1}{c}{0.84}\\
\hline
\multicolumn{1}{c}{$W+jets$} &\multicolumn{1}{c}{0.80}&\multicolumn{1}{c}{0.92}\\
\hline
\end{tabular}
}
\caption{\textit{Efficiencies of the neutrino reconstruction ($\epsilon_{\nu}$) and the top-tagging ($\epsilon_t$) procedures, for signal and backgrounds ($\sqrt{s}=14$ TeV).}}
\label{NeuTopEff}
\end{table}


Once we have recognized the $W$ and the $b$ that come from the decay of the reconstructed top, 
we are also able to tag the `other' (which are not part of the reconstructed top) $W$ and $b$ as 
the particles that form, with the reconstructed top $t$, our $Wtb$ final state\footnote{
Again, in the case where the `other' $W$ is a leptonically decayed $W$, we take both of the two leptonic $W$s, one for each of the two neutrinos we have reconstructed.}. 
This allows us to calculate the $M_{Wt}$ and $M_{Wb}$ distributions and to find the peaks from the heavy fermion resonances. 
In what follows, $W$ and $b$ will always denote the $W$ and the $b$ of the $Wtb$ final state, 
which are those coming from heavy vectors decay and not from the decay of the reconstructed top.
 Obviously, for the $G^* \to t\bar{t}$ component of the $G^*$ signal, these $W$ and $b$ particles also come from the decay of a top,
 which is the other not tagged top. This is also true for the $t\bar{t}$ component of the $WWbb$ background. 


\subsection{Preliminary cuts on $p_T$}\label{CUTzeroCost}

\begin{figure}[]
\mbox{\subfigure[]{\includegraphics[width=0.5\textwidth, angle=90]{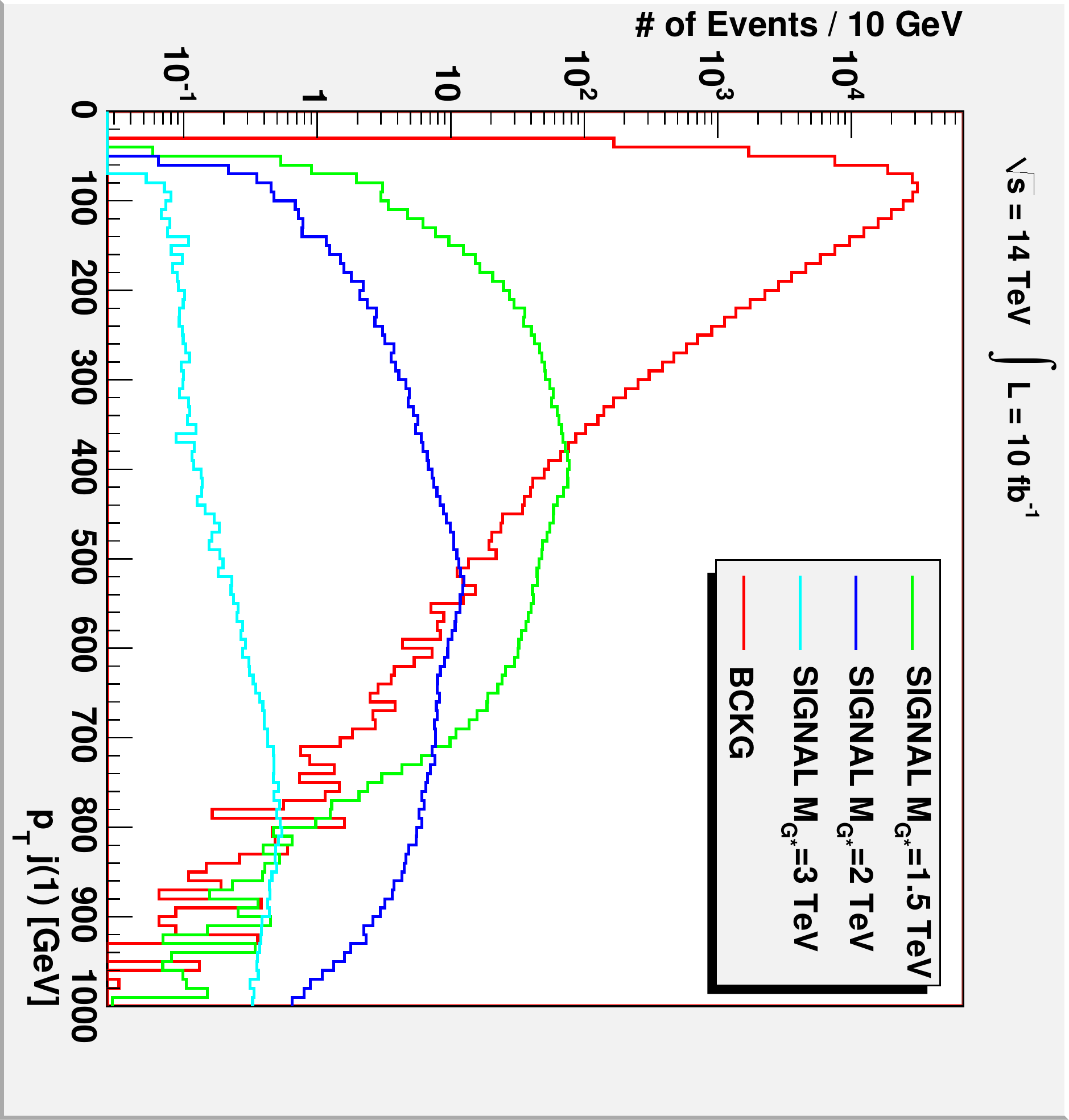}}
\subfigure[]{\includegraphics[width=0.5\textwidth, angle=90]{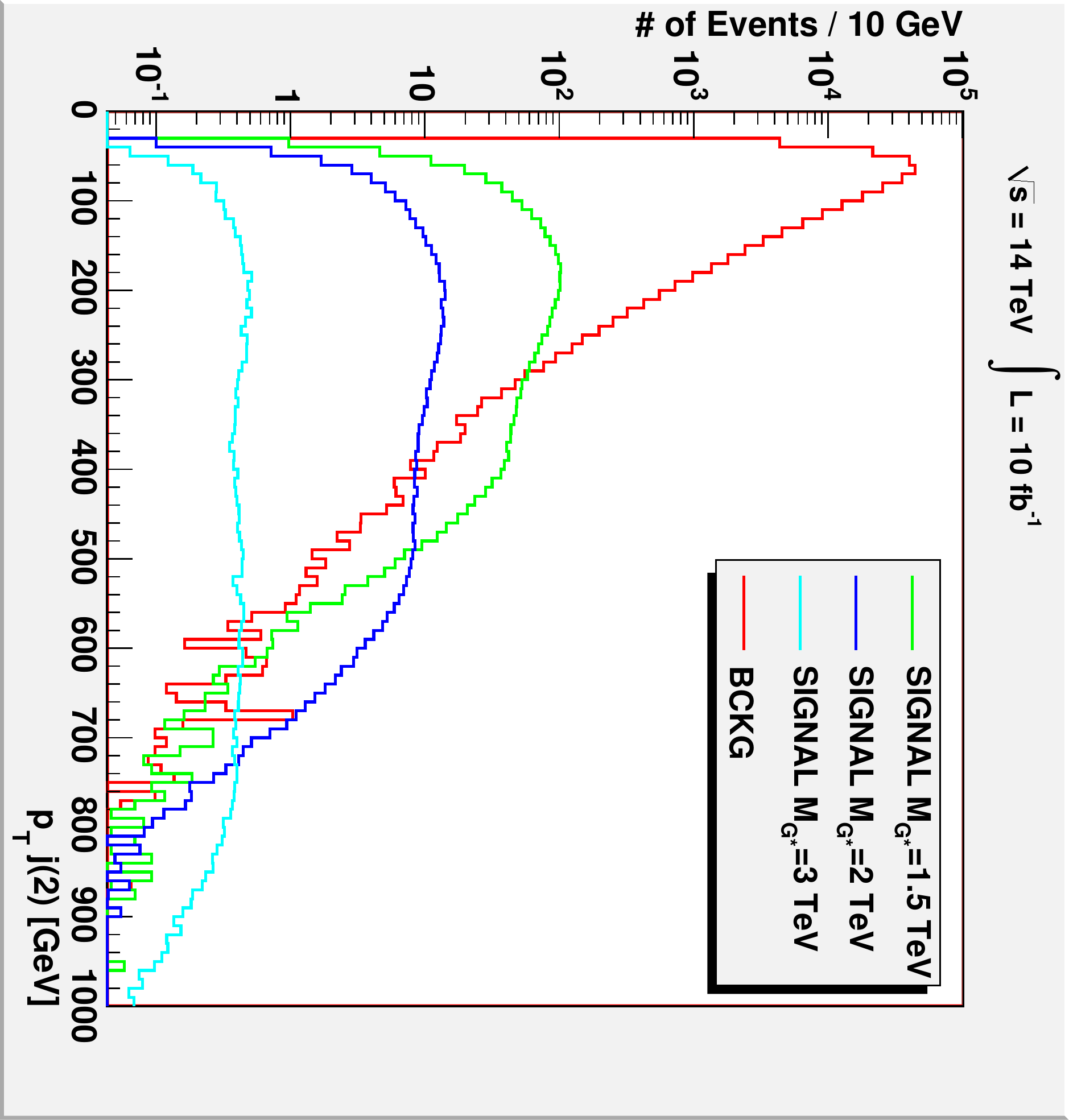}}
}\\
\mbox{\subfigure[]{\includegraphics[width=0.5\textwidth, angle=90]{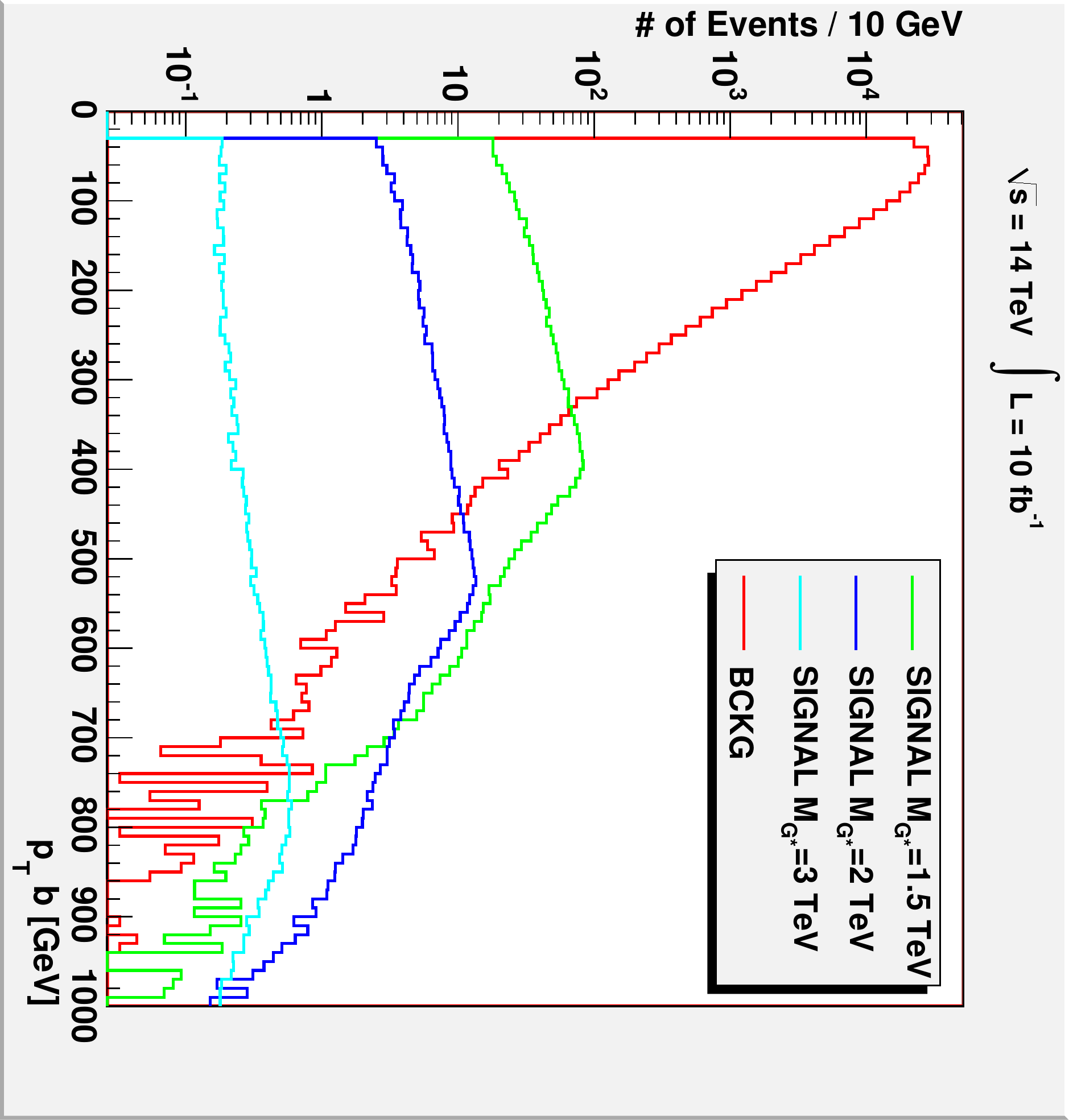}}
\subfigure[]{\includegraphics[width=0.5\textwidth, angle=90]{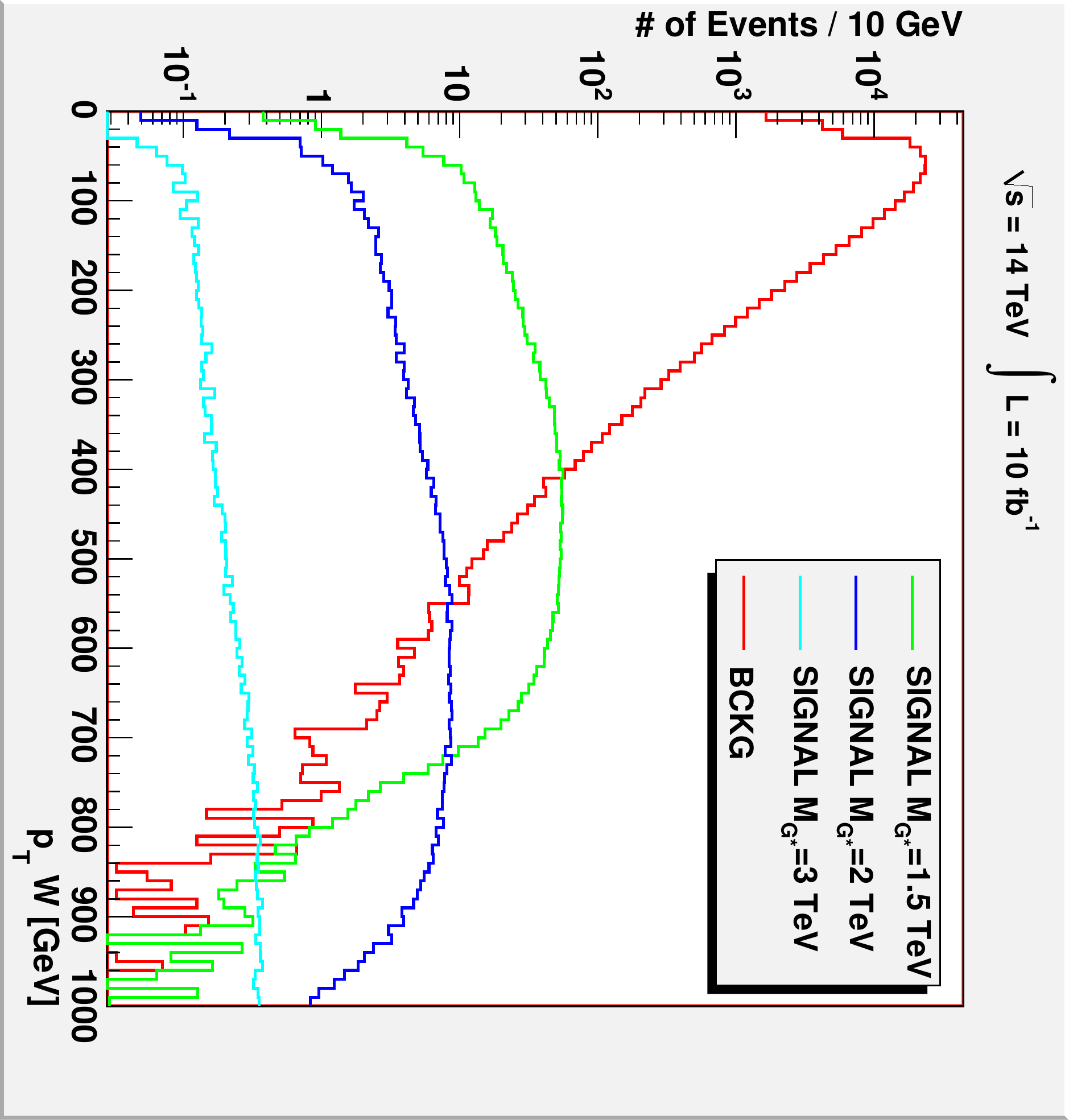}}
}
\centering
\mbox{\subfigure[]{\includegraphics[width=0.5\textwidth, angle=90]{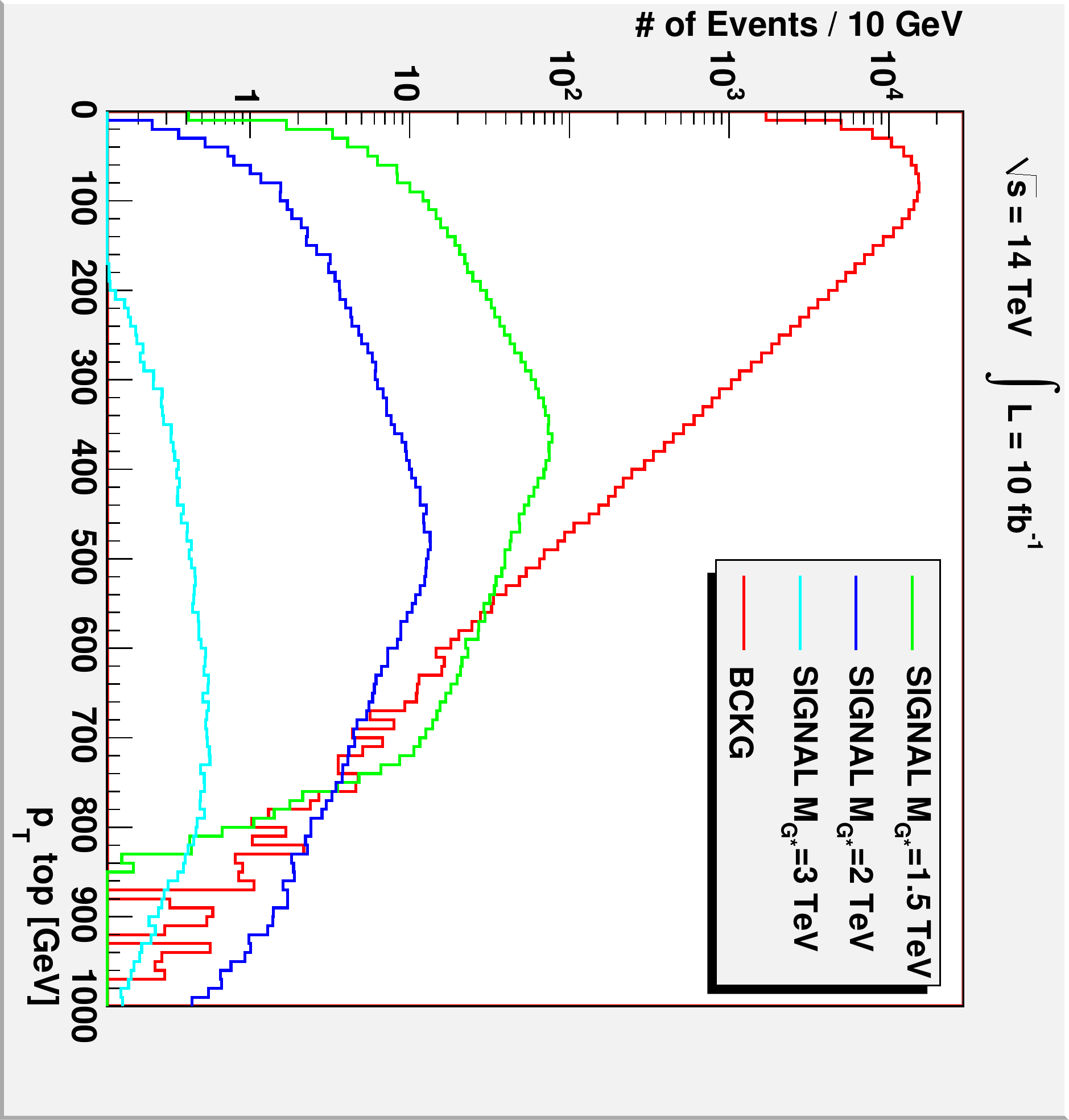}}}
\caption{\textit{Distribution of the $p_T$ of the hardest ($j(1)$) and the second hardest ($j(2)$) jet in the events 
and of the $W$, the top and the bottom in the reconstructed $Wtb$ final state. 
The red line denotes the distribution for the total background; in the same plot, we show the distributions for a signal with $M_{G^*}=1.5,2,3$ TeV.}}
\label{pT_distribution}
\end{figure}

At this point of the analysis, we can start to distinguish between the signal and the background. 
As we will show in the next section, it will be very effective for this purpose to look at the invariant mass distributions $M_{all}$, $M_{Wb}$ and $M_{Wt}$,
where, for the signal, we will recognize peaks in correspondence of the high mass values of the heavy colored vectors. \\
Before considering the invariant mass distributions, we can apply a preliminary selection on the events. 
We identify several cuts at `zero cost' for the signal; 
these are simple cuts on $p_T$ that throw out less than the $3\%$ of the signal events and that already give a significant contribution in reducing the background. 
This is a consequence of the fact that the signal final states are generally more energetic than the background ones, because they come from the decays of heavy particles, such as the heavy colored vectors.\\
We find that the $W$, the bottom and the top in the $Wtb$ final state of our signal, that we have reconstructed by using the procedure explained in the previous section,
are very energetic particles. Their transverse momentum is a good variable to exploit in the preliminary selection. 
It is useful as well to consider the $p_T$ of the hardest jet (light-jet or b-jet) and of the second hardest jet in the events.\\
We show in Fig. \ref{pT_distribution} the distributions of the $p_T$ of the $W$, of the top and of the bottom in the reconstructed $Wtb$ final state 
and of the hardest ($j(1)$) and the second hardest ($j(2)$) jet in the physical final state (\ref{final_states}).  
We show the distributions for the total background and for the signal referred to different $G^*$ mass values.\\
As expected, the signal with higher $G^*$ mass values has even more energetic final particles. 
Considering this feature, we will choose the values of the cuts so as to have an individual efficiency of $97\%$ on the signal at $M_{G*}=1.5$ TeV,
which corresponds to the less energetic case. Hence, we are sure that our cuts will not throw out more than the $3\%$ of the signal events, for all the $G^*$ masses above $1.5$ TeV. 

\begin{figure}[]
\centering
\includegraphics[width=0.5\textwidth,angle=90]{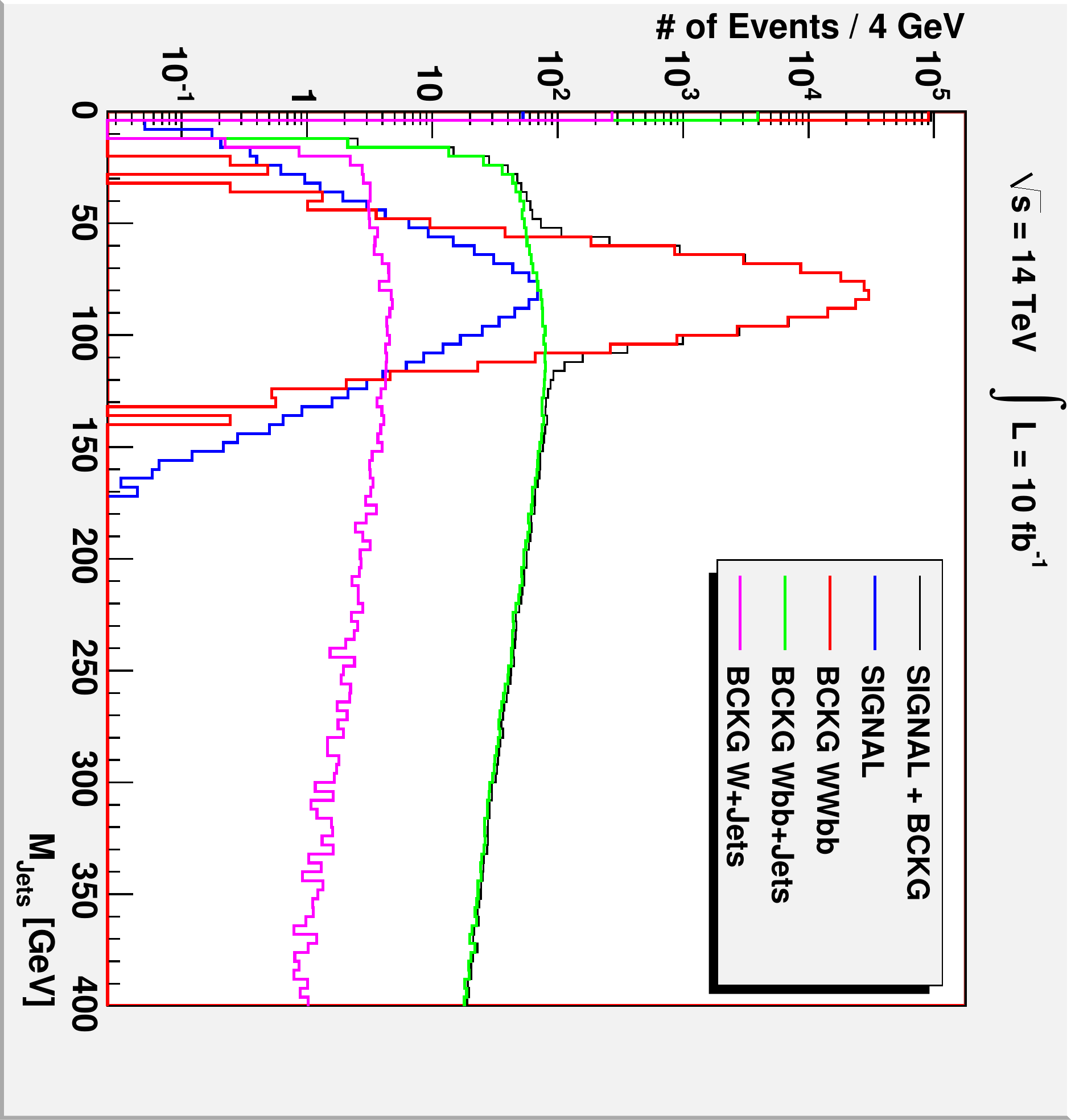}
\caption{\textit{Distribution of $M_{jets}$ for the signal with $M_{G*}=2$ TeV and for the different components of the background.}}
\label{Mjets}
\end{figure}

We include among the cuts at `zero cost' a cut on the invariant mass of the light jets, $M_{jets}<200$ GeV, 
which, for the signal and the $WWbb$ background, corresponds to a cut on the invariant mass of the hadronic $W$.
This cut is very effective to reduce the components of the $W+jets$ and $Wbb+jets$ backgrounds with high number of jets. 
These components have small cross section values at the beginning (as predicted by perturbation theory), but they could be not negligible after the selection we will do 
by considering essentially very energetic particles in the final state. 
The cut $M_{jets}<200$ GeV has an efficiency of $1.0$ for the signal and the $WWbb$ background; 
the efficiencies for the $Wbb+jets$ and the $W+jets$ backgrounds are, instead, of $0.72$ and of $0.84$ respectively. In particular, the $Wbb+jj$
sample is reduced by the $45\%$ and the $Wbb+jjj$ by the $87\%$. 
Therefore, This cut allows us to neglect the backgrounds which are formed by $Wbb$ plus more than three light jets in the final state.
We can neglect, as well, the backgrounds with $W$ and more than four light jets. \\
We show in Fig. \ref{Mjets} the $M_{jets}$ distribution for the signal with $M_{G*}=2$ TeV and for the different components of the background.
Since we are performing an analysis at parton level, we obtain a peak at $M_{jets}=0$, that comes from the backgrounds with only one light jet. 
The signal and the $WWbb$ background, as expected, have a resonance peak close to the hadronic $W$ mass, $M_{W}=80.4$ GeV. 
The components of the background with high number of light jets are distributed on high $M_{jets}$ values, therefore they are effectively reduced by the cut $M_{jets}<200$ GeV.\\
\noindent
The cuts at `zero cost' we impose after top tagging are the following:
\[
p_{Tj (1)} > 175\ GeV , \ p_{Tj (2)} > 85 \ GeV \ ,
\]
\[
 p_{T top} > 110 \ GeV  , \ p_{T W} > 110 \ GeV , \ p_{T b} > 70 \ GeV \ , 
\]
\begin{equation}
 M_{jets} < 200\  GeV 
\label{ZEROcut}
\end{equation}
 $j (1)$ denotes the hardest jet (light-jet or b-jet), $j (2)$ is the second hardest jet; 
$W$ and $b$ are not part of the tagged top.\\
The cuts are chosen so as to have an individual efficiency of $97\%$ on the signal at $M_{G*}=1.5$ TeV. \\

\begin{figure}[h]
\centering
\includegraphics[width=0.45\textwidth, angle=90]{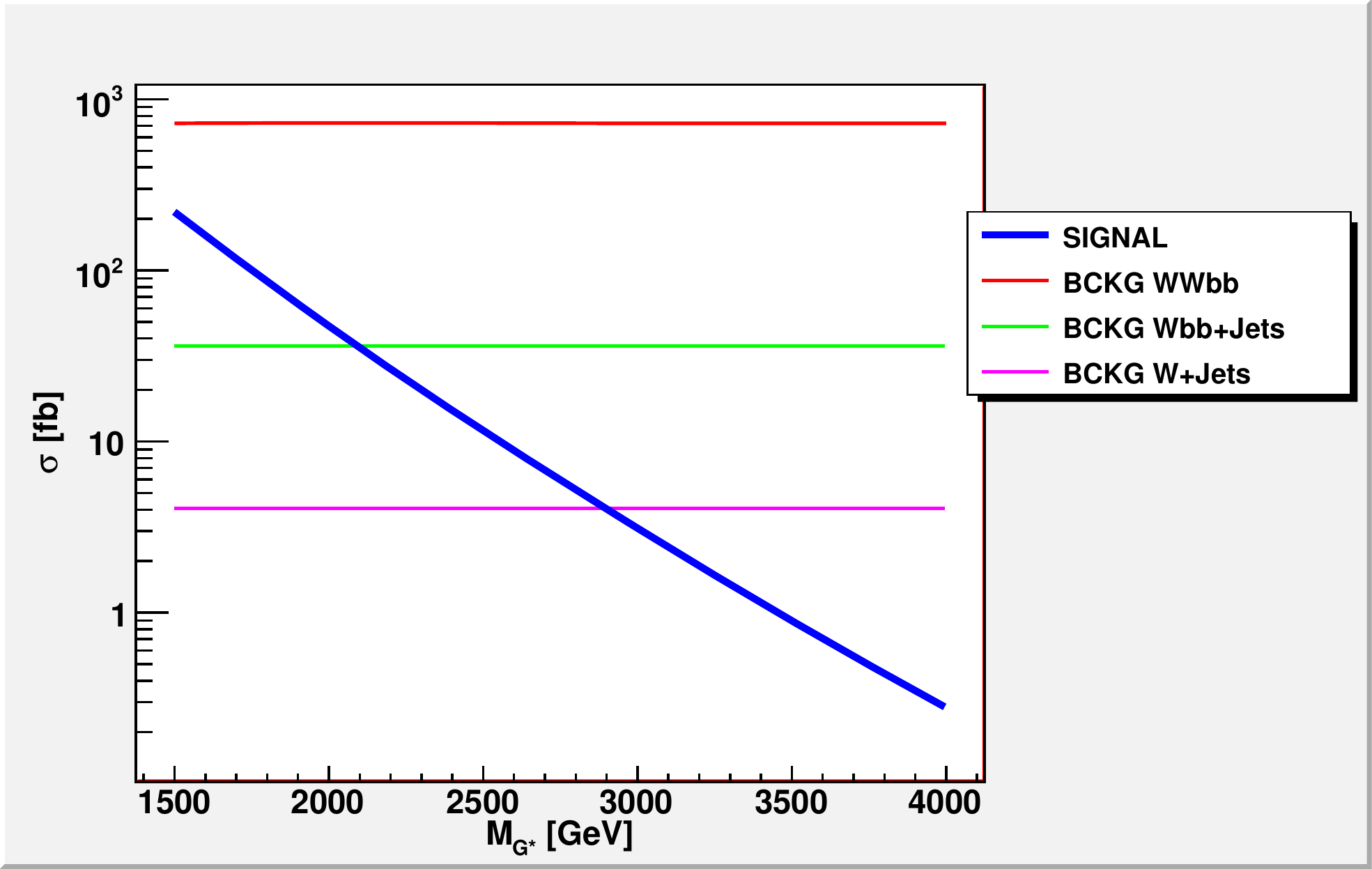}
\caption{\textit{Cross Section values after the cuts at `zero cost' (\ref{ZEROcut}) ($\sqrt{s}=14$ TeV).}
}
\label{fig_XSEC_AZerocut}
\end{figure}

We show in Tab. \ref{zeroCutEff} the efficiencies of the cuts on $p_T$ at `zero cost' (\ref{ZEROcut}), for signal and background.
After top tagging, more than the $\sim90\%$ of the signal events pass the zero cost cuts, while the number of events for the background reduces to the $\sim 3\%$ of its initial value 
\footnote{The cuts are a little bit less efficient for the $t\bar{t}$ component of the signal. Anyway, we can ignore this aspect, since our analysis will be focused on the $\psi\chi$ component, as we will further explain.}.\\
We show in Tab. \ref{XSEC_AZeroCut} (and in Fig. \ref{fig_XSEC_AZerocut}) the cross section values for signal and background that we have obtained after applying the cuts at `zero cost'.
 
\begin{table}[t]
\begin{tabular}[]{cccccccc}
 \multicolumn{2}{c}{}& \multicolumn{1}{c}{$p_{Tb}$} &\multicolumn{1}{c}{$p_{Ttop}$}&\multicolumn{1}{c}{$p_{TW}$} &\multicolumn{1}{c}{$p_{Tj(1)}$}&\multicolumn{1}{c}{$p_{Tj(2)}$}&\multicolumn{1}{c}{Combined Cuts}\\
\hline
\multicolumn{1}{c}{$M_{G*}=1.5$ TeV}&\multicolumn{1}{||c}{$\chi\psi$} &\multicolumn{1}{c}{0.99}&\multicolumn{1}{c}{0.97}&\multicolumn{1}{c}{0.98}&\multicolumn{1}{c}{0.98}&\multicolumn{1}{c}{0.98}&\multicolumn{1}{c}{}\\
\cline{2-7}
\multicolumn{1}{c}{}&\multicolumn{1}{||c}{$t\bar{t}$} &\multicolumn{1}{c}{0.85}&\multicolumn{1}{c}{0.98}&\multicolumn{1}{c}{0.89}&\multicolumn{1}{c}{0.87}&\multicolumn{1}{c}{0.91}&\multicolumn{1}{c}{}\\
\cline{2-7}
\multicolumn{1}{c}{}&\multicolumn{1}{}{} &\multicolumn{1}{c}{0.97}&\multicolumn{1}{c}{0.97}&\multicolumn{1}{c}{0.97}&\multicolumn{1}{c}{0.97}&\multicolumn{1}{c}{0.97}&\multicolumn{1}{c}{0.89}\\
\hline
\multicolumn{1}{c}{$M_{G*}=2$ TeV}&\multicolumn{1}{||c}{$\chi\psi$} &\multicolumn{1}{c}{1.0}&\multicolumn{1}{c}{0.98}&\multicolumn{1}{c}{0.99}&\multicolumn{1}{c}{1.0}&\multicolumn{1}{c}{0.99}&\multicolumn{1}{c}{}\\
\cline{2-7}
\multicolumn{1}{c}{}&\multicolumn{1}{||c}{$t\bar{t}$} &\multicolumn{1}{c}{0.84}&\multicolumn{1}{c}{0.97}&\multicolumn{1}{c}{0.89}&\multicolumn{1}{c}{0.87}&\multicolumn{1}{c}{0.90}&\multicolumn{1}{c}{}\\
\cline{2-7}
\multicolumn{1}{c}{}&\multicolumn{1}{}{} &\multicolumn{1}{c}{0.98}&\multicolumn{1}{c}{0.98}&\multicolumn{1}{c}{0.98}&\multicolumn{1}{c}{0.99}&\multicolumn{1}{c}{0.98}&\multicolumn{1}{c}{0.93}\\
\hline
\multicolumn{1}{c}{$M_{G*}=3$ TeV}&\multicolumn{1}{||c}{$\chi\psi$} &\multicolumn{1}{c}{1.0}&\multicolumn{1}{c}{0.99}&\multicolumn{1}{c}{1.0}&\multicolumn{1}{c}{1.0}&\multicolumn{1}{c}{0.99}&\multicolumn{1}{c}{}\\
\cline{2-7}
\multicolumn{1}{c}{}&\multicolumn{1}{||c}{$t\bar{t}$} &\multicolumn{1}{c}{0.81}&\multicolumn{1}{c}{0.94}&\multicolumn{1}{c}{0.83}&\multicolumn{1}{c}{0.77}&\multicolumn{1}{c}{0.85}&\multicolumn{1}{c}{}\\
\cline{2-7}
\multicolumn{1}{c}{}&\multicolumn{1}{}{} &\multicolumn{1}{c}{0.98}&\multicolumn{1}{c}{0.99}&\multicolumn{1}{c}{0.99}&\multicolumn{1}{c}{0.98}&\multicolumn{1}{c}{0.98}&\multicolumn{1}{c}{0.94}\\
\hline
\multicolumn{1}{c}{$M_{G*}=4$ TeV}&\multicolumn{1}{||c}{$\chi\psi$} &\multicolumn{1}{c}{1.0}&\multicolumn{1}{c}{0.99}&\multicolumn{1}{c}{1.0}&\multicolumn{1}{c}{1.0}&\multicolumn{1}{c}{1.0}&\multicolumn{1}{c}{}\\
\cline{2-7}
\multicolumn{1}{c}{}&\multicolumn{1}{||c}{$t\bar{t}$} &\multicolumn{1}{c}{0.78}&\multicolumn{1}{c}{0.89}&\multicolumn{1}{c}{0.75}&\multicolumn{1}{c}{0.65}&\multicolumn{1}{c}{0.77}&\multicolumn{1}{c}{}\\
\cline{2-7}
\multicolumn{1}{c}{}&\multicolumn{1}{}{} &\multicolumn{1}{c}{0.95}&\multicolumn{1}{c}{0.96}&\multicolumn{1}{c}{0.95}&\multicolumn{1}{c}{0.93}&\multicolumn{1}{c}{0.95}&\multicolumn{1}{c}{0.88}\\
\hline
\hline
\multicolumn{2}{c}{$WWbb$}&\multicolumn{1}{c}{0.55}&\multicolumn{1}{c}{0.47}&\multicolumn{1}{c}{0.27}&\multicolumn{1}{c}{0.082}&\multicolumn{1}{c}{0.32}&\multicolumn{1}{c}{0.032}\\
\hline
\multicolumn{2}{c}{$Wbb+jets$}&\multicolumn{1}{c}{0.46}&\multicolumn{1}{c}{0.32}&\multicolumn{1}{c}{0.38}&\multicolumn{1}{c}{0.21}&\multicolumn{1}{c}{0.37}&\multicolumn{1}{c}{0.034}\\
\hline
\multicolumn{2}{c}{$W+jets$}&\multicolumn{1}{c}{0.47}&\multicolumn{1}{c}{0.36}&\multicolumn{1}{c}{0.29}&\multicolumn{1}{c}{0.19}&\multicolumn{1}{c}{0.31}&\multicolumn{1}{c}{0.050}\\
\hline
\hline
\multicolumn{2}{c}{Total BCKG}&\multicolumn{1}{c}{0.54}&\multicolumn{1}{c}{0.46}&\multicolumn{1}{c}{0.28}&\multicolumn{1}{c}{0.089}&\multicolumn{1}{c}{0.32}&\multicolumn{1}{c}{0.032}\\
\end{tabular}
\caption{\textit{Efficiencies of the cuts on $p_T$ at `zero cost' (\ref{ZEROcut}), for signal and background ($\sqrt{s}=14$ TeV).}}
\label{zeroCutEff}
\end{table}

\begin{table}[t]
\subtable[SIGNAL]{
\begin{tabular}[]{cccc}
\multicolumn{1}{c}{}&\multicolumn{3}{||c}{ $\sigma\ [fb]$} \\
\hline
\hline
\multicolumn{1}{c}{$M_{G*}=1.5$ TeV}&\multicolumn{1}{||c}{$\chi\psi$} &\multicolumn{1}{c}{192}&\multicolumn{1}{c}{216}\\
\cline{2-3}
\multicolumn{1}{c}{}&\multicolumn{1}{||c}{$t\bar{t}$} &\multicolumn{1}{c}{23.7}&\multicolumn{1}{c}{}\\
\hline
\hline
\multicolumn{1}{c}{$M_{G*}=2$ TeV}&\multicolumn{1}{||c}{$\chi\psi$} &\multicolumn{1}{c}{42.7}&\multicolumn{1}{c}{46.6}\\
\cline{2-3}
\multicolumn{1}{c}{}&\multicolumn{1}{||c}{$t\bar{t}$} &\multicolumn{1}{c}{3.93}&\multicolumn{1}{c}{}\\
\hline
\hline
\multicolumn{1}{c}{$M_{G*}=3$ TeV}&\multicolumn{1}{||c}{$\chi\psi$} &\multicolumn{1}{c}{2.91}&\multicolumn{1}{c}{3.14}\\
\cline{2-3}
\multicolumn{1}{c}{}&\multicolumn{1}{||c}{$t\bar{t}$} &\multicolumn{1}{c}{0.225}&\multicolumn{1}{c}{}\\
\hline
\hline
\multicolumn{1}{c}{$M_{G*}=4$ TeV}&\multicolumn{1}{||c}{$\chi\psi$} &\multicolumn{1}{c}{0.246}&\multicolumn{1}{c}{0.278}\\
\cline{2-3}
\multicolumn{1}{c}{}&\multicolumn{1}{||c}{$t\bar{t}$} &\multicolumn{1}{c}{0.0318}&\multicolumn{1}{c}{}\\
\hline
\hline
\end{tabular}
}
\subtable[BACKGROUND]{
 \begin{tabular}[]{cccc}
\multicolumn{1}{c}{}&\multicolumn{3}{||c}{$\sigma\ [fb]$}\\
\hline
\hline
\multicolumn{1}{c}{$WWbb$}&\multicolumn{1}{||c}{} &\multicolumn{1}{c}{}&\multicolumn{1}{c}{724}\\
\hline
\hline
\multicolumn{1}{c}{$Wbb+jets$}&\multicolumn{1}{||c}{$Wbb+1J$} &\multicolumn{1}{c}{25.9}&\multicolumn{1}{c}{35.8}\\
\cline{2-3}
\multicolumn{1}{c}{}&\multicolumn{1}{||c}{$Wbb+2J$} &\multicolumn{1}{c}{9.30}&\multicolumn{1}{c}{}\\
\cline{2-3}
\multicolumn{1}{c}{}&\multicolumn{1}{||c}{$Wbb+3J$} &\multicolumn{1}{c}{0.627}\\
\cline{2-3}
\hline
\hline
\multicolumn{1}{c}{$W+jets$}&\multicolumn{1}{||c}{$W+3J$} &\multicolumn{1}{c}{2.90}&\multicolumn{1}{c}{4.05}\\
\cline{2-3}
\multicolumn{1}{c}{}&\multicolumn{1}{||c}{$W+4J$} &\multicolumn{1}{c}{1.15}&\multicolumn{1}{c}{}\\
\cline{2-3}
\hline
\hline
\multicolumn{1}{c}{Total BCKG}&\multicolumn{1}{||c}{} &\multicolumn{1}{c}{}&\multicolumn{1}{c}{764}\\
\hline
\hline
\end{tabular}
}
\caption{\textit{Cross Section values after the cuts at `zero cost' (\ref{ZEROcut}) ($\sqrt{s}=14$ TeV).}
}
\label{XSEC_AZeroCut}
\end{table}

\subsection{Discovery analysis - Invariant mass regions}
\begin{figure}[]
\mbox{\subfigure[\textit{Invariant mass distributions, $M_{all}$, $M_{Wt}$ and $M_{Wb}$ for the signal with $M_{G*}=2$ TeV ($M_{\tilde{T}}=M_{B}=1.33$ TeV).}]
{\includegraphics[width=0.47\textwidth, angle=90]{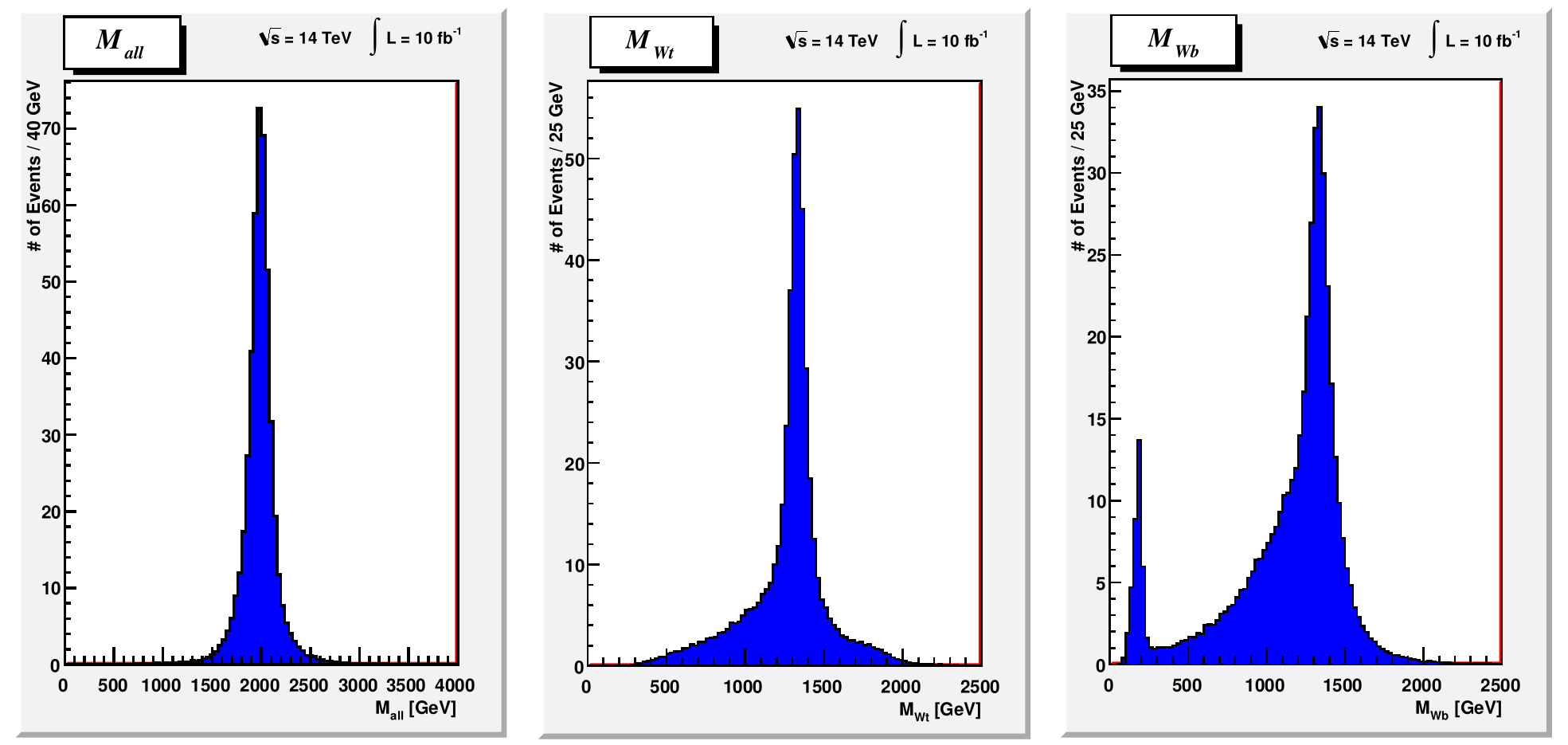}
\label{M20_invMass}}}
\mbox{\subfigure[\textit{Invariant mass distributions (in Log Scale), $M_{all}$, $M_{Wt}$ and $M_{Wb}$ for the total background and for the different components of the background.}]
{\includegraphics[width=0.50\textwidth, angle=90]{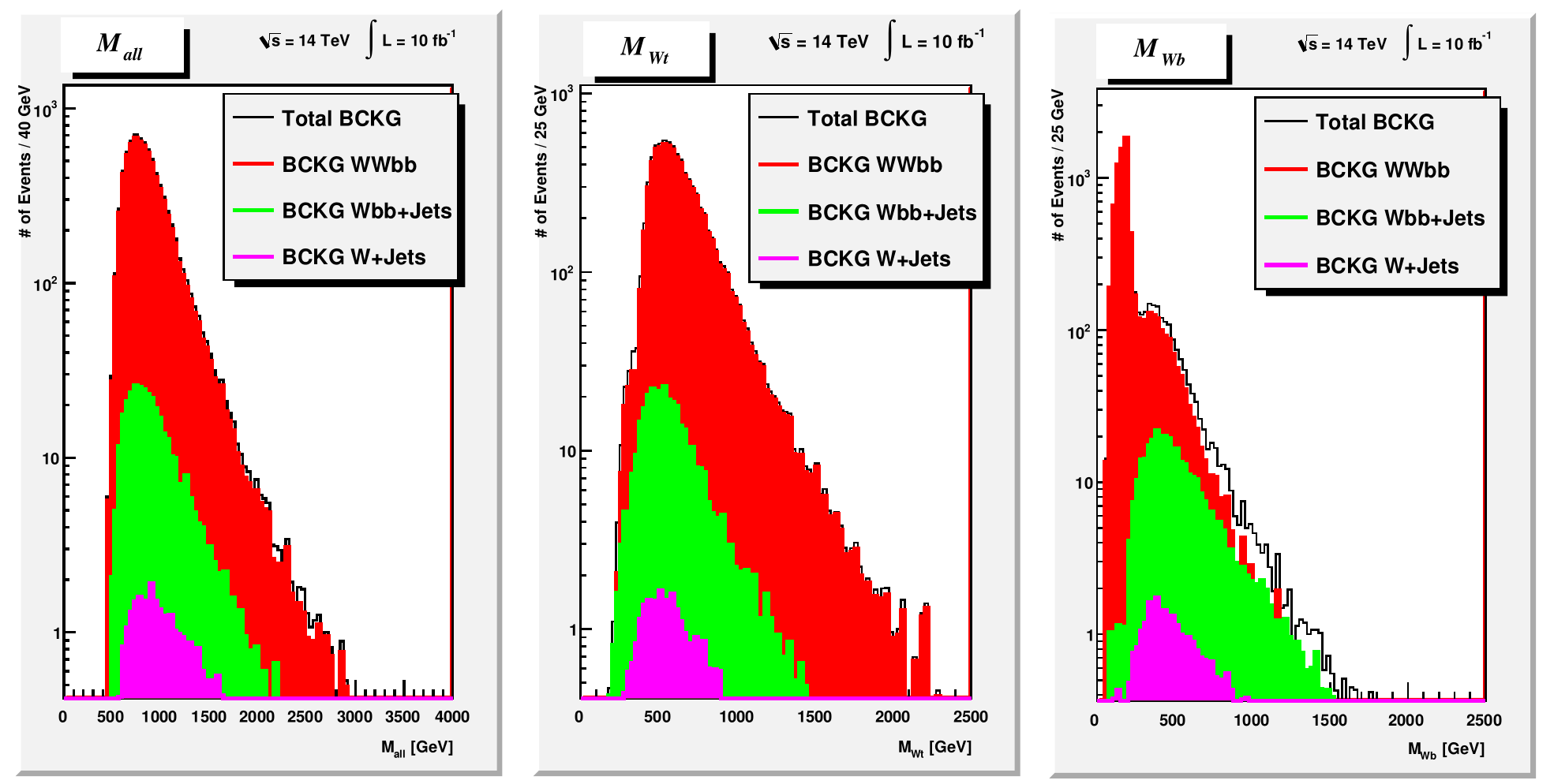}
\label{B_invMass}}}
\mbox{\subfigure[\textit{Invariant mass distributions (in Log Scale), $M_{all}$, $M_{Wt}$ and $M_{Wb}$ for the signal with $M_{G*}=2$ TeV plus the total background.}]
{\includegraphics[width=0.47\textwidth, angle=90]{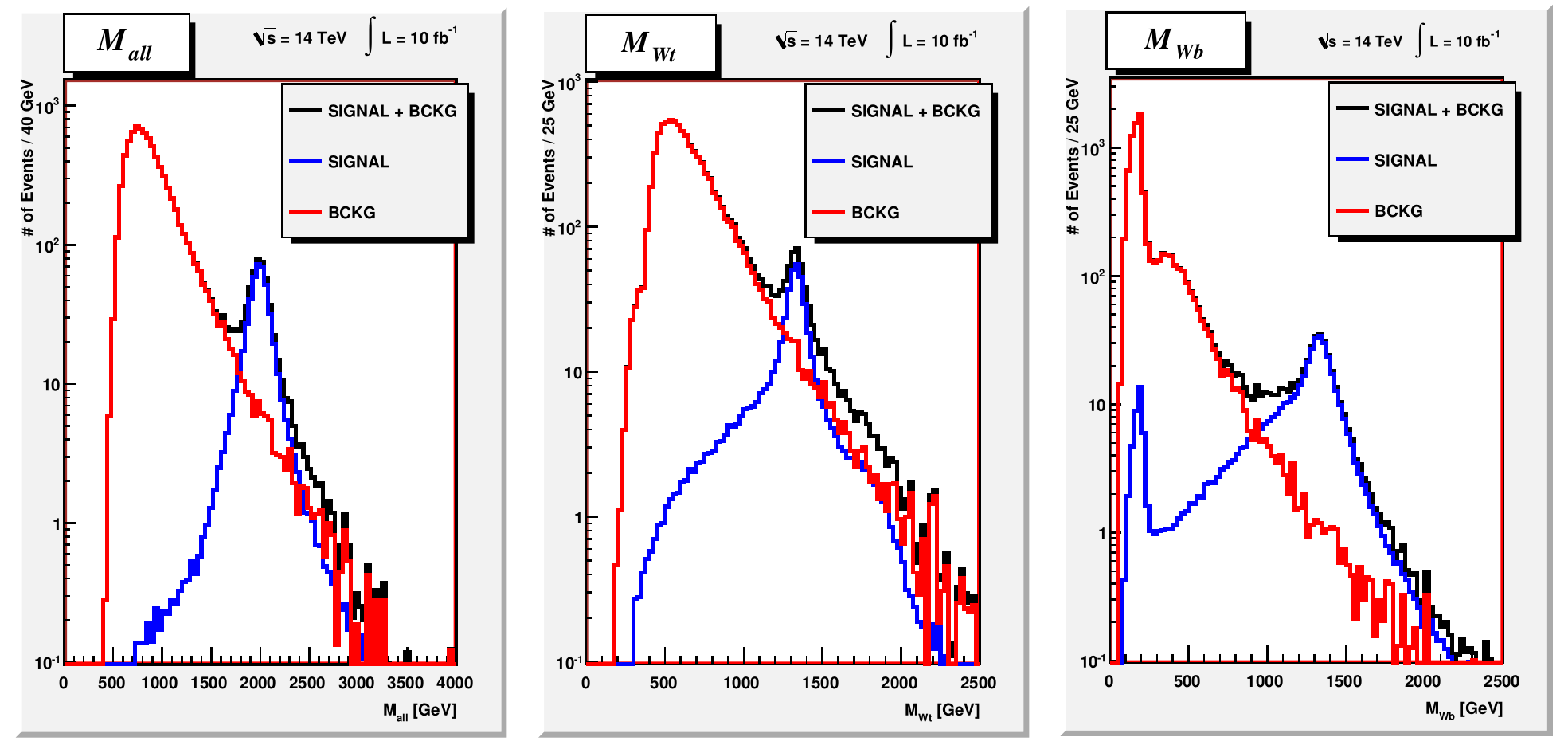}
\label{SandB_invMass}}}
\caption{\textit{Invariant Mass Distributions, $M_{all}$, $M_{Wt}$ and $M_{Wb}$, after the cuts at zero cost ($\sqrt{s}=14$ TeV).}}
\end{figure}

A very efficient way to distinguish between the signal and the background is looking at the 
invariant mass distributions: $M_{all}$, $M_{Wb}$ and $M_{Wt}$; 
where we can exploit the presence in the signal of heavy resonances, both from the $G^*$ and the heavy fermions.\\
We remind that $W$, $b$, $t$ denote the particles in the reconstructed $Wtb$ final state of our signal (Fig. \ref{wtb_S}): $t$ is the tagged top, $W$ and $b$ are not part of the tagged top.\\ 
The distribution of the signal events in the total invariant mass, $M_{all}$, will show a peak in correspondence of the $G^*$ mass, 
those in $M_{Wb}$ and in $M_{Wt}$ will show resonance peaks from the heavy partner of the top ($\tilde{T}$) 
and from the heavy partner of the bottom ($B$) respectively.\\
We have set for simplicity $M_{\tilde{T}}=M_{B}\ (=M_{G^*}/1.5)$; this should represent a quite realistic setting, 
since there is not a theoretical reason for a heavy fermion to be much heavier (or lighter) than the others.   
As consequence of the almost degeneracy in the heavy colored vector masses,
the component of the signal which is not resonant in the $M_{Wb}$ distribution, $G^*\to bB$, 
distributes quite close to the $\tilde{T}$ mass anyway. 
The same occurs for the $G^*\to t\tilde{T}$ component in the $M_{Wt}$ distribution.
We can check these features in Fig. \ref{M20_invMass}, where we show the invariant mass distributions, $M_{all}$, $M_{Wt}$ and $M_{Wb}$ for the signal with $M_{G*}=2$ TeV
(after the cuts at `zero cost'). In the $M_{Wb}$ distribution we can also recognize a bump near the top mass from the $G^*\to t\bar{t}$ component. 
The $M_{Wt}$ distribution shows a more narrow resonance than the one in $M_{Wb}$. This is due to the fact that the $M_{Wt}$ continuum, formed by the  
$G^*\to t\tilde{T}$ and the $t\bar{t}$ components, constitutes about one third of the $G^*$ signal and the most part of the signal comes from $G^*\to bB$. \\
The background distributes on low values of the invariant mass distributions, as Fig. \ref{B_invMass} shows. 
In particular, the $M_{Wb}$ distribution for the background has a high peak in correspondence of the top mass; 
the most part of the background, indeed, comes from $t\bar{t}$ events. 
For higher $M_{Wb}$ values, the $Wbb+jets$ and the $W+jets$ components of the background become relevant, as Fig. \ref{B_invMass} shows.\\
After the application of the cuts at `zero cost', we can already recognize signal excesses,
by looking at the invariant mass distributions, $M_{all}$, $M_{Wt}$ and $M_{Wb}$, as Fig. \ref{SandB_invMass} shows for a signal with $M_{G*}=2$ TeV.
An even clearer distinction between the signal and the background can be obtained with the following strategy.\\

\noindent
\textbf{The Main Strategy -}
We can clearly distinguish between the signal and the background
by considering 2D scatter plots of the $M_{all}$, $M_{Wt}$ and $M_{Wb}$ invariant mass distributions:
 \begin{center}
  $M_{all}$ $vs$  $M_{Wb}$ ,   $M_{Wt}$ $vs$ $M_{Wb}$  and  $M_{all}\  vs\ M_{Wt}$ .
 \end{center}
The background is predominantly distributed on low invariant mass values, 
 while the signal distributes close to the high values of the heavy colored vector masses. 
We can thus isolate an invariant mass region at low invariant mass values, that we treat as the `background region', 
and we can search for our heavy colored vector resonances in the remaining `signal region'. 
We have checked that once a `background region' is isolated, the excess of signal is evident in the invariant mass regions, 
also for the cases of lower signal cross sections, $M_{G^*}=3,4$ TeV (obviously, we need a sufficient integrated luminosity).
Once we have identified the signal excess, we can refine combined cuts in the invariant mass distributions to obtain the final discovery of the heavy colored vectors.\\ 

\begin{figure}[]
\centering
 \includegraphics[width=0.65\textwidth, angle=90]{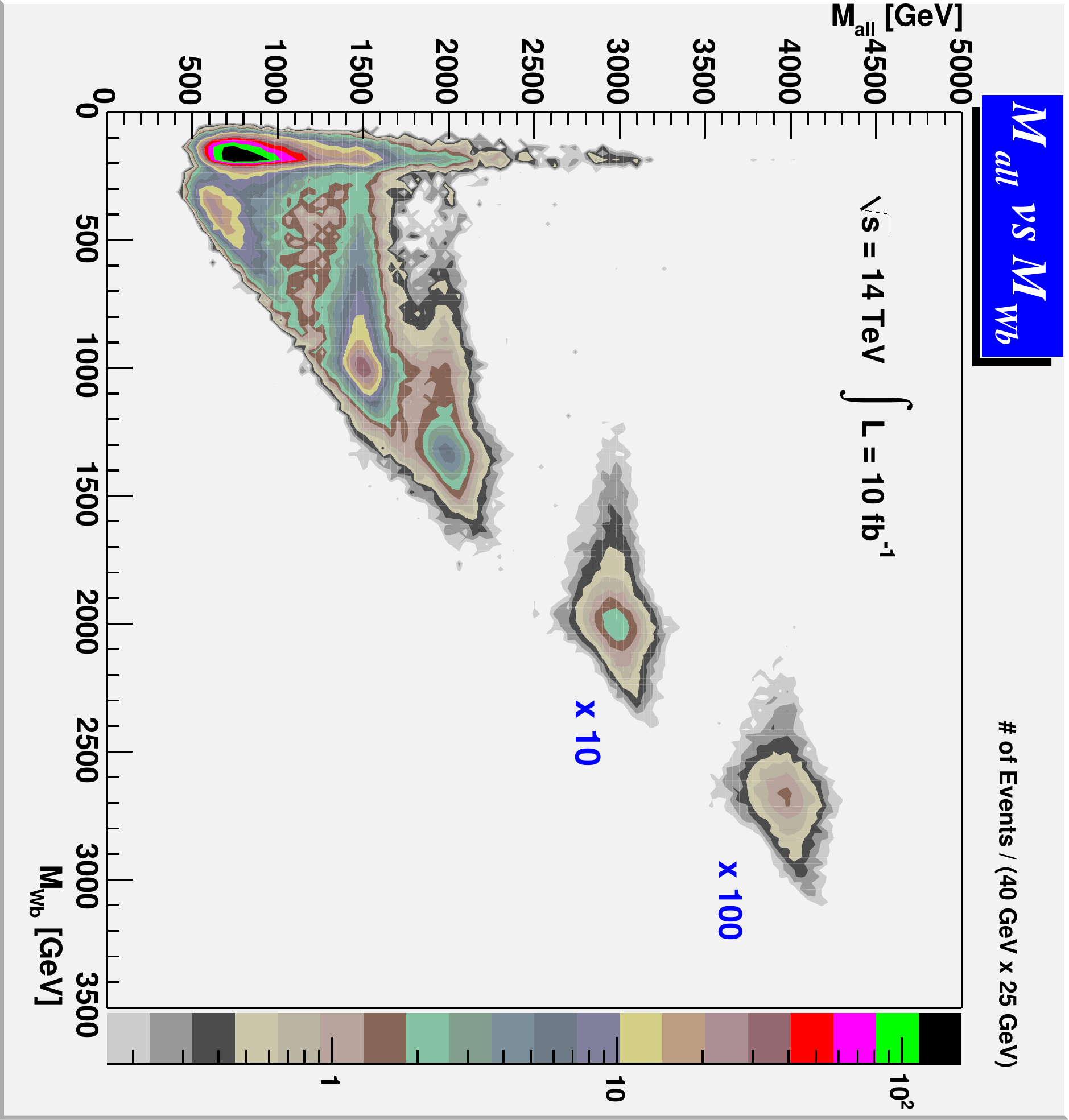}
\caption{\textit{contour plots of the 2D invariant mass distribution $M_{all}$ $vs$ $M_{Wb}$ for the background and for the signal with $M_{G*}=1.5,2,3,4$ TeV ($\sqrt{s}=14$ TeV). 
[The number of events for the signal with $M_{G*}=3(4)$ TeV has been multiplied by $10(100)$].} }
\label{Mtot_MWb}
\end{figure}

\begin{figure}[]
\centering
 \includegraphics[width=0.65\textwidth, angle=90]{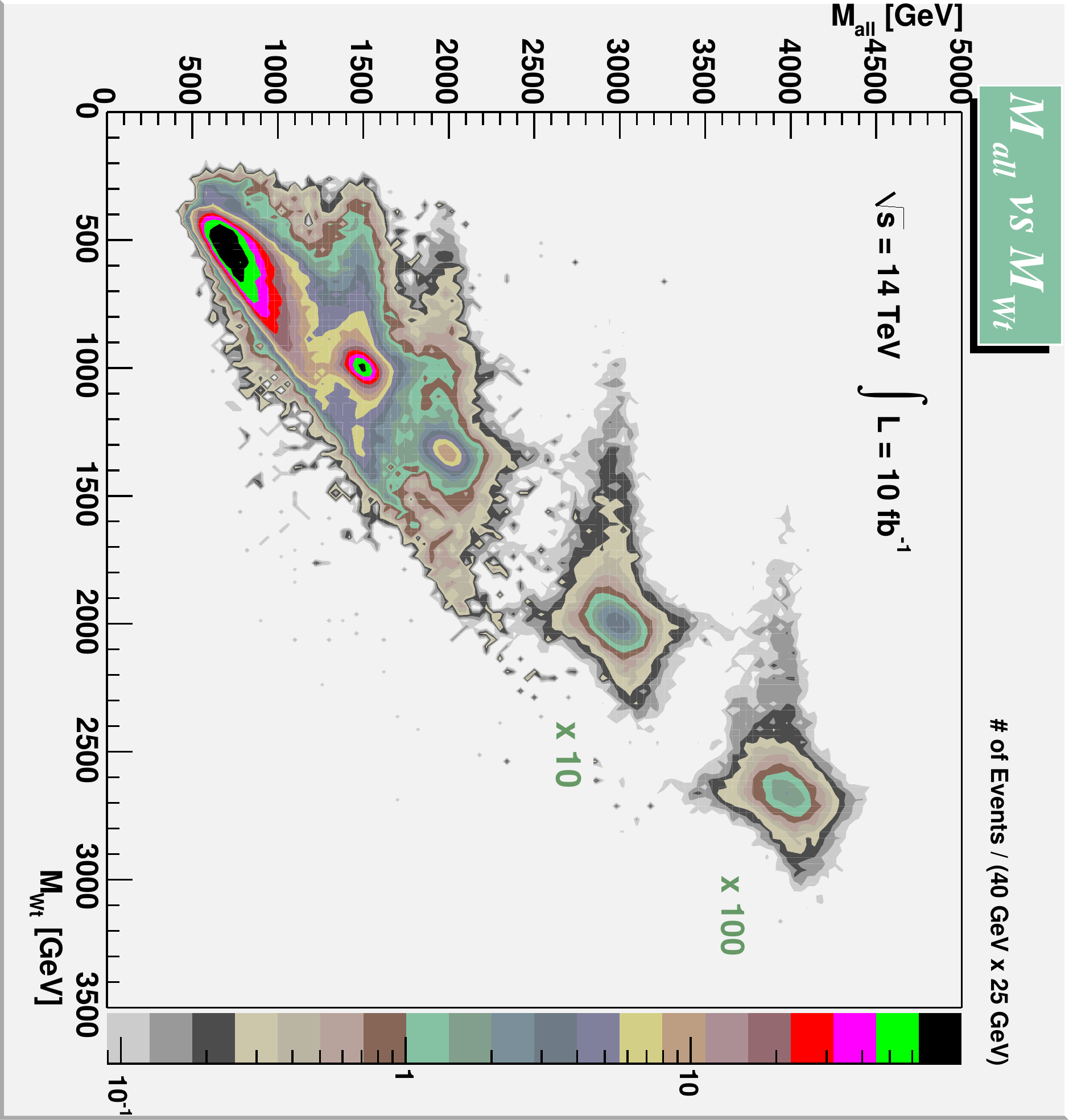}
\caption{\textit{contour plots of the 2D invariant mass distribution $M_{all}$ $vs$ $M_{Wt}$ for the background and for the signal with $M_{G*}=1.5,2,3,4$ TeV ($\sqrt{s}=14$ TeV). 
[The number of events for the signal with $M_{G*}=3(4)$ TeV has been multiplied by $10(100)$].} }
\label{Mtot_MWt}
\end{figure}

\begin{figure}[t]
\centering
 \includegraphics[width=0.65\textwidth, angle=90]{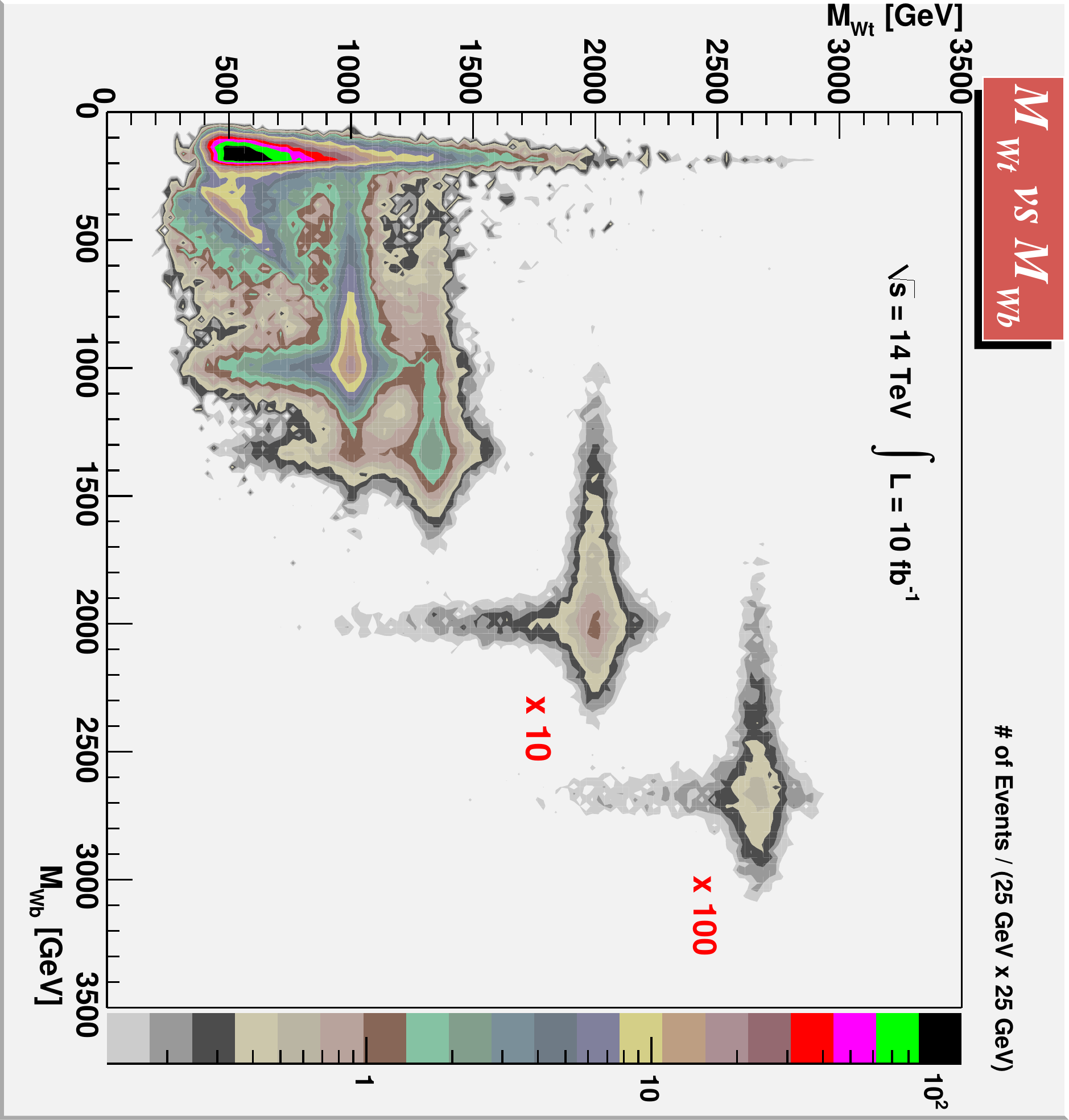}
\caption{\textit{contour plots of the 2D invariant mass distribution $M_{Wt}$ $vs$ $M_{Wb}$ for the background and for the signal with $M_{G*}=1.5,2,3,4$ TeV ($\sqrt{s}=14$ TeV). 
[The number of events for the signal with $M_{G*}=3(4)$ TeV has been multiplied by $10(100)$].} }
\label{MWt_MWb}
\end{figure}

We show in Fig.s \ref{Mtot_MWb}, \ref{Mtot_MWt} and \ref{MWt_MWb} the contour plots of the 2D invariant mass distribution $M_{all}$ $vs$ $M_{Wb}$, 
$M_{all}$ $vs$ $M_{Wt}$ and $M_{Wt}$ $vs$ $M_{Wb}$ respectively, for the background and for the signal with $M_{G*}=1.5,2,3,4$ TeV. 
These contour plots have been obtained from scatter plots of the expected number of events, after the `zero cost' cuts, for the background plus the different signals 
at an integrated luminosity of $10$ $fb^{-1}$ collected at the $14$ TeV LHC. 
The number of events for the signal with $M_{G*}=3(4)$ TeV has been multiplied by $10(100)$ in order to better visualize these signals together with
the background.\\
The contour plots show the expected features for the signal and the background: the background distributes predominantly on low invariant mass values and, 
in particular, it distributes in a region near the top mass in the $M_{Wb}$ invariant mass; 
the $\psi\chi$ signal distributes near the heavy colored vector masses; 
the $t\bar{t}$ component of the signal distributes on high invariant mass values (near the $G^{*}$ mass in $M_{all}$, near the $B$ mass in $M_{Wt}$)
except for the $M_{Wb}$ distribution, where it distributes near the top mass, as for the most part of the background.\\

Now we will follow the main strategy to discover our signal. \\
Since the most part of the signal lies near the small value of the top mass in the $M_{Wb}$ distribution, 
while the $\psi\chi$ signal distributes near the high value of the $\tilde{T}$ mass, a very efficient selection could be obtained by 
cutting the events with $M_{Wb}\lesssim m_{t}$. However, such a cut would throw out also the $t\bar{t}$ component of the signal.
In our first selection we will preserve this component by choosing a different cut in $M_{Wb}$.
This will allow us to discover with a unique analysis both the $\psi\chi$ and the $t\bar{t}$ components of the signal.
The simultaneous presence of the two types of the $G^*$ signal allows for a quick check on the model 
and also to extract in a simple way information on the top degree of compositeness, as we will explain in sec. \ref{tCOM}.
However, clearer discoveries (with higher $S/B$ ratios) can be achieved following
distinct strategies for the two types of the $G^*$ signal.
The $G^* \to t\bar{t}$ signal could be discovered apart from the $\psi\chi$ signal,
 following the specific analysis until now developed in the `ordinary' $G^*$ search at the LHC \cite{Agashe,Randall}. 
Therefore, our final analysis will be focused on the discovery for the $G^* \to \psi\chi$ signal. 
We will refine our cuts in the invariant mass distributions in order to optimize the discovery for the signal $G^* \to \psi\chi$. 
A strong point of our analysis, that makes it different from the previous searches for the $G^*$, will be the possibility to
effectively suppress the background by cutting the events with $M_{Wb}\lesssim m_{t}$. \\

\noindent
In order to discover the $G^*$ signal without throwing out the $t\bar{t}$ component, we derive the following cuts on the invariant mass distributions:\\

\noindent
For $M_{G^*}=1.5$ TeV: \\

\noindent
$M_{all}>1.3$ TeV and at least one of the conditions $M_{Wb}>0.8$ TeV, $M_{Wt}>0.8$ TeV respected. \\

\noindent
For $M_{G^*}=2$ TeV: 
\[
 M_{all}> 1.7\text{ TeV} \ \&\& \ ( M_{Wb}>1.1\  \text{TeV}\  ||\  M_{Wt}>1.1\ \text{TeV})
\]
For $M_{G^*}=3$ TeV: 
\[
 M_{all}> 2.7\text{ TeV} \ \&\& \ ( 1.4 \ \text{TeV} <M_{Wb}>2.6\  \text{TeV}\  ||\  M_{Wt}>1.4\ \text{TeV})
\]
For $M_{G^*}=4$ TeV: 
\begin{equation}
M_{all}> 3.6\text{ TeV} \ \&\& \ ( 2.4 \ \text{TeV} <M_{Wb}>3\  \text{TeV}\  ||\  M_{Wt}>2.4\ \text{TeV})
 \label{cut_preserve_tt}
\end{equation} 

\noindent
We have optimized the cuts to minimize the discovery luminosity.
 We first cut on the total invariant mass, then we throw out the region at low invariant mass values in the $M_{Wt}$ $vs$ $M_{Wb}$ distribution,
 by imposing inclusive conditions on $M_{Wb}$ and $M_{Wt}$.
This latter conditions preserve the $G^* \to t\bar{t}$ events, that distribute near the top mass in $M_{Wb}$ but on high $M_{Wt}$ values.\\
\noindent
In order to minimize the discovery luminosity, we find also useful to refine the $p_{T}$ cuts applied in (\ref{ZEROcut}). In particular, the 
cut on the hardest jet and on the $b$ (which is not part of the tagged top) give the highest significance values. 
For $M_{G^*}=1.5(2)$ TeV we refine the cut $p_{Tj(1)}>275 (400)$ GeV. 
For the cases of $M_{G^*}=3, 4$ TeV, we obtain a very high efficiency in the background suppression 
just after the application of the cuts in the invariant mass values, therefore we do not need a further refining. \\

In Table \ref{results} we show the final results of our analysis preserving the $G^*\to t\bar{t}$ component of the signal.
 We apply the cut on the invariant mass values in (\ref{cut_preserve_tt}) and the refining cuts in $p_{Tj(1)}$.
 $S/B$ denotes the signal over background ratio. 
$\mathcal{L}_{5 \sigma}$ denotes the integrated luminosity needed for a discovery at a significance of at least $5\sigma$. We also impose that 
at the integrated luminosity of $\mathcal{L}_{5 \sigma}$ at least ten events (both form signal or background) have passed all the cuts and are finally observed.
The statistical significance of the signal over the background is evaluated considering a Poisson distribution for the number of events passing all the cuts.
We define the minimum integrated luminosity required for a
discovery to be the integrated luminosity for which a goodness-of-fit test
of the SM-only hypothesis with Poisson distribution gives a p-value = $2.85\times 10^{-7}$,
 that corresponds to a $5\sigma$ significance in the limit of a gaussian distribution \cite{Stat_sign}.\\
 We show also the results for the analysis at $\sqrt{s}=7$ TeV. 
In this latter analysis we have applied the same strategy and the same cuts used for the case of $\sqrt{s}=14$ TeV, 
except for a little variation in the values of the `zero cost' cuts (\ref{7tev-ZEROcut}). 
The results for the simulation and the analysis at $\sqrt{s}=7$ TeV will be shown in detail in section \ref{7TEV}.\\
The results obtained show the possibility for a $G^*$ with a mass up to $\sim4$ TeV to be discovered at the $14$ TeV LHC.
The early stage of the LHC could discover a $G^*$ with a mass up to $\sim2$ TeV.
This result translates into the possibility for a heavy fermion, bottom or top partner, with a mass roughly 
included in a range $[M_{G^*}/2\ ,\ M_{G^*}]$ to be discovered at the LHC together with the $G^*$. 
We remind that the results shown refer to the setting (\ref{SetPar}), we are considering $\tan\theta_3=0.44$ and $s_R=0.6$. 
In section \ref{ParSpace} we will estimate the LHC discovery reach on the full parameter space: we will consider the variation of
 $M_{G*}$, $\tan\theta_3$ and $s_R$.
In section \ref{results} we will show how the clearness of the discovery cold be improved (we will obtain much higher $S/B$ ratios)
by refining an analysis specific for the $G^* \to \psi\chi$ channel.\\
 
\begin{figure}[]
\mbox{\subfigure[\textit{Final total invariant mass distribution for the signal with $M_{G*}=2$ TeV ($M_{\tilde{T}}=M_{B}=1.33$ TeV) plus the remaining background.}]
{\includegraphics[width=0.48\textwidth, angle=90]{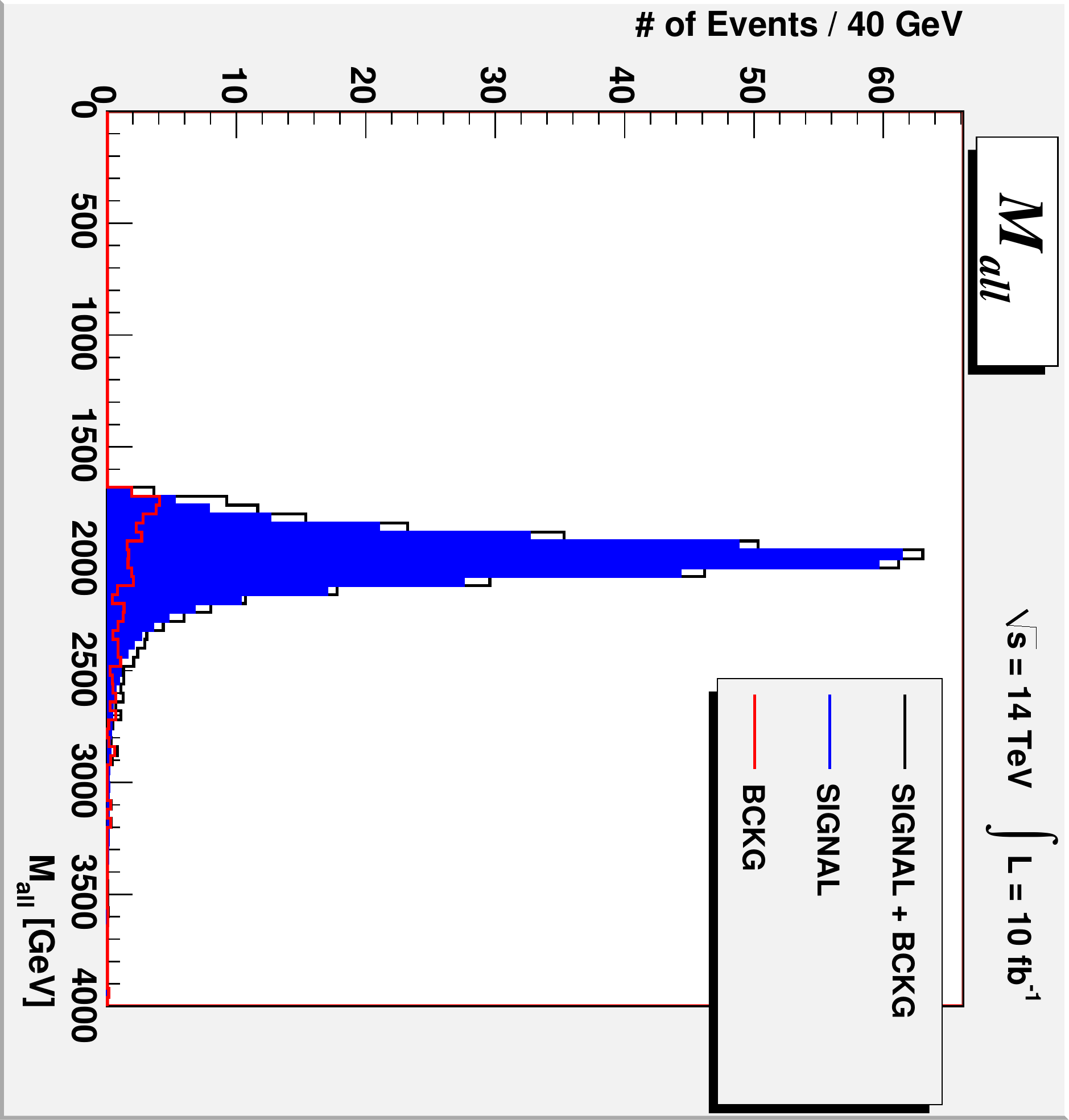}
\label{M20_all_tt}} \hspace{0.1cm}
\subfigure[\textit{Final $M_{Wt}$ distribution for the signal with $M_{G*}=2$ TeV ($M_{\tilde{T}}=M_{B}=1.33$ TeV) plus the remaining background.}]
{\includegraphics[width=0.48\textwidth, angle=90]{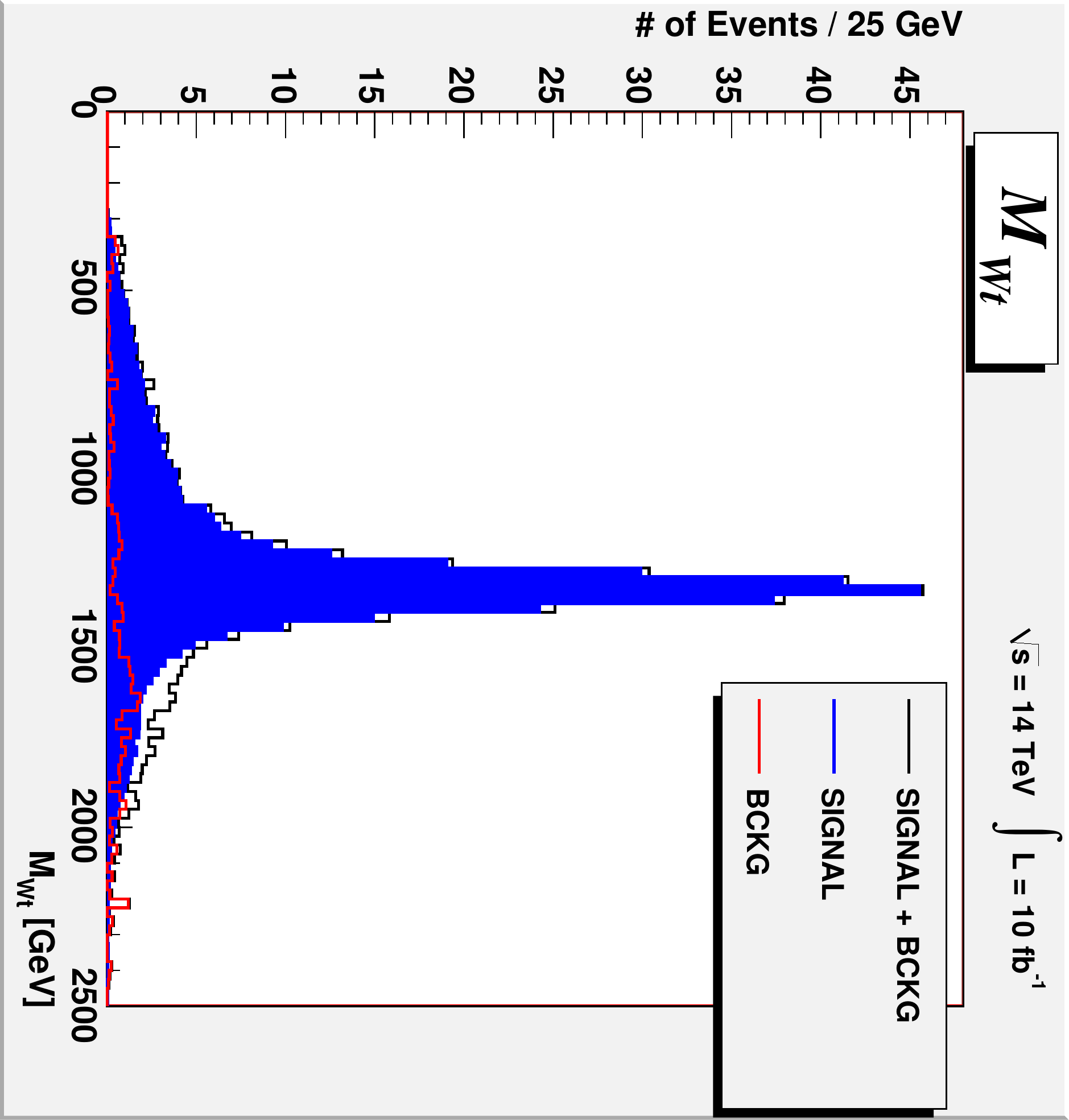}
\label{M20_wt_tt}}}\\
\centering
\mbox{\subfigure[\textit{Final $M_{Wb}$ distribution for the signal with $M_{G*}=2$ TeV ($M_{\tilde{T}}=M_{B}=1.33$ TeV) plus the remaining background.}]
{\includegraphics[width=0.52\textwidth, angle=90]{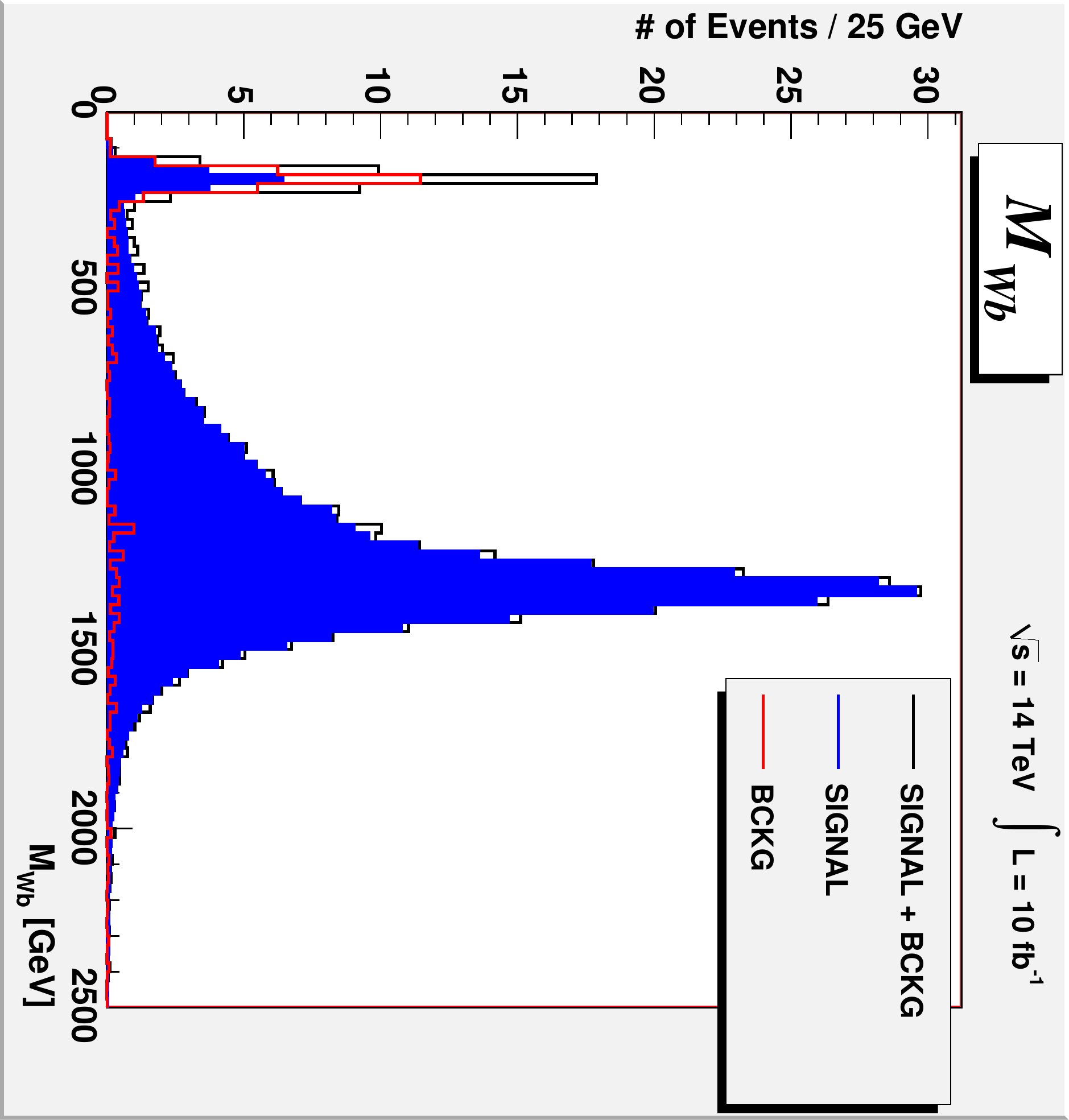}
\label{M20_wb_tt}}}
\caption{\textit{Final invariant mass distributions, $M_{all}$, $M_{Wt}$ and $M_{Wb}$, after the analysis preserving the $t\bar{t}$ component of the signal ($\sqrt{s}=14$ TeV).}}
\label{M20_with_tt}
\end{figure}

We show in Fig. \ref{M20_with_tt} the final invariant mass distributions, $M_{all}$, $M_{Wt}$ and $M_{Wb}$ for the signal with $M_{G*}=2$ TeV plus the remaining background.\\
We can clearly distinguish the resonances of the $G^*$, the $B$ and the $\tilde{T}$. 
As expected, the application of the cuts in \ref{cut_preserve_tt} has preserved the $G^* \to t\bar{t}$ component of the signal.
In the $M_{Wb}$ distribution we can distinguish a bump near the top mass from the $G^* \to t\bar{t}$ signal and also from the most part of the background.
The final $M_{Wb}$ distribution proves that the residual part of the background mainly distributes around the top mass. 
This clarify the effectiveness of imposing a cut $M_{Wb}\gtrsim m_t$ in the analysis we will perform for the $\psi\chi$ channel. \\
The final $M_{Wb}$ distribution in \ref{M20_wb_tt} could be useful to extract information on model parameters, such as the top degree of compositeness, 
as we are to explain in the following section. 

\begin{table}[]
\centering
 \begin{tabular}[]{ccccc}
&\multicolumn{2}{c}{ $\sqrt{s} = 14$ TeV } &\multicolumn{2}{c}{ $\sqrt{s} = 7$ TeV }\\
\hline
& $\mathcal{L}_{5 \sigma}$ & $S/B$ &$\mathcal{L}_{5 \sigma}$ & $S/B$\\
\hline
$M_{G^*}=1.5$ TeV & 57 pb$^{-1}$ & 6.2 & 0.40 fb$^{-1}$& 7.4 \\
$M_{G^*}=2$ TeV & 0.24 fb$^{-1}$ & 9.7 & 3.8 fb$^{-1}$& 6.6 \\
$M_{G^*}=3$ TeV & 4.0 fb$^{-1}$ & 5.4 & &  \\
$M_{G^*}=4$ TeV & 63 fb$^{-1}$ & 3.6 & &  \\
\end{tabular}\\
\caption{\textit{Results for the analysis preserving the $G^*\to t\bar{t}$ component of the signal. $\mathcal{L}_{5 \sigma}$ denotes the integrated luminosity needed for a $5\sigma$ discovery 
at the LHC, $S/B$ the Signal/Background ratio.}
}
\label{results}
\end{table}

\section{Information on the top degree of compositeness}\label{Hint_sR}

\begin{figure}[]
\centering
\includegraphics[width=0.65\textwidth]{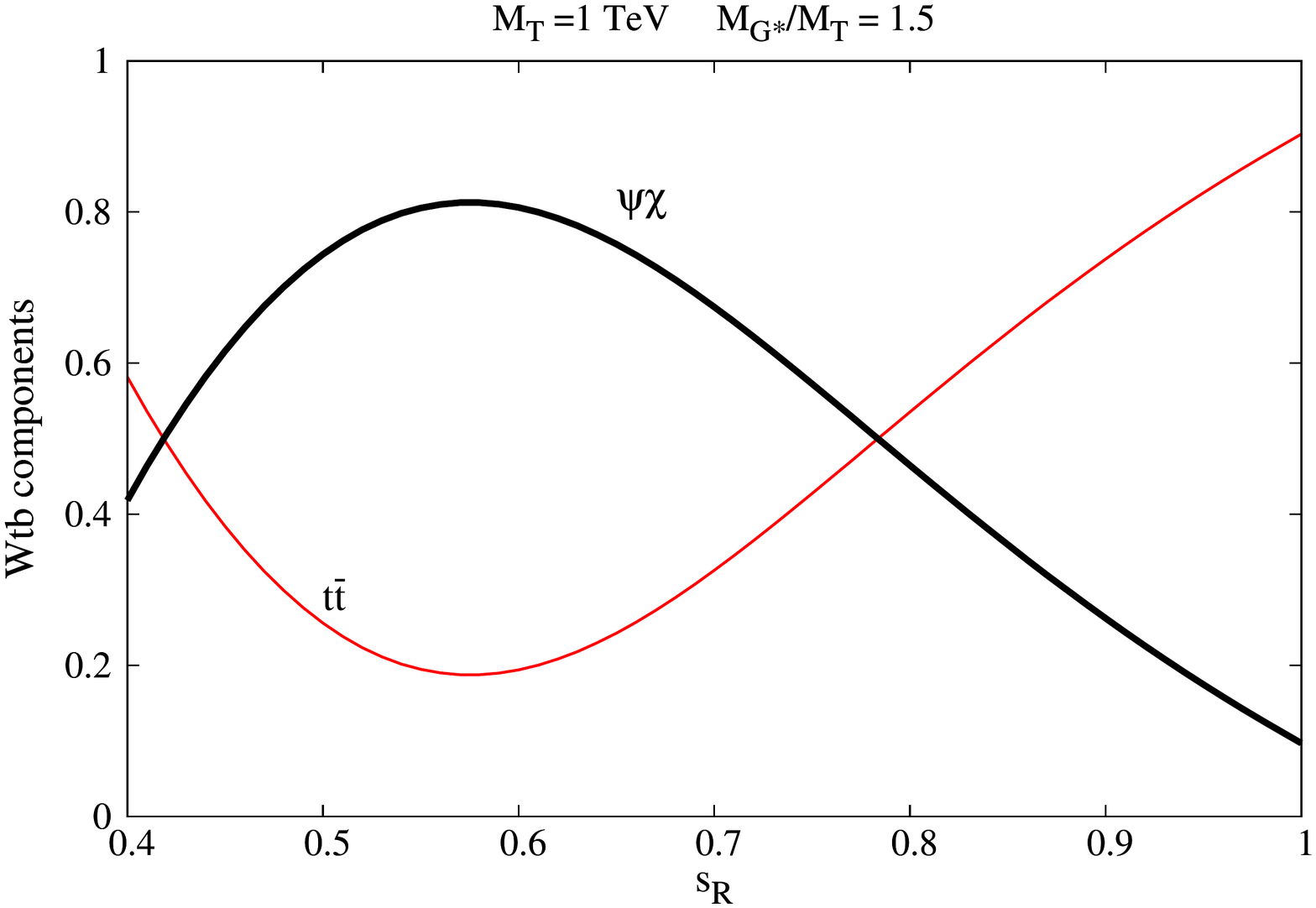}
\caption{\textit{Components of the $Wtb$ signal ($t\bar{t}$, red curve, $\psi\chi$, thick black curve), as function of the $t_R$ degree of compositeness ($s_R$).}
}
\label{Wtb_sR}
\end{figure}

The importance of the component $G^* \to t\bar{t}$ in the signal is related to the degree of compositeness of the top, $s_R$.
As we can check from the variation of the $G^*$ decay $BR$s with different $s_R$ values (Fig.s \ref{BR_sR1}, \ref{BR_sR08}, \ref{BR_sR06}), 
the $BR$ for the $G^*$ decay into top pairs increases when $s_R$ approaches its maximum value of $1$.
As a consequence, for larger $s_R$ values, the $BR$ for the $G^*$ decay into a SM fermion plus its heavy partner reduces.\\
In Fig. \ref{Wtb_sR} we show the result of the calculation of the relative amount of the components $G^* \to t\bar{t}$ and $G^* \to \psi\chi$ 
in our $G^* \to Wtb$ signal as functions of the top degree of compositeness. 
The result agrees with the expected trend of the $G^*$ $BR$s with larger $s_R$ values and it shows that the $t\bar{t}$ component 
slightly prevails on the $\psi\chi$ one even in the case of small $s_R$ values. This is a reasonable result since a small $s_R$ value corresponds to a large $s_1$ value, 
i.e. to a case of an almost fully composite left-handed top. The $BR$ for the $G^* \to t_L\bar{t}_L$ decay is therefore high in the case of a large $s_1$ value.
Fig. \ref{Wtb_sR} proves that the $\psi\chi$ component of the signal is maximum for intermediate top degree of compositeness, 
such as for the case $s_R=0.6$, that is the case we analyze, and it reduces slightly for small $s_R$ values and quite significantly for $s_R$ values close to $1$.
In this latter case, the number of events that lie near the top mass in the $M_{Wb}$ distribution should be larger than the one expected from Fig. \ref{Wtb_sR}
in the case $s_R=0.6$. The increase of this number, compared to the predicted value at $s_R=0.6$, could give an information on the true value of the top degree of compositeness.

An increase of the number of events near the top mass in the $M_{Wb}$ distribution should also occur, even though to a slighter extent, 
when $s_R$ approaches its minimum value (which is of about $0.4$ for $Y_*\sim 3$).
However, as we will explain in sec. \ref{tCOM}, the limit case of an almost fully composite left-handed top makes difficult the discovery of the $G^*$ both
in the $t\bar{t}$ and in the $\psi\chi$ channel. This is due to the predominance of the $G^*$ decays into pairs of custodians. 
These latter, indeed, become quite light in the limit of an almost fully composite left-handed top and can be produced in pairs in the decays of the $G^*$.
We point out that this limit case is quite disfavored by the electro-weak data \cite{pomarol_serra}, anyhow.\\

The plot in \ref{Wtb_sR} can be also read as an indication to follow one or the other analysis, the one for the $G^*$ discovery into the $t\bar{t}$ channel or that focused on the $G^* \to \psi\chi$ decay, depending on the expected $s_R$ value.

\section{Results for the $G^* \to \psi\chi$ search}\label{results}

The analysis performed until now has allowed to observe the $G^* \to \psi\chi$ and the $G^* \to t\bar{t}$ signals simultaneously, 
giving the opportunity to quickly test the model and to extract information on the top degree of compositeness.
However, as we already pointed out, the two components of the $G^*$ signal can be analyzed separately;
 the $G^* \to t\bar{t}$ signal can be more efficiently discovered by following the refining strategies in \cite{Agashe,Randall}, 
the $G^* \to \psi\chi$ signal, as well, can be more efficiently discovered\footnote{and by applying invariant mass cuts easier than those applied in the previous analysis (\ref{cut_preserve_tt})} by following the strategy that we are to present. \\
As advanced, a strong point of this strategy will be the possibility to
strongly suppress the background by cutting the events with $M_{Wb}\lesssim m_{t}$.\\

\textbf{Refining Strategy -} After the application of the cuts at `zero cost' (\ref{ZEROcut}), we consider again the invariant mass distributions, $M_{all}$, $M_{Wt}$ and $M_{Wb}$. 
The more efficient strategy that we find to discover the $G^* \to \psi\chi$ signal is to apply first a cut on the total invariant mass, $M_{all}$.
It is useful to choose a quite mild cut on $M_{all}$, in order to make an analysis quite independent on the specific values of the $G^*$ width 
(and therefore on the $\tan\theta_3$ parameter); the search strategy is finally optimized by considering cuts on the $M_{Wb}$ and the $M_{Wt}$ distributions.
The luminosity needed for the discovery can be minimized with a cut $M_{Wb}\gtrsim m_{t}$;
Once we have imposed a cut on $M_{Wb}$, we can further suppress the small remaining background by applying a cut on $M_{Wt}$. 
The $S/B$ ratio can then be maximized by optimizing the cut on $M_{Wt}$.
The optimized values of the $M_{Wb}$ and $M_{Wt}$ cuts can be chosen by looking at the 2D distribution $M_{Wt}$ $vs$ $M_{Wb}$, calculated after the application of the first cut on $M_{all}$.\\

\begin{figure}[]
\mbox{\subfigure[\textit{Background}]
{\includegraphics[width=0.48\textwidth, angle=90]{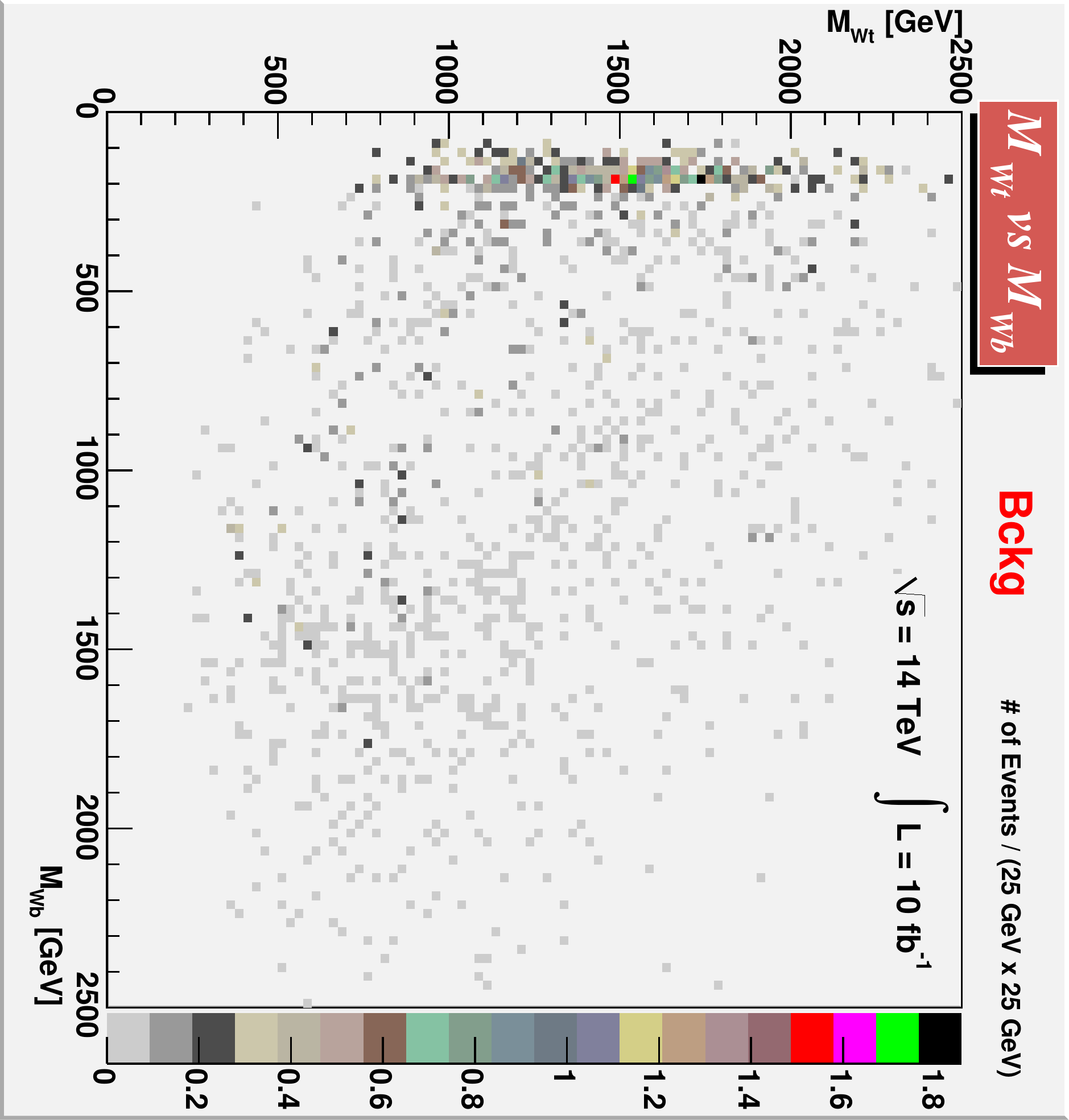}
}
\subfigure[\textit{Signal with $M_{G*}=2$ TeV ($M_T=1.33$ TeV)}]
{\includegraphics[width=0.48\textwidth, angle=90]{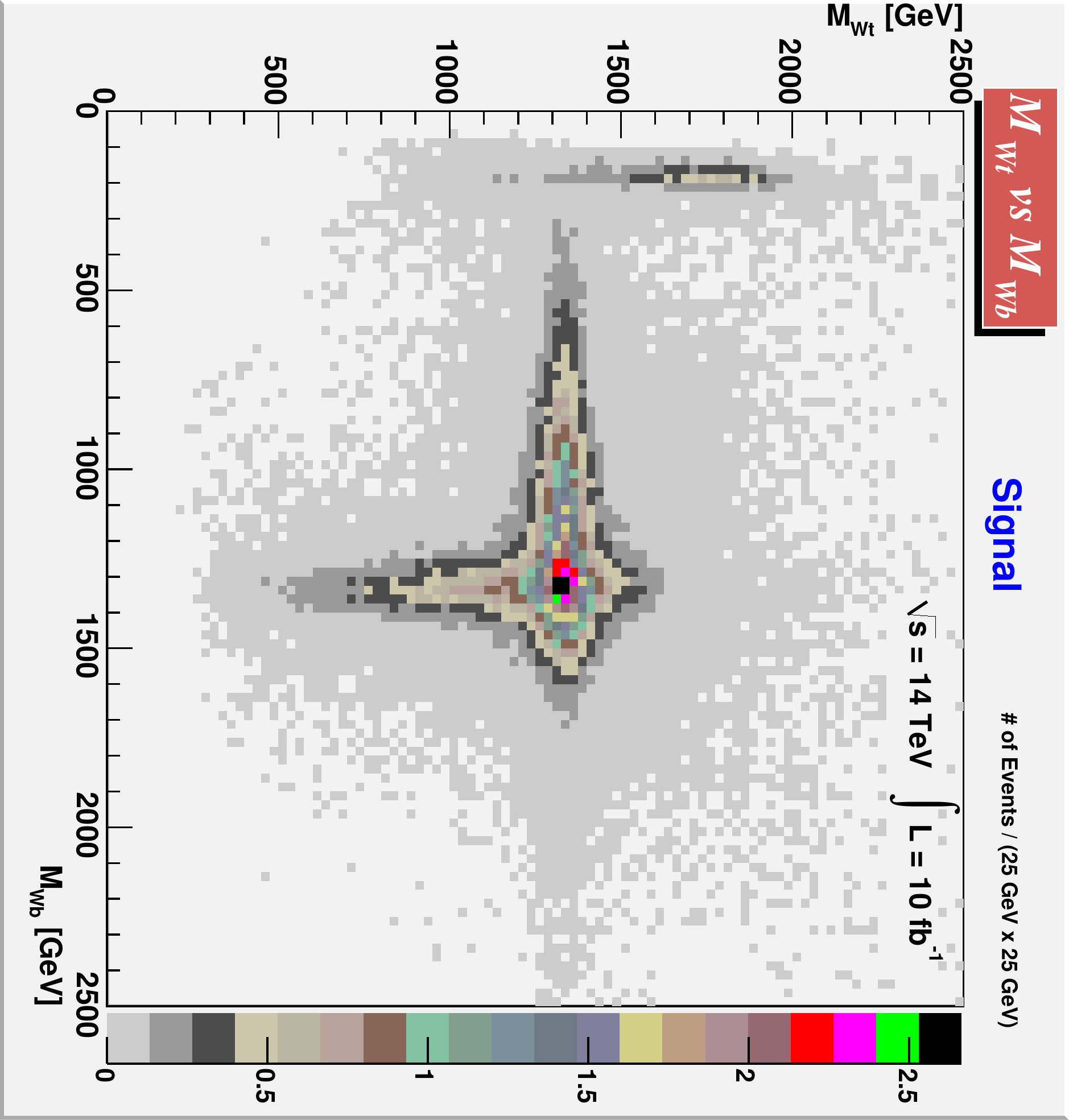}
}}
\caption{\textit{Scatter plots of $M_{Wt}$ $vs$ $M_{Wb}$ distribution for the background and for the signal with $M_{G*}=2$ TeV ($M_T=1.33$ TeV), 
after the application of a cut $M_{all}>1.7$ TeV ($\sqrt{s}=14$ TeV).}}
\label{MwtMwb_AcutMtot}
\end{figure}

In Fig. \ref{MwtMwb_AcutMtot} we show the $M_{Wt}$ $vs$ $M_{Wb}$ distribution for the background and for the signal with $M_{G*}=2$ TeV ($M_T=1.33$ TeV), 
after the application of a cut $M_{all}>1.7$ TeV. We can observe that the most part of the background distributes, as expected, near the top mass in $M_{Wb}$.
The cut $M_{Wb}>600$ GeV minimize the discovery luminosity for the signal with $M_{G*}=2$ TeV; 
after we have thrown out the region $M_{Wb}<600$ GeV, 
we can see that the residual part of the background can be efficiently suppressed by a cut on $M_{Wt}$, $M_{Wt}>1.1$ TeV, that preserves the most part of the signal events.\\


We derive the following cuts on the invariant mass distributions, optimized to minimize the integrated luminosity needed for the $G^* \to \psi\chi$ discovery:\\

\noindent
For $M_{G^*}=1.5$ TeV: 
\[
M_{all}>1.3 \text{ TeV} \ , \  M_{Wb}> 0.4\  \text{TeV}
\]
For $M_{G^*}=2$ TeV:
\[
 M_{all}> 1.7\text{ TeV} \ , \ M_{Wb}>0.6\  \text{TeV}
\]
For $M_{G^*}=3$ TeV: 
\[
 M_{all}> 2.5\text{ TeV} \ , \ M_{Wb}>0.6\  \text{TeV} \ , \ M_{Wb}>0.7\  \text{TeV}
\]
For $M_{G^*}=4$ TeV:
\begin{equation}
M_{all}> 3.2\text{ TeV} \ , \ M_{Wb}>0.7\  \text{TeV} \ , \ M_{Wt}>0.9\  \text{TeV}
 \label{cut_lum}
\end{equation} 
We denote this set of cuts by \textbf{SET (I)}.\\

\noindent
The signal over background ratio, $S/B$, for the cases $M_{G*}=1.5,2$ TeV, is then maximized by imposing the following cuts (after the `zero cost' cuts): \\

\noindent
For $M_{G^*}=1.5$ TeV: 
\[
M_{all}>1.3 \text{ TeV} \ , \  M_{Wb}> 0.4\  \text{TeV} \ , \  M_{Wt}> 0.9\  \text{TeV}
\]
For $M_{G^*}=2$ TeV:
\[
 M_{all}> 1.7\text{ TeV} \ , \ M_{Wb}>0.6\  \text{TeV} \ , \  M_{Wt}>1.1\  \text{TeV}
\]
For $M_{G^*}=3$ TeV: 
\[
 M_{all}> 2.5\text{ TeV} \ , \ M_{Wb}>0.6\  \text{TeV} \ , \  M_{Wt}> 0.7\  \text{TeV}
\]
For $M_{G^*}=4$ TeV:
\begin{equation}
M_{all}> 3.2\text{ TeV} \ , \ M_{Wb}>0.7\  \text{TeV} \ , \ M_{Wb}>0.9\  \text{TeV}
 \label{cut_SsuB}
\end{equation} 
We denote this set of cuts by \textbf{SET (II)}.\\

The SET (II) is given by the SET (I) plus a cut on $M_{Wt}$. For the cases $M_{G^*}=3,4$ TeV the cuts that minimize 
the discovery luminosity are also those that maximize the $S/B$ ratio.
For these cases of heavier $G^*$, $M_{G*}=3,4$ TeV, 
the cuts that optimize the discovery are basically those which preserve as much as possible of the signal events; 
the events of the signal from heavier composite resonances, indeed, distribute on very high value of the invariant masses, $M_{all}$, $M_{Wb}$, $M_{Wt}$; 
as a consequence the background is very suppressed by just mild cuts. We impose limit values that are faraway from the heavy resonance peaks; 
for example, the lower limit of $3.2$ TeV, imposed on the total invariant mass in the case of a signal with $M_{G*}=4$ TeV, corresponds to a cut $M_{all}\gtrsim (M_{G*}-4\Gamma(G*))$.  \\
We show in Tab. \ref{xsec_lum_disc} the cross section values for signals and backgrounds 
after the application of the cuts that minimize the discovery luminosity, SET (I) in (\ref{cut_lum}), and we show in Tab. \ref{xsec_SsuB} the cross section values for signals and backgrounds 
after the cuts that maximize $S/B$, SET (II) in (\ref{cut_SsuB}).\\
The statistical error associated to the cross section values has been calculated by considering a Poisson distribution for the number of events passing the cuts.
\footnote{We calculate the cross section after the application of a cut as
\[
 \sigma= \frac{n}{L}\ ,
\]
where $n$ is the number of *simulated* events that have passed the cut and $L$ is the integrated luminosity that we have reached in the *simulation*.\\
We consider a Poisson distribution for the true value of the number of events passing the cut, $\lambda$, given the observed number, $n$:
\[
 f(\lambda|n)=\frac{\lambda e^{-\lambda}}{n!}\ .
\]
$\lambda$ has a variance, $Var[\lambda]=n+1$; the variance we associate to the cross section is therefore
\[
 Var[\sigma]=\frac{n+1}{L^2} \ .
\]
\noindent
If, after the cuts, we obtain zero events in a particular sample, we fix an upper limit at $68\%$ C.L. on the cross section given by $\frac{1.1}{L^2}$. 
I thank C. Bini to having suggested this way to evaluate the statistical error.\\
When we sum over different cross section values, the error is summed in quadrature. 
We also consider that the cross section has to be a positive number; 
if in a particular sample we have obtained zero events and we have fixed an upper limit on the cross section of this sample, 
we do not add this error in the evaluation of the lower limit on the total cross section;
therefore, in these cases, we report an asymmetric error on the total cross section. 
} 

\begin{table}[]
\begin{tabular}[]{ccccc}
\multicolumn{1}{c}{$\sqrt{s}=14$ TeV}&\multicolumn{1}{||c}{SIGNAL}&\multicolumn{3}{|c}{BACKGROUND} \\
\multicolumn{1}{c}{SET (I)}&\multicolumn{1}{||c}{ $\sigma\ [fb]$}&\multicolumn{3}{|c}{ $\sigma\ [fb]$} \\
\hline
\hline
\multicolumn{1}{c}{$M_{G*}=1.5$ TeV} &\multicolumn{1}{||c}{$175.7 \pm 0.8$}&\multicolumn{1}{|c}{$WWbb$}&\multicolumn{1}{c}{$3.9 \pm 0.3$}&\multicolumn{1}{c}{$8.9 \pm 0.3$}\\
\cline{3-4}
\multicolumn{1}{c}{} &\multicolumn{1}{||c}{}&\multicolumn{1}{|c}{$Wbb+jets$}&\multicolumn{1}{c}{$4.1 \pm 0.1$}&\multicolumn{1}{c}{}\\
\cline{3-4}
\multicolumn{1}{c}{} &\multicolumn{1}{||c}{}&\multicolumn{1}{|c}{$W+jets$}&\multicolumn{1}{c}{$0.88 \pm 0.04$}&\multicolumn{1}{c}{}\\
\hline
\multicolumn{1}{c}{$M_{G*}=2$ TeV} &\multicolumn{1}{||c}{$39.5 \pm 0.2$}&\multicolumn{1}{|c}{$WWbb$}&\multicolumn{1}{c}{$1.0 \pm 0.2$}&\multicolumn{1}{c}{$2.4 \pm 0.2$}\\
\cline{3-4}
\multicolumn{1}{c}{} &\multicolumn{1}{||c}{}&\multicolumn{1}{|c}{$Wbb+jets$}&\multicolumn{1}{c}{$1.10 \pm 0.08$}&\multicolumn{1}{c}{}\\
\cline{3-4}
\multicolumn{1}{c}{} &\multicolumn{1}{||c}{}&\multicolumn{1}{|c}{$W+jets$}&\multicolumn{1}{c}{$0.31 \pm 0.01$}&\multicolumn{1}{c}{}\\
\hline
\multicolumn{1}{c}{$M_{G*}=3$ TeV} &\multicolumn{1}{||c}{$2.76 \pm 0.01$}&\multicolumn{1}{|c}{$WWbb$}&\multicolumn{1}{c}{$0.02 \pm 0.05$}&\multicolumn{1}{c}{$0.18 \pm 0.05$}\\
\cline{3-4}
\multicolumn{1}{c}{} &\multicolumn{1}{||c}{}&\multicolumn{1}{|c}{$Wbb+jets$}&\multicolumn{1}{c}{$0.09 \pm 0.02$}&\multicolumn{1}{c}{}\\
\cline{3-4}
\multicolumn{1}{c}{} &\multicolumn{1}{||c}{}&\multicolumn{1}{|c}{$W+jets$}&\multicolumn{1}{c}{$0.068 \pm 0.007$}&\multicolumn{1}{c}{}\\
\hline
\multicolumn{1}{c}{$M_{G*}=4$ TeV} &\multicolumn{1}{||c}{$0.231 \pm 0.001$}&\multicolumn{1}{|c}{$WWbb$}&\multicolumn{1}{c}{$<0.04$}&\multicolumn{1}{c}{$0.019 ^{+0.04}_{-0.007}$}\\
\cline{3-4}
\multicolumn{1}{c}{} &\multicolumn{1}{||c}{}&\multicolumn{1}{|c}{$Wbb+jets$}&\multicolumn{1}{c}{$0.007 ^{+0.008}_{-0.007}$}&\multicolumn{1}{c}{}\\
\cline{3-4}
\multicolumn{1}{c}{} &\multicolumn{1}{||c}{}&\multicolumn{1}{|c}{$W+jets$}&\multicolumn{1}{c}{$0.012 \pm 0.002$}&\multicolumn{1}{c}{}\\
\hline
\hline
\end{tabular}
\caption{\textit{Cross section values after the cuts that minimize the discovery luminosity, SET (I) in (\ref{cut_lum}). ($\sqrt{s}=14$ TeV)}
}
\label{xsec_lum_disc}
\end{table}

\begin{table}[]
\begin{tabular}[]{ccccc}
\multicolumn{1}{c}{$\sqrt{s}=14$ TeV}&\multicolumn{1}{||c}{SIGNAL}&\multicolumn{3}{|c}{BACKGROUND} \\
\multicolumn{1}{c}{SET (II)}&\multicolumn{1}{||c}{ $\sigma\ [fb]$}&\multicolumn{3}{|c}{ $\sigma\ [fb]$} \\
\hline
\hline
\multicolumn{1}{c}{$M_{G*}=1.5$ TeV} &\multicolumn{1}{||c}{$133.0 \pm 0.7$}&\multicolumn{1}{|c}{$WWbb$}&\multicolumn{1}{c}{$0.9 \pm 0.2$}&\multicolumn{1}{c}{$3.3 \pm 0.2$}\\
\cline{3-4}
\multicolumn{1}{c}{} &\multicolumn{1}{||c}{}&\multicolumn{1}{|c}{$Wbb+jets$}&\multicolumn{1}{c}{$1.97 \pm 0.09$}&\multicolumn{1}{c}{}\\
\cline{3-4}
\multicolumn{1}{c}{} &\multicolumn{1}{||c}{}&\multicolumn{1}{|c}{$W+jets$}&\multicolumn{1}{c}{$0.42 \pm 0.02$}&\multicolumn{1}{c}{}\\
\hline
\multicolumn{1}{c}{$M_{G*}=2$ TeV} &\multicolumn{1}{||c}{$32.3 \pm 0.2$}&\multicolumn{1}{|c}{$WWbb$}&\multicolumn{1}{c}{$0.13 \pm 0.07$}&\multicolumn{1}{c}{$0.82 \pm 0.08$}\\
\cline{3-4}
\multicolumn{1}{c}{} &\multicolumn{1}{||c}{}&\multicolumn{1}{|c}{$Wbb+jets$}&\multicolumn{1}{c}{$0.54 \pm 0.04$}&\multicolumn{1}{c}{}\\
\cline{3-4}
\multicolumn{1}{c}{} &\multicolumn{1}{||c}{}&\multicolumn{1}{|c}{$W+jets$}&\multicolumn{1}{c}{$0.15 \pm 0.01$}&\multicolumn{1}{c}{}\\
\hline
\multicolumn{1}{c}{$M_{G*}=3$ TeV} &\multicolumn{1}{||c}{$2.76 \pm 0.01$}&\multicolumn{1}{|c}{$WWbb$}&\multicolumn{1}{c}{$0.02 \pm 0.05$}&\multicolumn{1}{c}{$0.18 \pm 0.05$}\\
\cline{3-4}
\multicolumn{1}{c}{} &\multicolumn{1}{||c}{}&\multicolumn{1}{|c}{$Wbb+jets$}&\multicolumn{1}{c}{$0.09 \pm 0.02$}&\multicolumn{1}{c}{}\\
\cline{3-4}
\multicolumn{1}{c}{} &\multicolumn{1}{||c}{}&\multicolumn{1}{|c}{$W+jets$}&\multicolumn{1}{c}{$0.068 \pm 0.007$}&\multicolumn{1}{c}{}\\
\hline
\multicolumn{1}{c}{$M_{G*}=4$ TeV} &\multicolumn{1}{||c}{$0.231 \pm 0.001$}&\multicolumn{1}{|c}{$WWbb$}&\multicolumn{1}{c}{$< 0.04$}&\multicolumn{1}{c}{$0.019 ^{+0.04}_{-0.007}$}\\
\cline{3-4}
\multicolumn{1}{c}{} &\multicolumn{1}{||c}{}&\multicolumn{1}{|c}{$Wbb+jets$}&\multicolumn{1}{c}{$0.007 ^{+0.008}_{-0.007}$}&\multicolumn{1}{c}{}\\
\cline{3-4}
\multicolumn{1}{c}{} &\multicolumn{1}{||c}{}&\multicolumn{1}{|c}{$W+jets$}&\multicolumn{1}{c}{$0.012 \pm 0.002$}&\multicolumn{1}{c}{}\\
\hline
\hline
\end{tabular}
\caption{\textit{Cross section values after the cuts that maximize the $S/B$ ratio, SET (II) in (\ref{cut_SsuB}). ($\sqrt{s}=14$ TeV)}
}
\label{xsec_SsuB}
\end{table}

\section{Prospects for the discovery at the early LHC}\label{7TEV}

We repeat the analysis for the case $\sqrt{s}=7$ TeV, which is the center-of-mass energy of the current runs at the LHC. 
We can follow the same strategy adopted for the 
$\sqrt{s}=14$ TeV case, with only a small variation in the values of the cuts at `zero cost'. 
This is because the cuts are optimized to exploit the peculiar kinematics of the signal, 
and a change in the collider energy mainly implies a rescaling of the production cross sections of signal and background
via the parton luminosities, without affecting the kinematic distributions. \\

After we simulate at $\sqrt{s}=7$ TeV our signal events with MADGRAPH and the backgrounds using both MADGRAPH and MADEVENT\footnote{
as for the $\sqrt{s}=14$ TeV case, we have used the CTEQ6L1 pdf set. 
The samples for the $Wtb$ signal and the background $WWbb$ have been generated in MADGRAPH using the factorization and renormalization scales, 
 $Q=M_{G*}$ and $Q=\sqrt{M^2_W+\sum_b p^2_T}$ respectively. 
The samples for $Wbb+jets$ and $W+jets$ have been generated in ALPGEN using the scales $Q^2=M^2_W + p^{2}_{T,W}$.}, 
we impose the same acceptance cuts as for the case of $\sqrt{s}=14$ TeV (\ref{AcceptanceCut}). 
We show in Tab. \ref{7tevXSEC_AC-bTAG} the values of the cross sections for signals and backgrounds after the acceptance cuts and the b-tagging.
As for the case $\sqrt{s}=14$ TeV, we take into account a b-tagging efficiency of $60\%$ and a misidentification factor of $1/100$ for the light jets to be tagged as $b$ jets.\\

\begin{table}[t]
\subtable[SIGNAL]{
\begin{tabular}[]{ccccccc}
\multicolumn{1}{c}{}&\multicolumn{3}{||c}{ $\sigma\ [fb]$ After Acceptance Cuts} &\multicolumn{3}{||c}{ $\sigma\ [fb]$ After b-TAG}\\
\hline
\hline
\multicolumn{1}{c}{$M_{G*}=1.5$ TeV}&\multicolumn{1}{||c}{$\chi\psi$} &\multicolumn{1}{c}{82.5}&\multicolumn{1}{c}{98.8}&\multicolumn{1}{||c}{$\chi\psi$} &\multicolumn{1}{c}{29.8}&\multicolumn{1}{c}{35.7}\\
\cline{2-3}\cline{5-6}
\multicolumn{1}{c}{}&\multicolumn{1}{||c}{$t\bar{t}$} &\multicolumn{1}{c}{16.3}&\multicolumn{1}{c}{}&\multicolumn{1}{||c}{$t\bar{t}$} &\multicolumn{1}{c}{5.85}&\multicolumn{1}{c}{}\\
\hline
\hline
\multicolumn{1}{c}{$M_{G*}=2$ TeV}&\multicolumn{1}{||c}{$\chi\psi$} &\multicolumn{1}{c}{9.08}&\multicolumn{1}{c}{11.0}&\multicolumn{1}{||c}{$\chi\psi$} &\multicolumn{1}{c}{3.29}&\multicolumn{1}{c}{4.00}\\
\cline{2-3}\cline{5-6}
\multicolumn{1}{c}{}&\multicolumn{1}{||c}{$t\bar{t}$} &\multicolumn{1}{c}{1.96}&\multicolumn{1}{c}{}&\multicolumn{1}{||c}{$t\bar{t}$} &\multicolumn{1}{c}{0.714}&\multicolumn{1}{c}{}\\
\hline
\hline
\multicolumn{1}{c}{$M_{G*}=3$ TeV}&\multicolumn{1}{||c}{$\chi\psi$} &\multicolumn{1}{c}{0.106}&\multicolumn{1}{c}{0.271}&\multicolumn{1}{||c}{$\chi\psi$} &\multicolumn{1}{c}{0.0381}&\multicolumn{1}{c}{0.0975}\\
\cline{2-3}\cline{5-6}
\multicolumn{1}{c}{}&\multicolumn{1}{||c}{$t\bar{t}$} &\multicolumn{1}{c}{0.165}&\multicolumn{1}{c}{}&\multicolumn{1}{||c}{$t\bar{t}$} &\multicolumn{1}{c}{0.0594}&\multicolumn{1}{c}{}\\
\hline
\hline
\end{tabular}
}
\subtable[BACKGROUND]{
 \begin{tabular}[]{ccccccc}
\multicolumn{1}{c}{}&\multicolumn{3}{||c}{ $\sigma\ [pb]$ After Acceptance Cuts} &\multicolumn{3}{||c}{ $\sigma\ [pb]$ After b-TAG}\\
\hline
\hline
\multicolumn{1}{c}{$WWbb$}&\multicolumn{1}{||c}{} &\multicolumn{1}{c}{}&\multicolumn{1}{c}{13.5}&\multicolumn{1}{||c}{} &\multicolumn{1}{c}{}&\multicolumn{1}{c}{4.84}\\
\hline
\hline
\multicolumn{1}{c}{$Wbb+jets$}&\multicolumn{1}{||c}{$Wbb+1J$} &\multicolumn{1}{c}{0.584}&\multicolumn{1}{c}{0.865}&\multicolumn{1}{||c}{$Wbb+1J$} &\multicolumn{1}{c}{0.210}&\multicolumn{1}{c}{0.312}\\
\cline{2-3}\cline{5-6}
\multicolumn{1}{c}{}&\multicolumn{1}{||c}{$Wbb+2J$} &\multicolumn{1}{c}{0.281}&\multicolumn{1}{c}{}&\multicolumn{1}{||c}{$Wbb+2J$} &\multicolumn{1}{c}{0.102}&\multicolumn{1}{c}{}\\
\cline{2-3}\cline{5-6}
\hline
\hline
\multicolumn{1}{c}{$W+jets$}&\multicolumn{1}{||c}{$W+3J$} &\multicolumn{1}{c}{78.7}&\multicolumn{1}{c}{97.1}&\multicolumn{1}{||c}{$W+3J$} &\multicolumn{1}{c}{0.0188}&\multicolumn{1}{c}{0.0277}\\
\cline{2-3}\cline{5-6}
\multicolumn{1}{c}{}&\multicolumn{1}{||c}{$W+4J$} &\multicolumn{1}{c}{18.4}&\multicolumn{1}{c}{}&\multicolumn{1}{||c}{$W+4J$} &\multicolumn{1}{c}{0.00889}&\multicolumn{1}{c}{}\\
\cline{2-3}
\hline
\hline
\multicolumn{1}{c}{Total BCKG}&\multicolumn{1}{||c}{} &\multicolumn{1}{c}{}&\multicolumn{1}{c}{111}&\multicolumn{1}{||c}{} &\multicolumn{1}{c}{}&\multicolumn{1}{c}{5.18}\\
\hline
\hline
\end{tabular}
}
\caption{\textit{Cross Section values after the Acceptance Cuts (\ref{AcceptanceCut}) and the b-tagging ($\sqrt{s}=7$ TeV).}
}
\label{7tevXSEC_AC-bTAG}
\end{table}

\begin{table}[h!]
\subtable[SIGNAL]{
\begin{tabular}[]{cccc}
 \multicolumn{2}{c}{}& \multicolumn{1}{c}{$\epsilon_{\nu}$} &\multicolumn{1}{c}{$\epsilon_t$}\\
\hline
\multicolumn{1}{c}{$M_{G*}=1.5$ TeV}&\multicolumn{1}{||c}{$\chi\psi$} &\multicolumn{1}{c}{0.81}&\multicolumn{1}{c}{0.99}\\
\cline{2-4}
\multicolumn{1}{c}{}&\multicolumn{1}{||c}{$t\bar{t}$} &\multicolumn{1}{c}{0.82}&\multicolumn{1}{c}{1.0}\\
\hline
\multicolumn{1}{c}{$M_{G*}=2$ TeV}&\multicolumn{1}{||c}{$\chi\psi$} &\multicolumn{1}{c}{0.80}&\multicolumn{1}{c}{0.99}\\
\cline{2-4}
\multicolumn{1}{c}{}&\multicolumn{1}{||c}{$t\bar{t}$} &\multicolumn{1}{c}{0.80}&\multicolumn{1}{c}{1.0}\\
\hline
\multicolumn{1}{c}{$M_{G*}=3$ TeV}&\multicolumn{1}{||c}{$\chi\psi$} &\multicolumn{1}{c}{0.80}&\multicolumn{1}{c}{0.99}\\
\cline{2-4}
\multicolumn{1}{c}{}&\multicolumn{1}{||c}{$t\bar{t}$} &\multicolumn{1}{c}{0.82}&\multicolumn{1}{c}{1.0}\\
\hline
\end{tabular}
}
\subtable[BACKGROUND]{
\begin{tabular}[]{cccc}
 \multicolumn{1}{c}{}& \multicolumn{1}{c}{$\epsilon_{\nu}$} &\multicolumn{1}{c}{$\epsilon_t$}\\
\hline
\multicolumn{1}{c}{$WWbb$} &\multicolumn{1}{c}{0.81}&\multicolumn{1}{c}{1.0}\\
\hline
\multicolumn{1}{c}{$Wbb+jets$} &\multicolumn{1}{c}{0.79}&\multicolumn{1}{c}{0.92}\\
\hline
\multicolumn{1}{c}{$W+jets$} &\multicolumn{1}{c}{0.80}&\multicolumn{1}{c}{0.95}\\
\hline
\end{tabular}
}
\caption{\textit{Efficiencies of the neutrino reconstruction ($\epsilon_{\nu}$) and the top-tagging ($\epsilon_t$) procedures, for signal and backgrounds ($\sqrt{s}=7$ TeV).}}
\label{7tevNeuTopEff}
\end{table}

We then repeat the procedure explained in sec. \ref{reco} for the reconstruction of the neutrino and the tagging of the one top of our $Wtb$ signal.
We obtain efficiency values of the neutrino reconstruction ($\epsilon_{\nu}$) and the top-tagging ($\epsilon_t$), for signal and backgrounds,
 very similar to those obtained at $\sqrt{s}=14$ TeV (Tab. \ref{7tevNeuTopEff}). \\
 
We apply a preliminary selection of the events, by exploiting cuts at `zero cost' as the type of sec. \ref{CUTzeroCost}. 
The cuts at `zero cost' we impose after top tagging are the following:
\[
p_{Tj (1)} > 155\ GeV , \ p_{Tj (2)} > 75 \ GeV \ ,
\]
\[
 p_{T top} > 105 \ GeV  , \ p_{T W} > 90 \ GeV , \ p_{T b} > 65 \ GeV \ , 
\]
\begin{equation}
 M_{jets} < 200\  GeV \ .
\label{7tev-ZEROcut}
\end{equation}
The cuts are chosen so as to have an individual efficiency of $97\%$ on the signal at $M_{G*}=1.5$ TeV. \\
We show in Tab. \ref{7tevzeroCutEff} the efficiencies of the cuts on $p_T$ at `zero cost', for signal and background,
and in Tab. \ref{7tevXSEC_AZeroCut} the cross section values after we impose the cuts at `zero cost'.

\begin{table}[]
\begin{tabular}[]{cccccccc}
 \multicolumn{2}{c}{}& \multicolumn{1}{c}{$p_{Tb}$} &\multicolumn{1}{c}{$p_{Ttop}$}&\multicolumn{1}{c}{$p_{TW}$} &\multicolumn{1}{c}{$p_{Tj(1)}$}&\multicolumn{1}{c}{$p_{Tj(2)}$}&\multicolumn{1}{c}{Combined Cuts}\\
\hline
\multicolumn{1}{c}{$M_{G*}=1.5$ TeV}&\multicolumn{1}{||c}{$\chi\psi$} &\multicolumn{1}{c}{0.99}&\multicolumn{1}{c}{0.97}&\multicolumn{1}{c}{0.99}&\multicolumn{1}{c}{0.99}&\multicolumn{1}{c}{0.99}&\multicolumn{1}{c}{}\\
\cline{2-7}
\multicolumn{1}{c}{}&\multicolumn{1}{||c}{$t\bar{t}$} &\multicolumn{1}{c}{0.85}&\multicolumn{1}{c}{0.96}&\multicolumn{1}{c}{0.89}&\multicolumn{1}{c}{0.85}&\multicolumn{1}{c}{0.91}&\multicolumn{1}{c}{}\\
\cline{2-7}
\multicolumn{1}{c}{}&\multicolumn{1}{}{} &\multicolumn{1}{c}{0.97}&\multicolumn{1}{c}{0.97}&\multicolumn{1}{c}{0.97}&\multicolumn{1}{c}{0.97}&\multicolumn{1}{c}{0.97}&\multicolumn{1}{c}{0.90}\\
\hline
\multicolumn{1}{c}{$M_{G*}=2$ TeV}&\multicolumn{1}{||c}{$\chi\psi$} &\multicolumn{1}{c}{1.0}&\multicolumn{1}{c}{0.98}&\multicolumn{1}{c}{0.99}&\multicolumn{1}{c}{1.0}&\multicolumn{1}{c}{0.99}&\multicolumn{1}{c}{}\\
\cline{2-7}
\multicolumn{1}{c}{}&\multicolumn{1}{||c}{$t\bar{t}$} &\multicolumn{1}{c}{0.83}&\multicolumn{1}{c}{0.92}&\multicolumn{1}{c}{0.85}&\multicolumn{1}{c}{0.77}&\multicolumn{1}{c}{0.87}&\multicolumn{1}{c}{}\\
\cline{2-7}
\multicolumn{1}{c}{}&\multicolumn{1}{}{} &\multicolumn{1}{c}{0.97}&\multicolumn{1}{c}{0.97}&\multicolumn{1}{c}{0.97}&\multicolumn{1}{c}{0.97}&\multicolumn{1}{c}{0.97}&\multicolumn{1}{c}{0.91}\\
\hline
\multicolumn{1}{c}{$M_{G*}=3$ TeV}&\multicolumn{1}{||c}{$\chi\psi$} &\multicolumn{1}{c}{1.0}&\multicolumn{1}{c}{0.99}&\multicolumn{1}{c}{1.0}&\multicolumn{1}{c}{1.0}&\multicolumn{1}{c}{1.0}&\multicolumn{1}{c}{}\\
\cline{2-7}
\multicolumn{1}{c}{}&\multicolumn{1}{||c}{$t\bar{t}$} &\multicolumn{1}{c}{0.75}&\multicolumn{1}{c}{0.84}&\multicolumn{1}{c}{0.74}&\multicolumn{1}{c}{0.55}&\multicolumn{1}{c}{0.77}&\multicolumn{1}{c}{}\\
\cline{2-7}
\multicolumn{1}{c}{}&\multicolumn{1}{}{} &\multicolumn{1}{c}{0.85}&\multicolumn{1}{c}{0.90}&\multicolumn{1}{c}{0.84}&\multicolumn{1}{c}{0.73}&\multicolumn{1}{c}{0.86}&\multicolumn{1}{c}{0.63}\\
\hline
\hline
\multicolumn{2}{c}{$WWbb$}&\multicolumn{1}{c}{0.59}&\multicolumn{1}{c}{0.46}&\multicolumn{1}{c}{0.38}&\multicolumn{1}{c}{0.10}&\multicolumn{1}{c}{0.42}&\multicolumn{1}{c}{0.042}\\
\hline
\multicolumn{2}{c}{$Wbb+jets$}&\multicolumn{1}{c}{0.45}&\multicolumn{1}{c}{0.28}&\multicolumn{1}{c}{0.40}&\multicolumn{1}{c}{0.18}&\multicolumn{1}{c}{0.36}&\multicolumn{1}{c}{0.044}\\
\hline
\multicolumn{2}{c}{$W+jets$}&\multicolumn{1}{c}{0.47}&\multicolumn{1}{c}{0.34}&\multicolumn{1}{c}{0.34}&\multicolumn{1}{c}{0.19}&\multicolumn{1}{c}{0.35}&\multicolumn{1}{c}{0.054}\\
\hline
\hline
\multicolumn{2}{c}{Total BCKG}&\multicolumn{1}{c}{0.58}&\multicolumn{1}{c}{0.45}&\multicolumn{1}{c}{0.38}&\multicolumn{1}{c}{0.11}&\multicolumn{1}{c}{0.42}&\multicolumn{1}{c}{0.043}\\
\end{tabular}
\caption{\textit{Efficiencies of the cuts on $p_T$ at `zero cost' (\ref{7tev-ZEROcut}), for signal and background ($\sqrt{s}=7$ TeV).}}
\label{7tevzeroCutEff}
\end{table}

\begin{table}[]
\subtable[SIGNAL]{
\begin{tabular}[]{cccc}
\multicolumn{1}{c}{}&\multicolumn{3}{||c}{ $\sigma\ [fb]$} \\
\hline
\hline
\multicolumn{1}{c}{$M_{G*}=1.5$ TeV}&\multicolumn{1}{||c}{$\chi\psi$} &\multicolumn{1}{c}{22.2}&\multicolumn{1}{c}{25.4}\\
\cline{2-3}
\multicolumn{1}{c}{}&\multicolumn{1}{||c}{$t\bar{t}$} &\multicolumn{1}{c}{3.21}&\multicolumn{1}{c}{}\\
\hline
\hline
\multicolumn{1}{c}{$M_{G*}=2$ TeV}&\multicolumn{1}{||c}{$\chi\psi$} &\multicolumn{1}{c}{2.51}&\multicolumn{1}{c}{2.85}\\
\cline{2-3}
\multicolumn{1}{c}{}&\multicolumn{1}{||c}{$t\bar{t}$} &\multicolumn{1}{c}{0.342}&\multicolumn{1}{c}{}\\
\hline
\hline
\multicolumn{1}{c}{$M_{G*}=3$ TeV}&\multicolumn{1}{||c}{$\chi\psi$} &\multicolumn{1}{c}{0.0297}&\multicolumn{1}{c}{0.0486}\\
\cline{2-3}
\multicolumn{1}{c}{}&\multicolumn{1}{||c}{$t\bar{t}$} &\multicolumn{1}{c}{0.0189}&\multicolumn{1}{c}{}\\
\hline
\hline
\end{tabular}
}
\subtable[BACKGROUND]{
 \begin{tabular}[]{cccc}
\multicolumn{1}{c}{}&\multicolumn{3}{||c}{$\sigma\ [fb]$}\\
\hline
\hline
\multicolumn{1}{c}{$WWbb$}&\multicolumn{1}{||c}{} &\multicolumn{1}{c}{}&\multicolumn{1}{c}{167}\\
\hline
\hline
\multicolumn{1}{c}{$Wbb+jets$}&\multicolumn{1}{||c}{$Wbb+1J$} &\multicolumn{1}{c}{7.78}&\multicolumn{1}{c}{9.95}\\
\cline{2-3}
\multicolumn{1}{c}{}&\multicolumn{1}{||c}{$Wbb+2J$} &\multicolumn{1}{c}{2.17}&\multicolumn{1}{c}{}\\
\cline{2-3}
\hline
\hline
\multicolumn{1}{c}{$W+jets$}&\multicolumn{1}{||c}{$W+3J$} &\multicolumn{1}{c}{0.831}&\multicolumn{1}{c}{1.14}\\
\cline{2-3}
\multicolumn{1}{c}{}&\multicolumn{1}{||c}{$W+4J$} &\multicolumn{1}{c}{0.309}&\multicolumn{1}{c}{}\\
\cline{2-3}
\hline
\hline
\multicolumn{1}{c}{Total BCKG}&\multicolumn{1}{||c}{} &\multicolumn{1}{c}{}&\multicolumn{1}{c}{178}\\
\hline
\hline
\end{tabular}
}
\caption{\textit{Cross Section values after the cuts at `zero cost' (\ref{7tev-ZEROcut}) ($\sqrt{s}=7$ TeV).}
}
\label{7tevXSEC_AZeroCut}
\end{table}

The final discovery of our signal is then obtained by imposing the SET (I) of the cuts that minimize the integrated luminosity needed for the discovery (\ref{cut_lum})
and the SET (II) of the cuts that maximize the signal over background ratio (\ref{cut_SsuB}). 
We show in Tab.s \ref{xsec_lum_disc_7tev} and \ref{xsec_SsuB_7tev} the cross section values for signals and backgrounds 
after we impose the SET (I) and the SET (II) of cuts, respectively.

\begin{table}[]
\begin{tabular}[]{ccccc}
\multicolumn{1}{c}{$\sqrt{s}=7$ TeV}&\multicolumn{1}{||c}{SIGNAL}&\multicolumn{3}{|c}{BACKGROUND} \\
\multicolumn{1}{c}{SET (I)}&\multicolumn{1}{||c}{ $\sigma\ [fb]$}&\multicolumn{3}{|c}{ $\sigma\ [fb]$} \\
\hline
\hline
\multicolumn{1}{c}{$M_{G*}=1.5$ TeV} &\multicolumn{1}{||c}{$20.0 \pm 0.1$}&\multicolumn{1}{|c}{$WWbb$}&\multicolumn{1}{c}{$0.17 \pm 0.06$}&\multicolumn{1}{c}{$0.72 \pm 0.07$}\\
\cline{3-4}
\multicolumn{1}{c}{} &\multicolumn{1}{||c}{}&\multicolumn{1}{|c}{$Wbb+jets$}&\multicolumn{1}{c}{$0.43 \pm 0.03$}&\multicolumn{1}{c}{}\\
\cline{3-4}
\multicolumn{1}{c}{} &\multicolumn{1}{||c}{}&\multicolumn{1}{|c}{$W+jets$}&\multicolumn{1}{c}{$0.121 \pm 0.004$}&\multicolumn{1}{c}{}\\
\hline
\multicolumn{1}{c}{$M_{G*}=2$ TeV} &\multicolumn{1}{||c}{$2.28 \pm 0.01$}&\multicolumn{1}{|c}{$WWbb$}&\multicolumn{1}{c}{$0.06 \pm 0.04$}&\multicolumn{1}{c}{$0.15 \pm 0.04$}\\
\cline{3-4}
\multicolumn{1}{c}{} &\multicolumn{1}{||c}{}&\multicolumn{1}{|c}{$Wbb+jets$}&\multicolumn{1}{c}{$0.07 \pm 0.01$}&\multicolumn{1}{c}{}\\
\cline{3-4}
\multicolumn{1}{c}{} &\multicolumn{1}{||c}{}&\multicolumn{1}{|c}{$W+jets$}&\multicolumn{1}{c}{$0.029 \pm 0.002$}&\multicolumn{1}{c}{}\\
\hline
\multicolumn{1}{c}{$M_{G*}=3$ TeV} &\multicolumn{1}{||c}{$0.0301 \pm 0.0002$}&\multicolumn{1}{|c}{$WWbb$}&\multicolumn{1}{c}{$<0.03$}&\multicolumn{1}{c}{$0.007^{+0.03}_{-0.004}$}\\
\cline{3-4}
\multicolumn{1}{c}{} &\multicolumn{1}{||c}{}&\multicolumn{1}{|c}{$Wbb+jets$}&\multicolumn{1}{c}{$0.006 \pm 0.004$}&\multicolumn{1}{c}{}\\
\cline{3-4}
\multicolumn{1}{c}{} &\multicolumn{1}{||c}{}&\multicolumn{1}{|c}{$W+jets$}&\multicolumn{1}{c}{$0.0013 \pm 0.0005$}&\multicolumn{1}{c}{}\\
\hline
\hline
\end{tabular}
\caption{\textit{Cross section values after the cuts that minimize the discovery luminosity, SET (I) in \ref{cut_lum}. ($\sqrt{s}=7$ TeV).}
}
\label{xsec_lum_disc_7tev}
\end{table}

\begin{table}[]
\begin{tabular}[]{ccccc}
\multicolumn{1}{c}{$\sqrt{s}=7$ TeV}&\multicolumn{1}{||c}{SIGNAL}&\multicolumn{3}{|c}{BACKGROUND} \\
\multicolumn{1}{c}{SET (II)}&\multicolumn{1}{||c}{ $\sigma\ [fb]$}&\multicolumn{3}{|c}{ $\sigma\ [fb]$} \\
\hline
\hline
\multicolumn{1}{c}{$M_{G*}=1.5$ TeV} &\multicolumn{1}{||c}{$14.9 \pm 0.1$}&\multicolumn{1}{|c}{$WWbb$}&\multicolumn{1}{c}{$0.06 \pm 0.04$}&\multicolumn{1}{c}{$0.29 \pm 0.04$}\\
\cline{3-4}
\multicolumn{1}{c}{} &\multicolumn{1}{||c}{}&\multicolumn{1}{|c}{$Wbb+jets$}&\multicolumn{1}{c}{$0.19 \pm 0.01$}&\multicolumn{1}{c}{}\\
\cline{3-4}
\multicolumn{1}{c}{} &\multicolumn{1}{||c}{}&\multicolumn{1}{|c}{$W+jets$}&\multicolumn{1}{c}{$0.051 \pm 0.002$}&\multicolumn{1}{c}{}\\
\hline
\multicolumn{1}{c}{$M_{G*}=2$ TeV} &\multicolumn{1}{||c}{$1.85 \pm 0.01$}&\multicolumn{1}{|c}{$WWbb$}&\multicolumn{1}{c}{$0.02 \pm 0.03$}&\multicolumn{1}{c}{$0.06 \pm 0.03$}\\
\cline{3-4}
\multicolumn{1}{c}{} &\multicolumn{1}{||c}{}&\multicolumn{1}{|c}{$Wbb+jets$}&\multicolumn{1}{c}{$0.030 \pm 0.007$}&\multicolumn{1}{c}{}\\
\cline{3-4}
\multicolumn{1}{c}{} &\multicolumn{1}{||c}{}&\multicolumn{1}{|c}{$W+jets$}&\multicolumn{1}{c}{$0.013 \pm 0.001$}&\multicolumn{1}{c}{}\\
\hline
\multicolumn{1}{c}{$M_{G*}=3$ TeV} &\multicolumn{1}{||c}{$0.0301 \pm 0.0001$}&\multicolumn{1}{|c}{$WWbb$}&\multicolumn{1}{c}{$<0.03$}&\multicolumn{1}{c}{$0.007^{+0.03}_{-0.004}$}\\
\cline{3-4}
\multicolumn{1}{c}{} &\multicolumn{1}{||c}{}&\multicolumn{1}{|c}{$Wbb+jets$}&\multicolumn{1}{c}{$0.006 \pm 0.004$}&\multicolumn{1}{c}{}\\
\cline{3-4}
\multicolumn{1}{c}{} &\multicolumn{1}{||c}{}&\multicolumn{1}{|c}{$W+jets$}&\multicolumn{1}{c}{$0.0013 \pm 0.0005$}&\multicolumn{1}{c}{}\\
\hline
\hline
\end{tabular}
\caption{\textit{Cross section values after the cuts that maximize the $S/B$ ratio, SET (II) in \ref{cut_SsuB}. ($\sqrt{s}=7$ TeV).}
}
\label{xsec_SsuB_7tev}
\end{table}

\newpage
\section{Final results and LHC discovery reach in CHM parameter space}\label{ParSpace}

\begin{table}[h!]
\centering
 \scalebox{0.9}{\begin{tabular}[]{cccccccc}
\multicolumn{2}{c}{}&\multicolumn{3}{c}{ $\sqrt{s} = 14$ TeV } &\multicolumn{3}{c}{ $\sqrt{s} = 7$ TeV }\\
\cline{3-8}
\multicolumn{2}{c|}{} & $\mathcal{L}_{5 \sigma}$ & $S/B$ & $S/\sqrt{B}_{100}$ & \multicolumn{1}{|c}{$\mathcal{L}_{5 \sigma}$} & $S/B$ & \multicolumn{1}{c|}{$S/\sqrt{B}_{10}$}\\
\hline
\hline
\multicolumn{1}{|c}{$M_{G^*}=1.5$ TeV} & \multicolumn{1}{|c|}{SET (I)} & \textbf{54 pb}$\mathbf{^{-1}}$ & 20 & 589 & \multicolumn{1}{|c}{\textbf{0.48 fb}$\mathbf{^{-1}}$} & 28 & \multicolumn{1}{c|}{75}\\
\cline{2-8}
\multicolumn{1}{|c}{($M_{\chi}=1$ TeV)} & \multicolumn{1}{|c|}{SET (II)} & 73 pb$^{-1}$ & \textbf{40} & 732 & \multicolumn{1}{|c}{0.66 fb$^{-1}$} & \textbf{51} & \multicolumn{1}{c|}{87}\\
\hline
\hline
\multicolumn{1}{|c}{$M_{G^*}=2$ TeV} & \multicolumn{1}{|c|}{SET (I)} & \textbf{0.24 fb}$\mathbf{^{-1}}$ & 16 & 255 & \multicolumn{1}{|c}{\textbf{4.1 fb}$\mathbf{^{-1}}$} & 15 & \multicolumn{1}{c|}{19}\\
\cline{2-8}
\multicolumn{1}{|c}{($M_{\chi}=1.33$ TeV)} & \multicolumn{1}{|c|}{SET (II)} & 0.30 fb$^{-1}$ & \textbf{39} & 357 & \multicolumn{1}{|c}{5.2 fb$^{-1}$} & \textbf{31} & \multicolumn{1}{c|}{24}\\
\hline
\hline
\multicolumn{1}{|c}{$M_{G^*}=3$ TeV} & \multicolumn{1}{|c|}{SET (I)$\equiv$ (II)} & \textbf{3.4 fb}$\mathbf{^{-1}}$& \textbf{15} & 65 & \multicolumn{1}{|c}{$1.3\cdot 10^3$ fb$^{-1}$} & 4.3 & \multicolumn{1}{c|}{1.1}\\
\multicolumn{1}{|c}{($M_{\chi}=2$ TeV)} & \multicolumn{1}{|c|}{} & & & & \multicolumn{1}{|c}{} & & \multicolumn{1}{c|}{}\\
\hline
\hline
\multicolumn{1}{|c}{$M_{G^*}=4$ TeV} & \multicolumn{1}{|c|}{SET (I) $\equiv$ (II)} & \textbf{57 fb}$\mathbf{^{-1}}$ & \textbf{12} & \multicolumn{1}{c|}{17} \\
\multicolumn{1}{|c}{($M_{\chi}=2.67$ TeV)} & \multicolumn{1}{|c|}{} & & & \multicolumn{1}{c|}{}\\
\cline{1-5}
\cline{1-5}
\end{tabular}}\\
\caption{\textit{Final results for the discovery analysis of $G^* \to \chi\psi$. $\mathcal{L}_{5 \sigma}$ denotes the integrated luminosity needed for a $5\sigma$ discovery 
at the LHC, $S/B$ the Signal/Background ratio.
 $S/\sqrt{B}_{100(10)}$ represents the ratio between the number of signal events and the square root of the number of background events, at an integrated luminosity of $100(10)$ fb$^{-1}$.}
}
\label{final-results}
\end{table}

The final results of our analysis, optimized for the $G^* \to \chi\psi$ discovery, are summarized in Tab. \ref{final-results}. 
$S/B$ denotes the signal over background ratio. 
$\mathcal{L}_{5 \sigma}$ denotes the integrated luminosity needed for a discovery at a significance of at least $5\sigma$ 
\footnote{We take as value of the cross section the expected value plus one sigma. 
In this way, since for the cases of heavier $G^*$ the error associated with the cross section of the background is not negligible, 
we do a conservative estimate of the LHC reach.}. We impose that 
at the integrated luminosity of $\mathcal{L}_{5 \sigma}$ at least ten events (both form signal or background) have passed all the cuts and are finally observed.
 $S/\sqrt{B}_{100(10)}$ represents the ratio between the number of signal events and the square root of the number of background events, at an integrated luminosity of $100(10)$ fb$^{-1}$.
In cases where the number of signal and background events is quite high, $S/\sqrt{B}$ gives a good approximation of the statistical significance of the signal over the background.
We report in Tab. \ref{final-results} the $S/\sqrt{B}_{100(10)}$ values for the different signals, 
in order to compare our results with those from the analysis in the literature, as the one in Ref. \cite{Agashe}, 
that estimates the statistical significance using $S/\sqrt{B}$.\\


\begin{figure}[]
\mbox{\subfigure[\textit{Final total invariant mass distribution for the signal with $M_{G*}=2$ TeV ($M_{\tilde{T}}=M_{B}=1.33$ TeV) plus the remaining background.}]
{\includegraphics[width=0.5\textwidth, angle=90]{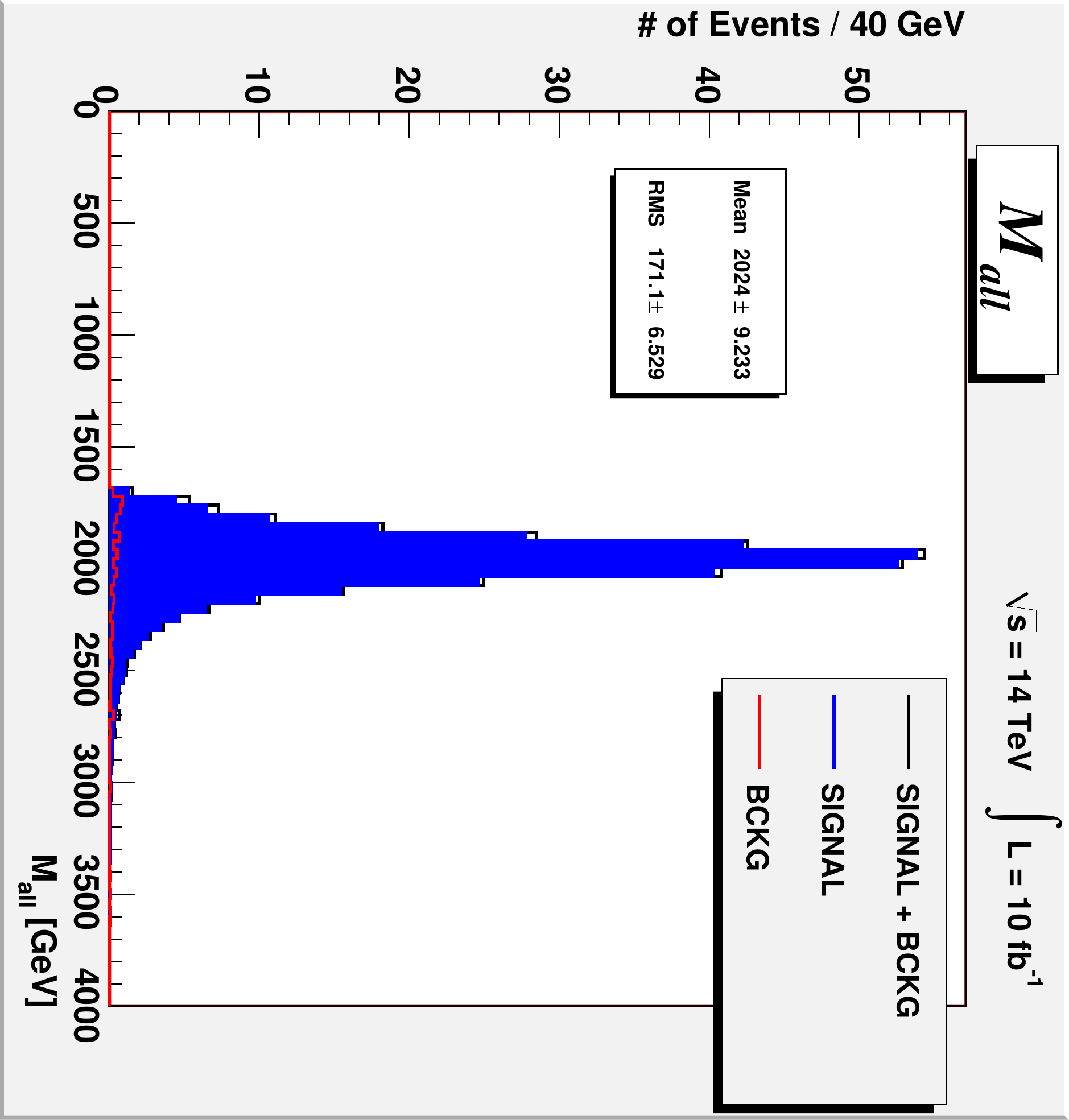}
\label{M20_all}} \hspace{0.1cm}
\subfigure[\textit{Final $M_{Wt}$ distribution for the signal with $M_{G*}=2$ TeV ($M_{\tilde{T}}=M_{B}=1.33$ TeV) plus the remaining background.}]
{\includegraphics[width=0.5\textwidth, angle=90]{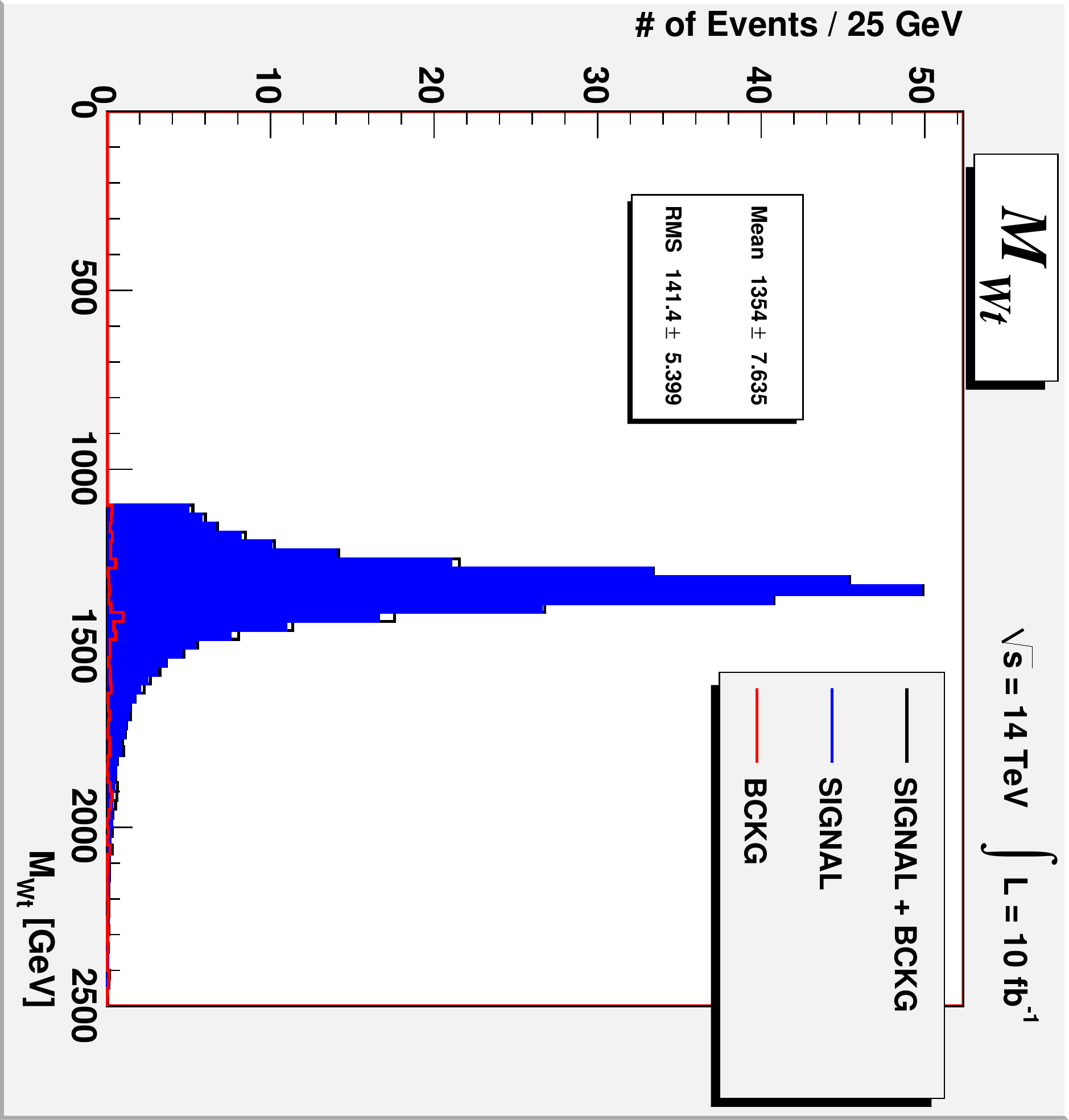}
\label{M20_wt}}}\\
\centering
\mbox{\subfigure[\textit{Final $M_{Wb}$ distribution for the signal with $M_{G*}=2$ TeV ($M_{\tilde{T}}=M_{B}=1.33$ TeV) plus the remaining background.}]
{\includegraphics[width=0.5\textwidth, angle=90]{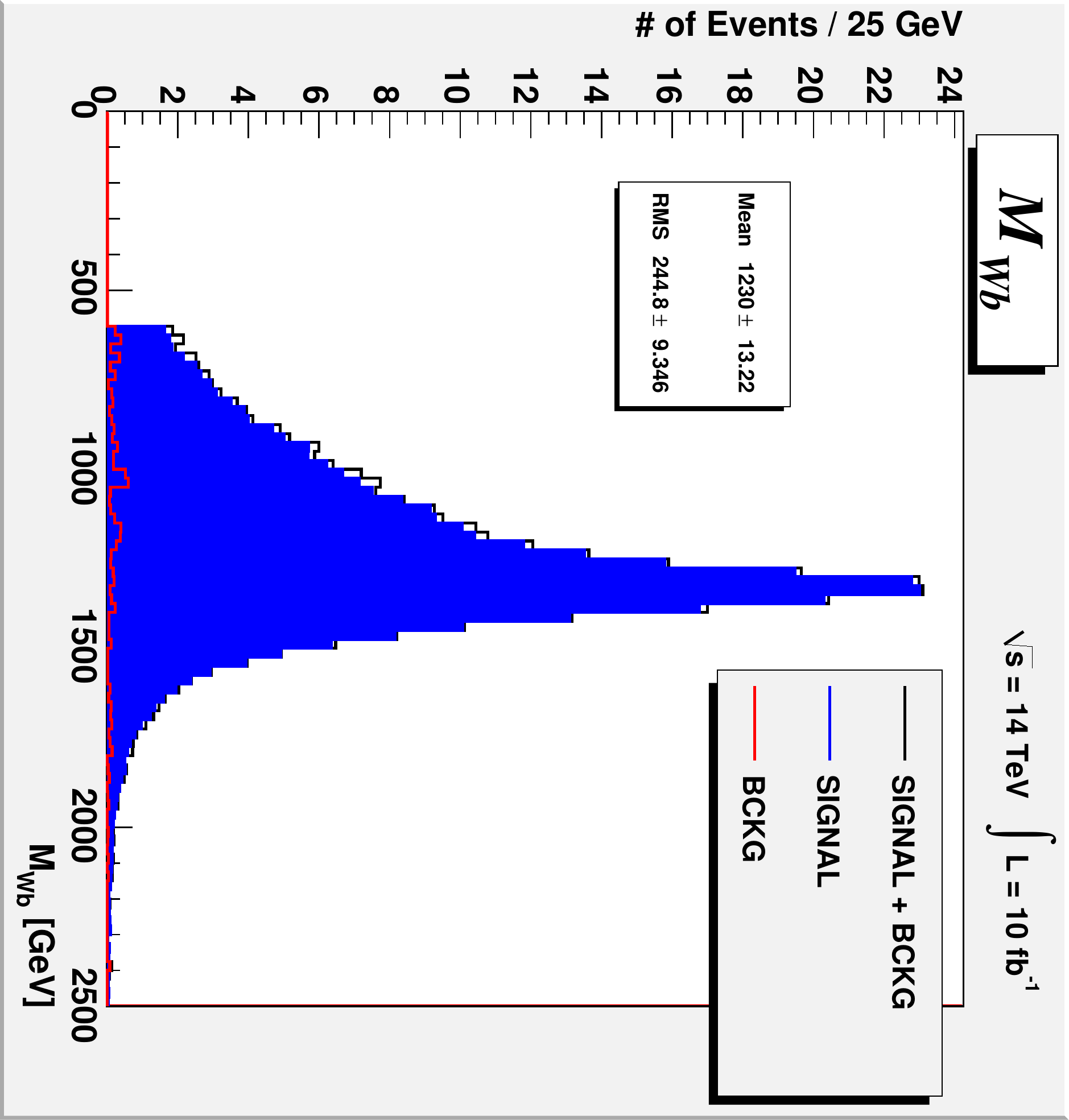}
\label{M20_wb}}}
\caption{\textit{Final invariant mass distributions, $M_{all}$, $M_{Wt}$ and $M_{Wb}$, after the analysis optimized for the $G^* \to \chi\psi$ discovery ($\sqrt{s}=14$ TeV).}}
\label{M20_fin}
\end{figure}

We show in Fig. \ref{M20_fin} the final invariant mass distributions, $M_{all}$, $M_{Wt}$ and $M_{Wb}$ for the signal with $M_{G*}=2$ TeV plus the remaining background.\\
We can clearly distinguish the resonances of the $G^*$, of the $B$ and of the $\tilde{T}$. \\

\noindent
The results obtained confirm those from the previous analysis preserving the $t\bar{t}$ component of the signal (Tab. \ref{results}):
a $G^*$ with a mass up to $\sim4$ TeV can be discovered at the $14$ TeV LHC and a $G^*$ with a mass up to $\sim2$ TeV can be discovered
in the early stage of the LHC, at $\sqrt{s}=7$ TeV. 
Nevertheless, the clearness of the discovery has been much improved compared to that from the analysis preserving $G^* \to t\bar{t}$;
we have obtained much higher $S/B$ ratios, up to a factor of $7$.\\
We can also compare our results with those in the literature, that refine the strategy for the $G^*$ discovery in the $t\bar{t}$ channel.
Ref. \cite{Agashe} searches for a Kaluza-Klein of the gluon in the channel $kkg \to t\bar{t}$; 
they consider $BR(kkg \to t_R\bar{t}_R)\simeq1$ and a coupling $g_{kkg u\bar{u}}=g_{kkg d\bar{d}}\simeq \frac{1}{5}g_{S}$ in the $kkg$ production. 
To compare their results with the ours, where we have $BR(G^* \to \psi\chi \to Wtb)\simeq 0.25$ and $g_{G* q\bar{q}}=0.44 g_{S}$ ($tg\theta_3=0.44$), 
we have to scale down by a factor of $\sim 1.2$ our signal cross sections and, as a consequence, our $S/B$ and $S/\sqrt{B}$ ratios 
\footnote{Indeed, because of the different value we used for $\tan\theta_3$, our signal cross sections are about five times larger than those expected at $\tan\theta_3=0.2$
(we are considering that the cross section for the $G^*$ production depends on $tg^2\theta_3$, see eq. (\ref{Gprod})). 
But, while Ref. \cite{Agashe} exploits the full $G^*$ production cross section, because it considers $BR(kkg \to t_R\bar{t}_R)=1$, 
in our assumptions (\ref{SetPar}), we are taking about the $25\%$ of the $G^*$ production cross section (see Fig.s \ref{BR_channels} and \ref{Wtb_sR}).
Therefore, we have to account for an overall scale factor of $\sim \left( \frac{0.44}{0.2}\right)^2 \cdot 0.25 =1.2$.}.
Ref. \cite{Agashe}, by exploiting peculiarities of the $kkg \to t_R\bar{t}_R$ signal like the high energy and the Left-Right asymmetry of the tops, 
obtains $S/B=2(1.6)$ and $S/\sqrt{B}_{100}=11(4.2)$ as result of the search for a $kkg$ with a mass of $3(4)$ TeV at the 14 TeV LHC. 
If we scale down by a factor of $1.2$ our $S/B$ and $S/\sqrt{B}_{100}$ ratios (in Tab. \ref{final-results}),
 we obtain values higher than those in \cite{Agashe};
for a $G^*/kkg$ with a mass of $3(4)$ TeV, our $S/B$ is higher by a factor of $\sim 6.3(6.3)$ and our $S/\sqrt{B}_{100}$ by a factor of $\sim 4.9(3.3)$, compared to those in \cite{Agashe}. 
We can thus affirm that the distinctive topology of the decay of a $G^*$ into one heavy fermion plus its SM partner clearly improves the LHC discovery
sensitivity to a heavy gluon.\\



\noindent
The results in Tab. \ref{final-results} refer to the setting (\ref{SetPar}), we have considered $\tan\theta_3=0.44$ and $s_R=0.6$. 
We want now to estimate the LHC discovery reach on the full parameter space, by taking into account a variation in $\tan\theta_3$ and $s_R$.\\
In order to evaluate the cross section for the signal at generic $\tan\theta_3$ and $s_R$ values, $\sigma_S (tg\theta_3, s_R)$, 
we assume the following relation between $\sigma_S (tg\theta_3, s_R)$ and the signal cross section at $\tan\theta_3=0.44$ and $s_R=0.6$, $\sigma_S (0.44, 0.6)$:
\begin{equation}
 \sigma_S (tg\theta_3, s_R) \simeq \left( \frac{tg\theta_3}{0.44}\right)^2 \frac{BR[G^*\to \psi\chi \to Wtb](tg\theta_3, s_R)}{BR[G^*\to \psi\chi \to Wtb](0.44, 0.6)} \ .
\label{Stheta3sR}
\end{equation} 

\begin{figure}[]
\centering
\includegraphics[width=0.5\textwidth]{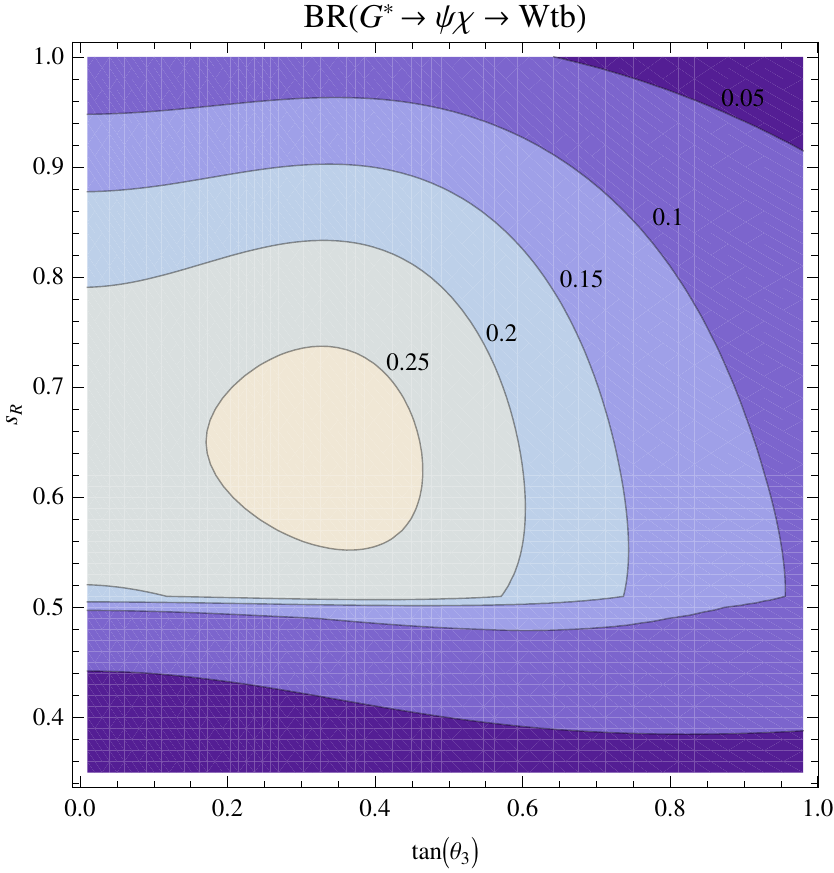}
\caption{\textit{Contour plot for the $BR(G^*\to \psi\chi \to Wtb)$ as a function of $\tan\theta_3$ and of the top degree of compositeness, $s_R$.}}
\label{BR_contour}
\end{figure}

\noindent 
We scale the signal cross section $\sigma_S (0.44, 0.6)$ by the factor
 $\left( \frac{tg\theta_3}{0.44}\right)^2$, taking into account the proportionality to $tg^2\theta_3$ of the cross section for the $G^*$ production (see eq. (\ref{Gprod})).
The ratio between the $BR$ for the $G^* \to \psi\chi$ decay in the $Wtb$ channel evaluated at generic $\tan\theta_3$ and $s_R$ values, $BR[G^*\to \psi\chi \to Wtb](tg\theta_3, s_R)$, 
and that evaluated at $\tan\theta_3=0.44$ and $s_R=0.6$ is calculated numerically by considering the formulas in sec. \ref{Sec_BRs} and \ref{BRsHf}. 
We show in Fig. \ref{BR_contour} the contour plot for $BR[G^*\to \psi\chi \to Wtb](tg\theta_3, s_R)$. 
We can see that, as expected, the $BR$ decreases for $s_R$ values close to $1$, 
as an effect of the predominance of the $G^*$ decay into top pairs and for $s_R$ close to its minimum value $s_R \sim 0.4$, where, 
as we will discuss in sec. \ref{tCOM}, the $G^*$ decays into pairs of custodians become dominant. 
The dependence of $BR[G^*\to \psi\chi \to Wtb]$ on $tg\theta_3$ is quite soft for moderate values of $tg\theta_3$; 
for larger $tg\theta_3$ values, instead, we can notice a decrease of the $BR$, that is a consequence of the enhancement of the $G^*$ 
decays into light jets.


\begin{figure}[]
\mbox{\subfigure[$s_R=0.6$]
{\includegraphics[width=0.4\textwidth, angle=-90]{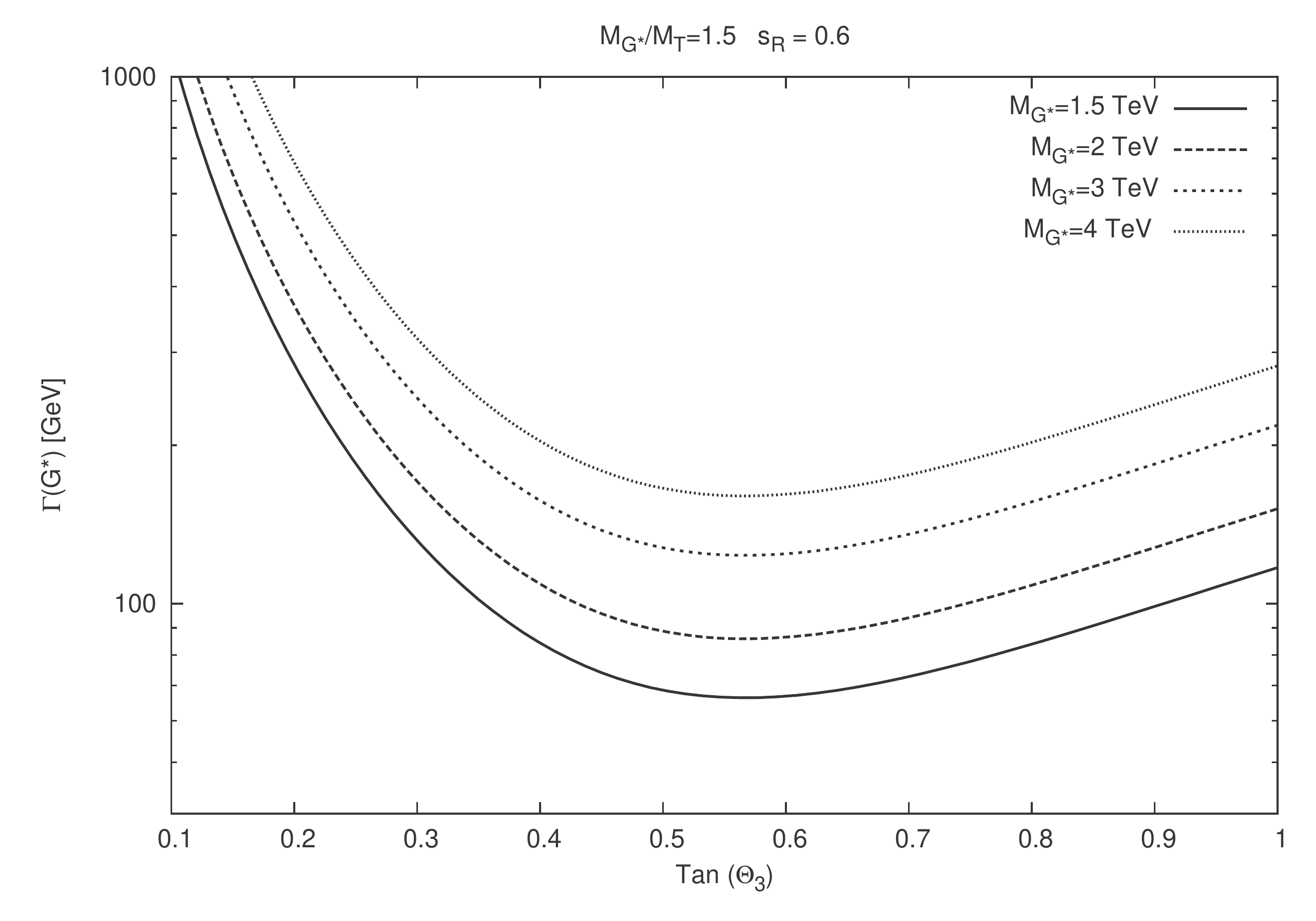}
}
\subfigure[$s_R=1$]
{\includegraphics[width=0.4\textwidth, angle=-90]{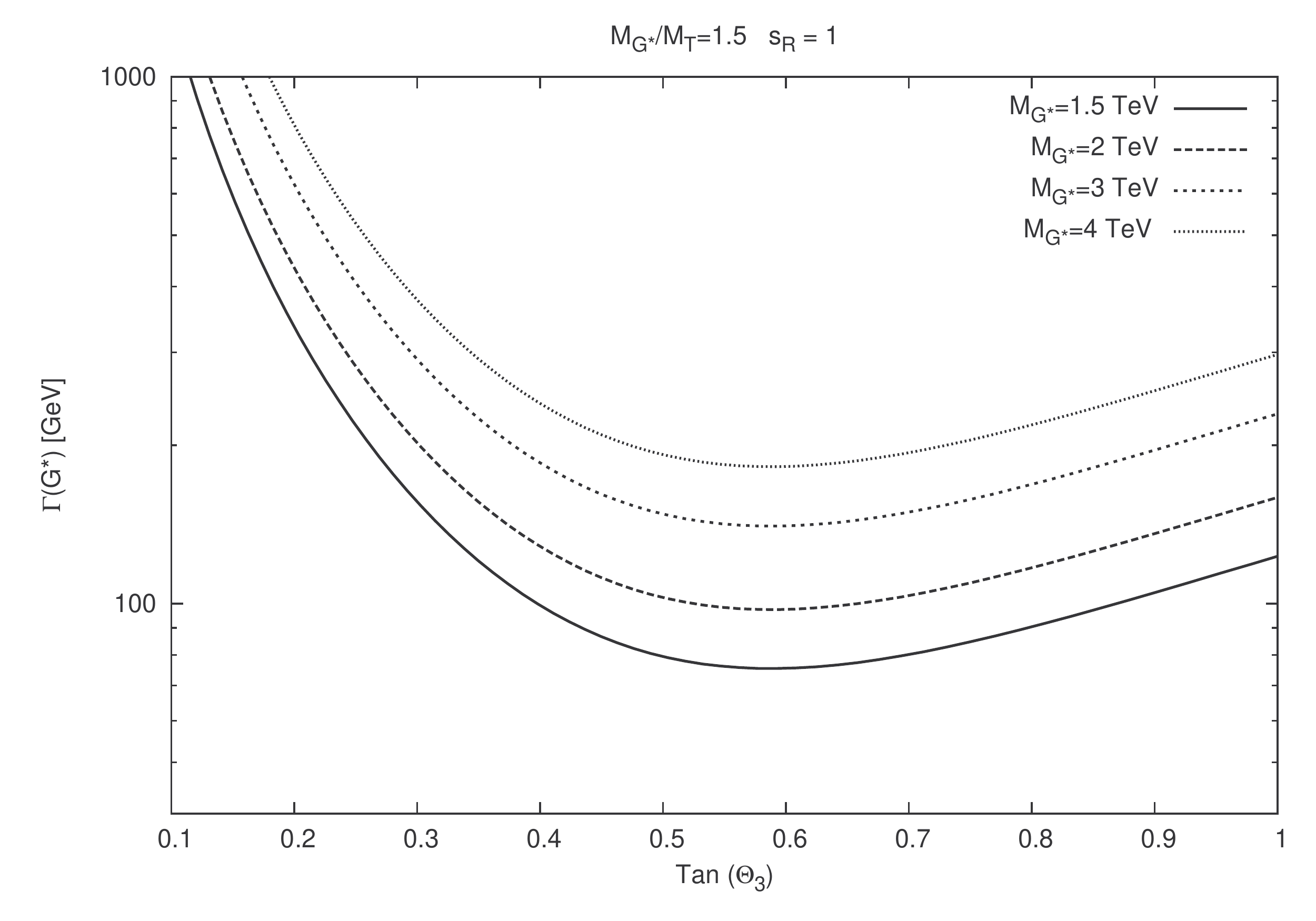}
}}
\caption{\textit{$G^*$ total decay width as function of $tg\theta_3$ for different $G^*$ masses.}}
\label{width_tantheta}
\end{figure}

\noindent
By taking into account the relation in (\ref{Stheta3sR}), we estimate the number of signal events for generic $\tan\theta_3$ and $s_R$ values 
at an integrated luminosity of $100$ fb$^{-1}$, 
for the analysis at $\sqrt{s}=14$ TeV, and at an integrated luminosity of $10$ fb$^{-1}$, 
for the analysis at $\sqrt{s}=7$ TeV. 
We take as starting values for the signal cross section, 
$\sigma_S (0.44, 0.6)$, those obtained from the analysis optimized for the $G^* \to \psi\chi$ search, 
after the application of the SET (I) of cuts that minimize the discovery luminosity 
(values in Tab. \ref{xsec_lum_disc}(\ref{xsec_lum_disc_7tev}) for $\sqrt{s}=14(7)$ TeV). \\
The number of background events at $100(10)$ fb$^{-1}$ is simply obtained by multiplying by $100(10)$ 
the cross section values of the background in Tab. \ref{xsec_lum_disc}(\ref{xsec_lum_disc_7tev}) 
\footnote{Again, we take as value of the cross section the expected value plus one sigma, in order to 
make a conservative estimate of the LHC reach.}.
This means that we are ignoring the variation with $tg\theta_3$ of the $G^*$ total decay width 
and we are assuming that the cuts on the total invariant mass that we have applied 
in the case $\tan\theta_3=0.44$ and $s_R=0.6$ can also be imposed in the case of generic $\tan\theta_3$ and $s_R$ values.  
This assumption is very reasonable since the cuts on $M_{all}$ that we have applied are quite mild, as we already pointed out.\\
The dependence of the $G^*$ total decay width on $tg\theta_3$ is shown in Fig. \ref{width_tantheta} for different $G^*$ masses 
and for $s_R=0.6$ and $s_R=1$ \footnote{In the calculation of the $G^*$ total decay width as function of $tg\theta_3$ we evaluate $\alpha_S$ at the scale of $M_{G*}$.}.
The previous claim can be checked by considering this graphic and the values of the $M_{all}$ cut in (\ref{cut_lum}).
Fig. \ref{width_tantheta} also shows that the $G^*$ total decay width is essentially independent of the value of $s_R$, 
we obtain very similar values for the $G^*$ total decay width at $s_R=0.6$ and at $s_R=1$. \\
In this analysis we are considering $Y_*$ fixed at the reference value, $Y_*=3$. Our results are indeed quite independent on $Y_*$, since 
this latter determines basically only the values of the heavy fermion widths. 
We checked that $\Gamma_{\chi}/ M_{\chi}$ remains small ($\lesssim 0.2$) up to $Y_*$ values of $\sim 6$. 
Moreover, we used cuts on $M_{Wb}$ and $M_{Wt}$ quite mild, much below the $M_{B}$ and $M_{\tilde{T}}$ mass values.\\

\begin{figure}[]
\centering
\mbox{\subfigure[$\sqrt{s}=14$ TeV]
{\includegraphics[width=0.8\textwidth]{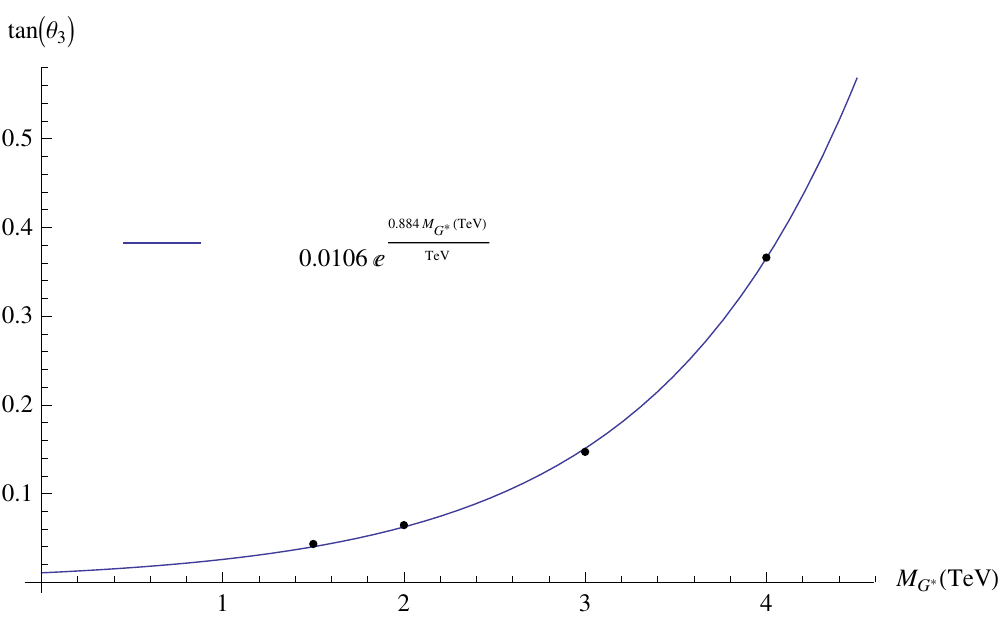}
}}
\mbox{
\subfigure[$\sqrt{s}=7$ TeV]
{\includegraphics[width=0.8\textwidth]{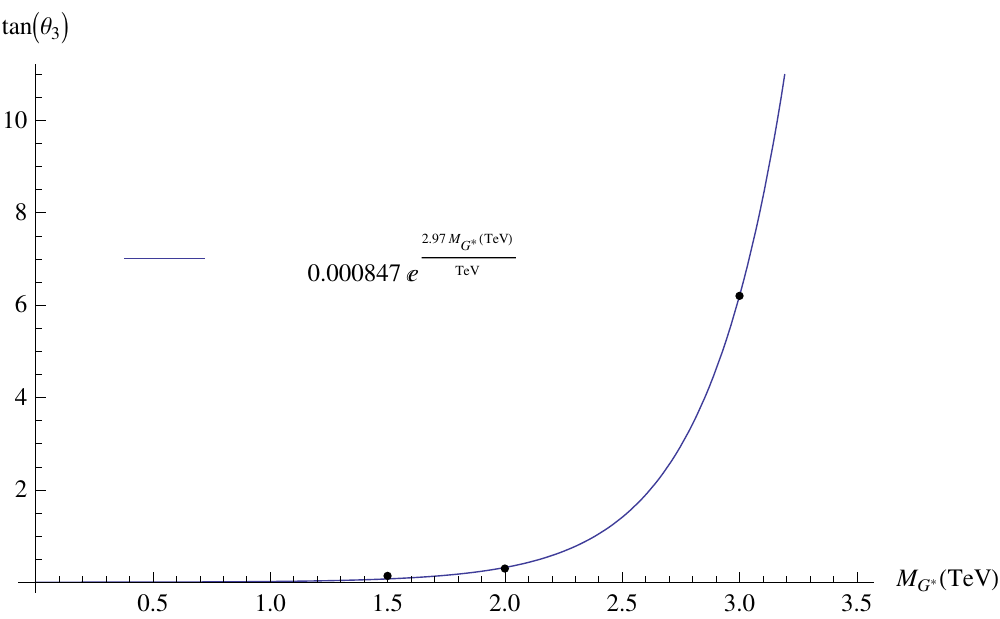}
}}
\caption{\textit{($M_{G*}$,$\tan\theta_3$) values  
for the $G^*$ discovery in the channel $G^* \to \psi\chi \to Wtb$ at an integrated luminosity of $100(10)$ fb$^{-1}$ 
at the $14(7)$ TeV LHC plus the respective fitting curve (\ref{eqFit}), for $s_R=0.6$.}}
\label{fit_reach}
\end{figure}

\noindent
At this point we are able to evaluate, for the different $G^*$ masses and for $s_R=0.6,0.8,1$, the values of $\tan\theta_3$ 
for which we can discover a heavy gluon in the channel $G^* \to \psi\chi \to Wtb$ at an integrated luminosity of $100(10)$ fb$^{-1}$ 
at the $14(7)$ TeV LHC.
 We claim the discovery when we reach a statistical significance of the signal over the background of $5\sigma$ 
\footnote{We remind that we are defining the minimum integrated luminosity required for a
discovery to be the integrated luminosity for which a goodness-of-fit test
of the SM-only hypothesis with Poisson distribution gives a p-value = $2.85\times 10^{-7}$,
 that corresponds to a $5\sigma$ significance in the limit of a gaussian distribution.}.
The values obtained are shown in Tab. \ref{values_CHM} 
(the values shown for $M_{G^*}=3$ TeV in the case $\sqrt{s}=7$ TeV have not physical meaning, since in our model $tg\theta_3\equiv\frac{g_{el}}{g_{com}}\lesssim 1$,
but they are useful to extrapolate the results to generic $M_{G*}$ values). \\
The values we have found are well fitted (Fig. \ref{fit_reach}) by exponential curves of the type:
\begin{equation}
 tg\theta_3= a e^{kM_{G*}} \ .
\label{eqFit}
\end{equation} 
We show in Tab. \ref{value_fit} the values of the $a$ and $k$ fit parameters, for the different $s_R$ values.  \\

\begin{table}[]
\centering
 \begin{tabular}[]{ccccccc}
\multicolumn{1}{c}{}&\multicolumn{3}{c}{ $\sqrt{s} = 14$ TeV } &\multicolumn{3}{c}{ $\sqrt{s} = 7$ TeV }\\
\cline{2-7}
\multicolumn{1}{c|}{} & $s_R=0.6$ & $s_R=0.8$ & $s_R=1$ & \multicolumn{1}{|c}{$s_R=0.6$} & $s_R=0.8$ & \multicolumn{1}{c|}{$s_R=1$}\\
\hline
\multicolumn{1}{|c|}{$M_{G^*}=1.5$ TeV} &  0.0432 & 0.0469 & 0.0846 & \multicolumn{1}{|c}{0.138} & 0.150 & \multicolumn{1}{c|}{0.264}\\
\hline 
\multicolumn{1}{|c|}{$M_{G^*}=2$ TeV}  & 0.0644 & 0.0700 & 0.126 & \multicolumn{1}{|c}{0.297} & 0.325 & \multicolumn{1}{c|}{0.770}\\
\hline
\multicolumn{1}{|c|}{$M_{G^*}=3$ TeV} & 0.147 & 0.159 & 0.280 & \multicolumn{1}{|c}{6.20} & 7.20 & \multicolumn{1}{c|}{11.6}\\
\hline
\multicolumn{1}{|c|}{$M_{G^*}=4$ TeV}  & 0.366 & 0.405 & \multicolumn{1}{c|}{0.833} \\
\cline{1-4}
\cline{1-4}
\end{tabular}\\
\caption{\textit{$tg\theta_3$ values for the $5\sigma$ discovery of $G^*\to\psi\chi\to Wtb$ at $100$($10$) fb$^{-1}$ at the 14(7) TeV LHC.}
}
\label{values_CHM}
\end{table}

\begin{table}[]
\centering
 \begin{tabular}[]{ccccc}
\multicolumn{1}{c}{}&\multicolumn{2}{c}{ $\sqrt{s} = 14$ TeV } &\multicolumn{2}{c}{ $\sqrt{s} = 7$ TeV }\\
\cline{2-5}
\multicolumn{1}{c|}{} & $a$ &  $k$ [TeV$^{-1}$] & \multicolumn{1}{|c}{$a$} & \multicolumn{1}{c|}{ $k$ [TeV$^{-1}$]}\\
\hline
\multicolumn{1}{|c|}{$s_R=0.6$} &  0.0106 & 0.884 & \multicolumn{1}{|c}{0.000847} & \multicolumn{1}{c|}{2.97}\\
\hline
\multicolumn{1}{|c|}{$s_R=0.8$}  & 0.0111 & 0.899  & \multicolumn{1}{|c}{0.000821} & \multicolumn{1}{c|}{3.03}\\
\hline
\multicolumn{1}{|c|}{$s_R=1$}  & 0.0146 & 1.01  & \multicolumn{1}{|c}{0.00370} & \multicolumn{1}{c|}{2.68}\\
\hline
\end{tabular}\\
\caption{\textit{$a$ and $k$ fit parameters (\ref{eqFit}), for the different $s_R$ values.}
}
\label{value_fit}
\end{table}

\begin{figure}[]
\mbox{\subfigure
{\includegraphics[width=1\textwidth]{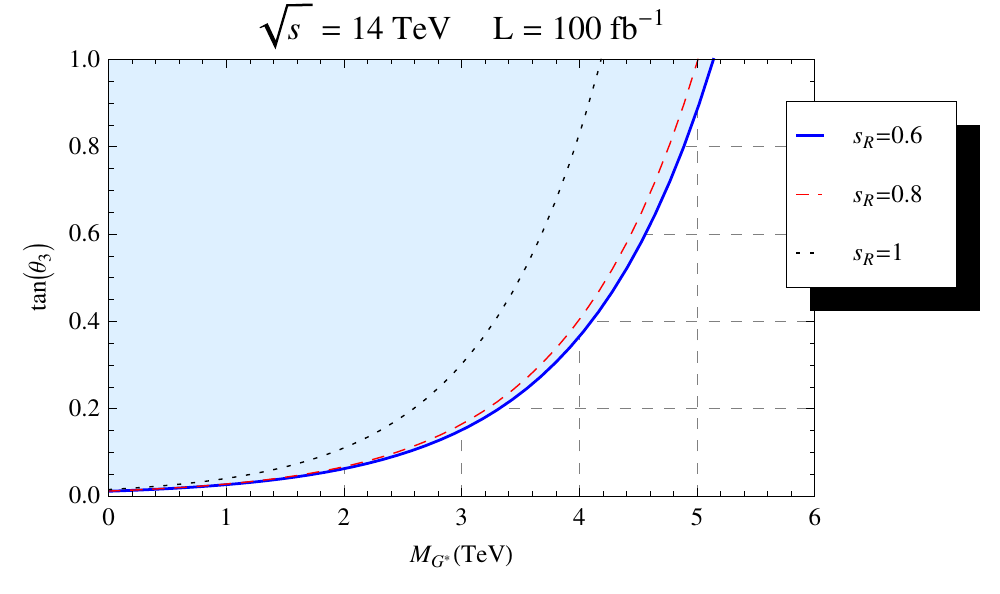}
\label{14tev-reach}}}\\
\mbox{\subfigure
{\includegraphics[width=1\textwidth]{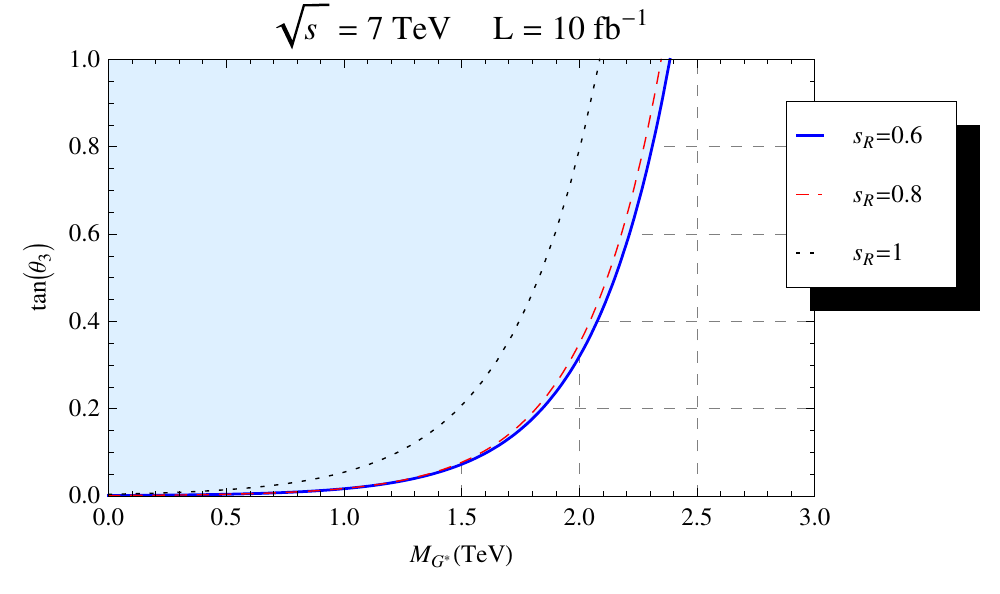}
\label{7tev-reach}}}
\caption{\textit{$14$ TeV and $7$ TeV LHC discovery reach on the parameter space of the Composite Higgs Model (TS5). 
The $5\sigma$ discovery region for heavy colored vectors in the channel $G^* \to \psi\chi\to Wtb$ extends above the blue curve in the case $s_R=0.6$,
above the dashed red curve for $s_R=0.8$ and above the dotted curve in the case $s_R=1$. 
The reach is calculated at an integrated luminosity of $100$ fb$^{-1}$ and $10$ fb$^{-1}$ for $\sqrt{s}=14$ TeV and $\sqrt{s}=7$ TeV respectively. 
}}
\label{CHM_reach}
\end{figure}

Fig. \ref{CHM_reach} summarizes our results and shows the LHC discovery reach on the parameter space of the Composite Higgs Model.
Fig. \ref{7tev-reach} shows the possibility for a heavy gluon with mass up to $\sim 2.4$ TeV to be discovered just in the early stage of the LHC, 
for quite large but theoretically allowed values of $tg\theta_3$. A $G^*$ with mass up to $\sim [1.8,2.2]$ TeV could be discovered 
for smaller and more theoretically motivated $tg\theta_3$ values, $tg\theta_3\in [0.2,0.6]$.
The reach will be much widened when LHC will be at its nominal design energy of $14$ TeV; 
a heavy gluon with mass up to $\sim 5$ TeV could be discovered at the 14 TeV LHC, as Fig. \ref{14tev-reach} shows.\\
The graphics in Fig. \ref{CHM_reach} can also be read as the possibility to exclude (in the hypothesis $M_T<M_{G*}<2M_T$) parts of the parameter space of the Composite Higgs Model, 
in the case of non-discovery. We can see that a significant part of the CHM parameter space can be already tested in the early runs of the LHC.\\


\begin{figure}[]
\centering
\mbox{
\includegraphics[width=0.45\textwidth, angle=90]{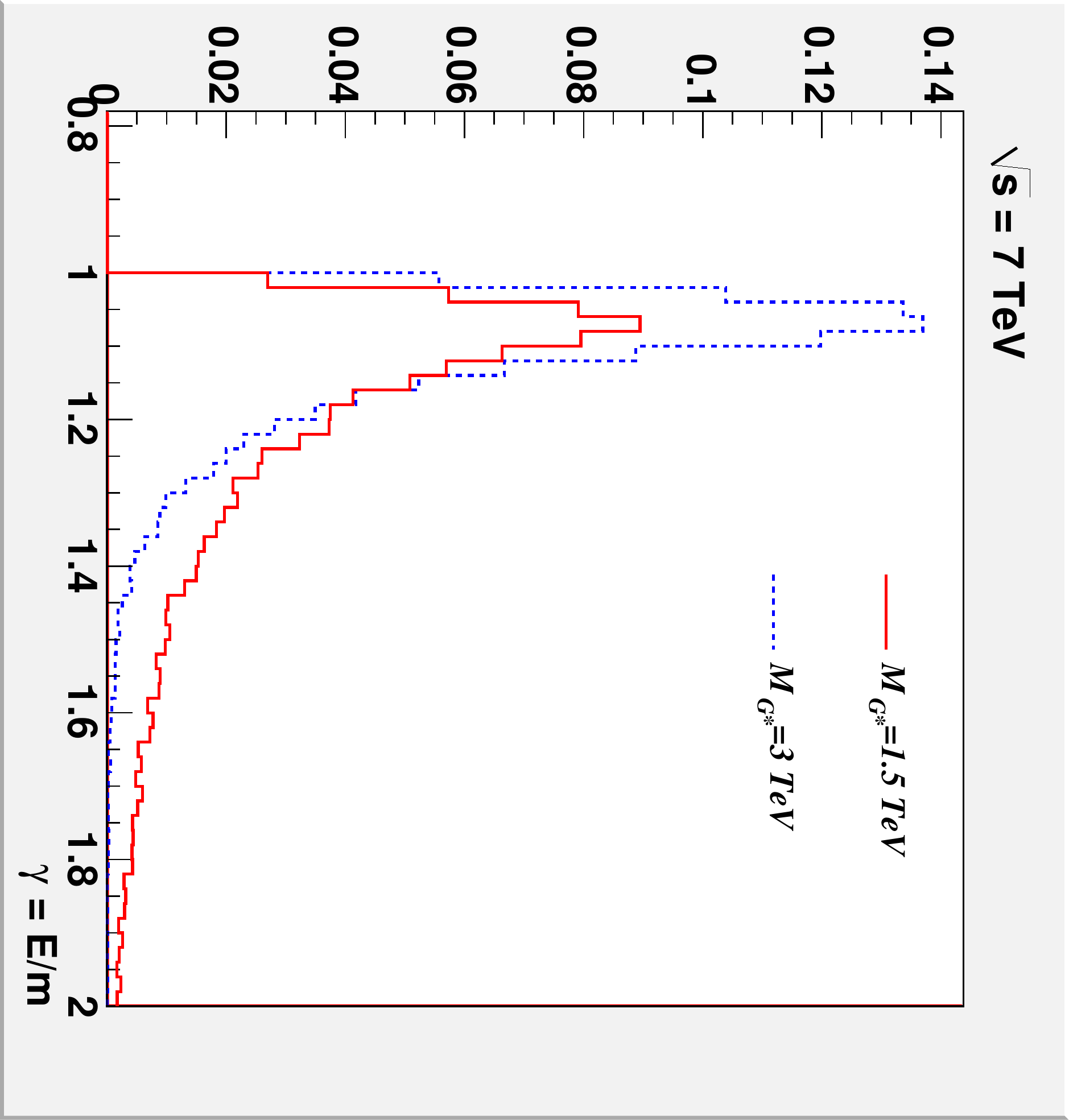}
\includegraphics[width=0.45\textwidth, angle=90]{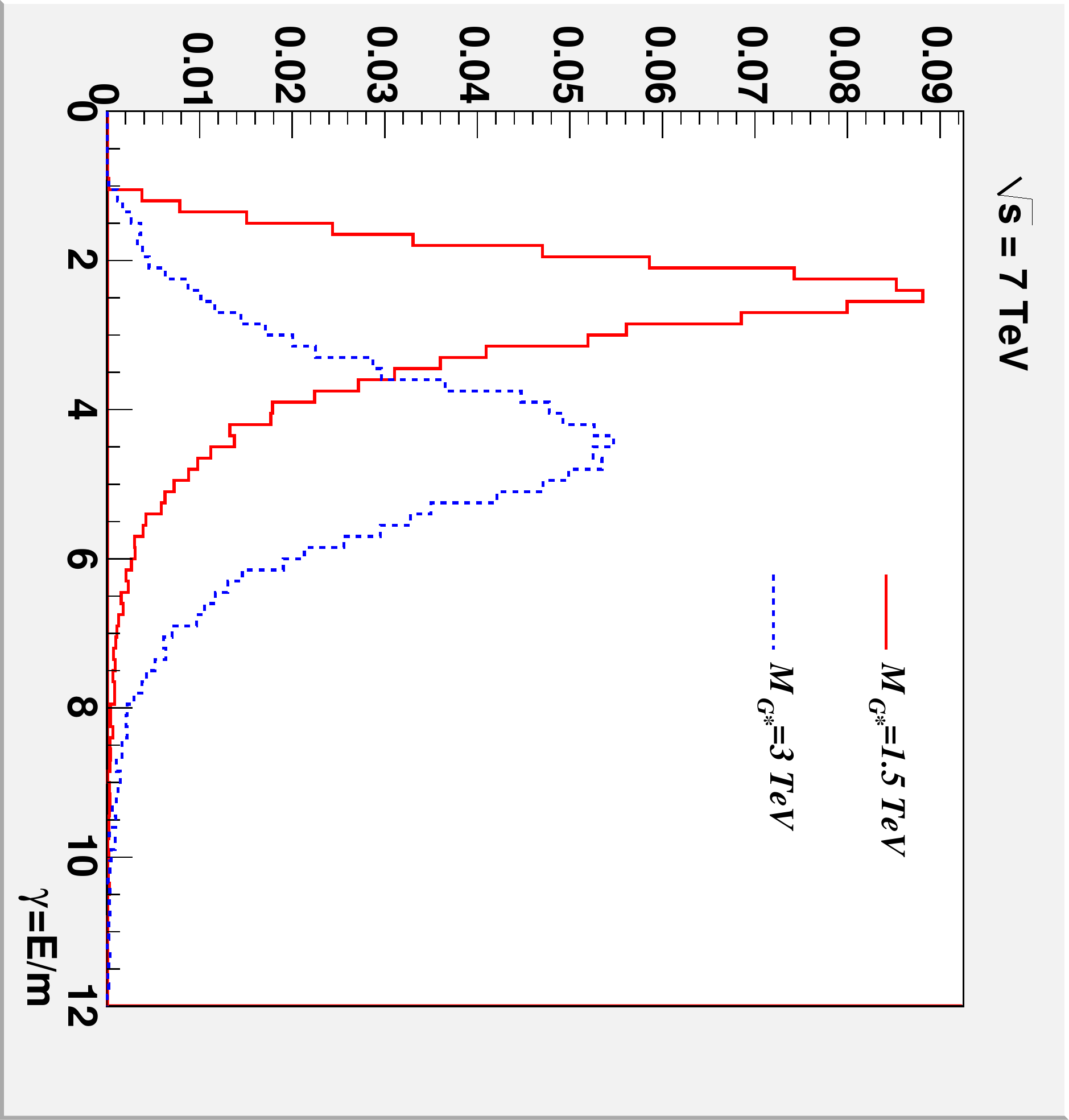}
}
\caption{\textit{Distribution of the boost factor $\gamma=E/m$ of $\tilde{T}$ (left plot) and of the top quark (right 
plot) in signal events $G^* \to \tilde{T}t$ with $M_{G*}=1.5$ TeV (continuous red line) and $M_{G*}=3$ TeV
(dashed blue line) for $\sqrt{s}=7$ TeV. The mass of the heavy fermion has been chosen such that
$M_{G*}/M_{\tilde{T}}=1.5$. Notice that this implies that $\tilde{T}$ is less boosted for $M_{G*}=3$ TeV than for 
$M_{G*}=1.5$ TeV. All the curves have been normalized to unit area.
}}
\label{boost}
\end{figure}

The current searches for a heavy gluon at the LHC are performed by analyzing heavy resonances decaying into top pairs \cite{LHC_tt}. 
They thus adopt (especially the CMS analysis) selection strategies which relies on a large boost of all the decay
products and which might have a poor efficiency on our topology of signal events. 
This can be explained by Fig. \ref{boost}, where we show the distribution of the boost factor $\gamma=E/m$ 
for the top and for the $\tilde{T}$ of the $G^*\to \tilde{T}t$ signal. We can see that while the SM (top or bottom) quark originating from the $G^*$ decay is always
highly boosted, for $M_{G*}/M_{\chi}=1.5$ the heavy quark is not.\\
Assuming, anyway, the same efficiency for the detection of a $G^*\to \psi\chi \to Wtb$ signal and of a $G^*\to \bar{t}t$ one, 
we find that the ATLAS analysis, performed using $200$ pb$^{-1}$ of $pp$ collisions at $\sqrt{s}=7$ TeV, 
is not yet sensitive to the parameter space of our two-site model. 
Considering a heavy gluon of $1.5$ TeV and the assumption $M_T<M_{G*}<2 M_{T}$
, we find that, 
in order for the $G^*\to t\bar{t}+\psi\chi\to Wtb$ signal cross section to lie in the range excluded by the analysis, 
we should have, for any possible $s_R$ value, a coupling $tg\theta_3 \gtrsim 2$, which is beyond the physical parameter space of the CHM; 
in CHM $tg\theta_3\equiv\frac{g_{el}}{g_{com}}\lesssim 1$. 
Obviously, with a higher luminosity of data collected the sensitivity can be improved. 
As we have shown in this project and as we want to 
stress here, it can be much improved by considering an analysis as the type presented in this project, focused on the channel $G^*\to\psi\chi$.\\ 

The LHC discovery/exclusion reach we have shown in Fig. \ref{CHM_reach} refers to the model TS5; however, it could be adapted with small variations to a
wider class of models (as to the Randall-Sundrum models or, more in general, to the models that predict the existence of heavy colored resonances).
Fig.s \ref{14tev-reach} and \ref{7tev-reach}, for example, can also be referred to the model TS10; 
the main difference lies in the case of a fully composite right-handed top ($s_R=1$). 
In this limit case, the search for a discovery in the TS10 should be less promising than in the TS5, as we will explain in section \ref{tCOM}.
Anyhow, as we have shown in the first part of the thesis, the case $s_R=1$ in the TS10 is quite disfavored by flavor observables (in particular by data from $b \to s\gamma$).


\section{Heavy fermions discovery}

\begin{figure}[h!]
\centering 
\includegraphics[width=0.65\textwidth]{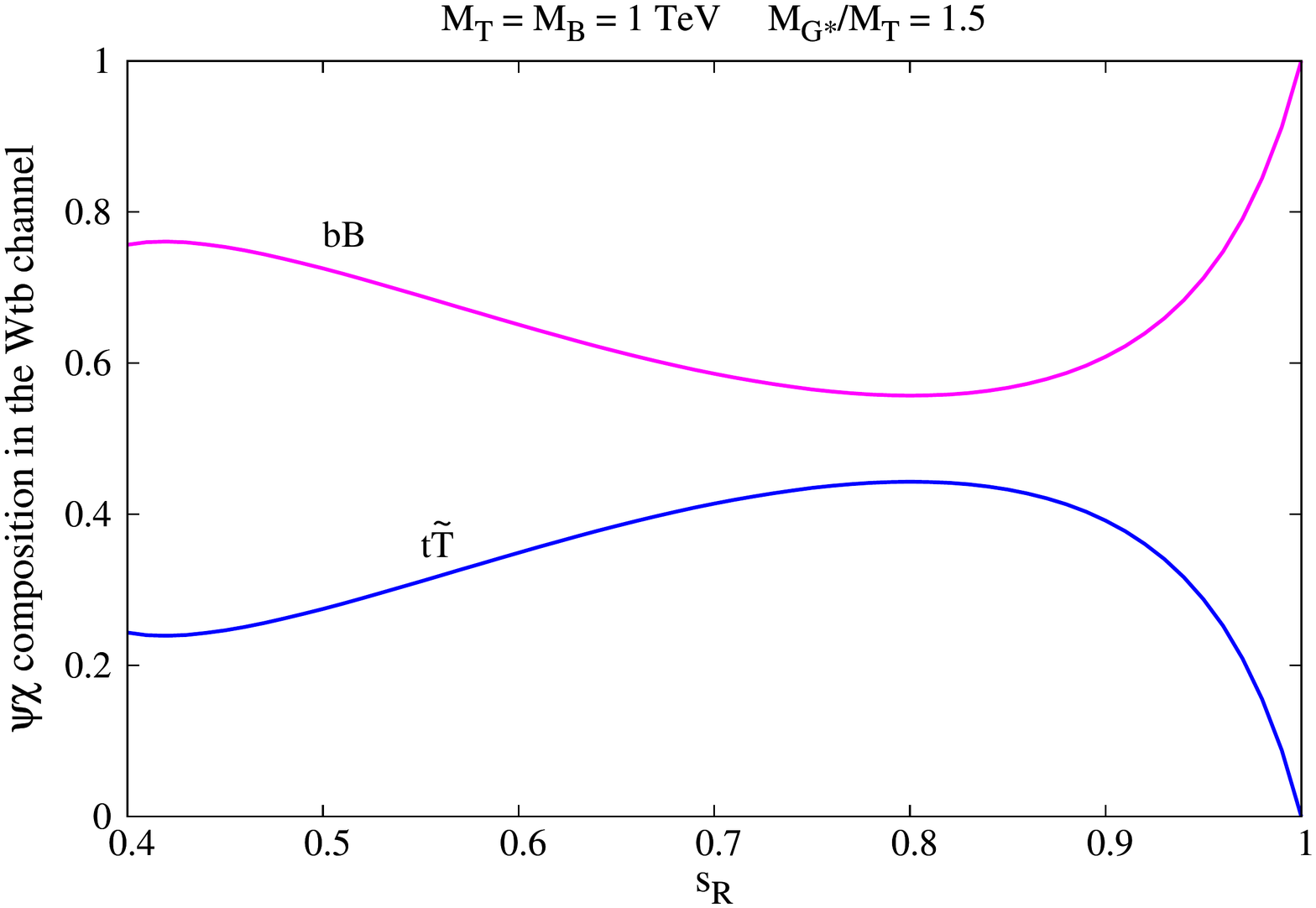}
\caption{\textit{ $\psi\chi$ composition in the $G^* \to \psi\chi \to Wtb$ channel as a function of $s_R$.}
}
\label{Hf_comp}
\end{figure}

The results in Tab. \ref{final-results} and in Fig. \ref{CHM_reach}, 
 that summarize the LHC potential to discover a heavy gluon with mass $M_{G*}$, can be translated into the possibility for a heavy fermion, bottom or top partner, 
with a mass roughly included in a range $[M_{G^*}/2\ ,\ M_{G^*}]$ to be discovered at the LHC together with the $G^*$. \\
Obviously, the cuts on the invariant mass distributions $M_{Wt}$ and $M_{Wb}$ should be adapted according to the real values of the heavy fermion masses;
the main strategy for the discovery, however, remains that we have followed in this analysis, where we made the simplifying assumption,  
$M_{B}=M_{\tilde{B}}=M_{\tilde{T}}=M_{G*}/1.5$.\\

The composition of the $\psi\chi$ signal depends on the top degree of compositeness. 
Therefore, the possibility for the discovery of one or the other type of heavy fermion (partner of the top or of the bottom) also depends on $s_R$. 
We show in Fig. \ref{Hf_comp} the $\psi\chi$ composition in the $G^* \to \psi\chi \to Wtb$ channel as a function of $s_R$. 
When $s_R\simeq 1$, $BR(G^*\to \tilde{T}t)\simeq 0$, therefore, in the case of a fully composite $t_R$, we cannot discover a $\tilde{T}$ in the $G^* \to \psi\chi$ channel. 
In the remaining cases ($s_R \lesssim 0.95$), we can see, instead, 
that the $\tilde{T}t$ part of the $\psi\chi$ signal remains close to (or above) its value at $s_R=0.6$ 
(for which we have shown the possibility for $\tilde{T}$ to be discovered together with the $G^*$), so we can discover both a $B$ and a $\tilde{T}$ in the $G^* \to \psi\chi$ channel.  \\

\begin{figure}[]
\centering
\mbox{
\subfigure[pair production]{
\includegraphics[width=0.35\textwidth]{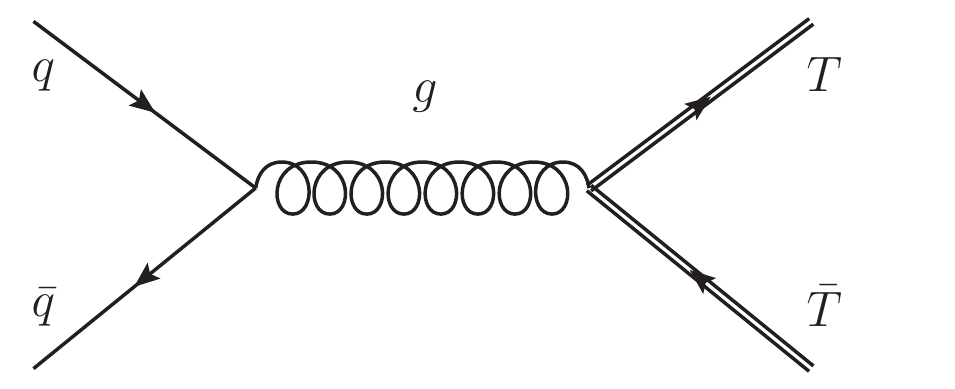}}
\subfigure[single production]{
\includegraphics[width=0.35\textwidth]{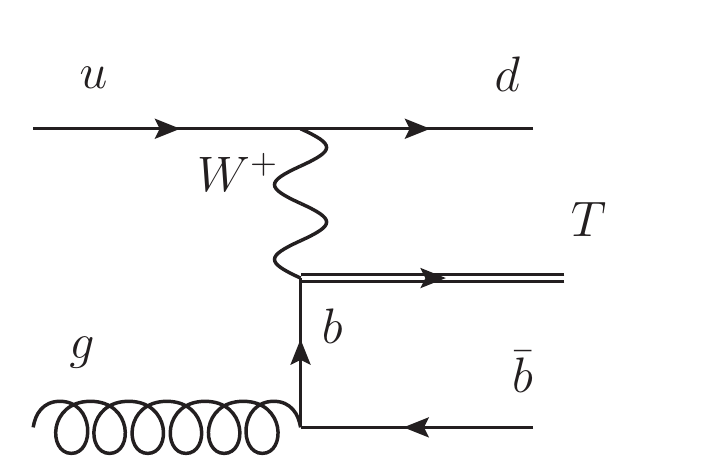}}
}
\caption{\textit{Diagrams for pair and single production of the heavy fermions.}
}
\label{pair_single}
\end{figure}

The typical channels considered for the search of heavy fermions at colliders are currently those of pair and single production (diagrams in Fig. \ref{pair_single}). 
Early searches for pair production of heavy fermions at CMS \cite{CMS_Hf} (with about $1$ fb$^-1$ of data collected) provide bounds on their masses, $M_{\chi}\gtrsim 495$ GeV, 
that are still below the range examined in this project (and also below the limit we have found from $b\to s\gamma$). 
Searches for single production at Tevatron \cite{single_tev}, as well, fix lower bounds on $M_{\chi}$ that are below $693$ GeV. 
We also point out that these latter analyses, contrary to what we have assumed in our models, consider sizable couplings between the heavy fermions and the SM quarks of the 1st and 2nd generation 
and, therefore, estimate signal cross sections higher than those predicted by CHM. 
An increasing of the integrated luminosity of the data collected at the LHC or, better, 
of the beams energy could bring to the discovery of heavy fermions with masses above $\sim 500$ GeV or, alternatively, to a strengthening of the bounds on $M_{\chi}$. 
We want to show here that the search for heavy fermions in the channel $G^* \to \psi\chi$, 
considered in our analysis, can improve the LHC sensitivity to heavy fermions\footnote{
the channel $G^* \to \psi\chi$ can improve the LHC discovery potential on the heavy fermions,
even though a non-discovery in this channel cannot imply a direct bound on the heavy fermions mass, 
since it can follow from the absence of a $G^*$ signal.}.\\
We can thus compare our results for the discovery of a $B$ heavy fermion with mass $\sim M_{G*}/1.5$ in the $G^*\to \psi\chi$ channel (Tab. \ref{final-results} and  Fig. \ref{CHM_reach}) 
with those recently obtained for the $B$ discovery at the $14$ TeV LHC, considering the typical single and pair production channels \cite{Wulzer}. 
Ref. \cite{Wulzer} finds a $S/B$ ratio of about $4(1.5)[0.9]$ and a discovery luminosity of $1.8(37)[420]$ fb$^{-1}$ 
\footnote{We report the $\mathcal{L}_{5\sigma}$ values calculated as in our analysis, considering the integrated luminosity for which a goodness-of-fit test
of the SM-only hypothesis with Poisson distribution gives a p-value = $2.85\times 10^{-7}$ and imposing that at least ten events are finally observed at $\mathcal{L}_{5\sigma}$. 
In Ref. \cite{Wulzer} $\mathcal{L}_{5\sigma}$ is evaluated in a less conservative way, 
as the luminosity at which $S/\sqrt{B}>5$ and at least $5$ signal events have been observed. 
Therefore \cite{Wulzer} shows different $\mathcal{L}_{5\sigma}$ values, $1.1(26)[327]$ fb$^{-1}$.} 
for a $B$ heavy fermion with a mass of $1(1.5)[2]$ TeV. 
We find, instead, that, if $M_B < M_{G^*}< 2 M_B$, a $B$ with a mass of $\sim 1(1.33)[2]$ TeV can be discovered in the $G^*\to \psi\chi$ channel, 
 considering our benchmark values $s_R=0.6$ and $tg\theta_3=0.44$, with a much higher $S/B$ ratio 
and a much lower discovery luminosity, $S/B=40(39)[15]$, $\mathcal{L}_{5\sigma}=0.054(0.24)[3.4]$ fb$^{-1}$. 
Higher $S/B$ ratios and lower discovery luminosities can be obtained even in the less optimistic cases of small $tg\theta_3$ values: 
for $tg\theta_3=0.2$ and a $B$ heavy fermion of $1$ TeV, for example, we find, by rescaling our signal cross section as in \ref{Stheta3sR}, $S/B\sim 8.3$ 
(about a factor of 2 higher than that in \cite{Wulzer}) and $\mathcal{L}_{5\sigma}\simeq0.35$ fb$^{-1}$ (about a factor of 5 lower than the discovery luminosity in \cite{Wulzer}). 
Notice also that the discovery reach can be improved. 
While, following the analysis in \cite{Wulzer}, a $B$ with a mass of $2$ TeV cannot be discovered 
with $300$ fb$^{-1}$ of integrated luminosity at the 14 TeV LHC, 
we find (fig. \ref{14tev-reach}) the possibility for a $B$ with mass up to $\sim3.4$ TeV to be discovered with $100$ fb$^{-1}$ 
in the channel $G^*\to \psi\chi$, for $tg\theta_3 \sim 1$, $s_R=0.6$ and $M_{G*}/M_{\chi}=1.5$.
The comparison proves that the $G^*\to \psi\chi$ channel is very promising also for the discovery of the heavy fermions 
and should be taken into account, together with the ordinary single and pair production channels, in the searches for heavy fermions at the LHC. 

\section{The composite top scenario}\label{tCOM}

\begin{figure}[h!]
\mbox{
\subfigure[$G^{*}$ decay Branching Ratios ($s_R=0.4$)]{
\includegraphics[width=0.52\textwidth]{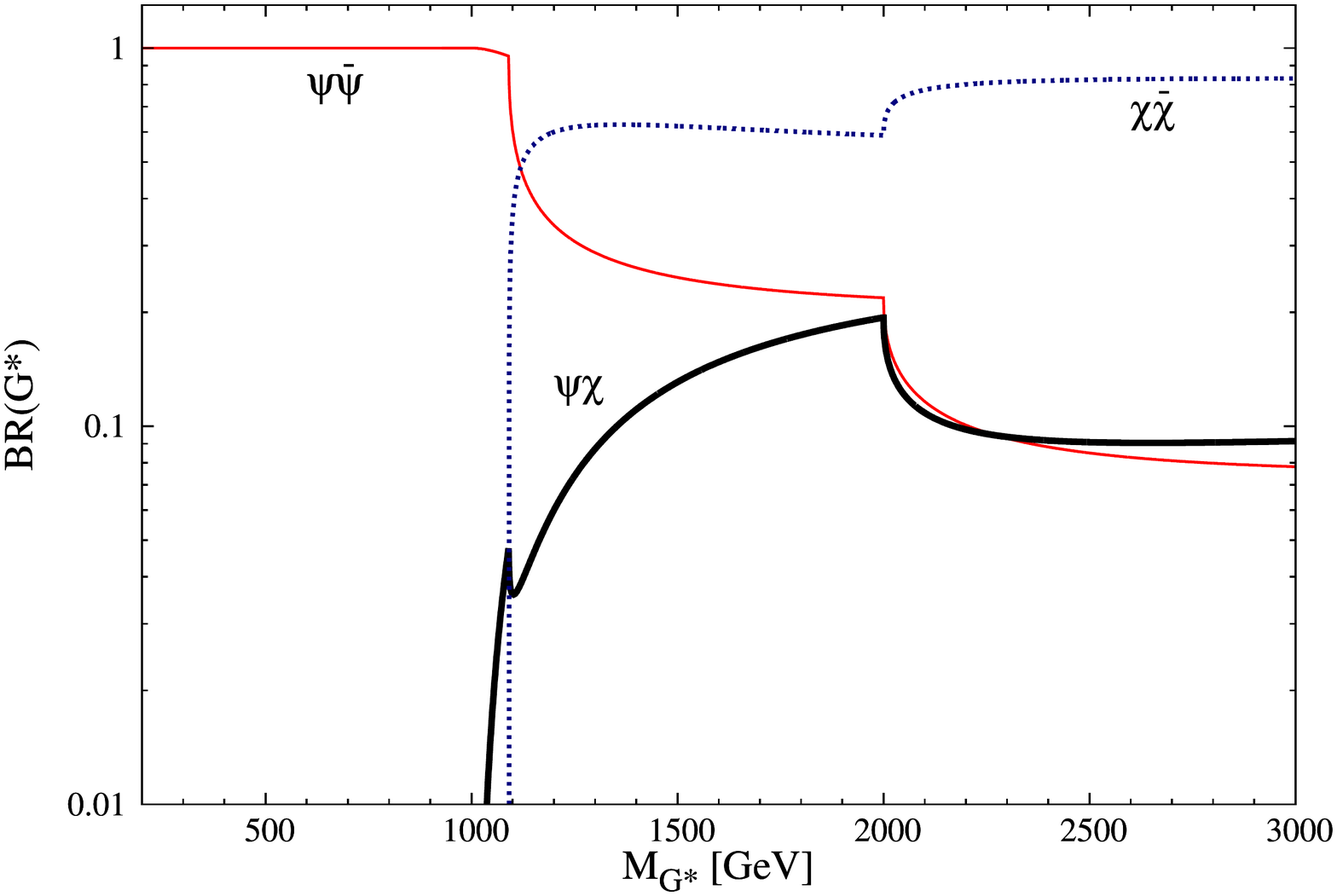}}
\subfigure[$G^{*}$ total decay width ($s_R=0.4$)]{
\includegraphics[width=0.52\textwidth]{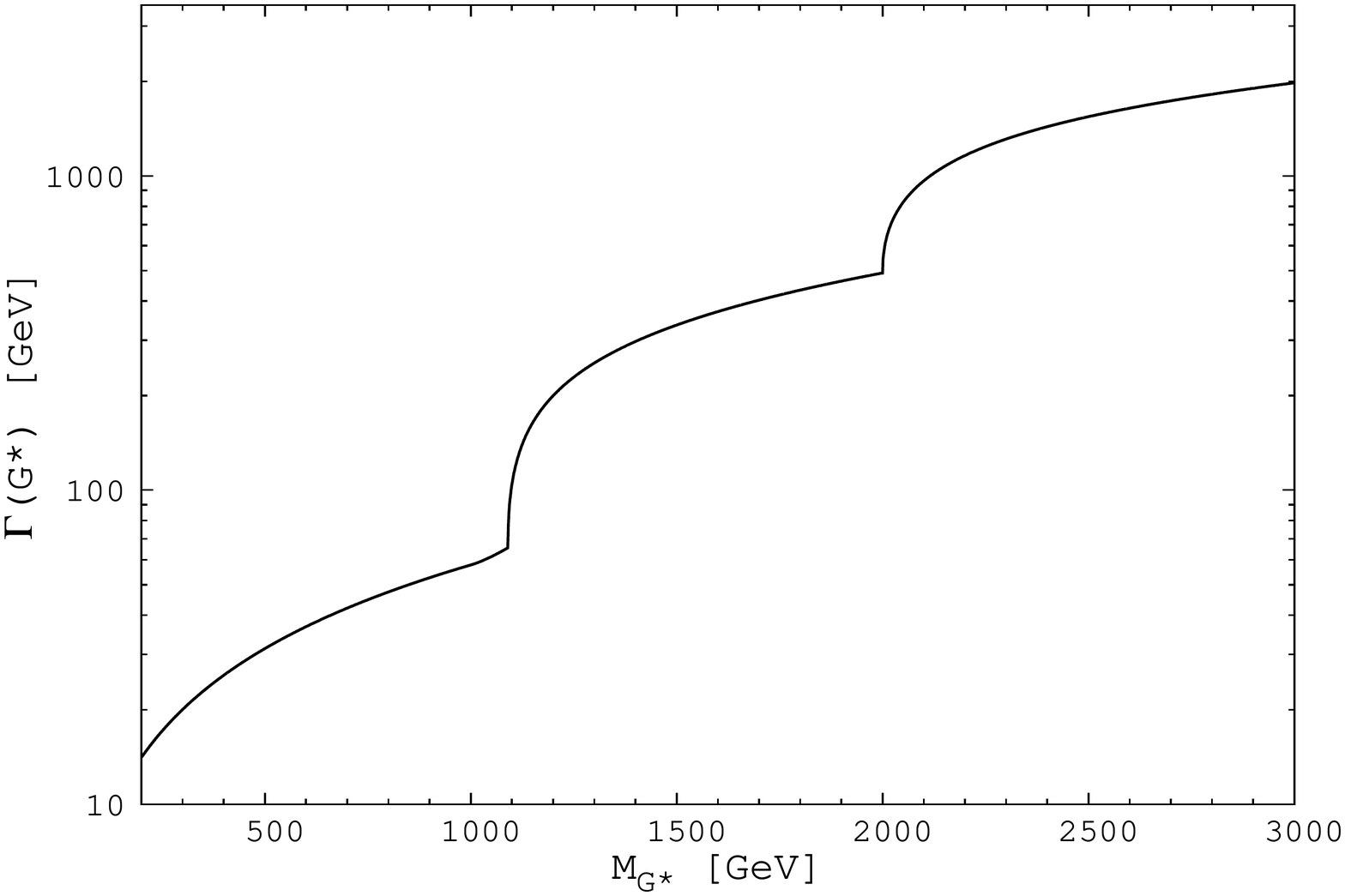}}
}
\caption{\textit{$G^{*}$ decay Branching Ratios and $G^{*}$ total decay width as functions of the $G^{*}$ mass in the TS5,
 with an almost fully composite $t_L$ ($s_R=0.4$).}
}
\label{BR_sR04}
\end{figure}

\begin{figure}[h!]
\centering 
\includegraphics[width=0.4\textwidth, angle=-90]{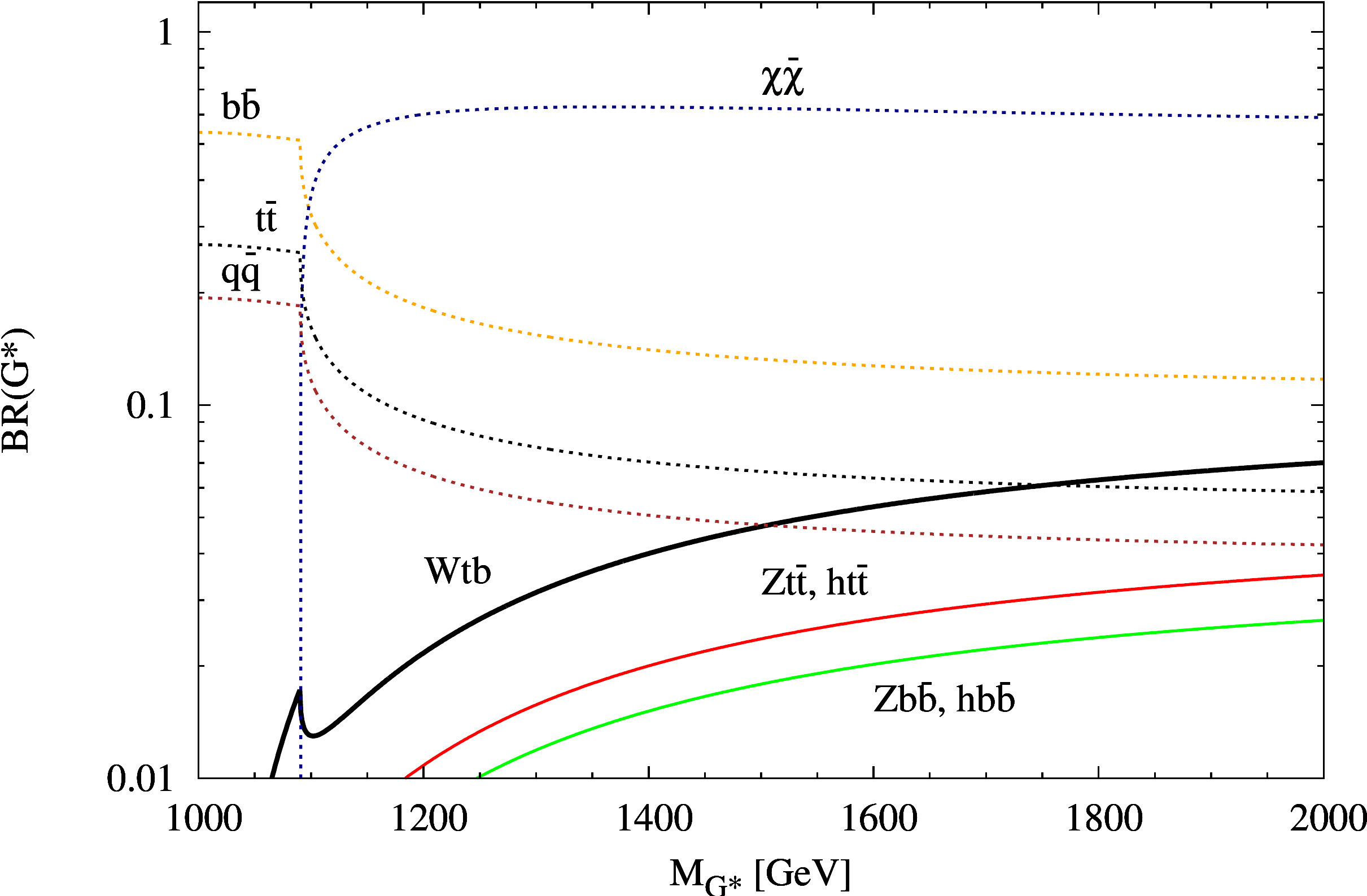}
\caption{\textit{BRs for the $G^*$ decays in the different search channels, considering the intermediate scenario ($M_T<M_{G*}<2 M_T$),
 with an almost fully composite $t_L$ ($s_R=0.4$).}
}
\label{BR_channels_sR04}
\end{figure}

The limit case of an almost fully composite left-handed top ($s_1 \simeq 1$, $s_R \simeq 0.4$)
\footnote{The limit value $s_R\sim 0.4$ is evaluated for $Y_*\sim 3$. 
Since, as we discussed, $s_1$ and $s_R$ are related to each other as $s_R=\frac{\sqrt{2}m_t}{Y_* v s_1}$, the specific $s_R$ values for which 
$s_1\to 1$ and the $G^*$ decays into custodian pairs become dominant depend on the value of $Y_*$.} 
makes difficult the discovery of the $G^*$ both
in the $t\bar{t}$ and in the $\psi\chi$ channel. This is due to the predominance of the $G^*$ decays into pairs of custodians.\\
In the TS5 the custodians $T_{2/3}$ and $T_{5/3}$ become light when the left-handed top becomes composite. 
As a consequence, for $s_R \simeq 0.4$, when the $G^{*}$ mass is above the threshold for the production of custodian pairs,
the decays into custodians become dominant (Fig.s \ref{BR_sR04} and \ref{BR_channels_sR04}).\\
 In this limit case, the search for the $G^*$ in the one heavy plus one SM fermion channel becomes less promising than in the `standard' scenario;
it could be interesting, instead, looking for the decays into custodian pairs.
These latter can decay into a $Z$ or a Higgs plus a top ($T_{2/3} \to Z(/h)t$) or into a $W$ plus a top ($T_{5/3} \to Wt$) 
and so they can give final states such as $WWtt$, $ZZtt$, $hhtt$ or $Zhtt$.\\
We point out that in the TS10 model there are also custodians ($\tilde{B}$, $\tilde{T}_{5/3}$ ...) that become light in the fully composite $t_R$ limit ($s_R=1$). 
Therefore, in the TS10, the $G^*$ discovery in the one heavy plus one SM fermion channel becomes quite difficult both 
in the fully composite $t_L$ and in the fully composite $t_R$ limit cases.
However, these limit cases seem to be quite disfavored by data; a fully composite $t_L$ is 
disfavored by electro-weak data \cite{pomarol_serra} both for the TS5 and the TS10, since it brings to too high values of the $S$ and $T$ parameters.
On the other hand, in the TS10, the opposite case of a fully composite $t_R$ is disfavored by the data from $b \to s\gamma$:
 as we have found in the first part of this project (sec. \ref{bsGamma}), in the case of a fully composite $t_R$, 
we obtain a strong constraint on the heavy fermion mass in the TS10, because of the absence of a $P_C$ protection to the effective $Wt_R b_R$ vertex.

\chapter{Conclusions}

Composite Higgs Models are among the compelling scenarios for physics beyond  
the Standard Model that can give an explanation of the origin of the EWSB
and that are going to be tested at the LHC.\\
In this thesis we have studied the phenomenology of this class of models. 
We have built two simple `two-site' models, the TS5 and the TS10, 
which incorporate a custodial symmetry and a $P_{LR}$ parity.\\
In the first part we have reconsidered the bounds on the CHM spectrum implied by flavor observables.
We have found in particular that the IR contribution to $b\to s \gamma$ induced by the flavor conserving effective vertex $W t_R b_R$ 
implies a robust Minimal Flavor Violating bound on the mass ($m_*$) of the new heavy fermions. 
The relevance of shifts to $W t_R b_R$ has been already pointed out in the literature, 
even though its importance in setting a bound on heavy fermion masses was unestimated in previous studies. 
We have also shown how this bound can be stronger in the case of the absence of a symmetry ($P_C$) protection to the effective $Wt_R b_R$ vertex. 
In particular, the constraints we have found are
\[m_* \gtrsim 1.4\ \text{TeV}
\]
in the TS5, and
\[m_* \gtrsim \frac{0.54}{\xi_{qL}}\ \text{TeV}\ , 
\]
where $\xi_{qL}$ denotes the $t_L$ degree of compositeness, in the TS10.\\
In addition to these bounds, we have calculated the constraints from the UV composite Higgs model contribution to $b \to s \gamma$.
Our results have shown that these bounds can be stronger than those from the IR contribution but they are model dependent; 
in particular they strongly depend on the assumptions made on the flavor structure of the composite sector. 
We have obtained an estimated limit
 \[
m_{*}\gtrsim (0.52)Y_{*}\ \text{TeV}
 \]
in a specific NP flavor scenario ($Y_{*}$ anarchic in the flavor space). \\ 
Even stronger bounds,
\[
m_{*}\gtrsim (1.3)Y_{*}\ \text{TeV} \ ,
\] 
can be obtained from $\epsilon^{'}/\epsilon_K$ 
but, again, they are model dependent and in principle could be loosened by acting on the NP flavor structure.
The lower bounds on $m_*$ we have found from $b\to s \gamma$, instead, are robust MFV bounds that cannot be evaded by acting on the NP flavor structure.\\
In the second part of the thesis we have studied in details the LHC phenomenology of the new heavy colored resonances. 
We have shown that heavy fermions have a great impact on the phenomenology of the heavy gluon; if the composite gluon is heavier than composite fermions,
 as flavor observables strongly suggest, 
the search in the channels where $G^*$ decays into one heavy fermion plus its Standard Model partner is very promising.\\
We have estimated the LHC sensitivity to the heavy colored vectors by analyzing these channels.
Our final results are summarized in Fig. \ref{CHM_reach} and show that a considerable part of the CHM parameter space can be already tested in the early stage of the LHC.\\
The LHC at $\sqrt{s}=7\,$TeV and with $10\,$fb$^{-1}$ of integrated luminosity 
should be able to discover a $G^*$ with mass in the range $M_{G^*} = 1.8 - 2.4$  for $\tan\theta_3 = 0.2-1$. 
On the other hand, by running at the design center-of-mass energy $\sqrt{s}=14\,$TeV, the LHC discovery reach extends to the mass range
$M_{G^*} = 3.3 - 5.2$  for $\tan\theta_3 = 0.2-1$ with an integrated luminosity $L =100\,$fb$^{-1}$.\\
Moreover we have found that the heavy-light decay topologies are very promising also for the discovery of the heavy fermions 
and should be taken into account, together with the ordinary single and pair production channels, in the searches for heavy fermions at the LHC.


\appendix 
\chapter{Two Site Models}
\section{TS5}
\label{TS5A}

Fermions rotate from the elementary/composite basis to the 'physical' light(SM)/heavy basis as
(we neglect $O\left(\Delta^{2}_{L2}\right)$ terms):

\begin{align}
\begin{split}\label{rotation_LL}
& \tan\varphi_{L1}=\frac{\Delta_{L1}}{M_{Q*}}\equiv \frac{s_{1}}{c_{1}}, \ \  s_{1}\equiv\sin\varphi_{L1} \  c_{1}\equiv\cos\varphi_{L1}\\
& s_{2}=\frac{\Delta_{L2}}{M_{Q'*}}\cos\varphi_{L1}\\ 
& s_{3}=\frac{\Delta_{L2}M_{Q'*}}{\Delta^{2}_{L1}+M^{2}_{Q*}-M^{2}_{Q'*}}\sin\varphi_{L1} \\
& \left\{\begin{array}{l}
	t_{L}=c_{1}t^{el}_{L}-s_{1}T^{com}_{L}-s_{2}T'^{com}_{L}\\
	T_{L}=s_{1}t^{el}_{L}+c_{1}T^{com}_{L}+s_{3}T'^{com}_{L} \\
        T'_{L}=\left(s_{2}c_{1}-s_{1}s_{3}\right)t^{el}_{L}-\left(s_{1}s_{2}+c_{1}s_{3}\right)T^{com}_{L}+ T'^{com}_{L} 
\end{array}  \right. \\
& \left\{\begin{array}{l}
	b_{L}=c_{1}b^{el}_{L}-s_{1}B^{com}_{L}-s_{2}B'^{com}_{L}\\
	B_{L}=s_{1}b^{el}_{L}+c_{1}B^{com}_{L}+s_{3}B'^{com}_{L} \\
        B'_{L}=\left(s_{2}c_{1}-s_{1}s_{3}\right)b^{el}_{L}-\left(c_{1}s_{3}+s_{1}s_{2}\right)B^{com}_{L}+B'^{com}_{L}
\end{array} \right.
\end{split}
\end{align}

\begin{align}
\begin{split}\label{rotation_LR}
& s_{4}=\Delta_{L2}\frac{\Delta_{L1}}{\Delta^{2}_{L1}+M^{2}_{Q*}-M^{2}_{Q'*}} \\
& \left\{\begin{array}{l}
	T_{R}=T^{com}_{R}+s_{4}T'^{com}_{R}\\
	T'_{R}=T'^{com}_{R}-s_{4}T^{com}_{R}
\end{array}  \right.  \ \ 
 \left\{\begin{array}{l}
	B_{R}=B^{com}_{R}+s_{4}B'^{com}_{R}\\
	B'_{R}=B'^{com}_{R}-s_{4}B^{com}_{R}  
\end{array} \right.
\end{split}
\end{align}

\begin{align}
\begin{split}\label{rotation_RR}
& \tan\varphi_{R}=\frac{\Delta_{R1}}{M_{\tilde{T}*}}\ \ s_{R}\equiv\sin\varphi_{R} \ \ c_{R}\equiv\cos\varphi_{R} \\
& \tan\varphi_{bR}=\frac{\Delta_{R2}}{M_{\tilde{B}*}}\ \ s_{bR}\equiv\sin\varphi_{bR} \ \ c_{bR}\equiv\cos\varphi_{bR} \\
& \left\{\begin{array}{l}
	t_{R}=c_{R}t^{el}_{R}-s_{R}\tilde{T}^{com}_{R}\\
	\tilde{T}_{R}=s_{R}t^{el}_{R}+c_{R}\tilde{T}^{com}_{R} 
\end{array}  \right. \ \ 
 \left\{\begin{array}{l}
	b_{R}=c_{bR}b^{el}_{R}-s_{bR}\tilde{B}^{com}_{R} \\ 
	\tilde{B}_{R}=s_{bR}b^{el}_{R}+c_{bR}\tilde{B}^{com}_{R}
\end{array} \right.
\end{split}
\end{align}

\noindent
Physical heavy fermion masses are related to the bare ones according to:
\begin{align}
\left\{\begin{array}{l}
M_{\tilde{T}}=\sqrt{M^{2}_{\tilde{T}*}+\Delta^{2}_{R1}}=\frac{M_{\tilde{T}*}}{c_{R}}\\
M_{\tilde{B}}=\sqrt{M^{2}_{\tilde{B}*}+\Delta^{2}_{R2}}=\frac{M_{\tilde{B}*}}{c_{bR}}\\
M_{T}=M_{B}=\sqrt{M^{2}_{Q*}+\Delta^{2}_{L1}}=\frac{M_{Q*}}{c_{1}} \\
M_{T5/3}=M_{T2/3}=M_{Q*} \\
M_{T'}=M_{B'}=\sqrt{M^{2}_{Q'*}+\Delta^{2}_{L2}}\simeq M_{Q'*}=M_{B-1/3}=M_{B-4/3} \end{array} \right.
\end{align}

\subsection{Yukawa Lagrangian}

In the elementary/composite basis the Yukawa Lagrangian reads:
\begin{eqnarray}	
\mathcal{L}^{YUK}=Y_{*U}Tr\left\{\bar{\mathcal{Q}}\mathcal{H}\right\}\tilde{T}+Y_{*D}Tr\left\{\bar{\mathcal{Q'}}\mathcal{H}\right\}\tilde{B}+h.c.\\ \nonumber
=Y_{*U}\left\{\bar{T}\phi^{\dag}_{0}\tilde{T}+\bar{T}_{2/3}\phi_{0}\tilde{T}+\bar{T}_{5/3}\phi^{+}\tilde{T}-\bar{B}\phi^{-}\tilde{T}\right\}\\ \nonumber
+Y_{*D}\left\{\bar{B}_{-1/3}\phi^{\dag}_{0}\tilde{B}+\bar{B'}\phi_{0}\tilde{B}+\bar{T'}\phi^{+}\tilde{B}-\bar{B}_{-4/3}\phi^{-}\tilde{B}\right\}+h.c.
\end{eqnarray}
\noindent
After field rotation to the mass eigenstate basis, before EWSB, $\mathcal{L}^{YUK}$ reads:

\begin{align}
\begin{split}\label{eq.Lagrange2}	
\mathcal{L}^{YUK}= & Y_{*U}c_{1}c_{R}\left(\bar{T}_{L}\phi^{\dag}_{0}\tilde{T}_{R}-\bar{B}_{L}\phi^{-}\tilde{T}_{R}\right)+Y_{*U}c_{R}\left(\bar{T}_{2/3L}\phi_{0}\tilde{T}_{R}+\bar{T}_{5/3L}\phi^{+}\tilde{T}_{R}\right)\\ 
& -Y_{*U}\left(s_{1}s_{2}+c_{1}s_{3}\right)c_{R}\left(\bar{T'}_{L}\phi^{\dag}_{0}\tilde{T}_{R}-\bar{B'}_{L}\phi^{-}\tilde{T}_{R}\right)-
 Y_{*U}s_{1}c_{R}\left(\bar{t}_{L}\phi^{\dag}_{0}\tilde{T}_{R}-\bar{b}_{L}\phi^{-}\tilde{T}_{R}\right)\\ 
& -Y_{*U}s_{R}\left(\bar{T}_{2/3L}\phi_{0}t_{R}+\bar{T}_{5/3L}\phi^{+}t_{R}\right)+Y_{*U}\left(s_{1}s_{2}+c_{1}s_{3}\right)s_{R}\left(\bar{T'}_{L}\phi^{\dag}_{0}t_{R}-\bar{B'}_{L}\phi^{-}t_{R}\right)\\ 
& -Y_{*U}c_{1}s_{R}\left(\bar{T}_{L}\phi^{\dag}_{0}t_{R}-\bar{B}_{L}\phi^{-}t_{R}\right)+Y_{*U}s_{1}s_{R}\left(\bar{t}_{L}\phi^{\dag}_{0}t_{R}-\bar{b}_{L}\phi^{-}t_{R}\right)\\ 
&+Y_{*U}\left(\bar{T}_{R}\phi^{\dag}_{0}\tilde{T}_{L}-\bar{B}_{R}\phi^{-}\tilde{T}_{L}\right)+Y_{*U}\left(\bar{T}_{2/3R}\phi_{0}\tilde{T}_{L}+\bar{T}_{5/3R}\phi^{+}\tilde{T}_{L}\right)\\ 
&-Y_{*U}s_{4}\left(\bar{T'}_{R}\phi^{\dag}_{0}\tilde{T}_{L}-\bar{B'}_{R}\phi^{-}\tilde{T}_{L}\right)\\ 
&+Y_{*D}c_{bR}\left(\bar{B}_{-1/3L}\phi^{\dag}_{0}\tilde{B}_{R}-\bar{B}_{-4/3L}\phi^{-}\tilde{B}_{R}\right)+Y_{*D}c_{bR}\left(\bar{B'}_{L}\phi_{0}\tilde{B}_{R}+\bar{T'}_{L}\phi^{+}\tilde{B}_{R}\right)\\ 
&-Y_{*D}s_{bR}\left(\bar{B}_{-1/3L}\phi^{\dag}_{0}b_{R}-\bar{B}_{-4/3L}\phi^{-}b_{R}\right)-Y_{*D}s_{bR}\left(\bar{B'}_{L}\phi_{0}b_{R}+\bar{T'}_{L}\phi^{+}b_{R}\right)\\ 
&-Y_{*D}s_{2}c_{bR}\left(\bar{b}_{L}\phi_{0}\tilde{B}_{R}+\bar{t}_{L}\phi^{+}\tilde{B}_{R}\right)+Y_{*D}s_{2}s_{bR}\left(\bar{b}_{L}\phi_{0}b_{R}+\bar{t}_{L}\phi^{+}b_{R}\right)\\ 
&-Y_{*D}s_{3}s_{bR}\left(\bar{B}_{L}\phi_{0}b_{R}+\bar{T}_{L}\phi^{+}b_{R}\right)+Y_{*D}s_{3}c_{bR}\left(\bar{B}_{L}\phi_{0}\tilde{B}_{R}+\bar{T}_{L}\phi^{+}\tilde{B}_{R}\right)\\
&+Y_{*D}\left(\bar{B'}_{R}\phi_{0}\tilde{B}_{L}+\bar{T'}_{R}\phi^{+}\tilde{B}_{L}\right)+Y_{*U}\left(\bar{B}_{-1/3R}\phi^{\dag}_{0}\tilde{B}_{L}-\bar{B}_{-4/3R}\phi^{-}\tilde{B}_{L}\right)\\ 
&+Y_{*D}s_{4}\left(\bar{B}_{R}\phi_{0}\tilde{B}_{L}+\bar{T}_{R}\phi^{+}\tilde{B}_{L}\right)+h.c. 
\end{split}
\end{align}
\noindent
After the EWSB top and bottom masses arise as:
\begin{equation}
	m_{t}=\frac{v}{\sqrt{2}}Y_{*U}s_{1}s_{R}
\label{tmass}
\end{equation}
\begin{equation}
	m_{b}=\frac{v}{\sqrt{2}}Y_{*D}s_{2}s_{bR} \ .
\label{bmass}
\end{equation}

\noindent
We have also electroweak mixings among fermions. The fermionic mass matrices for up and down states read, 
in the basis $\left(\bar{t}_L\ \bar{\tilde{T}}_L\ \bar{T}_{2/3L}\ \bar{T}_L\ \bar{T'}_L \right)$ 
$\left(t_R\ \tilde{T}_R\ T_{2/3R}\ T_R\ T'_R \right)$ for the up sector 
and in the basis $\left(\bar{b}_L\ \bar{\tilde{B}}_L\ \bar{B'}_L\ \bar{B}_{-1/3L} \ \bar{B}_L\right)$ 
$\left(b_R\ \tilde{B}_R\ B'_R\ B_{-1/3R} \ B_R\right)$ for the down-type fermions:

\begin{align}
\begin{split}\label{Mup}
&\mathcal{M}_{up}= \\
& \left(\begin{array}{ccccc}
	m_{t} & -Y_{*U}\frac{v}{\sqrt{2}}s_{1}c_{R} & 0 & 0& 0\\ 
	0 & M_{\tilde{T}} & Y_{*U}\frac{v}{\sqrt{2}} & Y_{*U}\frac{v}{\sqrt{2}}& -s_{4}Y_{*U}\frac{v}{\sqrt{2}}\\ 
 -Y_{*U}\frac{v}{\sqrt{2}}s_{R}& Y_{*U}\frac{v}{\sqrt{2}}c_{R} & M_{T2/3} & 0& 0\\ 
	-Y_{*U}\frac{v}{\sqrt{2}}c_{1}s_{R} &Y_{*U}\frac{v}{\sqrt{2}}c_{1}c_{R} & 0 & M_T & 0\\ 
	Y_{*U}\frac{v}{\sqrt{2}}\left(s_{1}s_{2}+c_{1}s_{3}\right)s_{R} & -Y_{*U}\frac{v}{\sqrt{2}}\left(s_{1}s_{2}+c_{1}s_{3}\right)c_{R}& 0& 0& M_{T'}
	\end{array}\right)
\end{split}
\end{align} 

\begin{align}
\begin{split}\label{Mdown}
&\mathcal{M}_{down}=\\ 
& \left(\begin{array}{ccccc}
	m_{b} & -Y_{*D}\frac{v}{\sqrt{2}}s_{2}c_{bR} & 0 & 0 & 0\\
	0 & M_{\tilde{B}} & Y_{*D}\frac{v}{\sqrt{2}} & Y_{*D}\frac{v}{\sqrt{2}} & Y_{*D}\frac{v}{\sqrt{2}}s_{4} \\
	-Y_{*D}\frac{v}{\sqrt{2}}s_{bR} & Y_{*D}\frac{v}{\sqrt{2}}c_{bR} & M_{B'} & 0 & 0\\
	-Y_{*D}\frac{v}{\sqrt{2}}s_{bR} & 	Y_{*D}\frac{v}{\sqrt{2}}c_{bR} & 0 & M_{B-1/3} & 0\\
		-Y_{*D}\frac{v}{\sqrt{2}}s_{3}s_{bR} & 	Y_{*D}\frac{v}{\sqrt{2}}s_{3}c_{bR} & 0 & 0 & M_B\\
	\end{array}\right)
\end{split}
\end{align} 
\newpage
\subsection{TS5 Lagrangian for $G^*$ and heavy fermion interactions}\label{Gstar_L}
The TS5 lagrangian (\ref{eq:Ltotal}) can be diagonalized, before EWSB, by performing the field rotations 
from the composite/elementary to the mass-eigenstate basis, 
that we have just shown, for the fermions, in (\ref{rotation_LL}), (\ref{rotation_LR}), (\ref{rotation_RR}); 
gluon and $G^*$ rotate as described in (\ref{GgMIX}). The diagonalized lagrangian reads

\begin{align}
\label{eq:Lrotated}
{\cal L} =& \, {\cal L}_{gauge}+ {\cal L}_{fermion}+ {\cal L}_{Higgs} \\[0.5cm]
\begin{split}   \label{eq:Lrotatedgauge}
{\cal L}_{gauge}= & -\frac{1}{4} G_{\mu\nu}G^{\mu\nu} \\[0.1cm]
 & +\frac{1}{2}\left(D_\mu G^*_\nu D_\nu G^*_\mu - D_\mu G^*_\nu D_\mu G^*_\nu \right)+\frac{1}{2}M^2_{G*}G^{* 2}_\mu 
     +\frac{ig_S}{2} \, G_{\mu\nu} \!\left[G^{*}_\mu,G^*_\nu \right] \\[0.1cm]
 & + 2i \, g_S \cot 2\theta_3 \, D_\mu G^*_\nu \left[G^{*}_\mu,G^*_\nu \right]+\frac{g^2_S}{4}\left(\frac{\sin^4\!\theta_3}{\cos^2\!\theta_3} 
    + \frac{\cos^4\!\theta_3}{\sin^2\!\theta_3}\right) \left[G^{*}_\mu,G^*_\nu \right]^2 
\end{split}  \\[0.5cm]
\begin{split}  \label{eq:Lrotatedfermion}
{\cal L}_{fermion}= 
       & \, \bar{q} \, i\!\Dslash q+ \bar{\psi}\, i\!\Dslash\psi+\bar{\chi}\left( i\!\Dslash - M_\chi\right)\chi \\[0.1cm]
       & -g_S \tan\theta_3 \, G^*_\mu \bar{q}\gamma^\mu q + g_S \left( \sin^2\!\varphi_{\psi} \cot\theta_3 
           -  \cos^2\!\varphi_{\psi}\tan\theta_3\right) G^*_\mu \bar{\psi}  \gamma^\mu \psi \\[0.1cm]
       & + g_S \, \frac{\sin\varphi_{\psi}\cos\varphi_{\psi}}{\sin\theta_3 \cos\theta_3} \, G^*_\mu \bar{\chi} \gamma^\mu \psi 
          +  g_S \left( \cos^2\varphi_{\psi} \cot\theta_3- \sin^2\varphi_{\psi} \tan\theta_3 \right) G^*_\mu \bar{\chi} \gamma^\mu \chi \\[0.1cm]
       & + h.c. + O(s_2) 
\end{split} 
\\[0.5cm]
\begin{split}  \label{eq:LrotatedHiggs}
 {\cal L}_{Higgs}= 
  & \, |D_\mu H|^2 -V(H) \\[0.1cm]
  & + Y_*  \cos\varphi_{L1} \cos\varphi_{tR} \, \bar{Q}_L\tilde{H} \tilde{T}_R + Y_* \cos\varphi_{tR} \, \bar{Q}_{uL}H\tilde{T}_R 
      -Y_*\sin\varphi_{L1}\cos\varphi_{tR}\,\bar{q}_L\tilde{H}\tilde{T}_R \\[0.1cm]
  & - Y_*\sin\varphi_{tR} \,\bar{Q}_{uL} H t_R -Y_*\cos\varphi_{L1}\sin\varphi_{tR}\,\bar{Q}_L\tilde{H}t_R 
     + Y_*\sin\varphi_{L1}\sin\varphi_{tR}\, \bar{q}_L\tilde{H}t_R \\[0.1cm]
  & - Y_* \left( s_2 \sin\varphi_{L1}  + s_3  \cos\varphi_{L1} \right)\left(\cos\varphi_{tR}\, \bar{Q}^{'}_L\tilde{H}\tilde{T}_R 
     - \sin\varphi_{tR}\, \bar{Q}^{'}_L\tilde{H}t_R \right)\\[0.1cm]
  & + Y_*\, \bar{Q}_R \tilde{H} \tilde{T}_L +Y_*\, \bar{Q}_u H \tilde{T}_L - s_4 Y_*\, \bar{Q}^{'}_R\tilde{H} \tilde{T}_L \\[0.1cm]
  & + Y_*\cos\varphi_{bR} \, \bar{Q}_{dL}\tilde{H}\tilde{B}_R+ Y_*\cos\varphi_{bR} \, \bar{Q}^{'}_{L}H\tilde{B}_R 
      - Y_* \sin\varphi_{bR} \,\bar{Q}_{dL}\tilde{H}b_R\\[0.1cm]
  & - Y_*\sin\varphi_{bR} \,\bar{Q}^{'}_{L}H b_R - Y_* s_2\cos\varphi_{bR}\, \bar{q}_{L} H \tilde{B}_R + Y_* s_2 \sin\varphi_{bR}\,\bar{q}_{L} H b_R \\[0.1cm]
  & - Y_* s_3\sin\varphi_{bR} \, \bar{Q}_{L} H b_R + Y_* s_3\cos\varphi_{bR} \,\bar{Q}_{L} H \tilde{B}_R\\[0.1cm]
  & + Y_* \, Q^{'}_R H\tilde{B}_L+ Y_* \, \bar{Q}_{dR}\tilde{H}\tilde{B}_L + Y_* s_4 \, \bar{Q}_R H \tilde{B}_L + h.c.
\end{split}
\end{align}

where $q = u,d,c,s$, $\psi = t_L, b_L , t_R, b_R$, and we have defined  $Q=(T,B)$, $Q'=(T',B')$, $Q_u=(T_{5/3}, T_{2/3})$, $Q_d=(B_{-1/3},B_{-4/3})$, 
$H=(\phi^{+}, \phi_0)$,  $\tilde{H}\equiv i\sigma^2 H^* = (\phi^\dag_0, -\phi^{-})$.  $\chi$ denotes any of the heavy fermions, except in the 
first term in the third line of eq.(\ref{eq:Lrotatedfermion}), where it denotes a top or bottom heavy partner, $T, B, \tilde T, \tilde B$.

\section{TS10}
\label{TS10A}

Fermions rotate from the elementary/composite basis to the 'physical' light(SM)/heavy basis as:
\begin{align}
\begin{split}\label{rotation_LL_TS}
& \tan\varphi_{L1}=\frac{\Delta_{L1}}{M_{Q*}}\equiv \frac{s_{1}}{c_{1}}\\  
& \left\{\begin{array}{l}
	t_{L}=c_{1}t^{el}_{L}-s_{1}T^{com}_{L}\\
	T_{L}=s_{1}t^{el}_{L}+c_{1}T^{com}_{L}
\end{array} \right.  \ \ 
 \left\{\begin{array}{l}
	b_{L}=c_{1}b^{el}_{L}-s_{1}B^{com}_{L}\\
	B_{L}=s_{1}b^{el}_{L}+c_{1}B^{com}_{L}  
\end{array} \right.
\end{split}
\end{align}

\begin{align}
\begin{split}\label{rotation_RR_TS}
& \tan\varphi_{R}=\frac{\Delta_{R1}}{M_{\tilde{Q}*}}\ \ s_{R}\equiv\sin\varphi_{R} \ \ c_{R}\equiv\cos\varphi_{R} \\
& \tan\varphi_{bR}=\frac{\Delta_{R2}}{M_{\tilde{Q}*}}\ \ s_{bR}\equiv\sin\varphi_{bR} \ \ c_{bR}\equiv\cos\varphi_{bR} \\
& \left\{\begin{array}{l}
	t_{R}=c_{R}t^{el}_{R}-s_{R}\tilde{T}^{com}_{R}\\
	\tilde{T}_{R}=s_{R}t^{el}_{R}+c_{R}\tilde{T}^{com}_{R} 
\end{array}  \right.  \ \ 
 \left\{\begin{array}{l}
	b_{R}=c_{bR}b^{el}_{R}-s_{bR}\tilde{B}^{com}_{R} \\ 
	\tilde{B}_{R}=s_{bR}b^{el}_{R}+c_{bR}\tilde{B}^{com}_{R}
\end{array} \right. 
\end{split}
\end{align}

\noindent
Physical heavy fermion masses are related to the bare ones as:
\begin{align}
\left\{\begin{array}{l}
M_{\tilde{T}}=\sqrt{M^{2}_{\tilde{Q}*}+\Delta^{2}_{R1}}=\frac{M_{\tilde{Q}*}}{c_{R}}\\
M_{\tilde{B}}=\sqrt{M^{2}_{\tilde{Q}*}+\Delta^{2}_{R2}}=\frac{M_{\tilde{Q}*}}{c_{bR}}\\
M_{\tilde{T}5/3}=M_{\tilde{T}'5/3}=M_{\tilde{T}'}=M_{\tilde{B}'}=M_{\tilde{Q}*}\\
M_{T}=M_B=\sqrt{M^{2}_{Q*}+\Delta^{2}_{L1}}=\frac{M_{Q*}}{c_{1}} \\
M_{T2/3}=M_{T5/3}=M_{Q*} 
\end{array} \right.
\end{align}\\

\noindent
In the elementary/composite basis the Yukawa Lagrangian reads:
\begin{equation}	
\mathcal{L}^{YUK}=+Y_{*}Tr\left\{\mathcal{H}\bar{\mathcal{Q}}\mathcal{\tilde{Q}'}\right\}+Y_{*}Tr\left\{\bar{\mathcal{Q}}\mathcal{H}\mathcal{\tilde{Q}}\right\} 
\end{equation}
\noindent

After field rotation to the mass eigenstate basis, before EWSB, $\mathcal{L}^{YUK}$ reads as in eq. (\ref{eq.Lagrange2_ts10}).\\

\noindent
After EWSB top and bottom masses arise as:

\begin{equation}
	m_{t}=\frac{v}{2}Y_{*}s_{1}s_{R}
\label{tmass_TS}
\end{equation}
\begin{equation}
	m_{b}=\frac{v}{\sqrt{2}}Y_{*}s_{1}s_{bR}
\label{bmass_TS}
\end{equation}

\noindent
The fermionic mass matrices for up and down states read, 
in the basis $\left(\bar{t}_L\ \bar{\tilde{T}}_L\ \bar{T}_{2/3L}\ \bar{T}_L\ \bar{\tilde{T}}'_L \right)$ 
$\left(t_R\ \tilde{T}_R\ T_{2/3R}\ T_R\ \tilde{T}'_R \right)$ for the up sector 
and in the basis $\left(\bar{b}_L\ \bar{\tilde{B}}_L\ \bar{\tilde{B}}'_L\ \bar{B}_L\right)$ 
$\left(b_R\ \tilde{B}_R\ \tilde{B}'_R\ B_R\right)$ for the down-type fermions:

\begin{align}
\mathcal{M}^{TS10}_{up}=\ Y_{*}\frac{v}{2}\left(\begin{array}{ccccc}
	\frac{m_{t}}{Y_{*}\frac{v}{2}} & -s_{1}c_{R} & 0 & 0& -s_{1}\\ 
	0 & \frac{M_{\tilde{T}}}{Y_{*}\frac{v}{2}} & -1& 1& 0\\ 
 s_{R} & -c_{R} & \frac{M_{T2/3}}{Y_{*}\frac{v}{2}} & 0& -1\\ 
	-c_{1}s_{R} & c_{1}c_{R} & 0 & \frac{M_{T}}{Y_{*}\frac{v}{2}} & c_{1}\\ 
	0 & 0 & -1& 1& \frac{M_{\tilde{T}'}}{Y_{*}\frac{v}{2}}
	\end{array}\right)
\label{Mup_mchm10}
\end{align}

\begin{align}
\mathcal{M}^{TS10}_{down}=\ Y_{*}\frac{v}{\sqrt{2}}\left(\begin{array}{cccc}
	\frac{m_{b}}{Y_{*}\frac{v}{\sqrt{2}}} & -s_{1}c_{bR} & -s_{1} & 0 \\
	0 & \frac{M_{\tilde{B}}}{Y_{*}\frac{v}{\sqrt{2}}} & 0 & 1\\
	0 & 0 &\frac{ M_{\tilde{B}'}}{Y_{*}\frac{v}{\sqrt{2}}} & 1 \\
	-c_{1}s_{bR} & 	c_{1}c_{bR} & c_{1} & \frac{M_B}{Y_{*}\frac{v}{\sqrt{2}}}\\
	\end{array}\right)
\label{Mdown_mchm10}
\end{align}

\begin{align}
\begin{split}\label{eq.Lagrange2_ts10}	
\mathcal{L}^{YUK}=\  & Y_{*}c_{1}c_{R}\frac{1}{\sqrt{2}}\left(\bar{T}_{L}\phi^{\dag}_{0}\tilde{T}_{R}-\bar{B}_{L}\phi^{-}\tilde{T}_{R}\right)-Y_{*}c_{R}\frac{1}{\sqrt{2}}\left(\bar{T}_{2/3L}\phi_{0}\tilde{T}_{R}+\bar{T}_{5/3L}\phi^{+}\tilde{T}_{R}\right)\\ 
& - Y_{*}s_{1}c_{R}\frac{1}{\sqrt{2}}\left(\bar{t}_{L}\phi^{\dag}_{0}\tilde{T}_{R}-\bar{b}_{L}\phi^{-}\tilde{T}_{R}\right)+Y_{*}s_{1}s_{R}\frac{1}{\sqrt{2}}\left(\bar{t}_{L}\phi^{\dag}_{0}t_{R}-\bar{b}_{L}\phi^{-}t_{R}\right)\\ 
& +Y_{*}s_{R}\frac{1}{\sqrt{2}}\left(\bar{T}_{2/3L}\phi_{0}t_{R}+\bar{T}_{5/3L}\phi^{+}t_{R}\right)
 -Y_{*}c_{1}s_{R}\frac{1}{\sqrt{2}}\left(\bar{T}_{L}\phi^{\dag}_{0}t_{R}-\bar{B}_{L}\phi^{-}t_{R}\right)\\ 
& +Y_{*}\frac{1}{\sqrt{2}}\left(\bar{T}_{R}\phi^{\dag}_{0}\tilde{T}_{L}-\bar{B}_{R}\phi^{-}\tilde{T}_{L}\right)-Y_{*}\frac{1}{\sqrt{2}}\left(\bar{T}_{2/3R}\phi_{0}\tilde{T}_{L}+\bar{T}_{5/3R}\phi^{+}\tilde{T}_{L}\right)\\ 
& +Y_{*}\left(\bar{T}_{5/3L}\phi^{\dag}_{0}\tilde{T}_{5/3R}-\bar{T}_{2/3L}\phi^{-}\tilde{T}_{5/3R}\right)+Y_{*}\left(\bar{T}_{5/3R}\phi^{\dag}_{0}\tilde{T}_{5/3L}- \bar{T}_{2/3R}\phi^{-}\tilde{T}_{5/3L}\right)\\
&-Y_{*}s_{1}c_{bR}\left(\bar{b}_{L}\phi_{0}\tilde{B}_{R}+\bar{t}_{L}\phi^{+}\tilde{B}_{R}\right)+Y_{*}s_{1}s_{bR}\left(\bar{b}_{L}\phi_{0}b_{R}+\bar{t}_{L}\phi^{+}b_{R}\right)\\ 
&-Y_{*}c_{1}s_{bR}\left(\bar{B}_{L}\phi_{0}b_{R}+\bar{T}_{L}\phi^{+}b_{R}\right)+Y_{*}c_{1}c_{bR}\left(\bar{B}_{L}\phi_{0}\tilde{B}_{R}+\bar{T}_{L}\phi^{+}\tilde{B}_{R}\right)\\
&+Y_{*}\left(\bar{B}_{R}\phi_{0}\tilde{B}_{L}+\bar{T}_{R}\phi^{+}\tilde{B}_{L}\right)+ Y_{*}\left(\bar{B}_{R}\phi^{\dag}_{0}\tilde{B}'_{L}+Y_{*}\bar{T}_{2/3R}\phi^{+}\tilde{B}'_{L}\right)\\
& Y_{*}\frac{1}{\sqrt{2}}\left(\bar{T}_{R}\phi^{\dag}_{0}\tilde{T}'_{L}+\bar{B}_{R}\phi^{-}\tilde{T}'_{L}\right)-Y_{*}\frac{1}{\sqrt{2}}\left(\bar{T}_{2/3R}\phi_{0}\tilde{T}'_{L}-\bar{T}_{5/3R}\phi^{+}\tilde{T}'_{L}\right)\\
& + Y_{*}c_{1}\frac{1}{\sqrt{2}}\left(\bar{T}_{L}\phi^{\dag}_{0}\tilde{T}'_{R}+\bar{B}_{L}\phi^{-}\tilde{T}'_{R}\right)-Y_{*}\frac{1}{\sqrt{2}}\left(\bar{T}_{2/3L}\phi^{\dag}_0\tilde{T}'_{R}-\bar{T}_{5/3L}\phi^{+}\tilde{T}'_{R}\right)\\
&- Y_{*}s_{1}\frac{1}{\sqrt{2}}\left(\bar{t}_{L}\phi^{\dag}_{0}\tilde{T}'_{R}+\bar{b}_{L}\phi^{-}\tilde{T}'_{R}\right)+Y_{*}\left(\bar{T}_{5/3R}\phi_{0}\tilde{T}'_{5/3L}-\bar{T}_{R}\phi^{-}\tilde{T}'_{5/3L}\right)\\
&+Y_{*}c_{1}\left(\bar{B}_{L}\phi^{\dag}_{0}\tilde{B}'_{R}-\bar{T}_{L}\phi^{-}\tilde{T}'_{5/3R}\right)-Y_{*}s_{1}\left(\bar{b}_{L}\phi^{\dag}_{0}\tilde{B}'_{R}-\bar{t}_{L}\phi^{-}\tilde{T}'_{5/3R}\right)\\
&+Y_{*}\bar{T}_{2/3L}\phi^{+}\tilde{B}'_{R}+Y_* \bar{T}_{5/3L}\phi_{0}\tilde{T}'_{5/3R} + h.c.
\end{split}
\end{align}
\noindent

\chapter{Constraints on heavy fermion masses}
\section{BOUND derivation}
\label{App_bound}

\[
 BR_{th}=(315 \pm 23)10^{-6}
\]
\[
 BR_{ex}=(355 \pm 24 \pm 9)10^{-6}
\]
\[
 \Gamma_{tot} \propto |\mathcal{C}_7(\mu_b)|^2+|\mathcal{C}^{'}_7(\mu_b)|^2\approx 
|\mathcal{C}^{SM}_7(\mu_b)+\mathcal{C}^{NP}_7(\mu_b)|^2+|\mathcal{C}^{'NP}_7(\mu_b)|^2
\]
\noindent
If we consider only the $\mathcal{C}_7$ contribution:
\[
 \frac{\Gamma_{tot}}{\Gamma_{SM}}=1+2\frac{Re(\mathcal{C}^{SM}_7(\mu_b)^{*}\mathcal{C}^{NP}_7(\mu_b))}{|\mathcal{C}^{SM}_7(\mu_b)|^2}+O(\Delta\mathcal{C}^2_7)
\]
For $\mu_b =5$ GeV, $\mu_W = M_W $, $\alpha_S =0.118$:\\

$C_7(\mu_b)=0.695 C_7(\mu_W)+0.086 C_8(\mu_W)-0.158 C_2(\mu_W)=-0.300$ \cite{Buras}. \\

\[
 \mathcal{C}^{NP}_7(\mu_b) = \left(\frac{\alpha_S (\mu_W)}{\alpha_S (\mu_b)} \right)^{\frac{16}{23}}\mathcal{C}^{NP}_7(\mu_w)= 0.695\ \mathcal{C}^{NP}_7(\mu_w)
\]
\noindent 
we obtain at $95 \%$ C.L.:
\[
 -0.0775 < \mathcal{C}^{NP}_7(\mu_w) < 0.0226
\]

\noindent
For $m_* = 1$ TeV:
\[
 \mathcal{C}^{NP}_7(\mu_W)=\left(\frac{\alpha_S (m_*)}{\alpha_S (m_t)} \right)^{\frac{16}{21}} \left(\frac{\alpha_S (m_t)}{\alpha_S (\mu_W)} \right)^{\frac{16}{23}}\simeq 0.79\ \mathcal{C}^{NP}_7(m_*)
\]
we obtain at $95 \%$ C.L.:
\[
 -0.0978 < \mathcal{C}^{NP}_7(m_*) < 0.0284
\]
\noindent 
If we consider only the $\mathcal{C}^{'}_7$ contribution:

\[
 \frac{\Gamma_{tot}}{\Gamma_{SM}} \simeq 1+\frac{|\mathcal{C}^{'NP}_7(\mu_b)|^2}{|\mathcal{C}^{SM}_7(\mu_b)|^2}
\]
\[
 C^{'}_7(\mu_b)\simeq C^{'NP}_7(\mu_b)=\left(\frac{\alpha_S (m_*)}{\alpha_S (m_t)} \right)^{\frac{16}{21}} \left(\frac{\alpha_S (m_t)}{\alpha_S (\mu_b)} \right)^{\frac{16}{23}} \mathcal{C}^{'NP}_7(m_*)\simeq 0.55\  \mathcal{C}^{'NP}_7(m_*)
\]
we obtain at $95 \% $ C.L.:

\[
 |\mathcal{C}^{'NP}_7(\mu_w)|<0.294
\]
\[
 |\mathcal{C}^{'NP}_7(m_*)|<0.372
\]

\section{CHM contribution to the effective coupling $Wt_Rb_R$ in the TS5}\label{wtrbr_calculation_TS5}
In this section (and in the next, for the TS10) we show in detail 
the calculation of the CHM contribution to the effective coupling $Wt_Rb_R$, $\mathcal{C}_R$.\\
\noindent
We show the expressions for the elementary/composite rotation angles and their values in the limit of a fully composite top. 
\begin{eqnarray*}
s_1=\frac{\Delta_{L1}}{\sqrt{M^{2}_{Q*}+\Delta^{2}_{L1}}}=\frac{\Delta_{L1}}{M_T}\\
 s_2=\frac{\Delta_{L2}}{M_{Q'*}}c_1\simeq \frac{\Delta_{L2}}{M_{T'}}c_1 \\
s_3=\frac{\Delta_{L2}M_{Q'*}}{M^{2}_{Q*}+\Delta^{2}_{L1}-M^{2}_{Q'*}}s_1\simeq s_2\frac{M^2_{T'}}{M^2_T-M^2_{T'}}\frac{s_1}{c_1}
\end{eqnarray*}
\\
$s_1$ limit values ($s^{MIN}_1\equiv$ fully composite $t_R$/ $s^{MAX}_1\equiv$ fully composite $t_L$):

\[
s^{MIN}_1=\frac{y_t}{Y_*}
\]
\[
s^{MAX}_1=\sqrt{1-\frac{y_b}{Y_*}}\simeq 1-\frac{1}{2}\frac{y_b}{Y_*}
\]
\[
s^{MIN}_{2}|_{s^{MAX}_1}\simeq\frac{y_b}{Y_*}
\]
\\
\ \\
In the TS5, we have two doublets of heavy fermions, one in the $X=2/3$ representation, $(T,B)$, and the other in the $X=-1/3$, $(T',B')$, 
that give a contribution to $\mathcal{C}_R$.\\
We calculate now the contribution from the first diagram ($T$, $B$ exchange). 
\begin{figure}[h!]
\centering
\scalebox{0.7}{\includegraphics{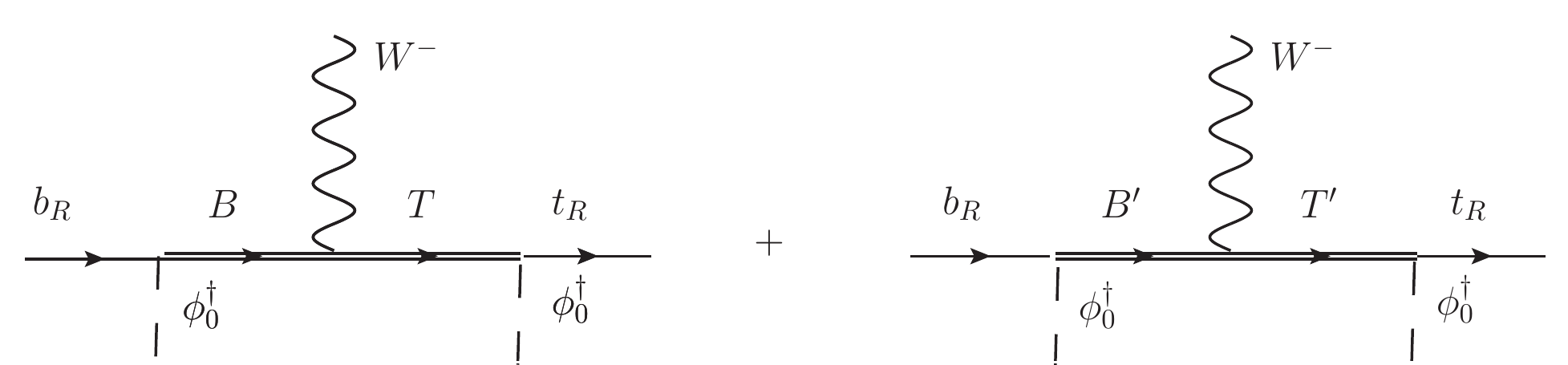}}
\caption{The two diagrams that give contribution to the effective vertex $W\bar{b}_{R}t_{R}$ in the TS5. 
In the TS10, only the first diagram gives a contribution to $W\bar{b}_{R}t_{R}$.}
\label{fig:wtRbR_dia}
\end{figure}
\\
\ \\
Couplings are as follows:
\[
 \alpha(t_R -T)=-Y^{U}_{*}c_1 s_R=-y_t\frac{c_1}{s_1}
\]

\begin{center}
\begin{footnotesize} $\ \ \overrightarrow{s_1\to s^{MIN}_1}\ \ -Y^{U}_{*}\sqrt{1-\frac{y^2_t}{Y^2_*}}\simeq -Y^U_*$\\
 $\ \ \overrightarrow{s_1\to s^{MAX}_1}\ \ -y_{t}\frac{\sqrt{\frac{y_b}{Y_*}}}{\sqrt{1-\frac{y_b}{Y_*}}}\simeq -y_t\sqrt{\frac{y_b}{Y_*}}$
\end{footnotesize}  
\end{center}

\[
 \alpha(b_R -B)=-Y^{D}_{*}s_3 s_{bR}=-y_b\frac{s_3}{s_2}=-y_b\frac{M^2_{T'}}{M^{2}_{T}-M^{2}_{T'}}\frac{s_1}{c_1}
\]

\begin{center}
\begin{footnotesize} $ \ \ \overrightarrow{s_1\to s^{MIN}_1}\ \ -y_b\frac{M^2_{T'}}{M^{2}_{T}-M^{2}_{T'}}\frac{y_t/Y_*}{\sqrt{1-(y_t/Y_*)^2}}\simeq -y_b\frac{y_t}{Y_*}\frac{M^2_{T'}}{M^{2}_{T}-M^{2}_{T'}}$\\
 $ \ \ \overrightarrow{s_1\to s^{MAX}_1}\ \ -y_b\frac{M^2_{T'}}{M^{2}_{T}-M^{2}_{T'}}\sqrt{\frac{Y_*}{y_b}}$
\end{footnotesize}  
\end{center}
\noindent
We obtain a contribution:
\[
 \mathcal{C}^{(I)}_R=\frac{\alpha(t_R -T)\alpha(b_R -B)}{M^2_T}=\frac{y_b y_t}{M^2_T}\frac{M^2_{T'}}{M^2_T-M^2_{T'}}
\]
\\
\ \\
Contribution from the second diagram ($T'$, $B'$ exchange):
\\
\ \\
Couplings:
\[
 \alpha(t_R -T')=Y^{U}_{*}(s_1s_2+c_1s_3)s_R = y_t(s_2+s_3\frac{c_1}{s_1})=y_ts_2\left(1+\frac{M^2_{T'}}{M^2_T-M^2_{T'}}\right) 
\]

\begin{center}
\begin{footnotesize} $\ \ \overrightarrow{s_1\to s^{MIN}_1}\ \ y_t\sqrt{1-\left(\frac{y_t}{Y_*}\right)^2 }\frac{\Delta_{L2}}{M_{T'}}\left(1+\frac{M^2_{T'}}{M^2_T-M^2_{T'}}\right)$\\
 $\ \ \overrightarrow{s_1\to s^{MAX}_1}\ \ y_t\frac{y_b}{Y_*}\left(1+\frac{M^2_{T'}}{M^2_T-M^2_{T'}}\right)$
\end{footnotesize}  
\end{center}

\[
 \alpha(b_R -B')=-Y^{D}_{*}s_{bR}=-y_b\frac{1}{s_2}
\]

\begin{center}
\begin{footnotesize} $ \ \ \overrightarrow{s_1\to s^{MIN}_1}\ \ -y_b\frac{M_{T'}}{\Delta_{L2}}\frac{1}{\sqrt{1-\left(\frac{y_t}{Y_*}\right)^2}}$\\
 $ \ \ \overrightarrow{s_1\to s^{MAX}_1}\ \ -Y_*$
\end{footnotesize}  
\end{center}
\noindent
We obtain for this contribution:
\[
 \mathcal{C}^{(II)}_R=\frac{\alpha(t_R -T')\alpha(b_R -B')}{M^2_{T'}}=-\frac{y_by_t}{M^2_{T'}}\frac{M^2_{T}}{M^2_T-M^2_{T'}}\ .
\]
\\
\ \\
By considering $\mathcal{C}^{(I)}_R$ and $\mathcal{C}^{(II)}_R$, we obtain the total contribution:
\[
 \mathcal{C}_R=-y_by_t\left(\frac{1}{M^2_T}+\frac{1}{M^2_{T'}} \right) 
\]

\section{Infrared bound in the TS10}\label{wtrbr_calculation_TS10}
$s_1$ limit values:

\[
s^{MIN}_1=\sqrt{2}\frac{y_t}{Y_*}
\]
\[
s^{MAX}_1=1
\]

\noindent
Couplings:
\[
 \alpha(t_R -T)=-Y^{U}_{*}c_1\frac{s_R}{\sqrt{2}}=-y_t\frac{c_1}{s_1}
\]

\begin{center}
\begin{footnotesize} $\ \ \overrightarrow{s_1\to s^{MIN}_1}\ \ \sim -\frac{1}{\sqrt{2}}Y_* $\\
 $\ \ \overrightarrow{s_1\to s^{MAX}_1}\ \ 0$
\end{footnotesize}  
\end{center}

\[
 \alpha(b_R -B)=-Y^{D}_{*}c_1s_{bR}=-y_b\frac{c_1}{s_1}
\]

\begin{center}
\begin{footnotesize} $\ \ \overrightarrow{s_1\to s^{MIN}_1}\ \ \sim -y_b\frac{1}{\sqrt{2}}\frac{Y_*}{y_t} $\\
 $\ \ \overrightarrow{s_1\to s^{MAX}_1}\ \ 0$
\end{footnotesize}  
\end{center}
\noindent
Contribution:
\[
 \mathcal{C}_R=\frac{\alpha(t_R -T)\alpha(b_R -B)}{M^2_T}=\frac{y_b y_t}{M^2_T}\frac{c^2_1}{s^2_1}
\]

\begin{center}
\begin{footnotesize} $\ \ \overrightarrow{s_1\to s^{MIN}_1}\ \ \sim \frac{1}{2}\frac{y_b}{y_t}Y^2_*\frac{1}{M^2_T} $\\
 $\ \ \overrightarrow{s_1\to s^{MAX}_1}\ \ 0$
\end{footnotesize}  
\end{center}
\noindent
Notice that in this model the heavy fermion couplings with Higgs vanish in the limit $s_1\to 1$. 
This implies that also $\mathcal{C}_R$ tends to zero in this limit.

\section{Numeric calculation of the Infrared bound}\label{numeric_bound}

\begin{figure}
\centering
\scalebox{1}{\includegraphics{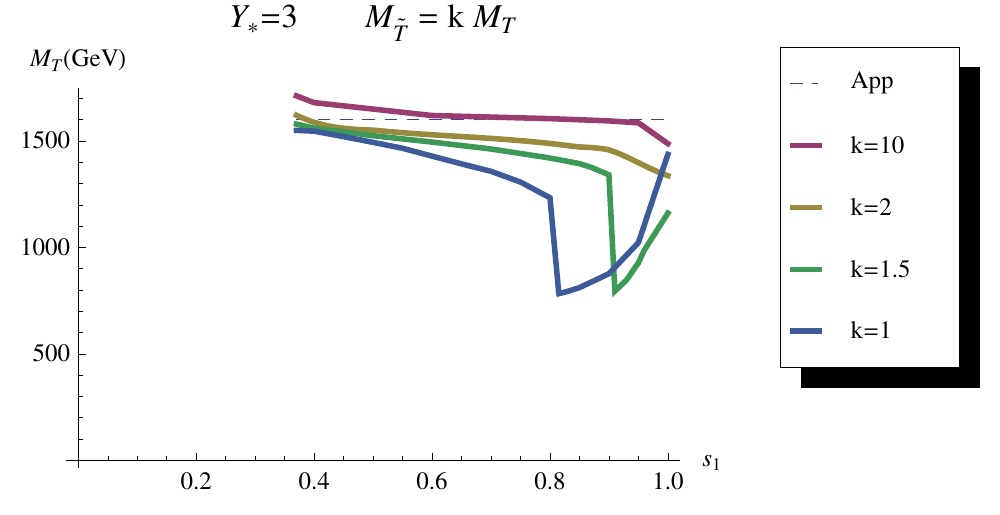}}
\caption{\textit{Bounds from $C^{CH-IR}_7$ in the TS5, obtained by analytic (dashed curve) and numeric (continuous curves) calculation.}}
\label{numeric}
\end{figure}

In this project we have evaluated the bounds on the mass of the bidoublet heavy fermions analytically, 
by considering an expansion in $x\equiv \frac{Y_{*}v}{\sqrt{2}m_{*}}$ and retaining only the $O(x)$ terms. 
This implies that we neglect $O(x^3)$ terms in the $\mathcal{C}_R$ contribution.
As we discussed in sec. \ref{IR_contr},
we expect that results from the numerical calculation do not differ more than O(1) from those obtained analytically. 
This is proved by the results of the numerical calculation of the bounds from $C^{CH-IR}_7$ in the TS5, shown in Fig. \ref{numeric}. 
The dashed curve denotes the analytical bound, obtained for $M_T=1.2 M_{T'}$. 
The continuous curves denote, for different values of the ratio between the doublet and the singlet heavy fermion masses, $k=M_{\tilde{T}}/M_T$, 
the bounds on $M_D$ obtained numerically. 
We can see that the numeric results are quite in agreement with the analytical ones, 
except for the case of an almost degeneracy between doublet and singlet heavy fermion masses and a quite composite $t_L$. 

\section{Charged Higgs ultraviolet contribution to $b\rightarrow s\gamma$ in the TS5}
\label{kchargedApp}
\begin{equation}
 \mathcal{H}^{eff}_{charged\ Higgs}=\frac{i\ e}{8\pi^{2}}\frac{(2\epsilon \cdot p)}{M^{2}_{w}}k_{charged}\left[V_{ts}\bar{b}(1-\gamma_{5})s+\frac{m_{s}}{m_{b}V_{ts}}\bar{b}(1+\gamma_{5})s \right] 
\end{equation} 
where
\begin{gather}
\nonumber
 k_{charged}\approx\sum^{4}_{i=1}\left(|\alpha^{(i)}_{1}|^{2}+|\alpha^{(i)}_{2}|^{2}\right)m_{b}\left(-\frac{2}{9}\right)\frac{M^{2}_{w}}{m^{2}_{*(i)}} + \\ \nonumber
\sum^{4}_{i=1}\left(\alpha^{(i)*}_{1}\alpha^{(i)}_{2}\right)m_{*(i)}\left(-\frac{5}{6}\right)\frac{M^{2}_{w}}{m^{2}_{*(i)}} \\ 
\label{kcharged}
\end{gather} 
the index $i$ runs over the four up-type heavy fermions of the model, $\mathbf{u}^{(i)}$, $m_{*(i)}$ denotes the physical mass of the the $\mathbf{u}^{(i)}$ heavy fermion
and the $\alpha^{(i)}_{1}$, $\alpha^{(i)}_{2}$ coefficients derive from the interactions:
\begin{equation}
 \mathcal{L}\supset \bar{\mathbf{u}}^{(i)}\left[\alpha^{(i)}_{1}(1+\gamma_{5})+\alpha^{(i)}_{2}(1-\gamma_{5})\right]bH^{+} + h.c.\ .
\end{equation} 
After the EWSB, we diagonalize the up-type quarks mass matrix of (\ref{Mup}) and the down-type one (\ref{Mdown}) perturbatively in $x\equiv\left(\frac{Y_{*}v}{\sqrt{2}m_{*}}\right)$, 
neglecting $O(x^{2})$. We find the following coefficients:
\begin{footnotesize}
\begin{gather}
\nonumber
 \alpha^{(\tilde{T})}_{1}= v Y^2_{*} s_1 s_2 s_{bR} \frac{M^2_T M^3_{\tilde{T}} + M^2_{T'} M^3_{\tilde{T}}-M^5_{\tilde{T}}+c_R M^3_T M^2_{T'} c_1}{4 M_T M_{\tilde{T}}(M^2_T-M^2_{\tilde{T}})(-M^2_{T'} + M^2_{\tilde{T}})c_1} \\ \nonumber
 \alpha^{(\tilde{T})}_{2}=\frac{Y_{*}s_1 c_R}{2 \sqrt{2}}\\ \nonumber
 \alpha^{(T)}_{1}= \frac{Y_{*}s_1 s_2 s_{bR}M^2_{T'}}{2 \sqrt{2}c_1 (M^2_{T'}-M^2_T)}\\ \nonumber
 \alpha^{(T)}_{2}= \frac{v Y^2_{*} s_1}{4}\left(\frac{c_R M_{\tilde{T}}+c_1 c^2_R M_T}{M^2_T-M^2_{\tilde{T}}}+\frac{c_1 s^2_R}{M_T}\right) \\  \nonumber
\alpha^{(T')}_{1}=-\frac{s_{bR}Y_{*}}{2 \sqrt{2}}\\ \nonumber
 \alpha^{(T')}_{2}= \frac{Y^2_{*}v s_2}{4}\frac{s^2_1 c_R M_{\tilde{B}}M_{\tilde{T}}M_T M^2_{T'}+s^2_1 c_1 c^2_R M_{\tilde{B}}M^2_T M^2_{T'}+c_1(M^2_{T'}-M^2_{\tilde{T}})\left(M^3_{T'}c_{bR}+M^2_T(s^2_1 s^2_R M_{\tilde{B}} -c_{bR}M_{T'}) \right) }{c_1 M_{\tilde{B}}M_{T'}(M^2_{T'}-M^2_T)(M^2_{T'}-M^2_{\tilde{T}})} \\ 
\label{alfaCharged}
\end{gather} 
\end{footnotesize}
the heavy fermion $T_{2/3}$ 
gives a contribution of $O(x^{2})$ to $k_{charged}$ and we can neglect it.\\
Considering the eq.(\ref{kcharged}) and the coefficients in (\ref{alfaCharged}), neglecting again $O(x^{2})$ terms, we obtain:
\[ k_{charged}=
\]
\begin{footnotesize}$-m_b M^2_W Y^2_*\frac{-15 M^2_T M_{T'} M^2_{\tilde{T}}\sqrt{1-s^2_{bR}} + M_{\tilde{B}} (15 M^2_{T'} M^2_{\tilde{T}} s^2_1 s^2_R + M^2_T (11 M^2_{T'}s^2_1 (-1 + s^2_R) + M^2_{\tilde{T}} (4 s^2_{bR} + 15 s^2_1 s^2_R)))}{144 M_{\tilde{B}} M^2_T M^2_{T'} M^2_{\tilde{T}}}$\end{footnotesize}\\
and:
 \begin{equation}
  k_{charged}\approx m_b M^2_W Y^2_*\frac{5}{48}\frac{1}{M_{B'}M_{\tilde{B}}}+O(s^2_1)+O(s^2_{bR}) \ ,
\label{kcharged_app2}
\end{equation} 
if we can neglect $O(s^{2}_{1})$.

\section{Ultraviolet contribution}\label{AppUVcontrib}
We find in the TS5:
\[
k_{neutral} = -m_b M^2_W Y^2_{*}\frac{1}{8}\left(\frac{c_{bR}}{M_{B'}M_{\tilde{B}}}-\frac{7}{18}\frac{s^{2}_{bR}}{M^2_{B'}}\right)= - m_b M^2_W Y^2_{*}\frac{1}{8}\frac{1}{M_{B'}M_{\tilde{B}}}+O(s^2_{bR})
\]

$k_{charged}=$\\

\begin{footnotesize}$-m_b M^2_W Y^2_*\frac{-15 M^2_T M_{T'} M^2_{\tilde{T}}\sqrt{1-s^2_{bR}} + M_{\tilde{B}} (15 M^2_{T'} M^2_{\tilde{T}} s^2_1 s^2_R + M^2_T (11 M^2_{T'}s^2_1 (-1 + s^2_R) + M^2_{\tilde{T}} (4 s^2_{bR} + 15 s^2_1 s^2_R)))}{144 M_{\tilde{B}} M^2_T M^2_{T'} M^2_{\tilde{T}}}$\end{footnotesize}
\[
=m_b M^2_W Y^2_*\frac{5}{48}\frac{1}{M_{B'}M_{\tilde{B}}}+O(s^2_1)+O(s^2_{bR})
\]
\noindent
and in the TS10:

\[
 k_{neutral}= m_b M^2_W Y^2_* \frac{7 M_T M^2_{T'} s^2_1 - 18 M_{\tilde{B}} M^2_{\tilde{B}'} \sqrt{1-s^2_1} + M^2_{\tilde{B}} (7 M_B s^2_1 -18 M_{\tilde{B}'} \sqrt{1-s^2_1})}{288 M^2_{\tilde{B}} M_B M^2_{\tilde{B}'}}+O(s_{bR})
\]
\[
 = - m_b M^2_W Y^2_{*}\frac{1}{16}\left( \frac{1}{M_{B}M_{\tilde{B}}}+\frac{1}{M_{B}M_{\tilde{B}'}}\right) +O(s^2_1)+O(s_{bR})
\]

\[
 k_{charged}= m_b M^2_W Y^2_* \left(\frac{5}{48}\frac{1}{M_{B}M_{\tilde{B}}}+\frac{5}{48}\frac{1}{M_{B}M_{\tilde{B}'}}+ \frac{5}{96}\frac{s^2_R}{M^2_{B}}\right) +O(s^2_1)+O(s^2_{bR})
\]




\backmatter

\end{document}